\documentclass[compsoc,conference,a4paper,10pt,times]{IEEEtran}

\usepackage[dvipsnames]{xcolor}

\usepackage[colorlinks=true,linkcolor=blue,breaklinks=True,citecolor=brown,urlcolor=blue]{hyperref}

\usepackage{dingbat}

\usepackage[most]{tcolorbox}

\usepackage{adjustbox}
\usepackage[linesnumbered,algo2e,ruled]{algorithm2e}
\usepackage{multirow}
\usepackage{booktabs,subcaption,dcolumn}
\usepackage{tikz}
\usepackage{enumitem}
\usepackage{tabularx}

\usepackage{pifont}
\usepackage{svg}
\usepackage{soul}
\usepackage[font=footnotesize]{caption}
\usepackage[noadjust]{cite}

\usepackage{amsmath}
\usepackage{amsfonts, amssymb}
\usepackage{colortbl}

\newcommand{\cmark}{{\ding{51}}}
\newcommand{\xmark}{{\ding{55}}}

\newcolumntype{?}{!{\vrule width 1.5pt}}

\newcommand{\textbox}[1]{
    \noindent\fbox{%
        \parbox{0.98\columnwidth}{%
            #1
        }%
    }
}

\newcommand{\tabvalue}[2]{\small{#1}\tiny{\text{$\pm$}$#2$}}

\newcommand{\tabvaluetemp}[1]{\small{#1}}

\newcommand\app[1]{{\texttt{\small{App.}}{\small #1}}}

\newcommand\smamath[1]{{\small $#1$}}
\newcommand\scmath[1]{{\scriptsize ${#1}$}}
\newcommand\smacal[1]{{\small $\mathcal{#1}$}}
\newcommand\sccal[1]{{\scriptsize $\mathcal{#1}$}}
\newcommand\smabb[1]{{\small $\mathbb{#1}$}}
\newcommand\scbb[1]{{\scriptsize $\mathbb{#1}$}}

\newcommand{\variable}[1]{{\footnotesize{\textsf{#1}}}}

\newcommand\revision[1]{%
  \bgroup%
  \hskip0pt\color{blue!80!black}%
  #1%
  \egroup
}

\newcommand\revisionGreen[1]{%
  \bgroup%
  \hskip0pt\color{green!50!black}%
  #1%
  \egroup
}

\newcommand\dataset[1]{{\fontfamily{pcr}\selectfont {\footnotesize #1}}}

\newcommand*\halfcirc[1][0.7ex]{%
  \begin{tikzpicture}
  \draw[fill] (0,0)-- (90:#1) arc (90:270:#1) -- cycle ;
  \draw (0,0) circle (#1);
  \end{tikzpicture}}

\DeclareUnicodeCharacter{0306}{ }

\newtheorem{definition}{\textsc{Def.}}

\newcommand{\defref}[1]{{\textsc{\footnotesize Def.~\ref{def:#1}}}}

\newtheorem{remark}{\textsc{Remark}}

\newtcolorbox{deftextbox}[1][]{%
    colback=black!5,
    colframe=black!5,
    notitle,
    sharp corners,
    borderline west={0pt}{0pt}{red!80!black},
    enhanced,
    breakable,
    left=0pt,
    right=0pt,
    top=0pt,
    bottom=0pt
    }
    
\newtcolorbox{tkwtextbox}[1][]{%
    colback=black!5,
    colframe=black!5,
    notitle,
    sharp corners,
    borderline west={0pt}{0pt}{green!80!black},
    enhanced,
    breakable,
    left=0pt,
    right=0pt,
    top=0pt,
    bottom=0pt
    }

\newtcolorbox{remtextbox}[1][]{%
    colback=black!5,
    colframe=black!5,
    notitle,
    sharp corners,
    borderline west={0pt}{0pt}{blue!80!black},
    enhanced,
    breakable,
    left=0pt,
    right=0pt,
    top=0pt,
    bottom=0pt
    }
    
\newcommand{\defboxN}[2]{
    \begin{deftextbox}
    \begin{definition}\label{def:#2} #1
            \end{definition}
    \end{deftextbox}
}

\newcommand{\takeawayN}[1]{
    \begin{tkwtextbox}
        \textbf{Takeaway}: {#1}
    \end{tkwtextbox}
}

\newcommand{\remarkN}[1]{
    \begin{remtextbox}
        \textbf{Remark}: {#1}
    \end{remtextbox}
}
\renewcommand{\figurename}{Fig.}

\AtBeginDocument{%
  \providecommand\BibTeX{{%
    \normalfont B\kern-0.5em{\scshape i\kern-0.25em b}\kern-0.8em\TeX}}}

\begin{document}

\title{SoK: Pragmatic Assessment of Machine Learning for\\Network Intrusion Detection}


\author{
 \IEEEauthorblockN{Giovanni Apruzzese, Pavel Laskov, Johannes Schneider\\}
\IEEEauthorblockA{\textit{Liechtenstein Business School -- University of Liechtenstein}\\
 \{giovanni.apruzzese, pavel.laskov, johannes.schneider\}@uni.li}\\
 }

\pagestyle{plain}
\maketitle

\begin{abstract}
Machine Learning (ML) has become a valuable asset to solve many real-world tasks. For Network Intrusion Detection (NID), however, scientific advances in ML are still seen with skepticism by practitioners.
This disconnection is due to the intrinsically limited scope of research papers, many of which primarily aim to demonstrate new methods ``outperforming'' prior work---oftentimes overlooking the practical implications for deploying the proposed solutions in real systems.
Unfortunately, the value of ML for NID depends on a plethora of factors, such as hardware, that are often neglected in scientific literature.

This paper aims to reduce the practitioners' skepticism towards ML for NID by \textit{changing} the evaluation methodology adopted in research. After elucidating which \textit{factors} influence the operational deployment of ML in NID, we propose the notion of \textit{pragmatic assessment}, which enable practitioners to gauge the real value of ML methods for NID. Then, we show that the state-of-research hardly allows one to estimate the value of ML for NID. As a constructive step forward, we carry out a pragmatic assessment. We re-assess existing ML methods for NID, focusing on the classification of malicious network traffic, and consider: hundreds of configuration settings; diverse adversarial scenarios; and four hardware platforms. Our large and reproducible evaluations enable estimating the quality of ML for NID. We also validate our claims through a user-study with security practitioners.
\end{abstract}
\vspace{-1em}

\section{Introduction}
\label{sec:introduction}

\noindent
Machine learning (ML) techniques have become an indispensable technology in many domains of computer science, such as computer vision~\cite{esteva2021deep, wu2019machine}, natural language processing~\cite{otter2020survey}, audio and speech recognition~\cite{amodei2016deep}, medical applications~\cite{litjens2017survey}, and increasingly in cybersecurity, e.g., malware analysis~\cite{ucci2019survey}, spam and phishing prevention~\cite{gangavarapu2020applicability}, as well as network intrusion detection (NID).
As stated by Arp et al. at the beginning of their paper: ``No day goes by without reading machine learning success
stories''~\cite{arp2022dos}.

However, deployment of ML methods in NID faces substantial skepticism~\cite{de2019information,sand2021soc,alahmadi202299} among practitioners---despite the fact that NID is one of the oldest applications of ML in cybersecurity~\cite{Kruegel:Anomaly, wang2006anagram, rieck2007language}. The main difficulty, as pointed out by Sommer and Paxson \cite{Sommer:Outside}, is that network environments exhibit ''immense variability''. Hence, most ML models for NID developed in research papers cannot be readily transferred into operational environments due to a large uncertainty about their genuine value. 
In the real world, what matters is not the improvements over prior work, but rather the balance between the performance and costs in routine deployment scenarios.

The main thesis of this paper is that \emph{research evaluations of ML in NID should account for pragmatic aspects of operational deployment}. 
We elucidate all such aspects by proposing the notion of ``pragmatic assessment'', whose goal is ensuring that practitioners have all the necessary information to determine whether a given ML solution is applicable to a given NID context. From the viewpoint of researchers, conducting such pragmatic assessments is challenging: almost every paper on ML in NID published in recent top-conferences has some shortcomings. However, as we will show in this work, \textit{it can be done}. As such, we endorse future efforts to adopt our proposed takeaways so as to facilitate the integration of ML research results into real network intrusion detection systems (NIDS).

\textsc{\textbf{Motivational Example.}}
Let us elucidate why the state-of-the-art of ML in NID is still at an early stage with respect to practical deployment. For this purpose, we compare NID with two popular domains in which ML has found applications: computer vision and malware analysis.

In \textbf{computer vision}, the evaluation methodology adopted in research is now standardized. Current benchmarks, e.g., \dataset{ImageNet}~\cite{deng2009imagenet}, were created before 2010 and are still widely used today---even in production~\cite{you2018imagenet}, because they contain data from the `real' world. Abundant literature\footnote{As of March 2022,~\cite{deng2009imagenet} has over 35K Google Scholar citations.} implicitly established reference standards,\footnote{E.g., \textit{ResNet}~\cite{he2016deep} models are known to consistently achieve very high \textit{accuracy} on \dataset{\scriptsize{ImageNet}}, which now represent the reference benchmark for computer vision (even on different datasets, e.g., \dataset{\scriptsize{CIFAR}}~\cite{mirzadeh2020improved}).} thereby removing the uncertainty on the real value of the findings obtained in research environments. 
A similar case can be said for \textbf{malware analysis}. After the release of \dataset{Drebin}~\cite{arp2014drebin} in 2014, containing real Android apps (benign and malicious), abundant\footnote{As of March 2022,~\cite{arp2014drebin} has over 2K citations on Google Scholar.} ML research has been carried out on \dataset{Drebin} (e.g.,~\cite{Demontis:Yes,li2021can, galvez2021less}). Agreeably, \dataset{Drebin} is not perfect (some papers found some `duplication' issues~\cite{irolla2018duplication}), and some malware families are not popular anymore. However, a crucial fact remains: malware is malicious \textit{everywhere} and \textit{everytime}~\cite{suarez2020eight}. Therefore, a proficient ML model trained on \dataset{Drebin} can be deployed on any system analyzing Android apps. For instance,~\cite{daoudi2021deep} show that the method originally proposed in~\cite{arp2014drebin} is effective even on (real!) apps collected in 2017--2019.

In contrast,\textbf{ ML research on NID is far from such maturity}. Among the root causes is the lack of (open-source) data that is representative of the real world. For instance, thousands of proposals were validated on the \dataset{KDD99} dataset, usually achieving near-perfect performance~\cite{mishra2018detailed}. However, the data in \dataset{KDD99} represents only a single network (from 1999), preventing to estimate the generalization capabilities of ML solutions~\cite{apruzzese2022cross}. We observe that, recently, more datasets are being publicly released (e.g.~\cite{sarhan2020netflow}).
Surprisingly, however, such `abundance' increased confusion. Consider, e.g., the recent findings of ML proposals evaluated on the popular \dataset{CICIDS17} dataset~\cite{Sharafaldin:Toward}. Specifically, we focus on~\cite{vinayakumar2019deep} and~\cite{pontes2021new}---both involving ML models based on diverse ML algorithms, including Decision Trees (DT) and Neural Networks (NN). While~\cite{vinayakumar2019deep} claims that NN are better than DT---as given by a superior F1-score (0.96 for NN, 0.95 for DT)---the opposite occurs in~\cite{pontes2021new}---with the DT reaching 0.99 F1-score, against the 0.96 of the NN. Moreover, it is misleading to only consider a single performance metric for NID: even a high F1 score may conceal a suboptimal false positive rate, making a given ML solution impractical for realistic deployments~\cite{arp2022dos}. To make all of this worse, recent efforts found that \dataset{CICIDS17} is flawed~\cite{engelen2021troubleshooting}. Finally, the authors of~\cite{andresini2021insomnia} showed that ML models trained on the (fixed) version of \dataset{CICIDS17} perform poorly against `unknown' attacks. The current situation of ML in NID can be summarized with a statement from Markus de Shon (who was Lead of the Detection Engineering at NetFlix): ``Application of ML in intrusion detection has been uneven at best, with deep and widespread (and generally justified) skepticism among subject matter experts''~\cite{de2019information}.

\textsc{\textbf{Contribution.}} 
Our aim is \textit{changing} the evaluation methodology adopted by research on ML for NID so as to remove the skepticism of practitioners towards the quality of scientific solutions. 
To reach our goal, we first summarize Machine Learning-based Network Intrusion Detection Systems (ML-NIDS). Then we provide four major contributions---each discussed in a dedicated section~(§) revolving around a given research question (RQ). 
\begin{itemize}
    \item The RQ of §\ref{sec:deployment} is: ``What are the \textit{factors}  taken into account by practitioners when developing ML-NIDS?'' To answer this RQ, we: (i)~elucidate the business relationships between the end-users of ML-NIDS and the developers of such ML-NIDS; (ii)~outline the challenges that such developers must face when devising their solutions; (iii)~present the factors that contribute to the real value of a ML-NIDS; and (iv)~validate our factors by directly asking the practitioners' opinion.
    
    \item The RQ of §\ref{sec:solution} is: ``What should research on ML in NID do to \textit{allow practitioners to estimate} the real value of the corresponding results?'' To answer this RQ, we: (i)~formalize our notion of a \textit{pragmatic assessment}; (ii)~explain how to conduct a pragmatic assessment through comprehensive guidelines.
    
    \item The RQ of §\ref{sec:sota} is: ``Does the state-of-the-art allow us to estimate the real value of ML methods for NID?'' To answer this RQ, we: (i)~review \textit{all} papers on ML-NIDS presented in top security conferences since 2017; (ii)~analyze to what extent they meet the conditions of pragmatic assessments; and (iii)~report the practitioners' viewpoint on the state of research.
    
    \item The RQ of §\ref{sec:demonstration} is: ``Can pragmatic assessments be done in research?''. We answer this RQ by performing the first pragmatic assessment of ML-NIDS. We do so through a large set of experiments focused on network traffic classification. Our evaluation reports the (statistically validated) performance of thousands of ML models, spanning across diverse datasets, algorithms, pipelines, and labeling budgets. Moreover, we perform our experiments on different platforms, and showcase the importance of \textit{hardware}---which is often overlooked in literature (and also in practice!).
\end{itemize}
We discuss our results and compare our paper with related work in §\ref{sec:discussion}.
To ensure reproducibility we release our code~\cite{pragmaticAssessment}: hence, our SoK can also serve as a benchmark for future studies. Due to the sheer size of our experimental campaign, the low-level details and results are provided in our code repository. Finally, we provide details on our survey with practitioners in the Appendix (\app{\ref{app:survey}}).

\section{Background and Problem Statement}
\label{sec:background}

\noindent
We first outline the general context of NIDS (§\ref{ssec:nids}). Then, we delve into the specific application of ML in NIDS (§\ref{ssec:ml-nids}), and explain how research on ML-NIDS is typically carried out (§\ref{ssec:research}). Finally, we elucidate the problem tackled by our SoK paper (§\ref{ssec:motivation}).

\subsection{Network Intrusion Detection Systems}
\label{ssec:nids}
\vspace{-0.5em}
\noindent
The security of IT systems spans over three tasks: prevention, detection, and reaction~\cite{yang2004security}. It is well-known that perfect prevention is unattainable, whereas the reaction phase implicitly assumes that the attack has already taken place. Hence, to minimize (or nullify) the damage resulting from a breach, a major role is played by the detection of cyber threats. In the case of network security, such a role is devoted to ``NIDS.'' Such term encompasses diverse meanings (e.g.,~\cite{liao2013intrusion, tsikoudis2014leonids}).
Let us provide the definition of NIDS adopted in our paper:

\defboxN{A NIDS is a \textit{system} that protects a \textit{network}, i.e., a set of (IT) systems that interact with each other.}{nids}

\noindent {\small (we refer the reader to the RFC~\cite{shirey2007internet} for the exhaustive definitions of ``network'' and ``system'')}

Since their conception~\cite{mukherjee1994network}, NIDS have undergone significant improvements. Initially, NIDS only analyzed data pertaining to network traffic, such as raw packet captures (PCAP), and the detection was performed by ``static'' methods, i.e., through human-defined ``signatures'' encoding patterns of known attacks. Due to the growing complexity of network environments as well as the appearance of adaptive attackers, however, static detection methods became infeasible. To cope with such a dynamic ecosystem, NIDS started to adopt automated detection techniques stemming from the data analytics domain, enabled by the availability of ``big data'' (potentially originating from diverse sources) and by advancements in computational power~\cite{chou2021survey}. Such data-driven techniques, which include (among others) ML methods, improved NIDS while reducing the burden on human operators. 

\begin{figure}[!htbp]
    \centering
    \includegraphics[width=0.7\columnwidth]{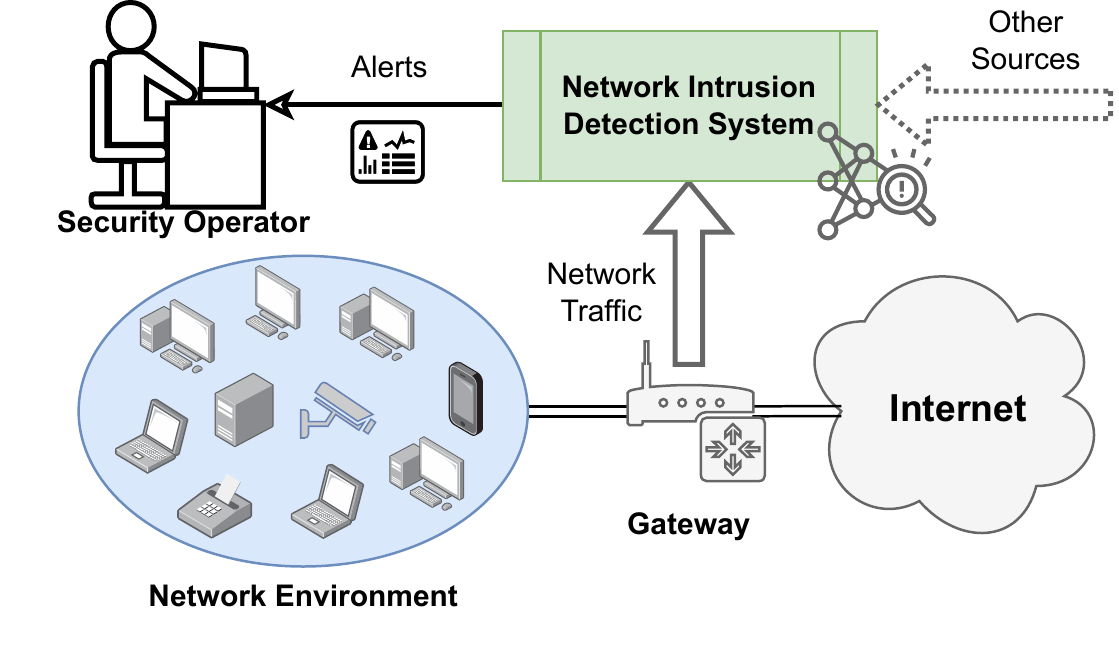}
    \caption{Typical deployment scenario of a NIDS.}
    \label{fig:network}
\end{figure}

We provide an illustration of the typical NIDS deployment in Fig.~\ref{fig:network}, where the NIDS monitors all communications performed by a given organization. The output of a NIDS is in the form of \textit{alerts} (which can be post-processed by dedicated modules, e.g.,~\cite{nadeem2021enabling}), which must be inspected and triaged by security operators. We stress that NIDS can be deployed anywhere in a given network, not just at the border (e.g.,~\cite{chaabouni2019network}). 

Current state-of-the-art NIDS---and related tools, such as SIEM, (e.g.,~\cite{radoglou2021spear, bryant2020improving})---may correlate information from various sources (e.g., whois geolocation~\cite{oprea2018made}, DNS logs~\cite{feng2017user}, or internal ACL), and can combine multiple detection approaches (e.g., either \textit{misuse}- or \textit{anomaly}-based~\cite{khraisat2019survey}) on diverse data-types~\cite{yehezkel2021network}. For example, a growing trend (even among practitioners~\cite{ucci2021near}) is analyzing NetFlows~\cite{Cisco:Flow}, i.e., high-level metadata summarizing the raw-communications between two endpoints~\cite{vormayr2020my}. We outline the advantages of NetFlow analyses in \app{\ref{sapp:netflow}}.

\subsection{Machine Learning and NIDS}
\label{ssec:ml-nids}
\vspace{-0.5em}
\noindent
A NIDS is a system that must orchestrate multiple components. Each component may consider diverse inputs, and its output may also be used by other components. All such components can adopt different analytical techniques, including those belonging to Machine Learning. 

The underlying principle of ML is to leverage the functionality of an ML \textit{model}. By applying a given ML \textit{algorithm} to some \textit{training} data, it is possible to develop a ML \textit{model} that can autonomously `predict' (new) data---e.g., determining whether an activity is legitimate or not. 
We provide our definition of a Machine Learning-based Network Intrusion Detection System (ML-NIDS):

\defboxN{An ML-NIDS is a NIDS that includes, among its components, a (trained) Machine Learning model.}{mlnids}

\begin{figure}[!htbp]
    \centering
    \includegraphics[width=\columnwidth]{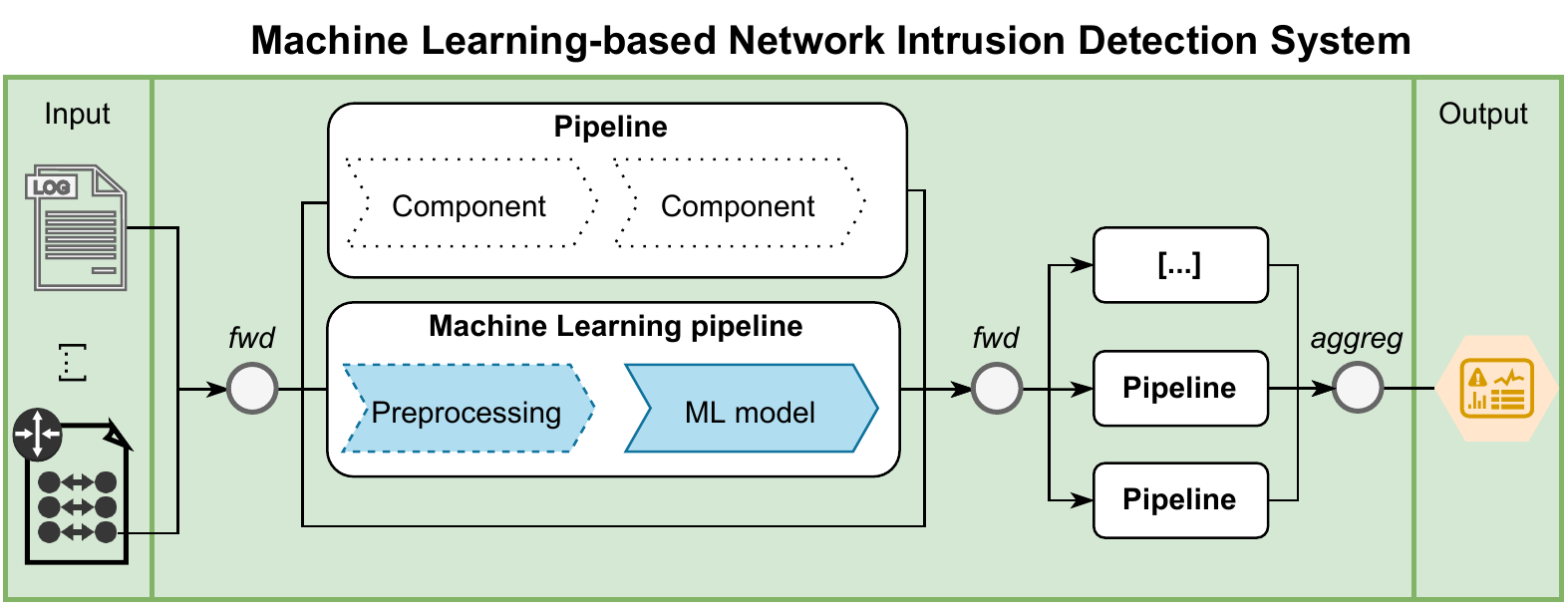}
    \caption{Architecture of an ML-NIDS.}
    \vspace{-1em}
    \label{fig:mlnids}
\end{figure}

\noindent To better understand \defref{mlnids}, we provide an exemplary architecture of an ML-NIDS in Fig.~\ref{fig:mlnids}. 
The ML-NIDS can receive different types of input data (either in batches or in real-time~\cite{corsini2021evaluation}), which are forwarded to specific \textit{pipelines}; such pipelines are made up of one or more \textit{components}, and analyze the given input(s). In the case of an ML-NIDS, at least one pipeline will include an ML model---typically preceded by a \textit{preprocessing} component tasked to transform the input data to a format accepted by the ML model (e.g., by extracting the relevant `features').
The output of all such pipelines can then be used as input to other pipelines (and respective components), which may leverage additional ML models (for the same, or a different task). All such results are then aggregated into the output of the NIDS (i.e., alerts).

Among the ML community, it is common to distinguish between \textit{supervised} and \textit{unsupervised} ML algorithms~\cite{Sommer:Outside}. The difference revolves around the notion of 

\noindent
``label'' which denotes the ground truth of a given sample. By providing such labels during the training stage of a supervised ML model, it is possible to `guide' the learning process and enable, e.g., classification tasks~\cite{joyce2021framework}. Obtaining such labels is, however, \textit{expensive} and often error-prone (as shown in~\cite{engelen2021troubleshooting}). 
We provide in \app{\ref{sapp:sup-unsup}} a more exhaustive description of supervised and unsupervised ML in the NID context---which is followed by an exemplary application of ML to detect malicious traffic (in \app{\ref{sapp:application}}).

As pointed out by many reviews (e.g.,~\cite{Buczak:Survey, Apruzzese:Deep, liu2019machine, ring2019survey}), the applications of ML for NID are highly successful. Accordingly, ML has been shown not only to automate crucial triaging operations~\cite{oprea2018made}, but also to exceed the detection capabilities of non-ML NIDS (e.g.,~\cite{Apruzzese:Periodic, Mirsky:Kitsune}). 

\subsection{ML-NIDS in Research}
\label{ssec:research}
\begin{figure}[!htbp]
    \centering
    \includegraphics[width=0.6\columnwidth]{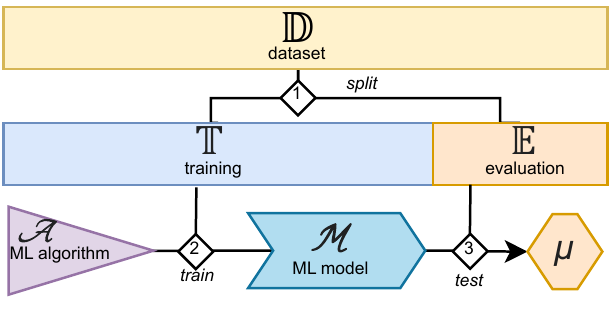}
    \caption{Typical ML workflow adopted in research.}
    \label{fig:typical}
\end{figure}

\vspace{-0.5em}
\noindent
Let us illustrate the common workflow adopted in research to assess ML-NIDS, schematically depicted in Fig.~\ref{fig:typical}. This workflow---typically borrowed from domains that are unrelated to cybersecurity---begins by acquiring a \textit{dataset}, \smabb{D}. Such \smabb{D} is divided into a \textit{train} and \textit{evaluation} (or ``test'') partition---\smabb{T} and \smabb{E} respectively---by following a given \textit{split} (e.g., 80:20, i.e., 80\% of \smabb{D} is put in \smabb{T}, and the remaining 20\% in \smabb{E}). Then, by using a given learning \textit{algorithm} \smacal{A} (e.g., DT) on \smabb{T}, a ML model \smacal{M} is developed: such \smacal{M} is then evaluated on \smabb{E}, and its quality is measured according to some performance \textit{metric}, \smamath{\mu} (e.g., F1-score, Accuracy). The intuition is that if \smamath{\mu} is `good enough' and `better' than existing proposals, then the respective research has achieved its purpose (e.g.~\cite{vinayakumar2019deep, pontes2021new}).

Despite being correct in principle, such a workflow has two intrinsic limitations from a practitioner's viewpoint.
\begin{enumerate}
    \item \textbf{The lack of an ``universal'' dataset for NID}. If such a dataset existed, it could be used in any assessment to generalize the performance of an ML-NIDS. However, the immense variability of networks~\cite{Sommer:Outside} makes creating such a dataset close to impossible (even security companies share such an opinion~\cite{van2022deepcase}). Furthermore, this problem prevents~\cite{van2022deepcase} a reliable `transferring' of ML models across different networks (in contrast, transfer is feasible in other domains in which ML has found applications~\cite{oliver2018realistic}). 
    
    \item \textbf{The primary focus is on the ML model,} \smacal{M}. Such \smacal{M}, however, is just a single component within the ML-NIDS (see \defref{mlnids}), which is a complex system. For instance, there are many elements that come both \textit{before} and \textit{after} \smacal{M}; moreover, \smacal{M} can be \textit{physically} deployed on devices mounting diverse \textit{hardware}.
    
\end{enumerate}
We claim that research papers can provide valuable insights for practitioners. In this paper, we accept the first problem (which cannot be solved today), and focus on rectifying the latter.
Specifically, we argue that practitioners are more interested in the (general) ML \textit{method} rather than in the (specific) ML \textit{model}. Hence, practitioners will appreciate if a research: accounts for the most likely scenarios to be faced by the ML-NIDS; and also allows to estimate the \textit{costs} required to sustain the ML-NIDS during its entire operational lifecycle~\cite{vishik2016key, wilson2014some}. 

\subsection{Skepticism of ML-NIDS Practitioners}
\label{ssec:motivation}
\vspace{-0.5em}
\noindent
According to a recent survey, over 75\% of companies employ ML solutions for network security~\cite{kshetri2021economics}. Most of such companies, however, \textit{delegate} their cybersecurity to third-party vendors~\cite{fischerHubner2021stakeholder}. Indeed, several commercial products for NID actively leverage ML (e.g.,~\cite{nguyen2018security, Darktrace:CyberML, Lastline:AI}). Yet, all such products adopt ML methods that are decades old and mostly in their unsupervised form (e.g., the one-class SVM of~\cite{ucci2021near} was proposed in 2002~\cite{tax2002one}). Simply put, the integration of research endeavours into operational environments is slow in the context of ML-NIDS. 

Such slow-pace stems from the skepticism~\cite{de2019information} of practitioners towards the `successes' of research papers. Such skepticism is well-founded: as we will show in our SoK, the current state of research hardly `complies' with the demands of professional ML-NIDS developers~(§\ref{sec:sota}). Indeed, our own survey (§\ref{ssec:opinion}) reveals that research papers---instead of providing answers---leave practitioners with \textit{uncertainty}, which can be summarized as: ``It works in \textit{your network}. But will it work equally well in \textit{my network}, and is it \textit{affordable} (now, and in the long-term)?''

\textbf{Our Goal.}
We firmly believe that the research community \textit{can} answer such a question. However, providing such an answer (\textit{which not necessarily needs to be always positive}~\cite{moore2010economics}) requires a radical change of the current assessment methodology---which should account for the necessities of real developers. To the best of our knowledge, such necessities have never been formalized in the context of ML-NIDS (related work is discussed in §\ref{ssec:related}). Therefore, we first elucidate all the \textit{factors} that practitioners must take into account whenever real deployment of ML in NID is considered. Then, we propose the notion of \textit{pragmatic assessment} which explains what research papers must do to satisfy the needs of practitioners. Finally, we \textit{perform the first} pragmatic assessment of ML in NID.
\section{Practical Deployment of ML in NID}
\label{sec:deployment}
\noindent
Our first contribution addresses the RQ: ``What are the \textit{factors} taken into account by practitioners when developing ML-NIDS?'' To answer this RQ, we must first elucidate the business perspective of ML-NIDS, and then describe the deployment challenges faced by developers when designing ML solutions for NID. 

\subsection{Business Perspective of ML for NID}
\label{ssec:business}
\begin{figure}[!htbp]
    \centering
    \includegraphics[width=0.99\columnwidth]{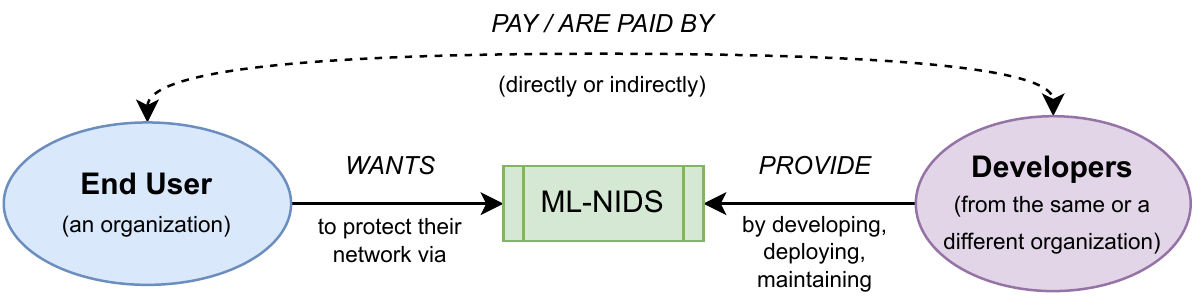}
    \caption{The business perspective of ML-NIDS.}
    \label{fig:business}
\end{figure}

\vspace{-0.5em}
\noindent
Consider an organization that uses a NIDS (which may or may not already leverage ML) to protect its network, and that wants to enhance such NIDS with a new ML solution for a given detection problem. To this purpose, the organization can develop the ML solution \textit{in-house}, or rely on \textit{commercial-off-the-shelf} (COTS) products~\cite{apruzzese2021modeling}. Let us elucidate the implications of these two use-cases.
\begin{itemize}
    \item \textbf{In-house.} The organization must first design the ML solution, which can be done either by replicating existing proposals or by devising an original one. Then, the organization must oversee the ML solution for its entire \textit{lifecycle}, which includes: data collection, preprocessing, and labeling (for both training and testing the ML model); development of the ML model (including repeated testings for parameter calibration); deployment of this ML model in the NIDS infrastructure; as well as any maintenance~\cite{paleyes2022challenges, pacheco2018towards}.
    \item \textbf{COTS.} The organization must choose among available products on the market the one that best fits their NIDS. Such a choice depends on the characteristics of a given COTS solution, as advertised by its \textit{vendor}. 
\end{itemize}
In both use cases, the deployment of a ML-NIDS entails two players: the \textit{end user} (i.e., the organization), and the \textit{developers} (i.e., either an external vendor, or the same organization). Such a relationship is represented in Fig.~\ref{fig:business}: The organization needs a solution according to its security strategy, and the developers provide a product to meet this demand. In any case, it is the \emph{developer} who has to make technical decisions and ensure the operational quality of the final product---during its entire lifecycle.

\remarkN{Real ML-NIDS require \textit{developers} who are responsible for the lifecycle of the ML model (Fig.~\ref{fig:business}).}

\vspace{-0.5em}

\subsection{Deployment Challenges of ML for NID}
\label{ssec:challenges}
\vspace{-0.5em}
\noindent
Three main challenges affect the real deployment of ML-NIDS: (1)~each network is unique, (2)~each network perpetually evolves over time, and (3)~the implicit presence of adversaries. These challenges (which contribute to the ``lack of an universal dataset'' mentioned in §\ref{ssec:research}) are emblematic of ML in NID and exist irrespective of who oversees the lifecycle of the ML-NIDS. Let us explain. 

\begin{enumerate}
    \item The \textbf{uniqueness of networks} has been pointed out in various works (old~\cite{Sommer:Outside} and recent~\cite{apruzzese2022cross}): some activities are legitimate in one network and illegitimate in another network. Hence, deployment of ML-NIDS requires training \textit{and} testing operations performed on data originating from the monitored network~\cite{apruzzese2022role}. 
    
    \item The \textbf{dynamic nature} of modern networks is another major hurdle for deployment of ML in NID. Every day, new hosts can appear or be removed; new services may be adopted; and new network segments may be attached---all of which may introduce new types of vulnerabilities. Such phenomena represent the well-known problem of ``concept drift''~\cite{jordaney2017transcend}. 
    
    \item The implicit \textbf{presence of adversaries} implies a different, more serious type of concept drift~\cite{andresini2021insomnia}. While the natural network evolution may be controllable to some degree\footnote{E.g., administrators know when organizations adopt new services.}, this is hardly the case for attackers who \textit{want} to evade a NIDS~\cite{corona2013adversarial}. Such adversaries are well motivated and may even devise evasion strategies that specifically target the ML model~\cite{apruzzese2021modeling}).
\end{enumerate}

\noindent
We observe that the last two challenges (intrinsic to most ML applications in cybersecurity~\cite{arp2022dos}) are unpredictable\footnote{E.g., even if administrators are aware of major changes, they do not know if such changes will impact the performance of their ML-NIDS.}, and hence cannot be solved \textit{during the development} of an ML model. Overcoming these challenges \textit{is possible} but requires re-assessments of the ML model \textit{after its deployment}. If the prediction quality of the ML model deteriorates, it must be updated or replaced. Such maintenance is indispensable for ML-NIDS, and accounting for its costs---unique to each network---is crucial for determining the pragmatic value of ML solutions for NID.

\remarkN{Real ML models for NID must be \emph{developed}, \emph{deployed}, and \emph{maintained} via periodic re-assessments. Such operations must be performed \emph{independently} for each network monitored by an ML-NIDS.}

\vspace{-0.5em}

\subsection{Factors affecting the real value of ML in NID}
\label{ssec:factors}
\vspace{-0.5em}
\noindent
The deployment challenges of ML for NID are well-known by practitioners, who must take into account several factors \textit{before} developing any ML model. We now answer our first RQ by connecting all the foundations described insofar and elucidate all such factors.

\textbf{Overview.}
The \textit{value} of any security solution can be expressed as the tradeoff between operational effectiveness and expenses. An ML method for NID is \textit{effective} if it yields an ML model \smacal{M} that exhibits, e.g., a high detection rate while raising few false alarms. The \textit{expenses} reflect all costs incurred during the lifecycle of \smacal{M}. We denote the value of an ML model as \smamath{\Psi}(\smacal{M}).
From the research perspective, \smamath{\Psi}(\smacal{M}) depends (at the high level) on the ML algorithm \smacal{A} and on the dataset \smabb{D} (§\ref{ssec:research}). However, from the practical viewpoint, \smabb{D} must be collected, and \smacal{M} must be deployed in the real NIDS. These operations introduce additional dependencies that crucially affect both the effectiveness and the expenses in \smamath{\Psi}(\smacal{M}). 

\textbf{Factors.}
We propose to formalize the dependencies contributing to \smamath{\Psi}(\smacal{M}) through the following five factors. 

\begin{itemize}
    \item \variable{Preprocessing} (\smacal{P}). There exist a plethora of mechanisms (each with its own operational costs) meant to transform raw data into the format accepted by an ML model \smacal{M}. These mechanisms affect the information included in \smabb{D} and utilized by \smacal{M} to make its predictions---hence influencing the effectiveness of \smacal{M}. Consider, for instance, the generation of NetFlows from PCAP, for which many tools are available---each having its own logic~\cite{vormayr2020my}: As shown in~\cite{sarhan2020netflow}, exactly the same raw data (in PCAP) yields different NetFlows (even if generated via similar tools), leading to ML models with different performance (we will also show this in our experiments).
    
    \item \variable{Data availability} (\smacal{D}). The quality of a given \smabb{D} is linked to its size and sample diversity, so that \smacal{M} can properly `learn' how to predict future data~\cite{vogelsang2019requirements}. However, obtaining such a \smabb{D} has a cost~\cite{apruzzese2022sok}, which is higher when \smacal{M} requires \textit{labelled} data for training\footnote{We observe that while \scbb{T} is required for supervised methods, a labelled \scbb{E} is always necessary to validate performance~\cite{yehezkel2021network}.}. Ground truth verification is costly and error-prone~\cite{van2022deepcase}, and it can lead to noisy samples~\cite{frenay2013classification}. For instance,~\cite{engelen2021troubleshooting} found many labelling issues in a well-known dataset for NID (the \dataset{CICIDS17}~\cite{Sharafaldin:Toward}). Finally, although some tools can (synthetically) generate malicious data (e.g., {\small CALDERA}~\cite{MITRE:URL}), some companies require several months to obtain a representative dataset of `normal' network activities (e.g., {\small CAIAC}~\cite{apruzzese2022role}).
    
    \item \variable{System Infrastructure} (\smacal{S}). Any \smacal{M} is just a single element within the NIDS, and hence its effectiveness depends on the NIDS infrastructure (§\ref{ssec:ml-nids}). The infrastructure determines, e.g., the type of data analyzed by \smacal{M}. For instance, the information included in the NetFlows analyzed by an \smacal{M} is dictated by the sensors deployed in the NIDS infrastructure. The infrastructure, furthermore, affects (i)~the type of decisions expected from \smacal{M}, (e.g., binary or multi-class classification); as well as (ii)~the logical arrangement of the individual decision units within the ML pipeline. For instance, a pipeline can include a standalone ML model, an ensemble of ML models, or a cascade of ML models (e.g.,~\cite{das2021network, Biggio:One, yu2018efficient}). We provide a schematic of an ML pipeline including a cascade of a binary and multi-class classifier in Fig.~\ref{fig:bmd}. 
    
    \item \variable{Hardware} (\smacal{H}). The \textit{detection} capabilities of a ML model are hardly affected by the computational resources available. However, hardware influence the \textit{runtime} for both the \textit{training} and the \textit{inference} stage of \smacal{M}. The former is necessary for the periodic re-training\footnote{Training-time is also crucial for fine-tuning: an optimal configuration will be found in less time for methods that are faster to train.} of \smacal{M}; the latter is crucial to determine where \smacal{M} can be physically deployed. Indeed, ML models for NID can be placed anywhere in a network~\cite{kshetri2021economics}, spanning from low-power IoT devices~\cite{da2019internet} to high-end computing platforms~\cite{kim2020ai}. 
    
    \item \variable{Unpredictability} (\smacal{U}). 
    It is impossible to know \emph{in advance} how the threat landscape and the network environment will evolve. Moreover, ML methods introduce further uncertainty by using randomized algorithms (e.g., Random Forests); but also because it is not possible to know a-priori how to collect a \smabb{T} that maximizes the effectiveness of \smacal{M} (and that does so in the long-term).
\end{itemize}
We can hence express the value of a ML method for NID as a function $f$ defined with the following equation (Eq.):
\begin{align}
    \label{eq:performance}
    \Psi(\mathcal{M}) = f(\mathcal{P}, \mathcal{D}, \mathcal{S}, \mathcal{H}, \mathcal{U}).
\end{align}
Because of \smacal{U}, we note that \smamath{\Psi}(\smacal{M}) is not deterministic.

We stress that all the factors above influence each other. For instance, \smacal{S} also implicitly affects \smacal{H}, but also \smacal{P}. Furthermore, ML solutions for NID should be continuously assessed (\smacal{U}), which requires both human and computational resources. For instance, updating \smacal{M} with new \smabb{T} may require additional labeling efforts~\cite{miller2016reviewer} (\smacal{D}); however, such retraining can be computationally expensive (\smacal{H}), and overlooking the training runtime can be detrimental~\cite{liu2016fp}.

\textbf{Practitioner Validation.}
We conducted a survey asking the opinion of practitioners on our proposed set of factors. Our population entails 12 practitioners with hands-on experience in ML and NID; overall, our participants work (or have worked) in the SOC of renown companies. (We provide all details in \app{{\small \ref{app:survey}}}.) 
The results of our survey are summarized in Table~\ref{tab:factors}, which reports the percentage of our interviewees that believed whether each of our factors was: ``not important''~({\small \textcircled{}}); ``important''~({\small \textbf{!}}); or ``crucial''~({\small \textcircled{\textbf{!}}}) for real deployments of ML-NIDS.

\begin{table}[!htbp]
    \centering
    \caption{Viewpoint of practitioners on our set of factors.}
    \resizebox{0.4\columnwidth}{!}{
        \begin{tabular}{c|c|c|c}
             \textbf{Factor} & \textcircled{} & \textbf{!} & \textcircled{\textbf{!}} \\
             \midrule
             \smacal{P} & 0\% & 9\% & 91\%\\
             \smacal{D} & 9\% & 18\% & 73\%\\
             \smacal{S} & 9\% & 27\% & 64\%\\
             \smacal{H} & 9\% & 64\% & 27\%\\
             \smacal{U} & 9\% & 18\% & 73\%\\ 
             \bottomrule
        \end{tabular}
    }
    \label{tab:factors}
\end{table}

\noindent
On average, 66\% of practitioners consider all our factors to be ``crucial'' for estimating the real value of a ML-NIDS. Interestingly, 0\% believe that preprocessing~(\smacal{P}) is ``not important,'' which was ranked as the most crucial factor by all our respondents. The unpredictability (\smacal{U}) and data availability~(\smacal{D}) are also deemed to be pivotal by 73\% of our population. The least relevant factor is hardware~(\smacal{H}), which is considered ``important'' by 64\%. However, as we will show, \smacal{H} can be the deciding factor to assert which ML solution is truly the best (§\ref{ssec:practical}).

\takeawayN{Our proposed five factors (\smacal{P, D, S, H, U}) are considered to be relevant for estimating the real value of ML in NID by most practitioners.}

\section{Pragmatic Assessment of ML-NIDS}
\label{sec:solution}
\noindent
We now address our second RQ:
``What should research on ML in NID do to \textit{allow practitioners to estimate} the real value of the proposed solutions?'' Indeed, practitioners must account for all the factors in Eq.~\ref{eq:performance}: they will not implement an ML method without knowing how much training data is required. They would also be reluctant to reproduce an ML method if it is not clear whether such a method is truly superior to existing solutions. Finally, an ML method for NID that has not been tested in an adversarial environment may contain security risks~\cite{Biggio:Wild}.

To answer our second RQ, we propose the following notion of \textit{pragmatic assessment} which draws on several past works from both the research (e.g.,~\cite{jenn2020identifying, arp2022dos, rimmer2022open}) and industrial (e.g.,~\cite{winter2021trusted, Europe:AI}) domains.

\defboxN{A \emph{pragmatic assessment} allows practitioners to assert the value of an ML method for NID iif:
\begin{itemize}
    \item the reported results are free of any experimental bias, and present high degree of confidence;
    \item the evaluation is carried out on testbeds resembling the (likely) operational scenarios of the NIDS;
    \item all requirements for developing the proposed ML method are clearly specified.
\end{itemize}
}{pragmatic}.

\vspace{-1em}
Let us explain how these three conditions can be met \textit{in research} and at a high level, starting from the last one.

\subsection{Development Requirements}
\label{ssec:transparency}
\vspace{-0.5em}
\noindent
A pragmatic assessment must transparently disclose \textit{all} information pertaining to the requirements for developing (and maintaining) a given solution. In the context of research on ML-NIDS, such information must include:
\begin{itemize}
    \item The schematic of the \textit{NIDS infrastructure} with respect to the proposed ML method (\smacal{S} in Eq.~\ref{eq:performance}). Such schematic must pinpoint `where' the corresponding ML model is meant to be deployed. Such information serves to establish: (i)~the function of the ML model; (ii)~which components/specifications are required to operate the ML model; and (iii)~whether additional components are required to post-process its output. 
    
    \item The \textit{hardware specifications} of the platforms used to train and test the ML model, which affect its runtime (\smacal{H} in Eq.~\ref{eq:performance}). Such specifications must include the RAM, the CPU (i.e., model, threads, maximum frequency) and---if necessary---the GPU. It is also important to report the CPU utilization during its runtime (i.e., how many threads were used, and at what frequency), because it plays a crucial role in the energy consumption. In particular, especially for the CPU, the \textit{exact} model must be reported\footnote{Note that CPUs can be under/overclocked and therefore exhibit different frequencies than those reported by their manufacturers~\cite{jalili2021cost}.}. For instance, stating that ``the CPU is an Intel Core i5'' (e.g.,~\cite{pereira2018dictionary}) is misleading because there are hundreds of such CPUs with significantly different performance: according to PassMark, an i5-470M is 35 times slower than a i5-12600KF~\cite{PassMark:URL}. To demonstrate the effects of `superficial' hardware specifications, we perform an original experiment §\ref{ssec:cpu}.
    
    \item The \textit{dataset composition} for both the training \smabb{T} and evaluation \smabb{E} partitions (\smacal{D} in Eq.~\ref{eq:performance}). Such information is crucial for supervised ML methods, as it allows determining the amount of labeled data necessary to develop the respective ML model. Such information, however, is also relevant for unsupervised ML algorithms, because even unlabelled data has a cost~\cite{apruzzese2022sok}. 
    
    \item The \textit{details of the ML method} used to develop the ML model. Such details include the feature set, the exact algorithm (e.g., DT) and its parameters, the task (e.g., binary or multi-class classification), and the design of its pipeline (e.g., stand-alone or ensemble). All such information contributes to \smacal{P} and \smacal{S} in Eq.~\ref{eq:performance}.
\end{itemize}
Finally, it is (obviously) desirable that the implementation code is openly released, and if the adopted dataset is publicly available. As stated by Lindauer et al.~\cite{lindauer2020best}, scientific reproducibility ``facilitates progress'': if the entire testbed is publicly accessible, then developers can determine if there are any similarities between the real and experimental environments---potentially enabling a direct transfer of the resulting ML model (if the environments are similar).

Reporting all the above-mentioned details also allows to roughly estimate the expenses for maintaining the ML solution (therefore accounting for part of \smacal{U} in Eq.\ref{eq:performance}).

\subsection{Likely Operational Scenarios}
\label{ssec:likely}
\vspace{-0.5em}
\noindent
Security systems must face real threats, hence pragmatic assessments must consider scenarios that are likely to occur in reality. 
To meet this condition, we propose three complementary use-cases that \textit{can} be taken into account in research on ML for NID. Given the lack of an universal dataset (§\ref{ssec:research}), our underlying intuition is to \textit{maximise the utility of a given dataset}. Doing this requires the researcher to use their domain expertise and `creativity'. 

\vspace{-0.5em}

\subsubsection{Closed and Open World}
\label{sssec:world}
It is not wrong to consider ``closed world'' scenarios, i.e., where the ML model expects each sample to resemble those seen during its training stage. However, ML methods should be assessed \textit{also} in ``open world'' scenarios~\cite{rimmer2022open}, due to the unpredictability of the threat landscape (\smacal{U} in Eq.~\ref{eq:performance}). Indeed, these are the scenarios that ML methods originally intended to address~\cite{Sommer:Outside}.
For unsupervised anomaly detection, open world scenarios are implicit: after learning a given concept of `normality', no pre-existing knowledge is required to detect anomalous behaviors (unsupervised methods have no notion of `classes'). In contrast, for classification problems (common in NID) assessing open world scenarios requires additional effort: Testing ML classifiers \textit{only} on a \smabb{E} having the exact same classes as \smabb{T} (closed world) prevents estimating any form of adaptability of the ML-NIDS. For a pragmatic assessment, the ML classifier should be evaluated \textit{also} on an \smabb{E} containing attacks different than those in \smabb{T} (open world). 

This can be done by (a)~injecting in \smabb{E} some malicious classes \textit{not included} in the original \smabb{D} -- e.g., by borrowing malicious samples from other datasets~\cite{pontes2021new}; or by entirely creating novel attack classes via, e.g.,~\cite{MITRE:URL} (as done in~\cite{bowman2020detecting}). Alternatively, it is possible to (b)~\textit{exclude} some malicious classes in \smabb{D} from being put in \smabb{T}, and put such classes in \smabb{E} instead. Both approaches are viable and can be combined in principle. However, as pointed out by Apruzzese et al.~\cite{apruzzese2022cross}, ``\textit{mixing data from different networks presents some fundamental issues}''. For instance, if two networks are considerably different then it is difficult to trust the resulting performance of an ML model. Therefore, mixing data from different networks should be done only after thorough topological analyses. 

\vspace{-0.5em}

\subsubsection{Static and Temporal Data Dependency}
\label{sssec:dependency}
ML methods were originally conceived by assuming the validity of the \textit{iid} principle, i.e., ``independent and identically distributed random variables''~\cite{dundar2007learning}. However, the iid principle does not always hold in network environments because the data (both benign and malicious) analyzed by a NIDS is likely to present temporal dependencies. As an example, a botnet-infected machine will first contact its CnC, and only afterwards it will execute the malicious commands received by the CnC. For this reason, it is recommended (e.g.,~\cite{arp2022dos}) to choose \smabb{E} so that its samples come `after' \smabb{T}. Investigating only this `temporal' case, however, prevents a generic assessment: the results will only resemble the `sequence' of the samples captured by a given \smabb{D}. Hence, to provide more general results, we propose to consider \textit{both} cases, i.e., by assuming that:
(a)~samples are all independent of each other; (b)~temporal dependencies may be present in the data stream.

Investigating both cases in research\footnote{We note, however, that investigating both cases may not be `universally' possible. Sequential ML methods that specifically look for temporal patterns (e.g.,~\cite{corsini2021evaluation}) implicitly assume the presence of temporal dependencies; whereas some datasets may simply not provide time-related information to investigate any form of temporal dependencies.} requires a dataset \smabb{D} containing time-related information. Assessing the `static' case is straightforward: it is sufficient to compose \smabb{T} and \smabb{E} by randomly sampling from \smabb{D}. On the other hand, for the `temporal' case, it is necessary to split \smabb{D} into \smabb{T} and \smabb{E} according to sensible temporal criteria. For instance, the split can be based on the \textit{timestamp} associated to each sample; it is also possible to choose as \smabb{E} the `last' portion of \smabb{D}, and use as \smabb{T} the `first' part (assuming that \smabb{D} is chronologically ordered).
Nevertheless, the \textit{time-gap} between \smabb{T} and \smabb{E} should not be overlooked. For example, the results can differ if only minutes pass between \smabb{T} and \smabb{E}, compared to when the gap is days or weeks.

 \vspace{-0.5em}

\subsubsection{Naive and Adaptive Adversaries}
\label{sssec:adversarial}
Security systems must always assume the presence of adversaries. Such adversaries can be `naive' and rely only on known offensive strategies (i.e., hoping to bypass an unpatched system). However, the most serious threats come from `adaptive' attackers who actively attempt to exploit the specific vulnerabilities of their target. In the case of ML methods, such vulnerabilities involve the so-called adversarial examples~\cite{Laskov:Practical}. After more than a decade of research demonstrating their effectiveness, it is paramount for pragmatic assessments to also consider such a threat. 

There are dozens of ways to bypass ML systems via adversarial examples~\cite{Apruzzese:Addressing} and considering all such ways is clearly infeasible since they are ultimately unpredictable (\smacal{U} in Eq.~\ref{eq:performance}). As stated by Biggio and Roli, priority should be given to the ``more likely threats''~\cite{Biggio:Wild}. The idea is endorsing \textit{defensive proactivity}: the developer evaluates an ML method in advance against the adaptive ``adversarial'' attacks that are more likely to occur in reality. 
To this end, it is crucial to consider adaptive attacks that conform to a \textit{threat models that are both viable and feasible}. We provide the following recommendations (extending those by~\cite{apruzzese2021modeling}) to facilitate the design of such threat models.
\begin{itemize}
    \item \textit{Adversarial Mindset}. Real attackers adopt a cost/benefit rationale~\cite{wilson2014some}: they will not launch attacks requiring huge resource investments---even if they are likely to succeed, there may be other targets (i.e., different from ML models) that yield a better `profit'.
    
    \item \textit{Consider the right ``Box''}. Adversarial ML threat models are often expressed with the notion of a ``box'' that identifies the system targeted by the attacker. In the case of ML methods for NID, the ``box'' is the entire NIDS---and not just the specific ML model. Hence, when considering a ``white-box'' attacker, such an attacker would have complete knowledge of the entire NIDS---i.e., a rather extreme circumstance, as such information is well-protected~\cite{apruzzese2021modeling}. For this reason, we recommend not to place ``white-box'' settings at top priority (contrarily to~\cite{arp2022dos}): such worst-case scenarios are feasible in general security, but not very likely against NIDS.\footnote{We argue that attackers with full knowledge of the whole NIDS would opt for more disruptive strategies than data perturbations.}
    
    \item \textit{Realizable Attacks}. Aside from conforming to the assumed threat model, the perturbation used to create an adversarial example should be physically realizable~\cite{tong2019improving}. This does not mean that it must be created in the ``problem-space''~\cite{Pierazzi:Intriguing}, as this may not be feasible\footnote{Complete realistic fidelity is almost impossible as it would require to reproduce the attacker's operations in the specific targeted network.} in research when operating on a pre-collected dataset \smabb{D}. Indeed, as observed by~\cite{apruzzese2021modeling}, even perturbations in the feature space can be realistic if the manipulation preserves the dependencies between features, and considers features on which a real attacker has some influence.\footnote{A very recent work~\cite{apruzzese2022spacephish} pointed out that attackers may even be able to directly control the feature representation of a given example.}
    
    \item \textit{Unbounded Perturbations}. Research on adversarial ML usually aims at devising minimal perturbations that are subject to self-imposed constraints (e.g., one pixel attacks~\cite{Su:One}). However, as also remarked by Carlini et al.~\cite{carlini2019evaluating} (and, more recently, also by~\cite{apruzzese2022position}), real attackers are not interested in `bounded' perturbations, as long as they achieve their goal (e.g., evading a security system). 
\end{itemize}
We make an important remark. Assessing the robustness to adversarial perturbations serves to gauge the vulnerabilities (or strengths) of an ML method \textit{before} its deployment. It is up to the end user of such an ML method to determine whether the envisioned threat deserves a dedicated treatment---which should be economically justified~\cite{moore2010economics}. 

\subsection{Unbiased and Statistically Validated Results}
\label{ssec:performance}
\vspace{-0.5em}
\noindent
The recent paper by Arp et al.~\cite{arp2022dos} provides sensible recommendations on how to conduct a meaningful evaluation of ML in cybersecurity. For instance, the base-rate fallacy should be considered, the right performance metrics should be measured, and comparisons should be made with the right baselines. All such guidelines are relevant for NIDS and must be followed also for our proposed pragmatic assessments. Such guidelines, however, lack a crucial piece: the performance of an ML method should be \textit{statistically validated}. The motivation is simple: to account for the (intrinsic) randomness of ML (\smacal{U} in Eq.~\ref{eq:performance}); and to mitigate the (intrinsic) sampling bias in \smabb{T} and \smabb{E}.

Such statistical validation is achieved by repeating the experiments\footnote{We stress that pragmatic assessments require such statistical validation for all the `likely' scenarios~(§\ref{ssec:likely}). For instance, the adversarial robustness should be repeated many times (as also recommended in~\cite{carlini2019evaluating}), each applying the same perturbation but to different samples.} for a sufficient amount of trials, whose focus is establishing the (unbiased) performance of the ML method---and not of a single ML model. Indeed, only by measuring the performance of a large `population' of ML models---all trained/tested in similar settings---it is possible to estimate the real value of the corresponding ML method. Moreover, large populations enable \textit{statistical comparisons}, a powerful tool for determining which ML method is truly the best. 
Carrying out comparisons that are statistically significant (i.e., assuming a target \smamath{\alpha < 0.05}), however, requires many trials. For instance, an ML method yielding an ML model with 0.992 accuracy cannot be claimed to be `better' than another ML method whose ML model exhibits 0.991 accuracy \textit{over a single trial}.
Therefore, in cases where two methods yield models with similar performance, a large amount of trials may be required\footnote{Some tests require a sample-size of at least 50~\cite{happ2019optimal}. However, the test may also be inconclusive: in this case, no claim can be made.}. We thus discourage relying just on cross-validation techniques, as they do not provide a sufficient amount of measurements for pragmatic assessments.\footnote{As an example, consider a \scbb{D} that is partitioned into \scbb{T} and \scbb{E} with an 80:20 split. Such a split allows to apply 5-fold cross validation, which produces only 5 results and hardly valid to determine whether an ML model is statistically better than another. In contrast, a more convincing and unbiased approach is to perform a large amount of trials by randomly sampling \scbb{T} and \scbb{E} from \scbb{D} many times (e.g., 50), each time with the same 80:20 split. Such an approach allows to compare two populations of 50 samples (via, e.g., a Welch's t-test~\cite{zimmerman1993rank}), enabling to derive sound conclusions on which ML method is better.}

\begin{figure}[!htbp]
    \centering
    \includegraphics[width=0.99\columnwidth]{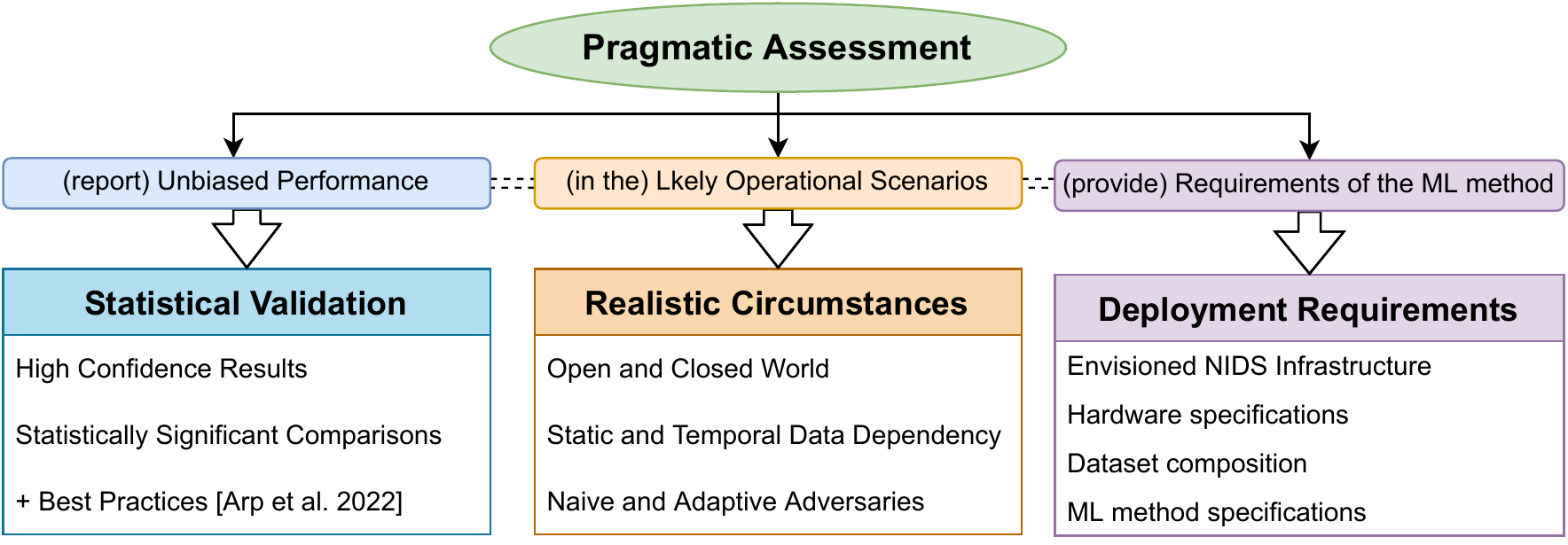}
    \caption{Characteristics of Pragmatic Assessments of ML in NID.}
    \label{fig:certification}
\end{figure}

\takeawayN{Accounting for all the factors contributing to the real value of ML for NID requires pragmatic assessments, summarized in Fig.~\ref{fig:certification}. Extensive information must be provided, diverse likely scenarios must be considered, and multiple trials must be made to provide statistically significant results.}

\vspace{-1em}

\subsection{Experiment: the importance of CPU specs}
\label{ssec:cpu}
\vspace{-0.5em}
\noindent
We perform a simple experiment to demonstrate the importance of reporting the \textit{complete} CPU specifications. 

\textbf{Objective.} We consider the simple task of measuring the runtime for training and testing an ML model on a given dataset. Specifically, we train and test a Decision Tree (DT) binary classifier on the \dataset{GTCS}~\cite{mahfouz2020ensemble} dataset (i.e., \smabb{D}); more details in \app{\ref{sapp:datasets}}. We randomly sample 80\% (i.e., \smabb{T}) of \smabb{D} to train the DT, and test it on the remaining 20\% (i.e., \smabb{E}). We repeat such experiments 10 times.

\textbf{Specifications.} We consider two different platforms, whose setup are nearly identical ``on the surface'': they both mount 8GB of DDR3 RAM (using the same frequencies), both run Windows 10 OS, and the experiments are done on the exact same version of Python and scikit-learn. 
The only difference is the exact model of the CPU: one is an Intel i5-4670, and the other is an Intel i5-430; both CPUs use their default clock speeds. Both training and testing the DT require only a single CPU core.

\textbf{Results.} On average, \textit{training} the DT on the i5-4670 requires 11.1s, but it takes 34.7s on the i5-430 (a 310\% increase). Whereas \textit{testing} requires an average of 0.39s on the i5-4670, and 1.38s on the i5-430 (a 350\% increase). Hence, reporting only a portion of the specifications (e.g., ``an Intel i5 CPU'') introduces a lot of uncertainty on the \textit{actual} performance of the final ML model.
\vspace{-0.5em}

\section{State-of-the-Art (in Research)}
\label{sec:sota}
\noindent
As a final motivation for this paper, we answer the following RQ: ``Does the state-of-the-art allow one to estimate the real value of ML methods for NID?'' We hence review recent literature to determine how much existing works `comply' to our notion of pragmatic assessment.

\vspace{0.5mm}
\textbox{
{\small \textbf{Disclaimer.} Similarly to~\cite{liao2021we, arp2022dos, apruzzese2022position}, the following analysis is not meant to invalidate previous works: ultimately, none of such works aimed at realistic deployment. Our intention is highlighting that the current evaluation protocol adopted in research papers can (and should) be improved. We provide a case-study describing some `practical redundancies' of a recent work (by the same authors of this SoK) in \app{\ref{sapp:flawed}}.}
}

\subsection{Methodology (literature review)}
\label{ssec:method}
\vspace{-0.5em}

\textbf{Scope and inclusion criteria.}
The amount of papers that propose to use ML for NID is off-the-charts. To perform a feasible but comprehensive analysis, we investigated all papers published in nine of the most reputable cybersecurity conferences.\footnote{We consider: IEEE SP and EuroSP; ACM CCS, AsiaCCS, ACSAC; NDSS and USENIX Security; as well as DIMVA and RAID.} 
For each venue, we investigated the proceedings from 2017 to 2021,\footnote{Some of these conferences still have to be held in 2022.} and selected all papers that fell within our scope. Such selection resulted in 30 papers,\footnote{We went through the proceedings four times over 5 months.} considering diverse types of networks (from enterprise~\cite{oprea2018made} to IoT~\cite{mudgerikar2019spion}) and cyber threats (from anomalous traffic~\cite{bortolameotti2017decanter} to APT~\cite{milajerdi2019holmes}, and even adaptive attacks~\cite{wang2021crafting}). Nonetheless, all such papers shared the same underlying assumption: the usage (and evaluation) of ML to detect `intrusions' in networks.\footnote{E.g., the ML-NIDS may analyze network data (e.g., NetFlows~\cite{oprea2018made}), or may account for data generated from an entire network (e.g., finding `anomalies' in the measurements of all sensors in a given network~\cite{erba2020constrained}). We do not consider ``malware detectors'' (analyzing, e.g., android apps~\cite{pendlebury2019tesseract} or javascript~\cite{fass2019hidenoseek} or PE files~\cite{aghakhani2020malware}) as ML-NIDS.}

\textbf{Analysis.} We inspected each selected paper from the perspective of our `pragmatic assessment' notion. Because each paper had different assumptions, we performed our analysis by asking ourselves six questions---each having a set of standardized answers ($\Rightarrow$). Specifically:
\begin{enumerate}
    \item ``Are the \textit{hardware} specifications clearly reported?''$\Rightarrow$Yes (\cmark); partially (\halfcirc, e.g., no details on CPU model); not provided (\xmark).
    
    \item ``Is the \textit{runtime} clearly specified?''$\Rightarrow$Yes (\cmark); only training time (\smabb{T}); only inference time (\smabb{E}); no (\xmark).
    
    \item ``Is the vulnerability to \textit{adaptive adversarial attacks} mentioned?''$\Rightarrow$Yes, and it is evaluated~(\cmark); yes, but only stated as a limitation~(\halfcirc); not mentioned (\xmark).
    
    \item ``Is the \textit{statistical significance} used to provide more convincing results?''$\Rightarrow$Yes (\cmark); no (\xmark).
    
    \item ``Is the training dataset ever changed to account for diverse \textit{data availability}?''$\Rightarrow$Yes (\cmark); no (\xmark).
    
    \item ``Is the evaluation done on (at least some) \textit{public data}?''$\Rightarrow$Yes (\cmark); yes, but it is not available today~(\halfcirc); no, such data has always been kept private~(\xmark). We also noted how many datasets were used.
\end{enumerate}
The results of such analysis are summarized in Table~\ref{tab:sota}. We also remark that we considered two additional criteria, namely: (i)~whether the ML models were tested only in a ``closed world'' setting; and (ii)~whether the paper considered different preprocessing operations. Such criteria are not included in Table~\ref{tab:sota} because the \textit{response was the same for all papers}, i.e.: all 30 papers evaluated their models (also) against unknown attacks (most of such papers are on anomaly detection, which implicitly assumes an ``open world'' setting); and none of the 30 papers considered different preprocessing mechanisms.

\begin{table}[!h]
    \centering
    \caption{State-of-the-Art: papers published since 2017 in top cybersecurity conferences that consider applications of ML linked with NID.}
    \resizebox{0.99\columnwidth}{!}{
        \begin{tabular}{c|c|c|c|c|c|c|c}
            \toprule
            \textbf{Paper} & Year & Hardware & Runtime & Adaptive & Stat. Sign. & Avail. & Pub. Data \\
            \midrule
            
            Bortolamelotti~\cite{bortolameotti2017decanter} & 2017 & \xmark & \xmark & \cmark & \xmark & \xmark & \xmark\ (1) \\
            Ho~\cite{ho2017detecting} & 2017 & \xmark & \xmark & \halfcirc & \xmark & \xmark & \xmark\ (1) \\
            Cho~\cite{cho2017viden} & 2017 & \xmark & \xmark & \cmark & \xmark & \xmark & \xmark\ (1) \\
            Siadati~\cite{siadati2017detecting} & 2017 & \xmark & \xmark & \halfcirc & \xmark & \xmark & \xmark\ (1) \\
            
            \midrule
            
            Oprea~\cite{oprea2018made} & 2018 & \xmark & \scbb{T} & \halfcirc & \xmark & \xmark & \xmark\ (1) \\
            Pereira~\cite{pereira2018dictionary} & 2018 & \halfcirc & \scbb{T} & \halfcirc & \xmark & \cmark & \halfcirc\ (1)\\
            Kheib~\cite{kneib2018scission} & 2018 & \xmark & \xmark & \halfcirc & \xmark & \xmark & \xmark\ (1) \\
            
            \midrule
            
            Araujo~\cite{araujo2019improving} & 2019 & \xmark & \scbb{E} & \xmark & \xmark & \cmark & \xmark\ (1) \\
            Mudgerikar~\cite{mudgerikar2019spion} & 2019 & \xmark & \cmark & \xmark & \xmark & \xmark & \xmark\ (1) \\
            Mirsky~\cite{Mirsky:Kitsune} & 2019 & \halfcirc & \cmark & \halfcirc & \xmark & \xmark & \cmark\ (1) \\
            Feng~\cite{feng2019systematic} & 2019 & \xmark & \xmark & \halfcirc & \xmark & \xmark & \cmark\ (2) \\
            Milajerdi~\cite{milajerdi2019holmes} & 2019 & \halfcirc & \cmark & \halfcirc & \xmark & \xmark & \cmark\ (1) \\
            Liu~\cite{liu2019log2vec} & 2019 & \halfcirc & \xmark & \halfcirc & \xmark & \xmark & \cmark\ (2) \\
            Du~\cite{du2019lifelong} & 2019 & \xmark & \scbb{T} & \halfcirc & \xmark & \xmark & \cmark\ (3) \\
            
            \midrule
            
            Erba~\cite{erba2020constrained} & 2020 & \halfcirc & \scbb{E} & \cmark & \xmark & \cmark & \cmark\ (2) \\
            Bowman~\cite{bowman2020detecting}& 2020 & \halfcirc & \scbb{E} & \xmark & \xmark & \xmark & \cmark\ (2) \\
            Leichtnam~\cite{leichtnam2020sec2graph} & 2020 & \halfcirc & \xmark & \xmark & \xmark & \xmark & \cmark\ (1) \\
            Singla~\cite{singla2020preparing} & 2020 & \xmark & \xmark & \xmark & \xmark & \cmark & \cmark\ (2) \\
            Han~\cite{han2020unicorn} & 2020 & \cmark & \cmark & \halfcirc & \xmark & \xmark & \cmark\ (2) \\
            Jan~\cite{jan2020throwing} & 2020 & \xmark & \xmark & \cmark & \cmark & \cmark & \xmark\ (1) \\
            
            \midrule
            
            Ghorbani~\cite{ghorbani2021distappgaurd} & 2021 & \cmark & \scbb{E} & \halfcirc & \xmark & \xmark & \xmark\ (1) \\
            Nabeel~\cite{nabeel2021cadue}& 2021 & \xmark & \xmark & \halfcirc & \xmark & \xmark & \xmark\ (1)\\
            Wang~\cite{wang2021crafting}& 2021 & \xmark & \scbb{E} & \cmark & \xmark & \xmark & \cmark\ (2) \\
            Piszkozub~\cite{piskozub2021malphase}& 2021 & \xmark & \xmark & \halfcirc & \xmark & \xmark & \halfcirc\ (2) \\
            Yuan~\cite{yuan2021recompose} & 2021 & \xmark & \xmark & \halfcirc & \xmark & \cmark & \cmark\ (1) \\
            Yang~\cite{yang2021cade}& 2021 & \xmark & \xmark & \halfcirc & \cmark & \xmark & \cmark\ (1) \\
            Barradas~\cite{barradas2021flowlens} & 2021 & \halfcirc & \cmark & \halfcirc & \xmark & \xmark & \cmark\ (1) \\
            Han~\cite{han2021deepaid} & 2021 & \cmark & \cmark & \cmark & \xmark & \cmark & \cmark\ (2) \\
            Liang~\cite{liang2021fare} & 2021 & \xmark & \scbb{T} & \halfcirc & \cmark & \cmark & \cmark\ (1)\\
            Fu~\cite{fu2021realtime} & 2021 & \halfcirc & \cmark & \cmark & \xmark & \xmark & \cmark\ (3) \\

            \bottomrule
        \end{tabular}
    }
    \label{tab:sota}
\end{table}

\vspace{-0.5em}

\subsection{Major Findings (and our interpretation)}
\label{ssec:findings}
\vspace{-0.5em}
\noindent
From Table~\ref{tab:sota}, we see that \textit{no one fits all}: despite being published in top conferences, no single paper allows to estimate the deployment value of the considered ML solutions. Nonetheless, we highlight some intriguing trends.

\textbf{\leftthumbsdown~~Only a snapshot.}  Most papers assess the quality of ML methods by training and testing the corresponding ML models on a single `snapshot'. For instance, such ML models are often evaluated only once, preventing to derive more general conclusions; it is concerning that the term `statistical significance' is mentioned only in 2 papers (i.e.,~\cite{jan2020throwing, yang2021cade}). 
Moreover, most papers (almost 70\%) do not vary the composition of their training dataset, preventing to estimate the value of the ML method when a company cannot afford to invest many resources in the data collection procedures. We acknowledge that some of these papers propose `unsupervised' ML techniques; however, even unlabelled data has a cost~\cite{apruzzese2022sok}. In addition, no paper considers different preprocessing mechanisms (§\ref{ssec:method}): we appreciate that most papers thoroughly describe the preprocessing operations of their solutions; however, such procedures (including all parameters) are never changed, preventing to determine their impact on the ML pipeline. Finally, most papers use a single dataset. 

\textbf{\leftthumbsdown~~Neglected Requirements.} Only three papers (i.e.,~\cite{han2021deepaid, han2020unicorn}) provide a holistic vision of the hardware and runtime requirements used to develop the corresponding ML models. For instance, the proposal in~\cite{liang2021fare} requires 2.5 hours to train, but no hardware information is provided. We find it concerning that even papers that specifically focus on IoT settings do not provide such details. For instance, the authors of E-Spion~\cite{mudgerikar2019spion} rightly state that ``E-Spion is specifically designed for resource-constrained IoT devices'': they do measure the CPU utilization, but without reporting \textit{which} CPU was used. Such an omission can be acceptable in research, but not when real deployments are considered.

\textbf{\leftthumbsup~~Smart Attackers.} On a positive note, the majority of papers considers an ``open world'' setting in which adversaries try to actively bypass the considered ML-NIDS. Some papers even evaluate the impact of adaptive attacks \textit{in addition} to measuring the performance in their absence---which is commendable. We remark that~\cite{erba2020constrained, wang2021crafting} specifically focus on such a threat, and hence have slightly diverse assumptions: for instance, not reporting the hardware or runtime is less of a problem for~\cite{erba2020constrained, wang2021crafting}. However, the lack of multiple trials ensuring statistically significant results is still an issue.

\subsection{Practitioners' Opinion}
\label{ssec:opinion}

\vspace{-0.5em}
\noindent
In our survey with practitioners, we also asked for their opinion on Table~\ref{tab:sota}. Specifically, \textit{after} asking the questions related to our factors (§\ref{ssec:factors}), we inquired whether the fact that some columns have many ``\xmark{}'' was: ``not very problematic''~({\small \textcircled{}}), ``problematic''~({\small \textbf{!}}), or ``very problematic''~({\small \textcircled{\textbf{!}}}). The results are shown in Table~\ref{tab:problematic}.   
Most practitioners (90\%) agree that the lack of statistically significant comparisons is ``very problematic.'' Moreover, 59\% believe that lack of data diversity is an issue. Perhaps surprisingly, 75\% can overlook the absence of evaluations against adaptive adversarial attacks. Finally, the lack of hardware specifications was also deemed to be not a crucial shortcoming---our evaluation will prove otherwise.

\begin{table}[!htbp]
    \centering
    \caption{Practitioners' opinion on the results displayed in Table~\ref{tab:sota}.}
    \resizebox{0.5\columnwidth}{!}{
        \begin{tabular}{c|c|c|c}
             \textbf{Column} & \textbf{\textcircled{}} & \textbf{!} & \textbf{\textcircled{\textbf{!}}} \\
             \midrule
             Hardware & 25\% & 75\% & 0\%\\
             Runtime & 0\% & 75\% & 25\%\\
             Adversarial & 8\% & 67\% & 25\%\\
             Stat. Sign. & 0\% & 10\% & 90\%\\
             Avail. & 16\% & 42\% & 42\%\\ 
             Pub. Data & 0\% & 41\% & 59\%\\ 
             \bottomrule
        \end{tabular}
    }
    \label{tab:problematic}
\end{table}

Nonetheless, at the end of our questionnaire we posed one last question to our interviewees: ``In general, do you think that research papers facilitate the practitioners' job in determining the \textit{real value} of the proposed ML methods?'' The answers were enlightening: 92\% are ``uncertain'', whereas 8\% are ``left with more questions than answers after reading a research paper''.

\takeawayN{Despite abundant work proposing ML methods for NID, the state-of-the-art the art does not allow practitioners to determine the real value of existing ML solutions. We attempt to change the current evaluation protocols with our proposed \textit{pragmatic assessment} notion---which \textit{can be done}, as we will now show.}
\noindent
(Given our findings, we wondered: ``did the situation change in {\small 2022}?'' We investigate this question in \app{\ref{sapp:2022}})
\section{Demonstration of a Pragmatic Assessment}
\label{sec:demonstration}

\noindent
To bridge the gap between research and practice, we now focus on our last RQ: ``Can pragmatic assessments be done in research?'', and make a constructive step towards the integration of state-of-the-art ML methods into real NIDS. Specifically, our goal is threefold:
 ({\small \textbf{i}})~Demonstrate that our guidelines \textit{can} be followed in research experiments;
 ({\small \textbf{ii}})~showcase an exemplary case-study of ML for NID, malicious NetFlow classification, wherein we \textit{pragmatically assess} existing ML methods;
 ({\small \textbf{iii}})~provide \textit{statistically validated results} for future studies, by publicly disclosing the complete details and low-level implementation.

Our evaluation is massive, hence the complete details are reported in the Appendix (and repository~\cite{pragmaticAssessment}). Here, we summarize our testbed~(§\ref{ssec:setup}), present some original results~(§\ref{ssec:results}), and derive practical considerations~(§\ref{ssec:practical}).

\subsection{Experimental Setup}
\label{ssec:setup}
\vspace{-0.5em}
\noindent
Our evaluation revolves around the well-known problem of malicious NetFlow classification, which can be done via ML.\footnote{Out of the 30 papers in Table~\ref{tab:sota}, 16 use NetFlow-related data:~\cite{bortolameotti2017decanter,oprea2018made,araujo2019improving,Mirsky:Kitsune,liu2019log2vec,leichtnam2020sec2graph,singla2020preparing,wang2021crafting,ghorbani2021distappgaurd,piskozub2021malphase,yuan2021recompose,yang2021cade,barradas2021flowlens,liang2021fare,fu2021realtime,han2021deepaid}.} We chose this problem because it allows one to devise diverse ML pipelines. Indeed, NetFlow is generated by preprocessing raw PCAP data; moreover, \textit{detecting} malicious NetFlows can be seen either as a binary or multi-class classification problem (because a sample can belong to diverse malicious classes). Such a problem can be tackled through diverse ML pipelines, e.g., it is possible to create an \textit{ensemble} of `specialized' binary classifiers (each trained on a subset of the available data---similarly to~\cite{Mirsky:Kitsune}); but it is also possible to create a cascade of a binary and multi-class classifier: the former determines whether a NetFlow is benign or malicious, and the latter infers the specific class of a malicious NetFlow, e.g., a DDoS or a Botnet (a schematic of such `cascade' is shown in Fig.~\ref{fig:bmd}). Moreover, many (labeled) datasets are publicly available, ensuring scientific reproducibility. 

These characteristics enable a broad coverage of use-cases.
In particular, we consider \textit{thousands} of different configurations, which vary depending on the following:
\begin{itemize}
    \item \textbf{Source Dataset (5)}: \dataset{CTU13}, \dataset{NB15}, \dataset{UF-NB15}, \dataset{CICIDS17}, \dataset{GTCS}. Each of these datasets is created via a different NetFlow tool: Argus, nProbe, Zeek, FlowMeter. An overview of these datasets is in Table~\ref{tab:datasets}, while more details are in \app{\ref{sapp:datasets}}.
    
    \item \textbf{Data Availability for training (4)}: \textit{Abundant} (80\% of \smabb{D}), \textit{Moderate} (40\%), \textit{Scarce} (20\%), \textit{Limited} (only 100 samples per class in \smabb{D}). Refer to \app{\ref{sapp:availability}}.
    
    \item Size of the \textbf{feature set (2)}: \textit{Complete} (i.e., using all features provided by the NetFlow tool) or \textit{Essential} (using only half of such features). Refer to \app{\ref{sapp:features}}.
    
    \item \textbf{ML Pipeline (6)}: a single \textit{binary detector} (\variable{BD}); a single \textit{multi-class detector} (\variable{MD}); a \textit{cascade} of \variable{BD} and \variable{MD} (\variable{BMD}); as well as three ensembles which vary depending on how the output is determined: via a \textit{logical or} (\variable{ED-o}), through \textit{majority voting} (\variable{ED-v}), or via a \textit{stacked} classifier (\variable{ED-s}). Refer to \app{\ref{sapp:pipeline}}.
    
    \item \textbf{ML Algorithm (4)}: Random Forest (RF), Logistic Regression (LR), Histogram Gradient-boosting (HGB), Decision Tree (DT). Refer to \app{\ref{sapp:algorithms}}.
    
    \item \textbf{Hardware specifications (6)}: a high-end computing appliance, a workstation, a common desktop, an old laptop, a virtual machine with reduced capabilities, and a Raspberry Pi 4B. Refer to \app{\ref{sapp:hardware}}.
\end{itemize}
Each combination can be seen as an unique ML-NIDS, which is assessed against: \textit{known} (by testing on the same attacks seen at training), \textit{unknown} (by testing on attacks not seen during training), and \textit{adversarial} attacks (based on~\cite{Apruzzese:Evading}, as they are feasible and hence likely to occur~\cite{apruzzese2021modeling}). A detailed description of all these distinct operational scenarios' is provided in \app{\ref{app:scenarios}}.
For each ML-NIDS, we compute the true and false positive rate (\smamath{tpr} and \smamath{fpr}); accuracy (\smamath{Acc}, but only for multi-classification tasks); and runtime (for both training and testing).

\vspace{-0.5em}

\begin{table}[!htbp]
    \centering
    \caption{Summary of the datasets of our experimental evaluation.}
    \label{tab:datasets}
    \resizebox{0.9\columnwidth}{!}{
        \begin{tabular}{c|c|c|c|c|c}
            
            \begin{tabular}{c} Dataset\\Name \end{tabular} & 
            \begin{tabular}{c} Benign\\Samples \end{tabular} &
            \begin{tabular}{c} Malicious\\Samples \end{tabular} &
            \begin{tabular}{c} Attack\\Classes \end{tabular} &
            Features &
            \begin{tabular}{c} NetFlow\\Software \end{tabular} \\
            \toprule
            
            \dataset{CTU13}~\cite{Garcia:CTU} & 16.7M & 403K & 6 & 30 & Argus~\cite{Argus:URL} \\
            \dataset{NB15}~\cite{UNSWNB15:Dataset} & 2.2M & 105K & 7 & 45 & Zeek~\cite{Zeek:URL} \\
            \dataset{UF-NB15}~\cite{sarhan2020netflow} & 2.3M & 78K & 7 & 40 & nProbe~\cite{nProbe:URL}\\
            \dataset{CICIDS17}~\cite{Sharafaldin:Toward} & 1.6M & 433K & 9 & 76 & FlowMeter~\cite{engelen2021troubleshooting} \\
            \dataset{GTCS}~\cite{mahfouz2020ensemble} & 140K & 378K & 4 & 80 & FlowMeter~\cite{lashkari2017characterization} \\
            
            \bottomrule
        \end{tabular}
    }
    \vspace{-0.5em}
\end{table}

To provide statistically significant results and remove any bias, we repeat all our experiments (both training and testing) multiple times, specifically: 1000 times for the \textit{limited} data availability (as there is a high chance of bias), and 100 times for the three other availability settings. Such repetitions are done by randomly sampling \smabb{T} from \smabb{D} according to the data availability setting; whereas \smabb{E} is always chosen by randomly selecting 20\% of the available samples of each class available in a given \smabb{D}. Moreover, we always follow the ``dos'' proposed by Arp et al.~\cite{arp2022dos}. (Our evaluation is \textit{fair}: for each trial, we train all our models on the same \smabb{T}, and evaluate them on the same \smabb{E}.)

Finally, we also perform an extra set of experiments in which \smabb{T} and \smabb{E} are chosen by taking the temporal domain into account, i.e.: \smabb{E} contains only the `last' 20\% samples of a given dataset, and \smabb{T} contains the `first' samples.

\remarkN{our evaluation is massive, and is due to our goal of providing a benchmark for future studies. A single research paper needs not to perform an evaluation of the same magnitude as the one in this SoK.
}

\vspace{-1em}

\subsection{Main Results (Quantitative Analysis)}
\label{ssec:results}
\vspace{-0.5em}
\noindent
Let us discuss the results of detectors using HGB, since it is a very recent algorithm for NID. Here, we aggregate the results of \textit{all} datasets, and we focus on the \textit{detection} performance on the \textit{high-end} platform. Fine-grained results are in \app{\ref{app:benchmark}}, which reports the \textit{multi-classification} performance, and the runtime on different hardware.

\vspace{-1em}

\subsubsection{Baseline Performance}
\label{sssec:baseline}
\begin{figure}[!htbp]
    \centering
    \begin{subfigure}{0.25\columnwidth}
        \centering
        \includegraphics[width=\linewidth]{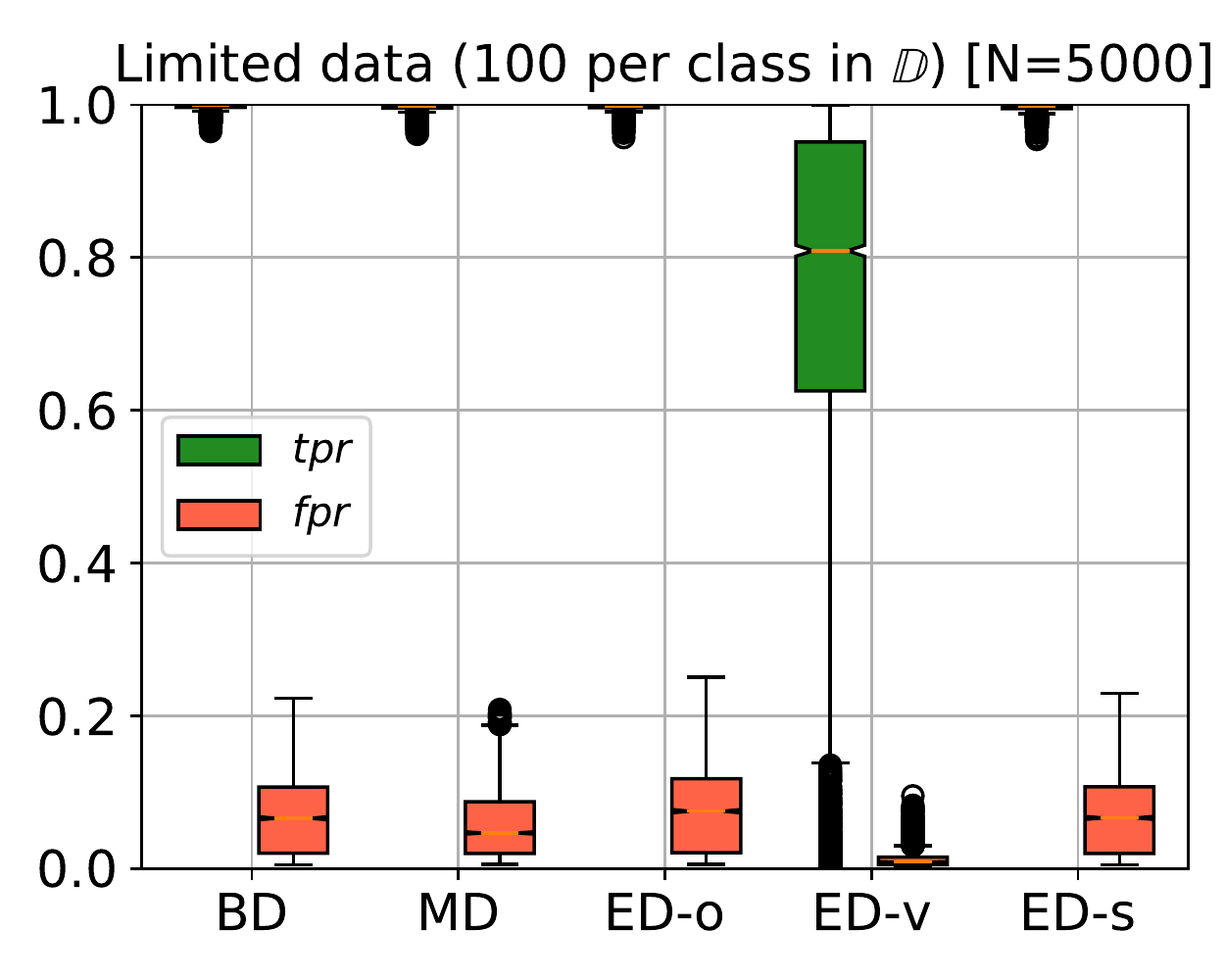}
        \label{sfig:baseline_limited}
    \end{subfigure}\hfill%
    \begin{subfigure}{0.25\columnwidth}
        \centering
        \includegraphics[width=\linewidth]{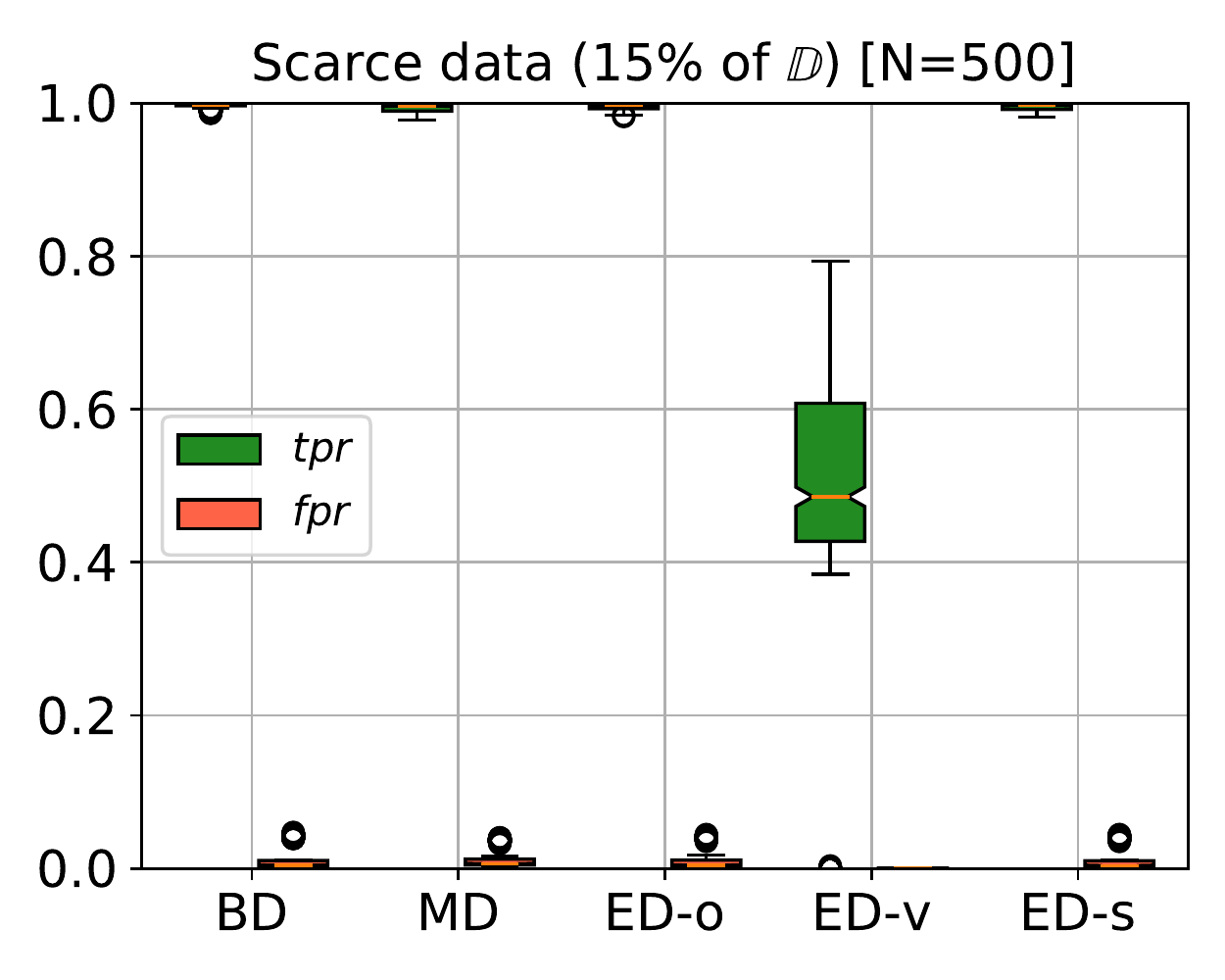}
        \label{sfig:baseline_scarce}
    \end{subfigure}\hfill%
    \begin{subfigure}{0.25\columnwidth}
        \centering
        \includegraphics[width=\linewidth]{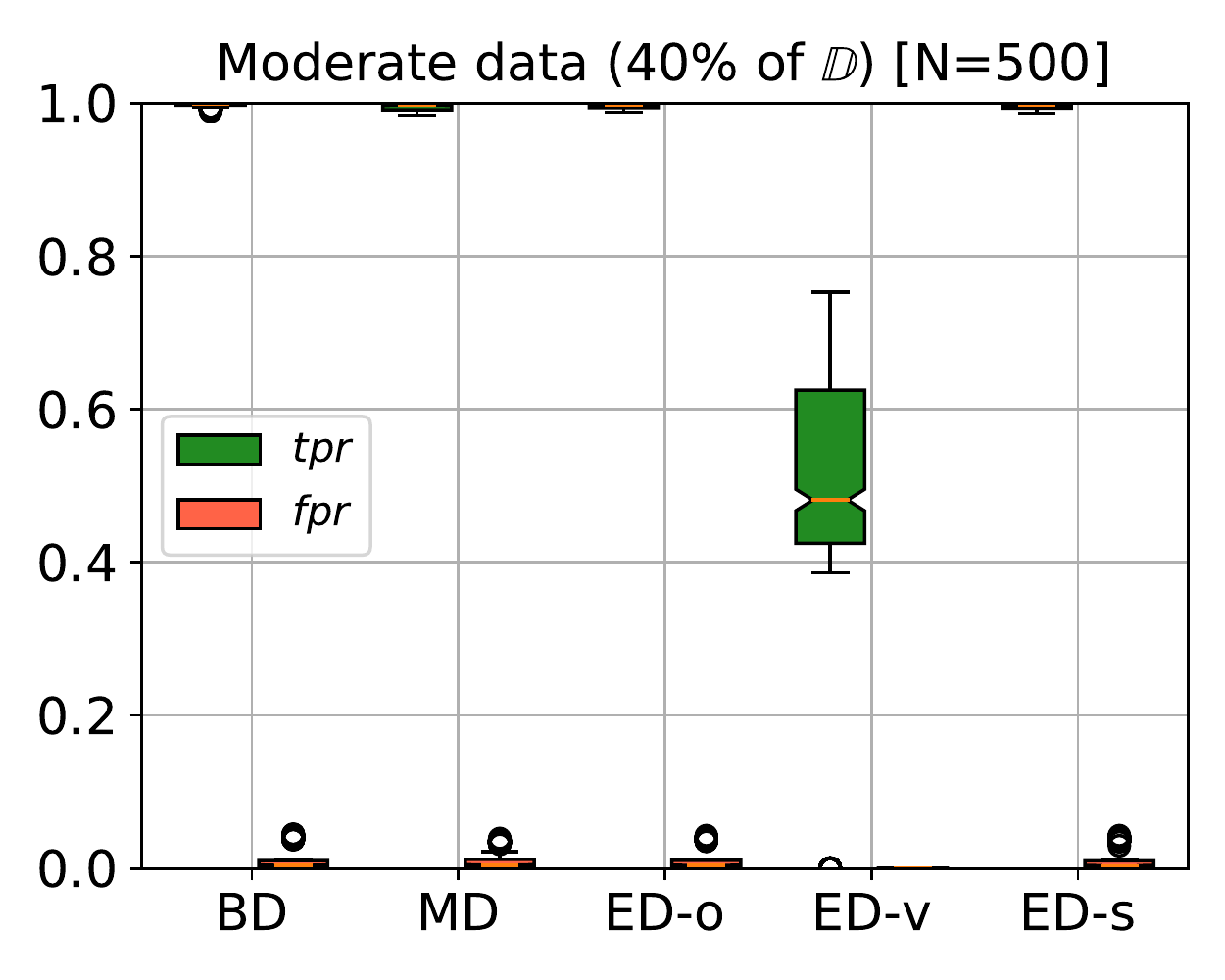}   
        \label{sfig:baseline_moderate}
    \end{subfigure}\hfill%
    \begin{subfigure}{0.25\columnwidth}
        \centering
        \includegraphics[width=\linewidth]{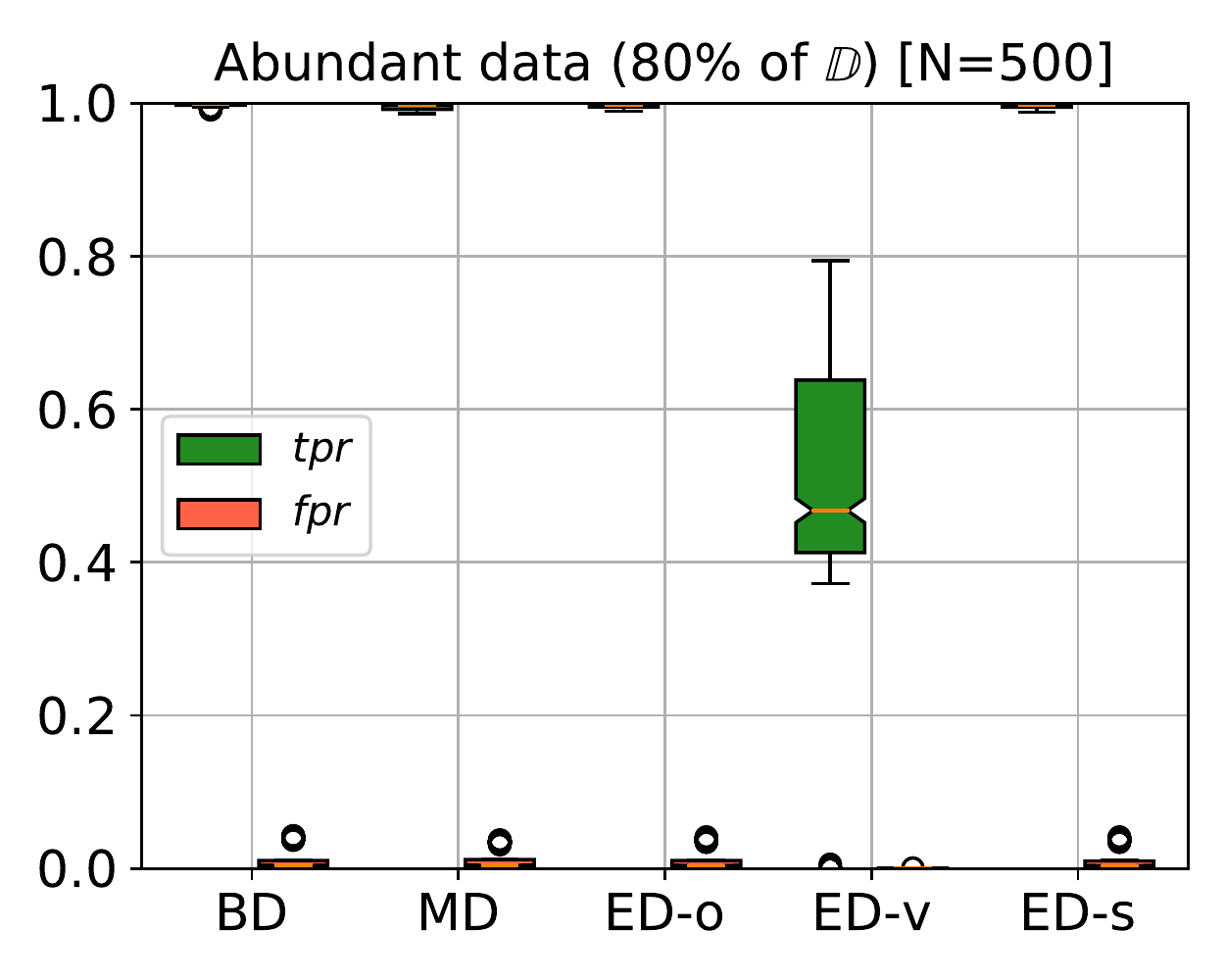}
        \label{sfig:baseline_abundant}
    \end{subfigure}
    \vspace{-1em}
    \caption{Baseline Performance.}
    \label{fig:baseline}
\end{figure}

We report in Fig.~\ref{fig:baseline} the boxplots showing the \smamath{tpr} and \smamath{fpr} of our detectors for increasing (left to right) data availability settings. We can see that detectors using \variable{ED-v} exhibit the worst \smamath{tpr} but the best \smamath{fpr}, which is understandable because they require multiple classifiers to agree on the maliciousness of a sample. In contrast, the other detectors appear to have comparable performance. We find it intriguing that \variable{MD} detectors appear to be effective even using a very limited amount of labels (see rightmost plot in Fig.~\ref{fig:baseline}).

\vspace{-1em}
\subsubsection{Detection of unknown attacks}
\label{sssec:unknown}
\begin{figure}[!htbp]
    \centering
    \begin{subfigure}{0.25\columnwidth}
        \centering
        \includegraphics[width=\linewidth]{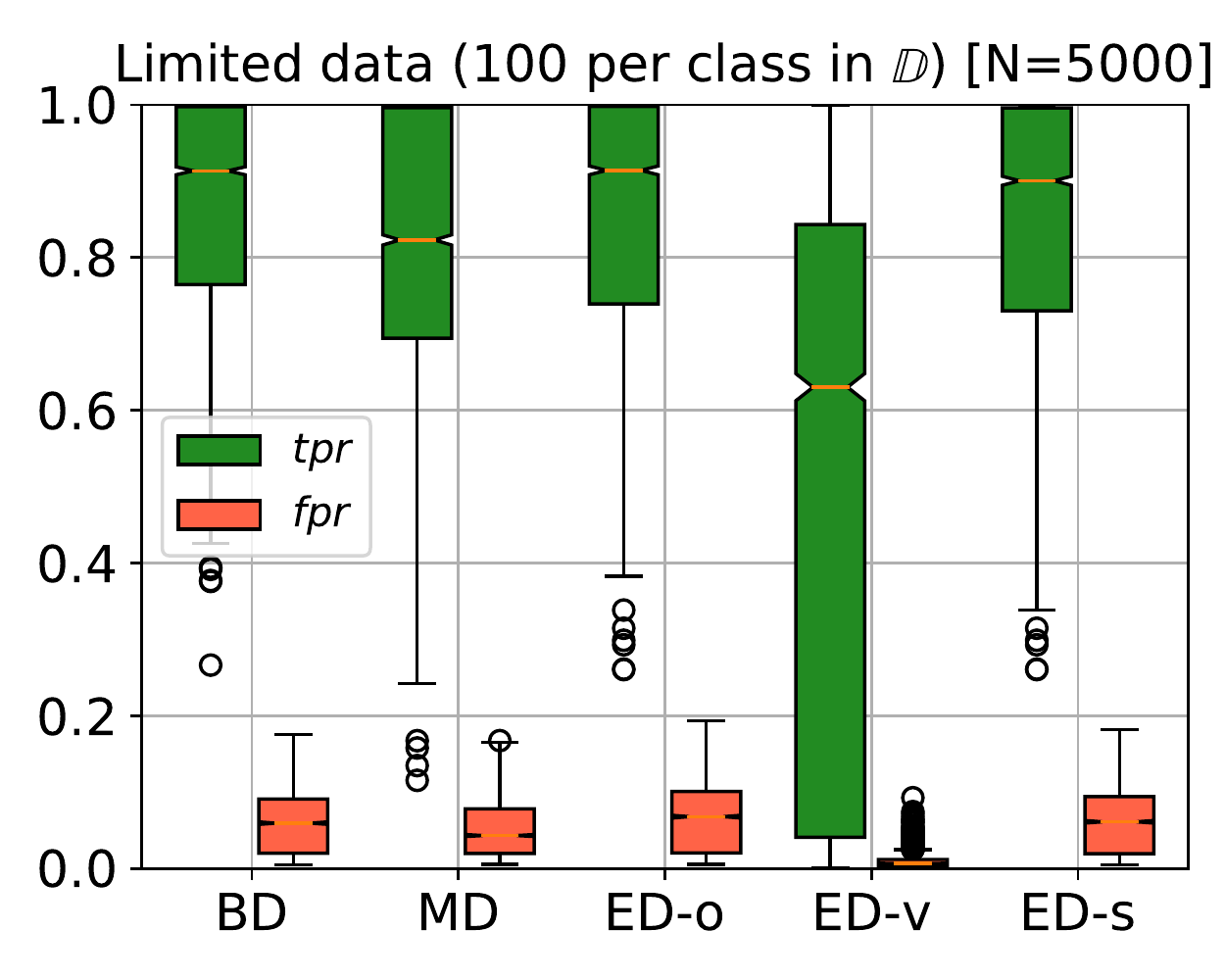}
        \label{sfig:unknown_limited}
    \end{subfigure}\hfill%
    \begin{subfigure}{0.25\columnwidth}
        \centering
        \includegraphics[width=\linewidth]{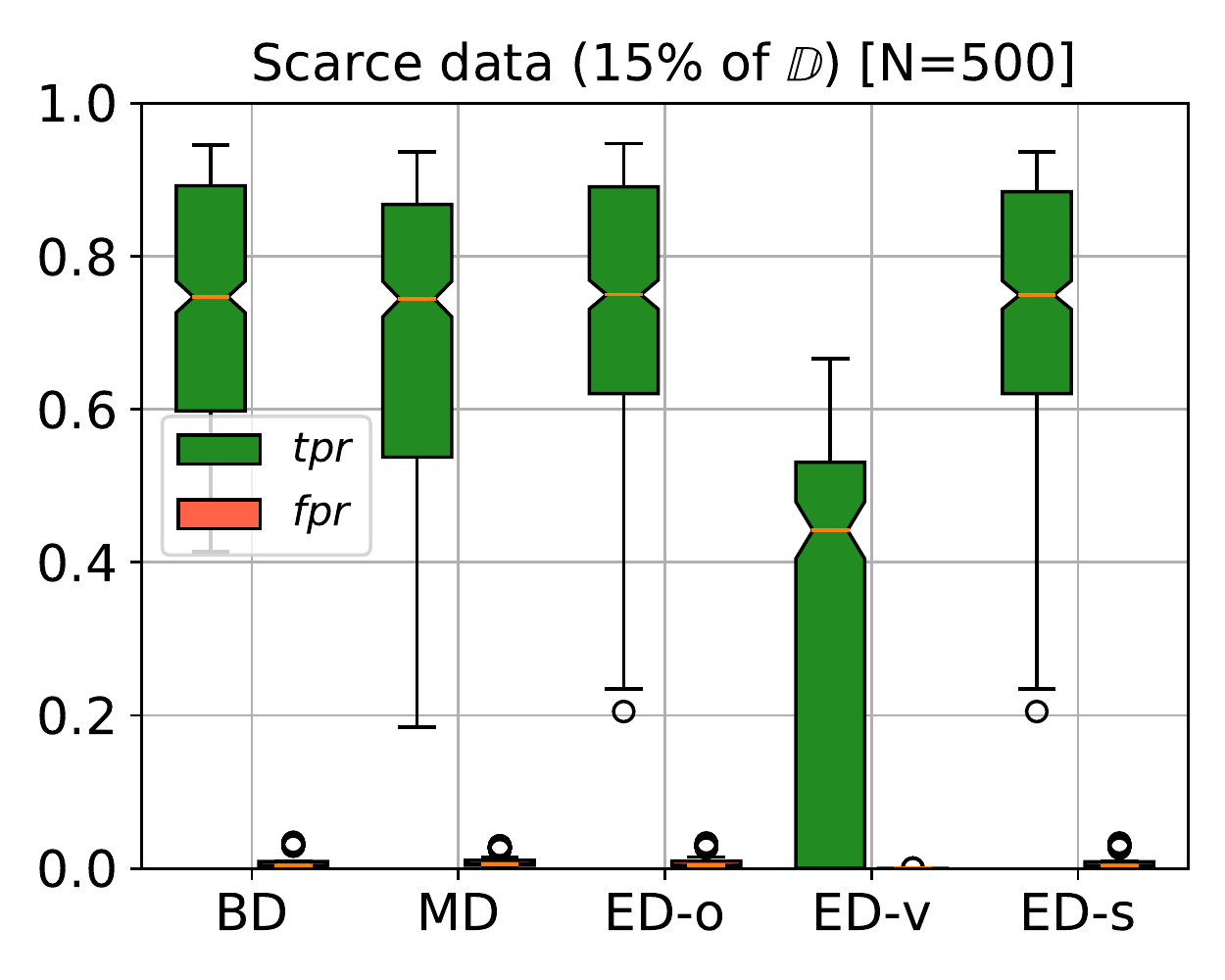}
        \label{sfig:unknown_scarce}
    \end{subfigure}\hfill%
    \begin{subfigure}{0.25\columnwidth}
        \centering
        \includegraphics[width=\linewidth]{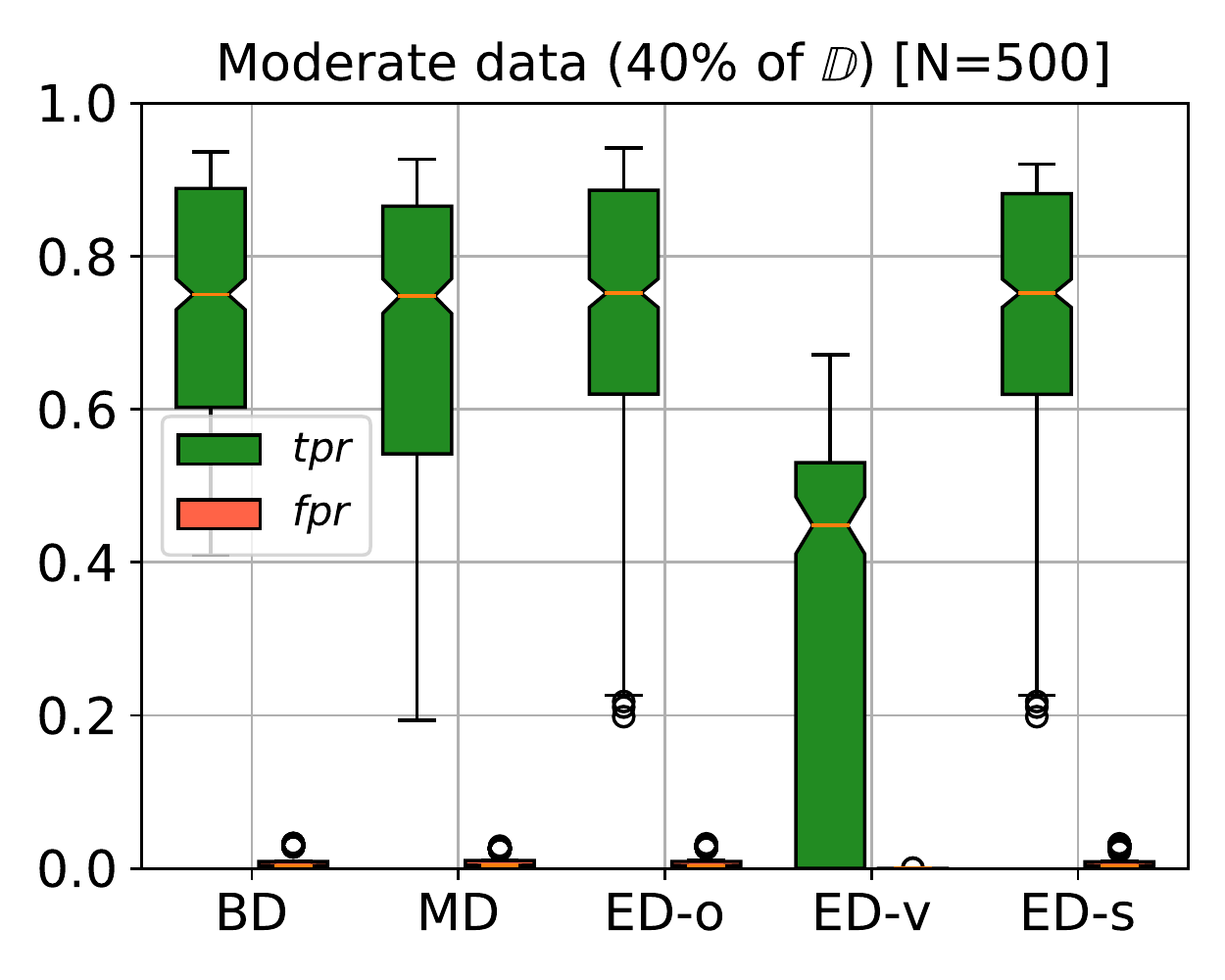} 
        \label{sfig:unknown_moderate}
    \end{subfigure}\hfill%
    \begin{subfigure}{0.25\columnwidth}
        \centering
        \includegraphics[width=\linewidth]{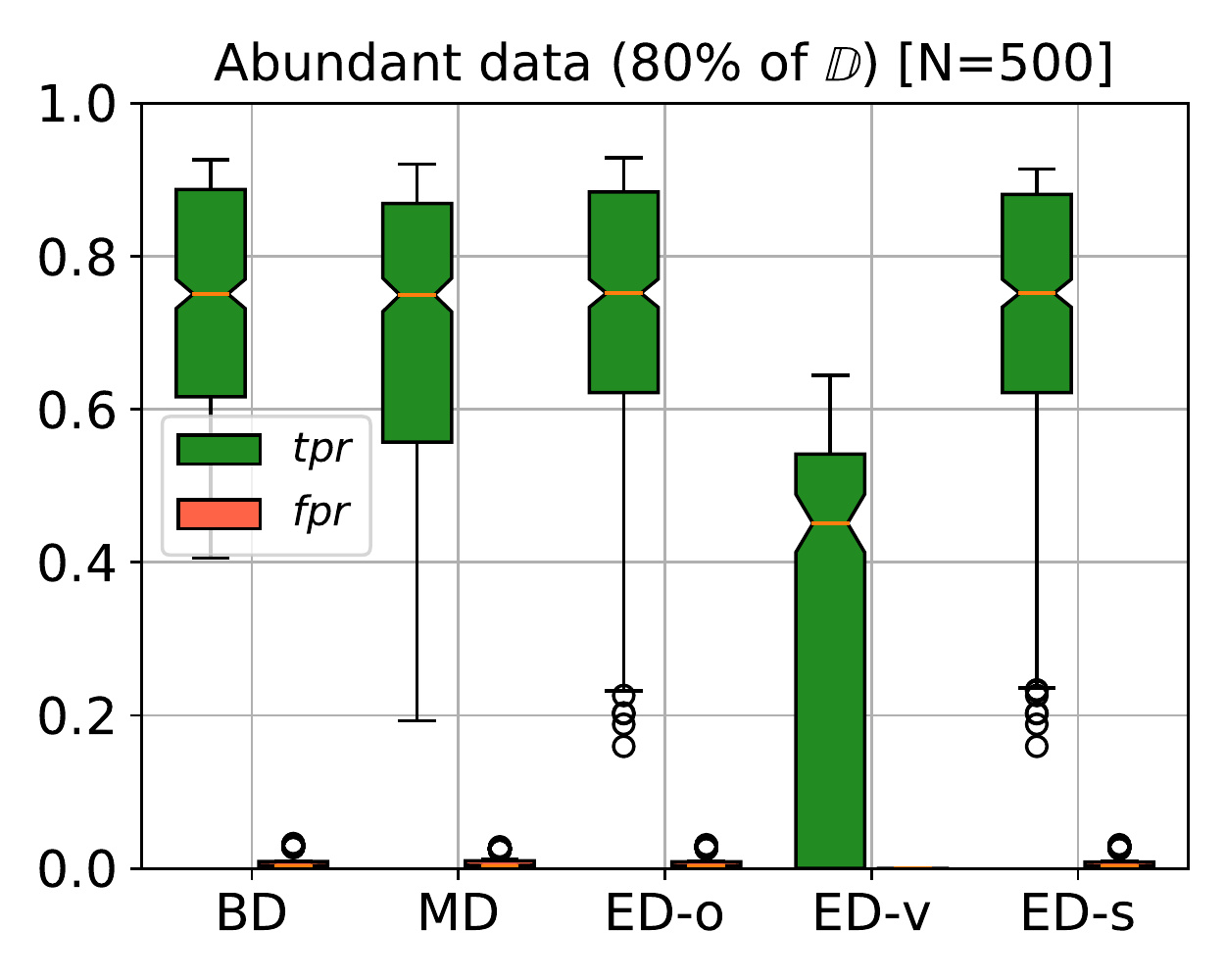}
        \label{sfig:unknown_abundant}
    \end{subfigure}
    \vspace{-1em}
    \caption{Detection of unknown attacks.}
    \label{fig:unknown}
\end{figure}

We report in Fig.~\ref{fig:unknown} the performance against unknown attacks---which is computed by excluding one malicious class from a given \smabb{T}, re-training all the involved ML models on such new \smabb{T}, and testing them on the benign portion of \smabb{E} (for the \smamath{fpr}), and on the `excluded' malicious class (and then averaging the resulting \smamath{tpr}). From Fig.~\ref{fig:unknown} (which has the same structure as Fig.~\ref{fig:baseline}) we can see that the \smamath{tpr} decreases, which is expected because the attacks are unknown. The detectors based on \variable{BD} appear to be the most robust. It is intriguing that the best results are achieved in the \textit{Limited} data availability setting. Such phenomenon can be explained by the fact that training on few samples allows ML models to generalize better on `unseen' classes.

\vspace{-1em}

\subsubsection{Adversarial Robustness}
\label{sssec:robustness}
\begin{figure}[!htbp]
    \centering
    \begin{subfigure}{0.25\columnwidth}
        \centering
        \includegraphics[width=\linewidth]{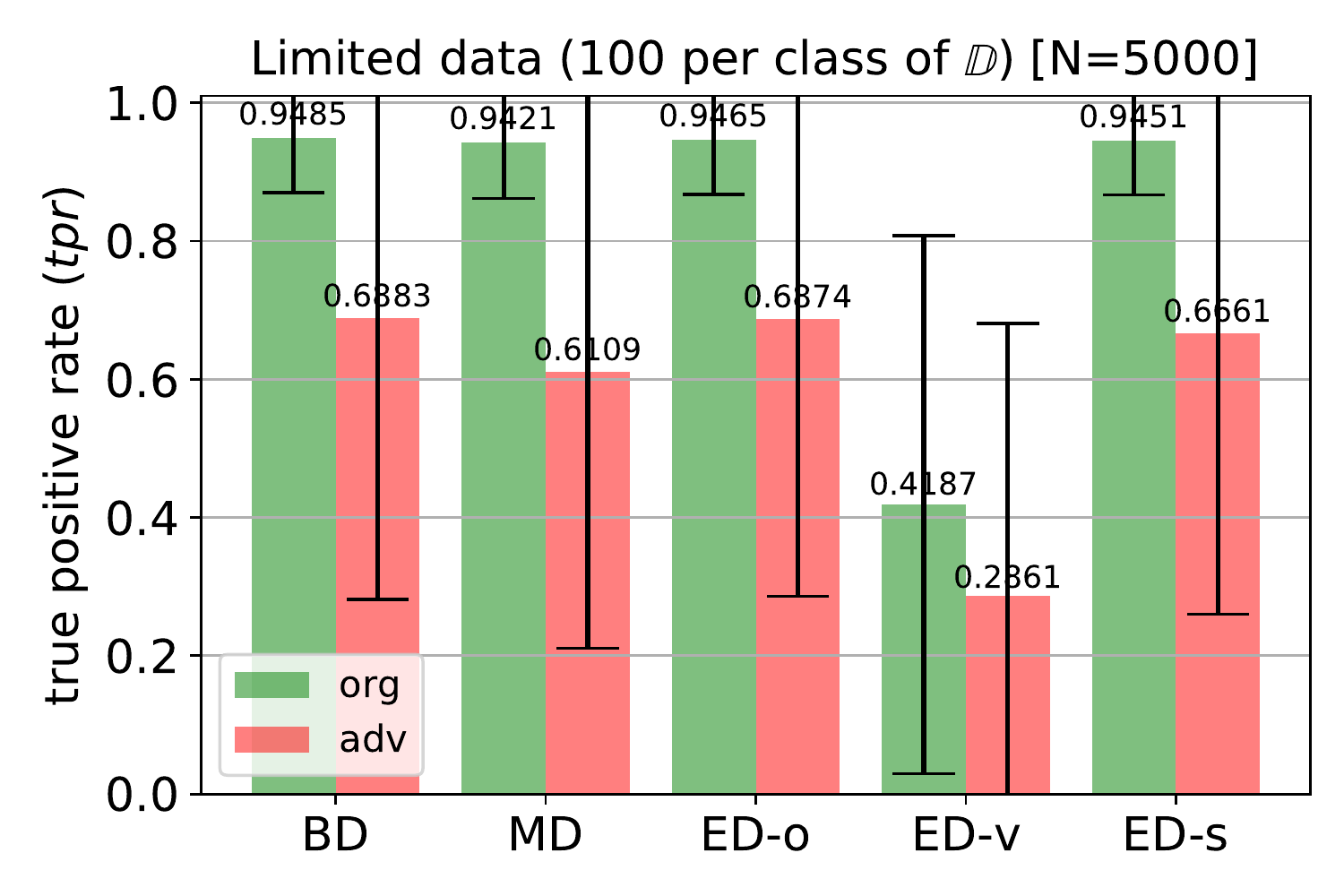}
        \label{sfig:robustness_limited}
    \end{subfigure}\hfill%
    \begin{subfigure}{0.25\columnwidth}
        \centering
        \includegraphics[width=\linewidth]{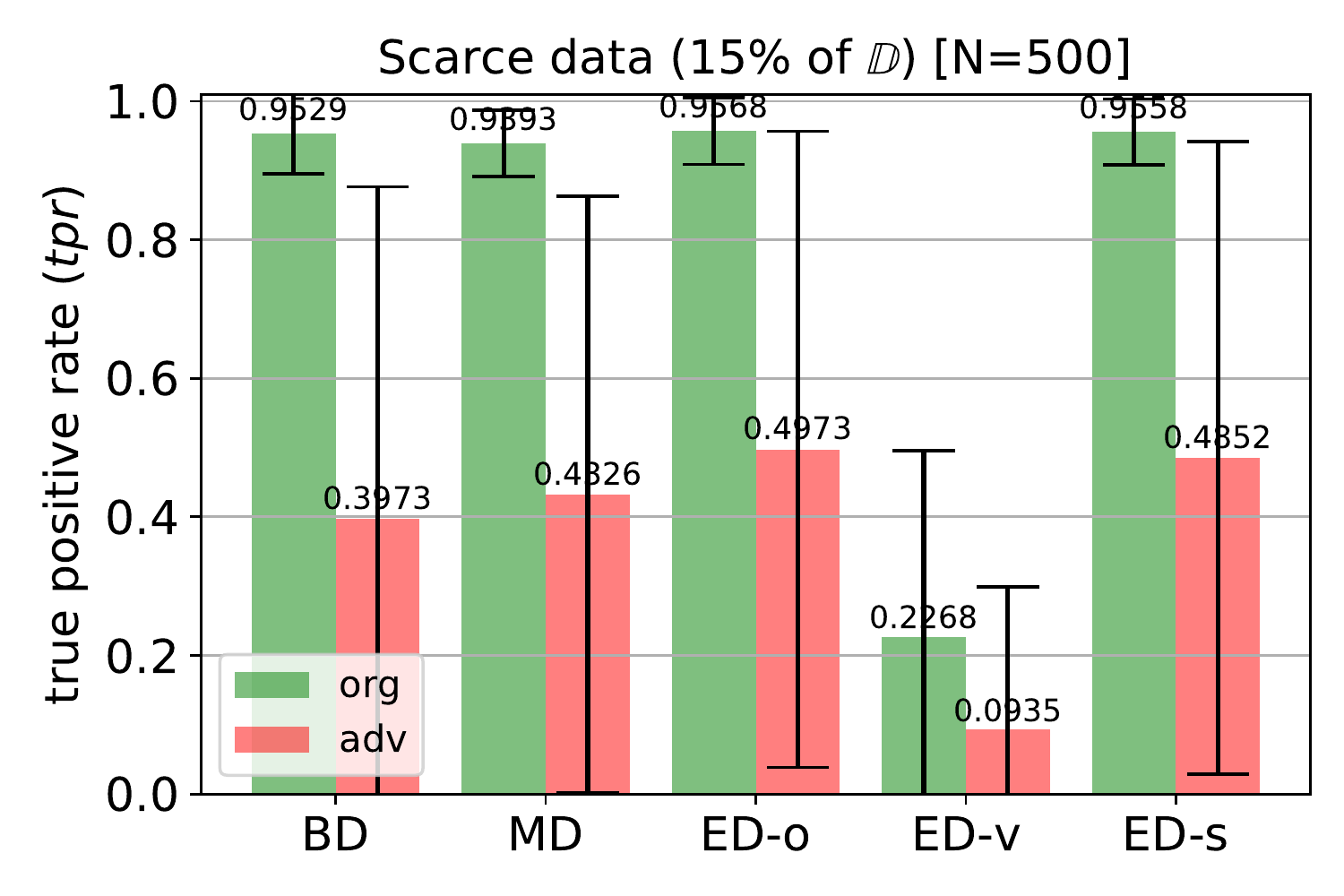}
        \label{sfig:robustness_scarce}
    \end{subfigure}\hfill%
    \begin{subfigure}{0.25\columnwidth}
        \centering
        \includegraphics[width=\linewidth]{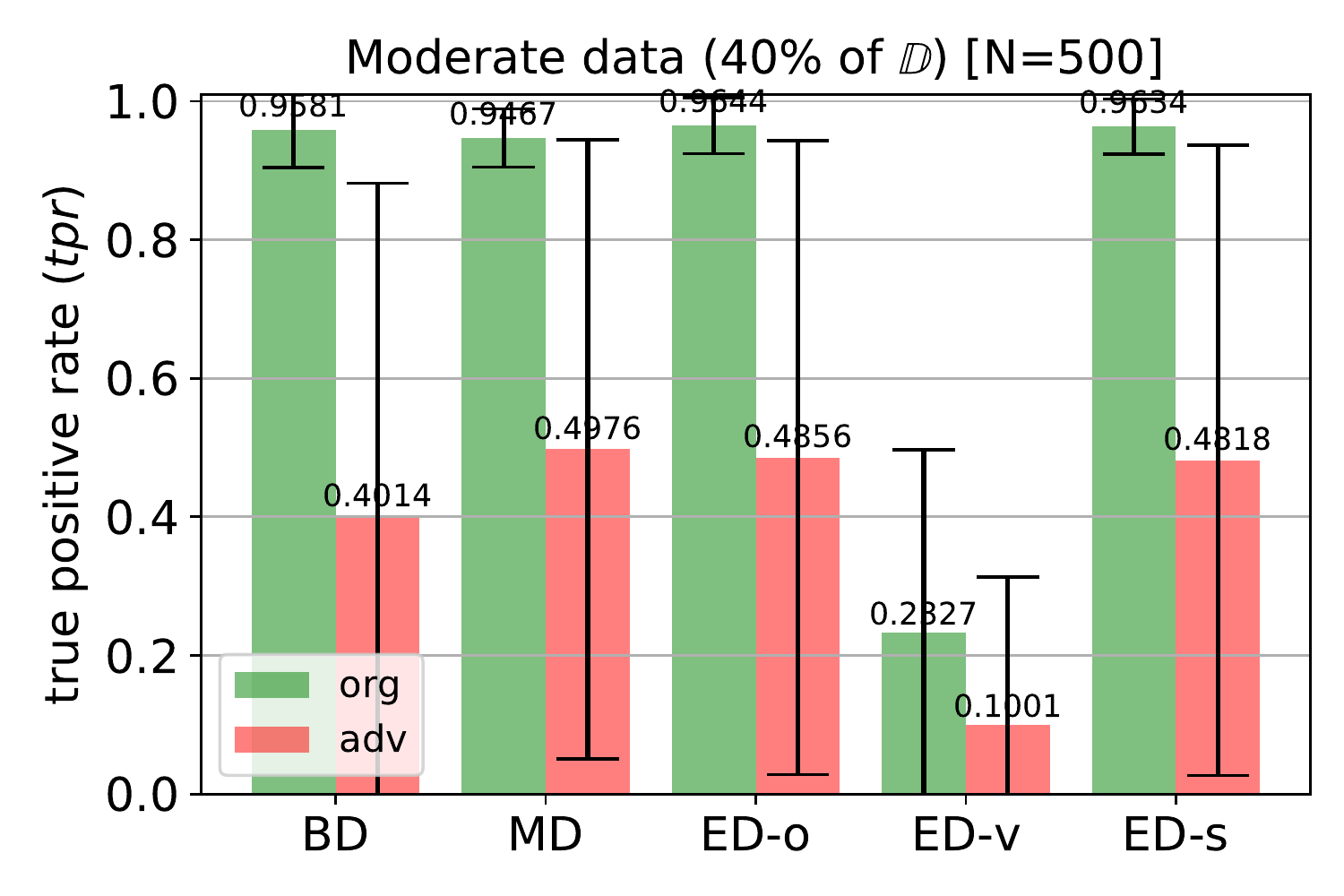}   
        \label{sfig:robustness_moderate}
    \end{subfigure}\hfill%
    \begin{subfigure}{0.25\columnwidth}
        \centering
        \includegraphics[width=\linewidth]{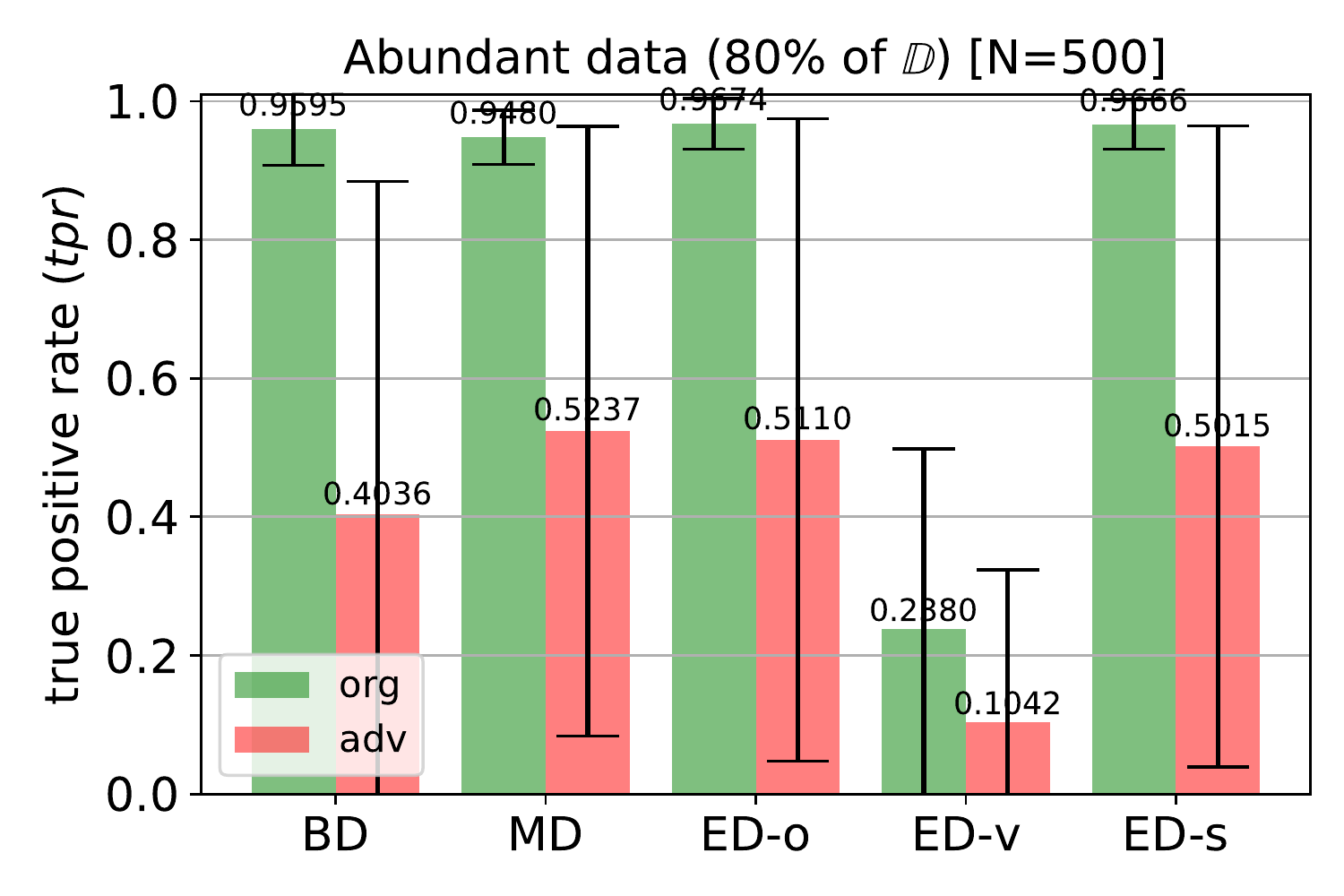}
        \label{sfig:robustness_abundant}
    \end{subfigure}
    \vspace{-1em}
    \caption{Robustness to adversarial attacks.}
    \label{fig:robustness}
\end{figure}

We measure the robustness of our detectors against the evasion attacks proposed in~\cite{Apruzzese:Evading}. The results are shown in Fig.\ref{fig:robustness}, reporting the \smamath{tpr} both before (green bars) and after (red bars) the application of the adversarial perturbations for all the detectors and for increasing amounts of training data (left to right). From these results, we can see that our detectors are \textit{more robust} when they are trained with \textit{less data}: indeed, the red bars in the leftmost graph are always higher than those in the other graphs (a similar phenomenon as the one in Fig.~\ref{fig:unknown}). In particular, we observe that \variable{BD} is the most resilient detector when using limited data, but the weakest (aside from \variable{ED-v}) when more data is available for training.

\vspace{-1em}

\subsubsection{Runtime}
\label{sssec:runtime}
\begin{figure}[!htbp]
    \centering
    \begin{subfigure}{0.25\columnwidth}
        \centering
        \includegraphics[width=\linewidth]{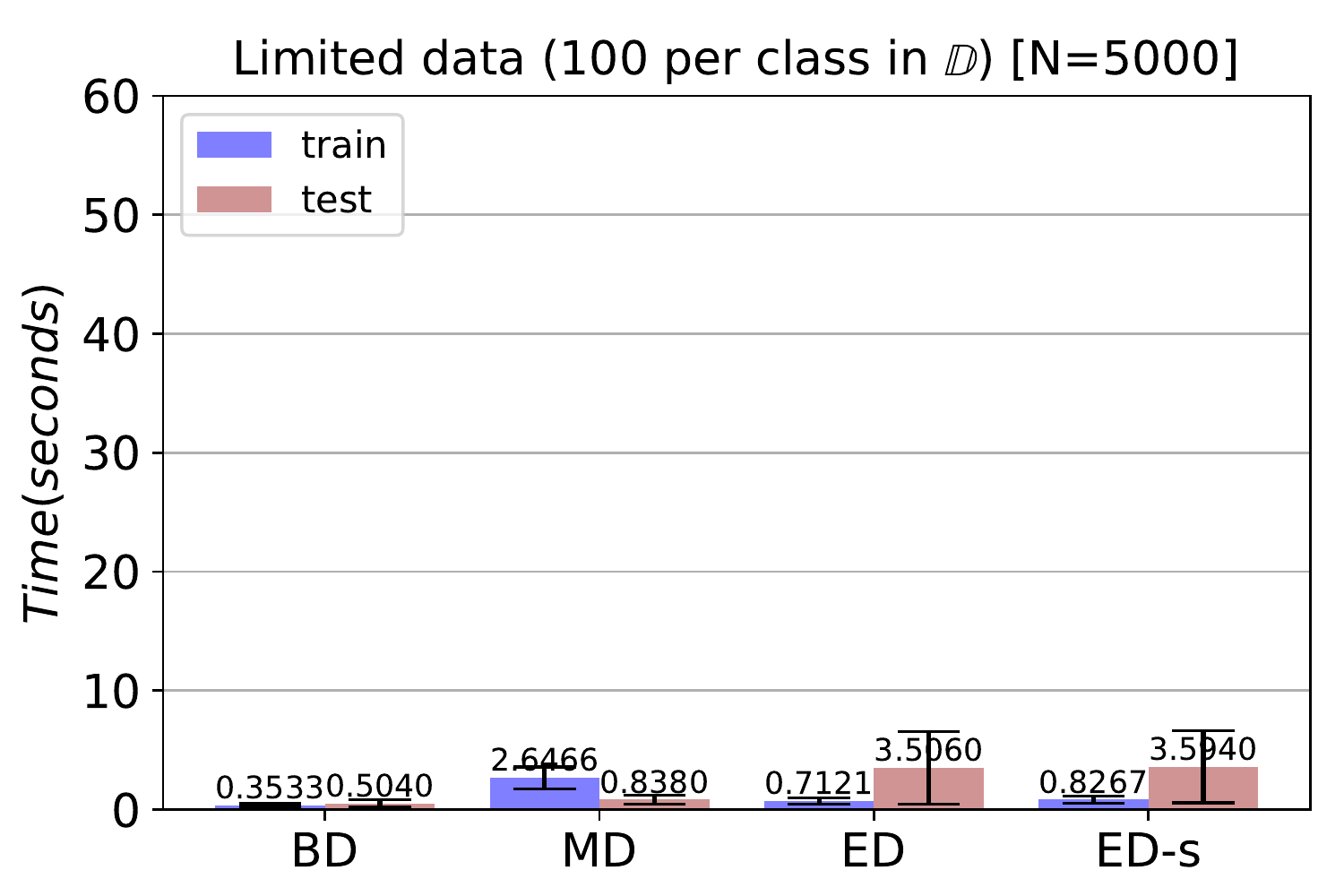}
        \label{sfig:runtime_limited}
    \end{subfigure}\hfill%
    \begin{subfigure}{0.25\columnwidth}
        \centering
        \includegraphics[width=\linewidth]{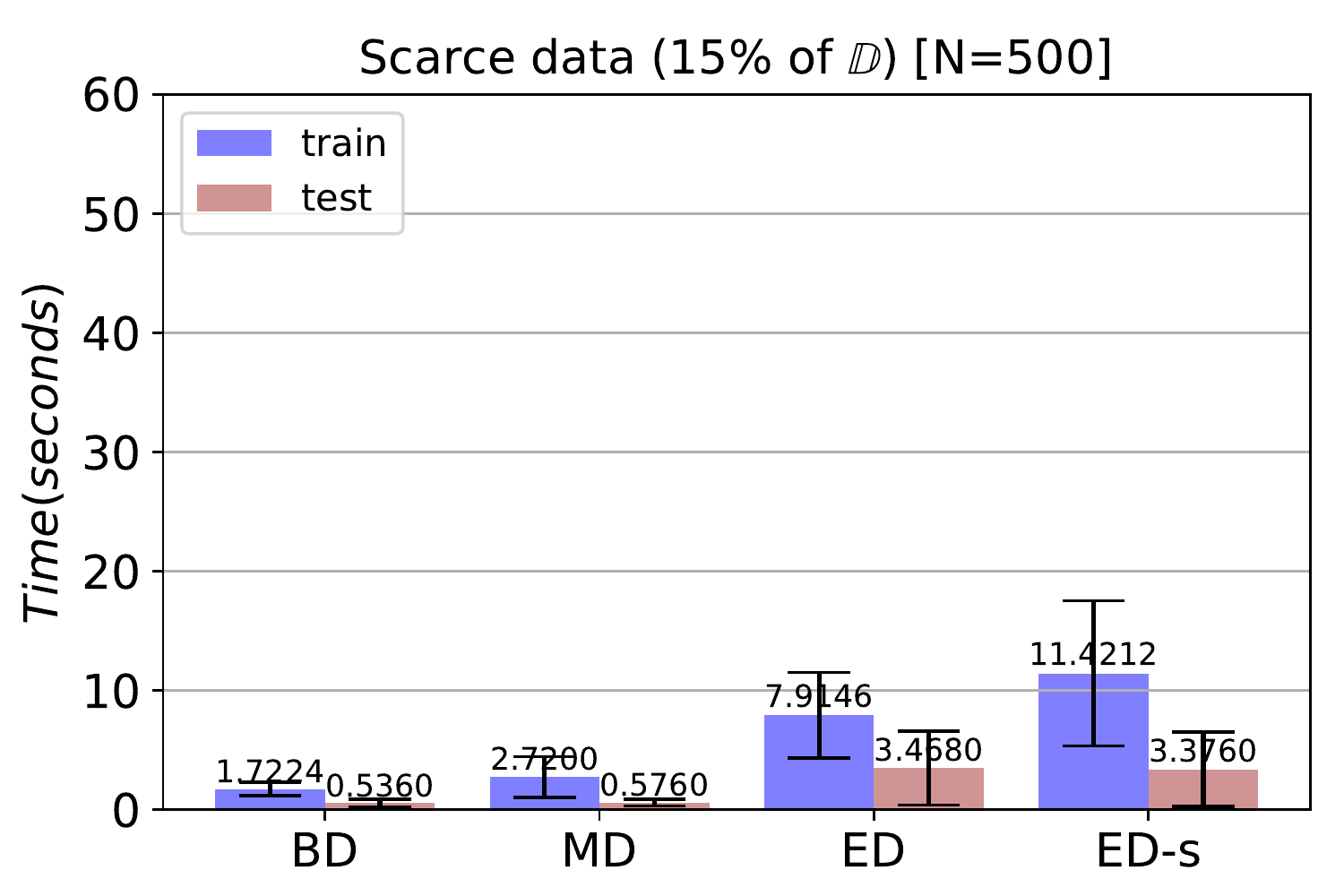}
        \label{sfig:runtime_scarce}
    \end{subfigure}\hfill%
    \begin{subfigure}{0.25\columnwidth}
        \centering
        \includegraphics[width=\linewidth]{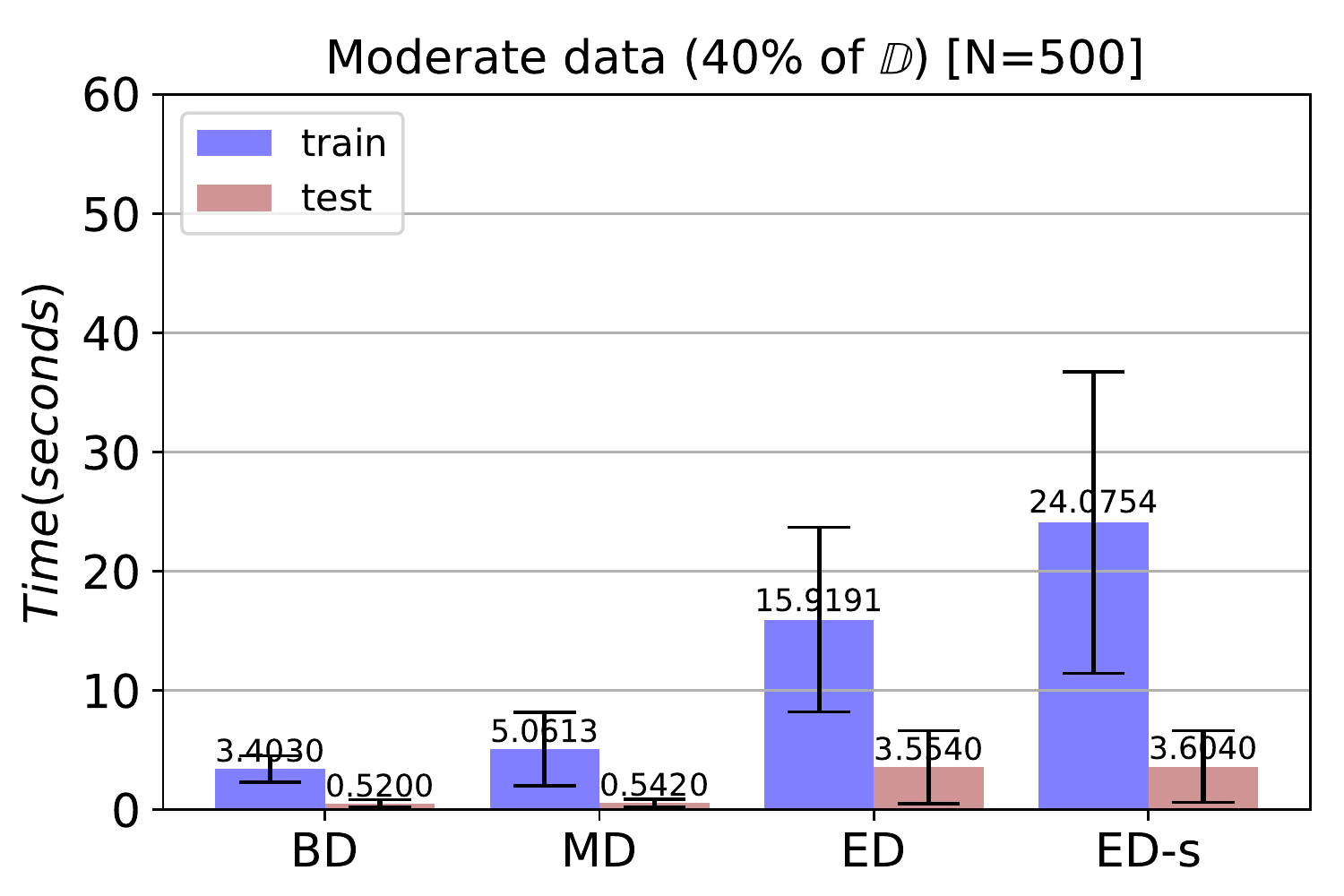} 
        \label{sfig:runtime_moderate}
    \end{subfigure}\hfill%
    \begin{subfigure}{0.25\columnwidth}
        \centering
        \includegraphics[width=\linewidth]{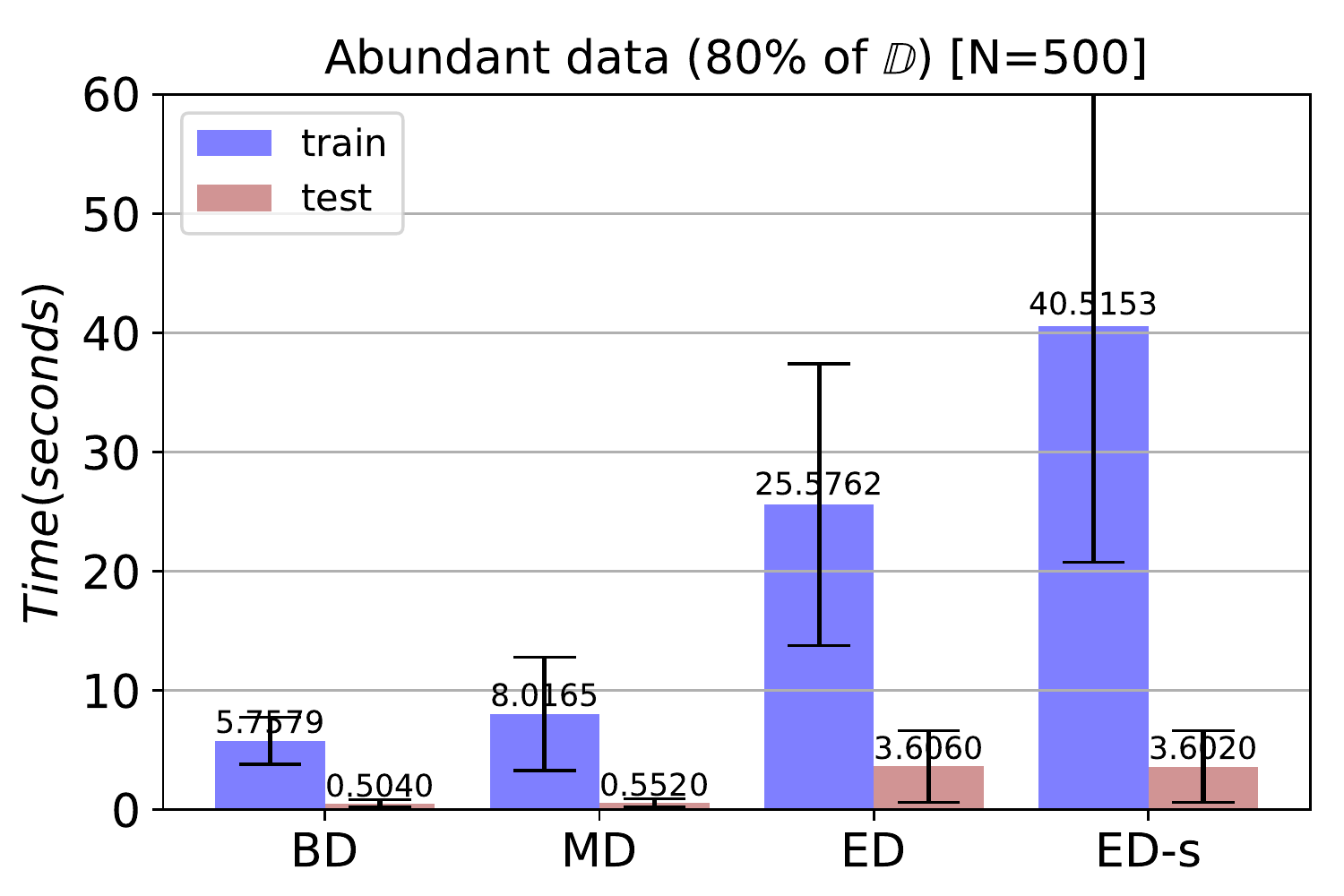}
        \label{sfig:runtime_abundant}
    \end{subfigure}
    \caption{Runtime (on the \textit{high-end} platform).}
    \label{fig:runtime}
\end{figure}

We report the operational runtime (as measured on the high-end platform) for all our detectors (\variable{ED} includes both \variable{ED-o} and \variable{ED-v}) in Fig.~\ref{fig:runtime}. Each plot (related to a specific data availability setting) reports the time (in seconds) for training (blue bars) and testing (brown bars) the respective detectors.
We can see that, on limited data, training is computationally less expensive than testing. Moreover, we also observe that training the ensembles is much more resource intensive than the simple \variable{MD} and \variable{BD} (the latter being the `cheapest' to train). More details are in \app{\ref{sapp:highend}}).

\vspace{-1em}

\subsection{Practical Considerations}
\label{ssec:practical}
\vspace{-0.5em}
\noindent
By inspecting all our results (\app{\ref{app:benchmark}}), we conclude that the ``no free lunch'' statement~\cite{wolpert1995no} is, once again, correct. For instance, methods based on HGB can have a slightly superior performance than, e.g., RF (cf. the \smamath{fpr} and \smamath{tpr} of \variable{BD} using the complete feature set with limited training data on \dataset{GTCS} in Table~\ref{tab:gtcs_baseline_static}). However, the HGB is worse in ``open world'' scenarios (Table~\ref{tab:gtcs_open_static})---but it has a lower runtime (cf. Tables~\ref{tab:gtcs_training} and~\ref{tab:gtcs_testing}). In this cases, it is up to practitioners to decide which method to deploy in their NIDS. Our pragmatic assessment, which reports both the effectiveness (i.e., \smamath{tpr}, \smamath{fpr}, or \smamath{Acc}) and expenses (i.e., all the requirements as well as the operational runtime) enables practitioners to make informed decisions.
Let us use our experiments to draw some practical considerations.

\textbf{The power of statistical comparisons.}
By performing a massive amount of trials, it is possible to carry out statistical tests that can be used to infer which ML method truly `outperforms' other competitors.\footnote{In real scenarios, even a \scmath{0.0001\%} can be significant: a single false negative can compromise a system, whereas the \scmath{fpr} must be close to 0.} For instance, by looking at the multi-classification tables (\app{\ref{sapp:classification}}), we can see that the \smamath{Acc} of \variable{BMD} and \variable{MD} tend to be close (e.g., on \dataset{GTCS} using HGB with Limited data, \variable{BMD} has 0.994 while \variable{MD} 0.982). With a Welch's t-test, we find that the resulting \smamath{p}-value is less than 0.0001 (i.e., below the usual target \smamath{\alpha=0.05}), which statistically proves that \variable{BMD} is better than \variable{MD} (accuracy-wise). Despite being powerful, such verifications are underused in NID literature (§\ref{sec:sota}). 

\textbf{Hardware can determine the winner.} Let us recall our motivational example (§\ref{sec:introduction}). By looking at the detection results in the closed world scenario using the complete feature set on \dataset{CICIDS17} (Table~\ref{tab:ids17_baseline_static}), we can see that---with the exception of LR methods---almost all detectors (\variable{BD}, \variable{MD}, \variable{ED-s} and \variable{ED-o}) achieve near-perfect performance in the abundant data availability setting, with \smamath{tpr} close to 1, and \smamath{fpr} close to 0; even a statistical test cannot determine the best method. In these cases, the `winner' can be determined by looking at the runtime in Table~\ref{tab:ids17_training}. We can see that the fastest method to train is the \variable{BD} using HGB, which requires 9s. However, HGB uses \textit{all 36 cores} of the high-end platform: in contrast, training the \variable{BD} using DT requires 25s but by using only one core. From a CPU utilization perspective, the HGB is 13 times slower than DT. Hence, our takeaway is that the `best' ML method for NID on \dataset{CICIDS17} uses DT as ML algorithm.\footnote{We reached out to some of the respondents of our survey (after they filled the questionnaire), and told them about such a finding: this made them change their mind on the importance of hardware.}

\textbf{Small or Big Data?}
Some intriguing results have been obtained by ML models trained with limited amount of labeled data (see the leftmost plots in Figs.~\ref{fig:baseline} to~\ref{fig:robustness}). Some of our ML models exhibited similar \smamath{tpr} as those trained with a considerably higher budget, but they had a higher \smamath{tpr} against both unknown and adversarial attacks (but at the expense of a slightly higher \smamath{fpr} against `known' attacks). 
This finding is noteworthy, as it may help in demystifying the necessity of having training datasets that count millions of samples. To quote a recent statement by Andrew Ng: ``\textit{Collecting more data often helps, but if you try to collect more data for everything, that can be a very expensive activity}''~\cite{andrew2021unbiggen}. We hence endorse development of ML methods that require smaller training datasets.

\textbf{Concrete use-case.}
Suppose an organization wants to deploy an ML method in their NIDS for identifying malicious NetFlows. The organization can compare their own network environment with those captured by our five considered datasets, and see whether there exist any similarities between our testbed and their real network. Suppose that the organization finds some similarities with the networks contained in \dataset{NB15} and \dataset{UF-NB15}: at this point, the organization can determine whether the NetFlow tool used by their NIDS is compatible to those used in \dataset{NB15} and \dataset{UF-NB15} (potentially by also considering the Essential and Complete feature set considered in our experiments). For example, if the organization has already a NetFlow tool using nProbe, then such an organization can almost directly transfer our ML methods trained on \dataset{UF-NB15} onto their NIDS. Otherwise, the organization needs to manually deploy nProbe into their NIDS first (which requires some expenses). In the (likely) chance that the organization finds no similarities with their own network and those captured by our chosen datasets, such an organization can choose to develop the solution that best fits their necessities, e.g., by choosing the one that provides the best performance while requiring the least amount of labels. 
\section{Discussion and Related Work}
\label{sec:discussion}
\noindent
We present some intrinsic difficulties of pragmatic assessments, perform some reflective exercises on our findings, and compare our paper with related literature.

\subsection{Challenges of Pragmatic Assessments}
\label{ssec:limitations}
\vspace{-0.5em}
\noindent
A research paper that fulfills each criteria in \defref{pragmatic} would be appreciated by practitioners. However, while some conditions are easy to meet, others are more difficult. Let us discuss some (current and future) challenges, so as to clarify the function of pragmatic assessments. 

\textbf{Statistical Significance.}
Obtaining results that are devoid of bias requires to perform multiple randomized trials. The experiments carried out in this paper required \textit{weeks} of computations---some of which are performed on expensive hardware. Furthermore, some ML methods are rooted on the existence of temporal patterns among data (e.g.,~\cite{corsini2021evaluation}): in these cases, performing many trials for statistically significant comparisons requires to either split the original \smabb{D} into different subsets or use completely different \smabb{D}. 
Therefore, we acknowledge that pragmatic assessments are not simple---which explains the situation portrayed in Table~\ref{tab:sota}. They are, however, doable: for instance, Liang et al.~\cite{liang2021fare} performed more than 50 trials for some of their experiments. Nonetheless, in some cases (i.e., if the results are `notably' different) only few trials are sufficient: the crux is reporting \textit{how many} trials have been performed. Finally, we encourage future works to rely on statistical tests when claiming that a given ML method ``outperforms the state-of-the-art.''

\textbf{Shortage of Public Data.}
A well-known problem in NID is the lack of datasets usable for research purposes~\cite{apruzzese2022role}. Such a lack makes it impossible for scientific papers to \textit{exactly} replicate the (real) network environment in which the proposed ML method can be deployed. Therefore, a pragmatic assessment is meant to ``allow practitioners\footnote{We stress that such an ``estimate'' is outside the scope of a research paper, since it can only be done by the developers of real products.} to estimate the real value of an ML method for NID,'' and not to ``ensure that every ML method for NID is deployed in practice.'' Indeed, the latter requires researches to evaluate their ML-NIDS in every possible network environment, which is clearly unfeasible. Nonetheless, future endeavours should attempt to evaluate their ML methods on diverse datasets---which is important to practitioners (§\ref{ssec:opinion}). We outline the opinion of practitioners on NID datasets in \app{\ref{sapp:pradata}}.

\textbf{Concept Drift and Explainability.}
A pragmatic assessment should not aim at investigating the robustness of an ML method to the concept drift problem.\footnote{This requires the researcher to know---in advance---whether a given dataset contains instances of such drift, which may not be the case. Simultaneously, concept drift is unpredictable and it is not known a priori whether it will occur or not. Hence, results derived from `synthetic' testbeds are questionable due to such unpredictability (\sccal{U} in Eq.~\ref{eq:performance}).} 
Indeed, robustness to concept drift can be realistically assessed only \textit{after} the deployment of an ML model; in contrast, the goal of a pragmatic assessment is to guide decision making \textit{before} such deployment. Nevertheless, we acknowledge that some ML methods can better deal with concept drift~\cite{lu2018learning}, such as lifelong learning (e.g.,~\cite{pendlebury2019tesseract}), or those methods that present a high explainability~\cite{andresini2021insomnia, meske2022}. In particular, we mention that the participants of our survey commented that providers of security solutions should favor methods that are ``explainable to their clients''. Unfortunately, it is well-known that the decisions of ML models are difficult to interpret~\cite{bhatt2020explainable}. Hence, we cannot put the ``explainability'' into our proposed factors, as it would be unfeasible to fulfill by research.

\subsection{Reflections and Recommendations}
\label{ssec:reflections}
\vspace{-0.5em}

\textbf{Feasibility and Sweet Spots.} 
We provide some recommendations that can maximize the pragmatic value of research without requiring extensive effort. We focus on those aspects that apply to ``any'' paper on ML-NIDS.
\begin{itemize}
    \item \textit{Experimental details.} Providing all details (§\ref{ssec:transparency}) of the testbed (including hardware) is straightforward. The only issue are page limitations: in these cases, researchers can provide a link to supplementary files (but we also endorse editors and organizers to accept longer papers during the peer-review).
    \item \textit{Performance.} As recommended by Arp et al.~\cite{arp2022dos}, at least two `classification' performance metrics should be computed (we used \smamath{tpr} and \smamath{fpr}), which is trivial to accomplish. Moreover, measuring the runtime (for both training and testing) is also straightfoward and requires just few lines of code (plus, it helps in devising a sound and efficient experimental workflow).
    \item \textit{Testbed variety.} Typically, a research paper on ML-NIDS requires to evaluate (i)~the proposed method, and (ii)~a suitable baseline for comparison---both of which should be assessed in the same settings.\footnote{Hence, we reiterate that it is not necessary to consider hundreds of combinations (as we did in our demonstration).} However, we endorse papers that assume ML-NIDS requiring large \smabb{T} to also assess cases entailing a `very small' \smabb{T} (some real product require months of data collection before they can be deployed~\cite{apruzzese2022role}). Doing this is feasible since the training time is shorter, and the `smaller' \smabb{T} can be generated as a subset of the `larger' \smabb{T} (but in both cases, \smabb{E} should be the same).\footnote{Even if the performance with the `small' \scbb{T} is subpar, it would not subtract to the paper's contribution (as long as it is sensible to assume that the proposed ML-NIDS requires large \scbb{T}).}
    \item \textit{Repetitions (supervised ML).} As can be seen from our evaluation, when using `large' \smabb{T} the performance does not change substantially (see the distribution of \smamath{fpr} and \smamath{tpr} in §\ref{ssec:results} for the Scarce, Moderate and Abundant data); hence, for these cases, we recommend at least 3x3 repetitions (i.e., changing \smabb{E} and \smabb{T} three times each). However, when considering small \smabb{T} (see the results for the limited data in §\ref{ssec:results}) the performance can greatly vary; hence, for these cases, we recommend at least 10x10 repetitions. We stress that the training time for the Limited data was significantly inferior than for all the other cases (refer to Fig.~\ref{fig:runtime}), hence such a higher amount of repetitions should be feasible to perform.\footnote{We believe our proposed ``repetition sweet-spots'' to be feasible to integrate in any ML-NIDS paper; however, a paper can provide a valid scientific contribution even without following our recommendations.}
\end{itemize}
Simply put, meeting the requirements for our pragmatic assessment is well within the reach of most researchers. 

\textbf{The role of our factors.} We discuss the relevance of our factors (Eq.~\ref{eq:performance}) by using our experiments (§\ref{sec:demonstration}):
\begin{itemize}
    \item \smacal{P} can be observed by comparing the results on \dataset{NB15} and \dataset{UF-NB15} (e.g., Table~\ref{tab:nb15_open_static} and Table~\ref{tab:ufnb15_open_static}), because these datasets contain the exact same raw data, but the NetFlow tool (i.e., the preprocessing) is different. E.g., the \variable{MD} using HGB with scarce data is robust against our adversarial attacks on \dataset{NB15} (0.96 \smamath{tpr}), but the same method on \dataset{UF-NB15} is very weak (0.47 \smamath{tpr}).
    \item \smacal{D} can be observed from any table (e.g., Table~\ref{tab:gtcs_baseline}) as the performance clearly changes as \smabb{T} increases.
    \item \smacal{S} can be observed by comparing the multi-classification results of any table (e.g., Table~\ref{tab:gtcs_multi}), as the performance of \variable{MD} and \variable{BMD} differs due to different pipelines (cf. our remark on statistical significance); but also by comparing the results of different algorithms in any table.
    \item \smacal{H} is shown by Table~\ref{tab:hardware}, as the runtime changes up to 400\% under diverse hardware settings.
    \item \smacal{U} is highlighted by the great variance of results achieved across our entire evaluation, which confirms the role played by randomness.
\end{itemize}
The unpredictability \smacal{U} is also implicit: we cannot foresee what is going to happen \textit{after} any ML model is deployed.

\textbf{User-study: Limitations.}
Our questionnaire (see \app{\ref{app:survey}}) resembles that of \textit{structured interviews} (used also, e.g., by~\cite{fischerHubner2021stakeholder}), thereby allowing to derive \textit{quantitative} results, while protecting our participants against possible NDA violations~\cite{opdenakker2006advantages}.
Such a design choice was chosen because our goal is to \textit{validate the importance of our proposed factors} (§\ref{ssec:factors}), and to get the opinion of practitioners on the \textit{current state-of-research} (§\ref{sec:sota}). 
Although our closed-questions could introduce some form of bias, we remark that (i)~each question had a `negative' answer; and (ii)~in some cases, the viewpoint of our population went against our theses.
We acknowledge that our questionnaire could have been formulated in an `open question' format; however, such a design choice could also be affected by bias, since we ultimately had to interpret the (unstructured) answers we received and map them to our proposed factors. 
We therefore acknowledge that some practitioners may have some priorities that are complementary to our factors. To account for such a limitation, we invited our respondents to give us some feedback \textit{after} they filled their questionnaire, thereby allowing us to derive additional insight (discussed in \app{\ref{sapp:analysis}}).

\subsection{Related Work}
\label{ssec:related}
\vspace{-0.5em}
\noindent
Let us compare our paper with prior literature. We stress that our focus is on ML for NID, and we do not claim generality over different domains. Nonetheless, we discuss how our pragmatic assessment can be extended to other security applications of ML in \app{\ref{sapp:extensions}}.

\textbf{Technical papers.}
Taken individually, most papers on ML-NIDS have different goals than ours. The authors of~\cite{binbusayyis2019identifying} aim to `outperform the state-of-the-art'; those of~\cite{andresini2021insomnia} focus on concept drift, which is unpredictable hence impossible to detect \textit{before} the deployment of a ML solution (as explained in §\ref{ssec:challenges}).
Pendlebury et al.~\cite{pendlebury2019tesseract} aim to eliminate experimental bias, but ultimately consider a different security problem (i.e., malware analysis). An intriguing research area focuses on \textit{privacy} of ML (e.g.,~\cite{jayaraman2019evaluating}), which is complementary to our goal.
Finally, a significant number of papers perform evaluations on outdated datasets (e.g., the \dataset{NSL-KDD}~\cite{iwendi2020use}), which makes the corresponding results of questionable value for modern and realistic deployments. Others present uncertainties due to overlooking some factors that real developers must take into account (cf. §\ref{sec:sota}). In contrast, our testbed involves recent datasets (including their `fixed' version~\cite{engelen2021troubleshooting}), increasing the realistic fidelity of our experiments. 
Due to (i)~the broad combination of use-cases, (ii)~the hundreds of trials to remove bias, and (iii)~the consideration of many likely deployment scenarios, our evaluation enables a fair and statistically validated benchmark of existing ML methods for NID---benefiting both research and practice.

\textbf{Reviews and Surveys.}
Some reviews tackle the entire cybersecurity domain (e.g.,~\cite{leszczyna2021review}) and do not delve into the specificity of ML; or focus on trustworthy ML development, but not from the perspective of NID (e.g.,~\cite{xiong2022towards}). Some papers focus on `deployment' challenges of ML, such as:~\cite{baier2019challenges} and~\cite{jenn2020identifying}, which are both very generic and do not focus on networked systems;~\cite{paleyes2022challenges}, which does not have any form of practitioner validation, nor systematically explains how research can fulfill their needs; and~\cite{pacheco2018towards}, which is on network applications, but not specific of cybersecurity and thus do not consider the presence of malicious entities---which are intrinsic of NID. A recent paper~\cite{alahmadi202299} interviewed 21 SOC analysts but neither proposes nor empirically evaluates any solution that can meet practitioners' needs from the researcher perspective; and do not focus on ML (only 10\% of their population uses ML!).
We also mention~\cite{damiani2020certified} and~\cite{jiang2018trust}, which propose `certification' of ML \textit{models}---which is not relevant for NID research, whose focus is on the ML \textit{method} (due to the impossibility of reliably transferring ML models across network environments~\cite{apruzzese2022cross}).
More related works provide a broad overview (e.g.~\cite{apruzzese2022role}) or highlight the issues (e.g.,~\cite{Apruzzese:Deep}) of ML for generic cybersecurity tasks; others may focus on a single aspect of ML for NID, such as the architecture of an ML-NIDS (e.g.,~\cite{giacinto2005network}), the role of features (e.g.,~\cite{das2021network}), the impact of unlabelled data~\cite{apruzzese2022sok}, or the weakness to adaptive ``adversarial'' attacks (e.g.,~\cite{apruzzese2021modeling}). 
Our paper \textit{extends} all such works by providing original takeaways---some of which are overlooked, or even contrast those by past work. We provide in \app{\ref{sapp:comparison}} an in-depth comparison of this SoK with the (closest) related work by Arp et al.~\cite{arp2022dos} (presented at USENIX Security'22).

\textbf{Summary.} 
To the best of our knowledge, no paper: (i)~elucidates the factors contributing to the real value of ML for NID, and (ii)~explains how research can account for such factors; and then (iii)~demonstrates how to do this in practice through a statistically-validated re-assessment of hundreds of diverse ML-NIDS; and (iv)~performs a user study with practitioners to validate its major claims.

\section{Conclusion}
\label{sec:conclusion}
\noindent
The integration of ML methods proposed in research into operational NIDS is progressing at a slow pace, due to the (justified) skepticism of developers towards the results reported in scientific literature. 
Our SoK paper aims to rectify this problem by changing the existing evaluation methodology adopted in this research domain. We do this by proposing the notion of \textit{pragmatic assessments}, whose objective is allowing practitioners to estimate the operational effectiveness and required expenses related to the entire lifecycle of a ML method for NID. After presenting irrefutable evidence that prior research does not allow to estimate the real value of ML for NID, we perform the first pragmatic assessment. Our massive evaluation represents a benchmark for future research, but is also useful for practitioners who can ascertain the real value of existing ML methods.

One may ask: ``\textbf{Must any future research paper perform a pragmatic assessment to be considered a significant contribution?}'' Our answer is a clear ``\textbf{no}'': a paper that does not meet all requirements of a pragmatic assessment can still be useful \textit{for research}. Indeed, we acknowledge that pragmatic assessments are tough to carry out. However, as we showed, they \textit{can be done}. Hence, we endorse future work to improve their evaluations by embracing our guidelines and using our resources.

\paragraph{\textsc{\textbf{Ethical Statement}}}
Our institutions do not require a formal IRB approval for carrying out the research presented in this paper. During our efforts, we always adhered to the Menlo report~\cite{bailey2012menlo}. Our experiments do not raise any ethical concern (they are a re-assessment of prior work). Our survey with practitioners was done so as to preserve the anonymity of our respondents---which is why we cannot disclose any further information about our population.
All our participants were informed that their responses would have been used for research. Furthermore, all our participants know the identity and the contact details of the authors of this paper, which they can use to explicitly request their responses to be deleted.

\paragraph{\textsc{\textbf{Acknowledgements}}}
We would like to thank: the Program Committee of EuroS\&P'23 and NDSS'23 for the constructive comments that improved this paper immensely; the practitioners who contributed to our user-study; and the Hilti Corporation for funding.

\appendices

\section{Additional Background and Use-cases}
\label{app:examples}
\noindent
We provide some supplementary descriptions and examples to facilitate the understanding of our paper.

\subsection{NetFlow-based analyses (and tools)}
\label{sapp:netflow}
\vspace{-0.5em}

\textbf{Problem.} Analyzing the \textit{all} raw-data generated by modern networks is problematic~\cite{vormayr2020my}, due to the sheer size of full packet captures (PCAP). Indeed, performing deep packet inspection (DPI) is computationally demanding (in terms of processing and storaging), besides also raising~\cite{deng2017commoner} privacy concerns\footnote{Encryption may solve the issue, but makes DPI challenging~\cite{kim2021p2dpi}.}. 

\textbf{Solution.} To make automated analyses of network data more feasible, a convenient alternative is to analyze high-level summaries of the communications between two endpoints, commonly referred to as \textit{NetFlows}. A NetFlow can be roughly expressed as the following tuple:
\begin{equation}
\label{exp:flow}
\resizebox{0.9\columnwidth}{!}{
$
    \text{NetFlow}\!=\!{\text{\small(\textit{srcIP}, \textit{dstIP}, \textit{srcPort}, \textit{dstPort}, \textit{proto}, \textit{startTime} \textit{endTime}, ...)}}
    $,
    }
\end{equation}
where the last three dots can include any element that relates to the other fields (e.g., the amount of bytes transferred in the `flow'). Compared to traditional PCAP, NetFlows present several advantages. For instance, the PCAP version of the \dataset{CICIDS17} dataset is of 50GB, whereas its NetFlow version requires just 1GB~\cite{engelen2021troubleshooting}. Such low requirements makes NetFlow viable for real time analyses, and the implicit lack of privacy issues is appreciated in commercial products  as well as for research---including (e.g.,~\cite{vinayakumar2019deep, pontes2021new, sarhan2020netflow}), but not limited to (e.g.,~\cite{Apruzzese:Pivoting, vormayr2020my, chuah2021challenges}), ML-specific proposals.

\textbf{Variants.} Initially introduced by CISCO in 1996~\cite{hofstede2014flow}, the concept of NetFlow has evolved substantially over the years. For instance, besides being available by default on most CISCO routers, it is possible to generate NetFlows via open-source software tools, such as Argus~\cite{Argus:URL}, nProbe~\cite{nProbe:URL}, or CIC-FlowMeter~\cite{lashkari2017characterization} (including its fixed version by Engelen et al.~\cite{engelen2021troubleshooting}). We even mention that Zeek~\cite{Zeek:URL} (formerly BroIDS), among the leading tools for network monitoring\footnote{Zeek also provides additional logging tools (e.g.,~\cite{bowman2020detecting}), and is frequently mentioned in research papers (e.g.,~\cite{leichtnam2020sec2graph,van2022deepcase,wolsing2022ipal,dodia2022exposing}).}, has implemented its own variant of NetFlows, named ``connection logs''~\cite{Zeek:connlog}. In our evaluation, we will consider all these variants.

\vspace{1em}

\textbox{
{\small \textbf{Disclaimer:} NetFlow represent a cost-effective solution to perform data-driven analyses for NID. However, despite being used in both research and practice, NetFlows are not a panacea~\cite{chuah2021challenges,bowman2020detecting}, and we invite future work to explore also other data-types (or create new ones!).}
}

\subsection{Supervised vs Unsupervised ML}
\label{sapp:sup-unsup}
\vspace{-0.5em}
\noindent
It is common to distinguish between \textit{supervised} and \textit{unsupervised} ML algorithms~\cite{Sommer:Outside}. Supervised algorithms require the training data to be provided with \textit{labels} that denote the ground truth of each sample, and are hence suited to `specific' tasks (e.g., distinguishing benign from malicious samples). In contrast, unsupervised algorithms do not have such requirement, making them applicable only to more `generic' tasks (e.g., grouping similar data): indeed, without ground truth labels it is not possible to `supervise' what the ML model is actually learning~\cite{joyce2021framework}. 

At a high-level, both supervised and unsupervised ML algorithms are applicable to either misuse or anomaly detection approaches. For instance, labelled data can serve as a guide to produce the signatures (for misuse detection, e.g.,~\cite{pontes2021new}) or to establish the notion of normality (for anomaly detection, e.g.,~\cite{ma2016supervised}); at the same time, the signatures can be determined by extracting some rules after clustering (e.g.~\cite{nappa2014cyberprobe}), while the normality can be determined from the clusters with most data points (e.g.,~\cite{syarif2012unsupervised}).

\subsection{An use-case of (supervised) ML in NID}
\label{sapp:application}
\begin{figure}[!htbp]
    \centering
    \includegraphics[width=\columnwidth]{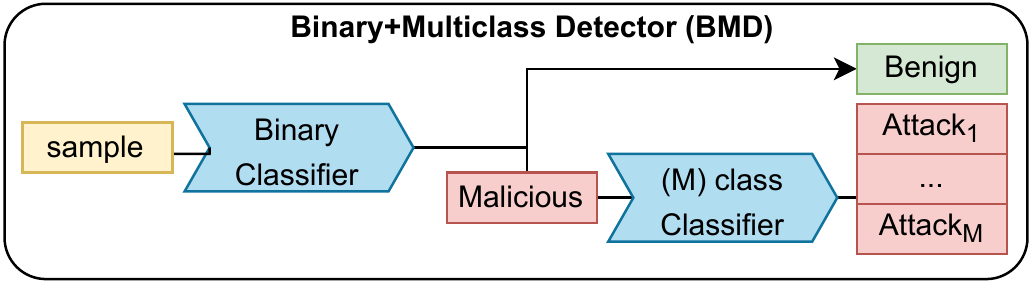}
    \caption{An ML pipeline representing a detector by cascading a 2- and M-class classifier. The Binary classifier first analyzes a sample, predicting whether such sample is benign or malicious. If the sample is malicious, then it is forwarded to a M-class classifier that determines the specific malicious class (out of M possible classes).}
    \label{fig:bmd}
\end{figure}

\vspace{-0.5em}
\noindent
Consider a NIDS that includes a ML model analyzing NetFlows. Such ML-NIDS will receive the raw network traffic from the gateway (Fig.~\ref{fig:network}). Such data is in PCAP format, and not usable by the ML model: hence, the PCAP is preprocessed into NetFlows (e.g., by using Argus~\cite{Argus:URL}), and the resulting NetFlows are sent as input to the ML model. The output of such ML model can be further utilized, e.g., by an additional ML model. For instance, it is possible to create a cascade of a binary and multiclass classifier (depicted in Fig.~\ref{fig:bmd}): the first ML model determines whether a NetFlow is benign or malicious, and the second ML model analyzes only the malicious outputs---according to the $M$ `known' classes. 

The initial PCAP can also be analyzed via traditional signature-based approaches (but in separate pipelines).

\subsection{A `practically redundant' ML-NIDS}
\label{sapp:flawed}
\vspace{-0.5em}
\noindent
We present a case-study of a `redundant' ML-NIDS adopted in a recent paper in a high-quality journal,~\cite{apruzzese2020deep}. Our objective is showcasing the immaturity of related research from the perspective of operational deployment.\footnote{To avoid `pointing-the-finger', we observe that some of the authors of~\cite{apruzzese2020deep} are shared with those of this SoK; nonetheless, we note that the methodology adopted in~\cite{apruzzese2020deep} derives from others peer-reviewed papers.} 

\begin{figure}[!htbp]
    \centering
    \includegraphics[width=0.6\columnwidth]{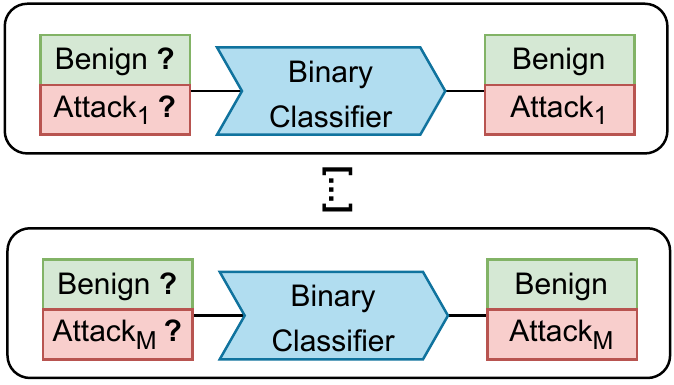}
    \caption{Exemplary design of a `redundant' ensemble of ML models for NID (used in~\cite{apruzzese2020deep}). Each classifier is trained on a specific attack (out of M); however, each classifier receives only samples that are either benign, or belong to the specific attack that the classifier can recognize.}
    \label{fig:edf}
\end{figure}

In~\cite{apruzzese2020deep}, an ML-NIDS is first developed, and then assessed in adversarial scenarios. The evaluation is based on the \dataset{CTU13} dataset, which contains malicious samples belonging to 5 different botnet families. The adversarial attacks are carried out by applying small perturbations to the malicious samples of each family: the ML-NIDS is then tested on such adversarial samples.
From a `research' perspective, such methodology is \textit{correct}, because the goal in~\cite{apruzzese2020deep} was the assessment of adversarial attacks. However, from a `practical' perspective, the ML-NIDS considered in~\cite{apruzzese2020deep} is \textit{redundant} due to a questionable architectural design (schematically depicted in Fig.~\ref{fig:edf}). Indeed, such ML-NIDS consists in an ensemble of ML models, with the logic that each model is dedicated to a specific family; however, each model of the ensemble is tested \textit{only} on the samples of its specific family. Hence, the ML models of~\cite{apruzzese2020deep} can only be viable if the NIDS knows---in advance!---which `attacks' should be forwarded to the ML model(s), therefore defeating the entire purpose of using ML to detect an attack. 

Furthermore, the ML-NIDS analyzes samples that are either benign, or correspond to one among 5 botnet families, i.e., a ``closed world'' setting. For instance, how would the ML models in~\cite{apruzzese2020deep} behave on samples that are generated via different malicious activities (e.g., a brute-force attack)? 
Finally, all the experiments in~\cite{apruzzese2020deep} are performed by attacking only a single ML model (based on RF). Hence, the effectiveness of the resulting attacks is questionable: what if the attacked ML model was slightly different from the one considered in the evaluation? We acknowledge that some adversarial attacks can be ``transferred''~\cite{Demontis:Adversarial}, but such operations do not guarantee the same degree of effectiveness. By performing many trials, all such uncertainties could be removed.

In the pragmatic assessment carried out in this SoK, we do not make any of such `redundancies', and hence our results have a higher practical value.

\section{Survey with Practitioners}
\label{app:survey}
\noindent
A crucial contribution of our paper is the survey we carried out with real practitioners. Our aim was to substantiate two of our major claims: (i)~whether our factors were truly relevant for practitioners; and (ii)~whether practitioners truly see research proposals of ML for NID with skepticism---and, if yes, what are some possible reasons.

Let us explain how we performed our survey, which is rooted in \textit{fairness} and \textit{transparency} to minimize bias.

\subsection{Selection of Participants}
\label{sapp:participants}
\vspace{-0.5em}

\textbf{Eligibility Criteria.}
Our goal was collecting the opinion of `practitioners' in the context of Machine Learning-based Network Intrusion Detection. For our survey, such ``practitioners'' entail people that have (had) first-hand experience with such technologies in the industrial sector. In-line with what we described in §\ref{ssec:business}, we hence focused on people who either work, or have worked for, companies that either: (a)~provide cybersecurity to third-parties, e.g., by monitoring the networks of their clients via ML-NIDS; or (b)~manage their own cybersecurity, e.g., they have a section entirely devoted to developing ML-NIDS that protect the network of the entire organization. Furthermore, since we were also interested in collecting meaningful opinions on the current state-of-research, our participants had to have some connection with the research domain (most co-authored peer-reviewed publications).

\textbf{Population.}
Overall, we reached out and found agreements with a total of 12 `practitioners'. To prevent bias, the companies for which our practitioners work (or have worked) are all different. To provide comprehensive and diverse opinions, we did not set ourselves any boundary to either the \textit{location} of the company (some are from the USA, some are based in the EU), or in its \textit{size} (some have dozens of employees, some are world-leaders in cybersecurity). Although our population may appear small, we stress that the corresponding companies have clients distributed everywhere in the World. 
To ensure fairness, all our interviewees were unaware of the \textit{specific} research we were carrying out; and none of the authors of this SoK had ever asked the opinion of the respondents of our survey \textit{beforehand}.\footnote{In other words, we did not `cherry pick' people that we knew would confirm our claims (some responses go against some claims!)} Finally, also for fairness, we reached out to our population by sending a generic email, stating that ``we want to collect the opinion of practitioners on ML-NIDS about the current state of research and practical deployment of such technologies.''

\subsection{Survey Design}
\label{sapp:questionnaire}
\vspace{-0.5em}
\noindent
We carry out our survey through an \textit{online questionnaire} having 13 questions with fixed answers.\footnote{We created a copy of our questionnaire for reviewing purposes, accessible at this link: \url{https://forms.gle/TxfwmAqG7zi5WCsZ9}}

\subsubsection{Organization}
\label{ssapp:organization}
The 13 questions were distributed into four `pages'~({\small P}), each with a specific purpose:
\begin{enumerate}[label={\small P{{\arabic*}})}]
    \item Introducing the questionnaire to the participant, and determining their suitability for our questions.
    \item Collecting the opinion on our proposed \textit{factors}.
    \item Collecting the opinion on the situation of Table~\ref{tab:sota}.
    \item Collecting the opinion on the state of research.
\end{enumerate}
Aside from {\small P1}, all the questions in the other pages had three possible answers, which can be summarized as: ``yes''; ``yes, but''; and ``no''.

To prevent `snooping bias'~\cite{arp2022dos}, the questionnaire was designed so that participants could not see the questions of a given page until they answered the previous ones. We gave the possibility of participants of not answering questions, because some participants may not have had the expertise to answer all of them.
Once they submitted their answers, their response was recorded and no changes could be made. There was no time limit for any question. 

We distributed the questionnaire to our participants (after reaching an agreement) via email, which included the link to our questionnaire. We asked each participant to provide us some form of confirmation that they submitted their answers---this was necessary to avoid cases in which a participant filled the questionnaire more than once.

\subsubsection{Questions}
\label{ssapp:questions}

Only one question was asked in {\small P1} and {\small P4}: in {\small P1}, we asked whether the company of the participant had a connection with ML and NID; such a question acted as a form of verification (a `negative' answer would terminate the survey);
in {\small P4}, we asked the simple question reported (verbatim) at the end of §\ref{ssec:opinion}. 

In contrast, {\small P2} and {\small P3} had 5 and 6 questions, respectively. 
In {\small P2}, we: considered each of our proposed five factors (§\ref{ssec:factors}); provided a brief explanation of such factor; and then asked ``how important'' such factor was for the respondent.
In {\small P3}, we first displayed an anonimysed (author names were hidden) version of Table~\ref{tab:sota}, and briefly explained what each column represented. Then, for each of the six columns in table, we asked ``how problematic'' it was that such a column had a certain amount of \xmark{}.

\subsection{Analysis and Feedback}
\label{sapp:analysis}
\vspace{-0.5em}
\noindent
After filling the questionnaire, some of our respondents gave us some \textbf{feedback}, which we now summarize.
\begin{itemize}
    \item \textit{``It depends!''} Many respondents commented that they felt the urge to answer all questions with ``it depends''. We were expecting this, which is why we did not include such a possibility in our questions: all participants would have chosen that option. 
    
    \item \textit{``I did not expect that!''} Some respondents stated that hardware is often not a concern in operational environments, because computational resources are abundant. We responded to them by showing some of our results, and they changed their mind: apparently, they did not expect that some ML methods may exhibit similar detection performance, while requiring substantially different time to train or test.
    
    \item \textit{``It must be explainable!''} Some respondents commented that their clients always ask for ``reasons why something (bad) happened,'' thereby inducing security providers to favor `explainable' ML methods. We were aware of the importance of this
    `factor' (as also evidenced in~\cite{alahmadi202299}) but we could not include it in our list because it would be unfeasible for \textit{any} research to provide an exhaustive answer for practical purposes---at least today~\cite{bhatt2020explainable} (even~\cite{alahmadi202299} argues that explanations are client-specific!).
\end{itemize}
Let us provide some \textbf{additional information}.
\begin{itemize}
    \item \textit{Timeline.} The first response was registered at the end of Jun. 2022, and the last at the start of Oct. 2022. 
    
    \item \textit{Missing answers.} One of our respondents did not answer any of the questions in {\small P2}, whereas two respondents skipped the ``Stat. Sign.'' question in {\small P3}.
    
    \item \textit{Length.} Filling the questionnaire required \smamath{\sim}20 mins.
\end{itemize}
Our repository~\cite{pragmaticAssessment} also includes some code-snippets providing a breakdown of the answers received.

\section{Complementary Analyses}
\label{app:complementary}
\noindent
We provide in this Appendix some additional considerations that further enrich the contributions of this SoK.

\subsection{State of Research (in 2022)}
\label{sapp:2022}
\vspace{-0.5em}
\noindent
In §\ref{sec:sota} we presented the state-of-research from 2017 until 2021. This was because we carried out our survey with practitioners in Summer'22 (i.e., 2022 was still ongoing). However, at the time of writing, all venues considered in our analysis have been held also in 2022: we find instructive to analyze also this year to see if there are any `improvements' w.r.t. the situation portrayed in §\ref{sec:sota}.

\textbf{Methodology.} We perform the exact same analysis described in §\ref{sec:sota}, but by considering the proceedings of 2022. We repeated our analysis twice, between Feb. and March 2023. We identified 16 papers, reported in Table~\ref{tab:sota2022}. Altogether, these papers have various goals related to ML-NIDS (e.g., evaluating novel attacks~\cite{erba2022assessing}, or proposing explainability methods for ML~\cite{jacobs2022ai}).
Similarly to §\ref{sec:sota}, all these papers consider a single preprocessing mechanism (the only exceptions are:~\cite{apruzzese2022sok,dhooge2022establishing}, which consider data generate via different NetFlow tools); and consider open-world settings (aside from~\cite{apruzzese2022sok}, wherein the evaluation represents a closed-world setting). 

\begin{table}[!h]
    \centering
    \caption{State-of-the-Art (2022): papers published in top cybersecurity conferences that consider applications of ML linked with NID.}
    \resizebox{0.99\columnwidth}{!}{
        \begin{tabular}{c|c|c|c|c|c|c|c}
            \toprule
            \textbf{Paper} & Year & Hardware & Runtime & Adaptive & Stat. Sign. & Avail. & Pub. Data \\
            \midrule

            Apruzzese~\cite{apruzzese2022sok} & 2022 & \cmark & \scbb{T} & \xmark & \cmark & \cmark & \cmark\ (3) \\ 
            Arp~\cite{arp2022dos} & 2022 & \xmark & \xmark & \halfcirc & \xmark & \xmark & \cmark\ (1) \\ 
            D'hooge~\cite{dhooge2022establishing} & 2022 & \xmark & \xmark & \xmark & \xmark  & \cmark & \cmark\ (8)\\ 
            Dodia~\cite{dodia2022exposing} & 2022 & \xmark & \xmark & \xmark & \cmark & \xmark & \cmark\ (1) \\ 
            Erba~\cite{erba2022assessing} & 2022 & \xmark & \xmark & \cmark & \xmark & \xmark & \cmark\ (1)\\ 
            Feng~\cite{feng2022cj} & 2022 & \cmark & \cmark & \halfcirc & \xmark & \cmark & \cmark\ (1) \\ 
            Fu~\cite{fu2022encrypted} & 2022 & \cmark & \scbb{E} & \halfcirc & \xmark & \xmark & \cmark\ (2) \\ 
            Jacobs~\cite{jacobs2022ai} & 2022 & \xmark & \xmark & \xmark & \xmark & \xmark & \cmark\ (6)\\ 
            King~\cite{king2022euler} & 2022 & \cmark & \cmark & \xmark & \xmark & \cmark & \cmark\ (3) \\ 
            Landen~\cite{landen2022dragon} & 2022 & \xmark & \scbb{T} & \cmark & \xmark & \cmark & \xmark\ (1) \\ 
            Sharma~\cite{sharma2022lumos} & 2022 & \halfcirc\ & \xmark & \halfcirc & \xmark & \xmark & \xmark\ (1) \\ 
            Tekiner~\cite{tekiner2022lightweight} & 2022 & \cmark & \scbb{E} & \cmark & \cmark & \cmark & \cmark\ (3) \\ 
            Van Ede~\cite{van2022deepcase} & 2022 & \cmark & \cmark & \cmark & \xmark & \cmark & \cmark\ (1)\\ 
            Wang~\cite{wang2022enidrift} & 2022 & \cmark & \cmark & \cmark & \xmark & \cmark & \cmark\ (1) \\ 
            Wang~\cite{wang2022maddc} & 2022 & \xmark & \xmark & \xmark & \xmark & \xmark & \cmark\ (3) \\ 
            
            Wolsing~\cite{wolsing2022ipal} & 2022 & \xmark & \xmark  & \xmark & \xmark & \xmark & \cmark\ (3)\\ 
            \bottomrule
        \end{tabular}
    }
    \label{tab:sota2022}
\end{table}

\textbf{Improvements.}
By looking at Table~\ref{tab:sota2022}, we observe an improvement w.r.t. Table~\ref{tab:sota}
Notably, we appreciate the utilization of more public datasets (which we believe stems from the increased release of open NID datasets\footnote{Papers using NetFlows:~\cite{apruzzese2022sok,arp2022dos,dhooge2022establishing,dodia2022exposing,feng2022cj,fu2022encrypted,jacobs2022ai,king2022euler,wang2022enidrift,wang2022maddc}.}) and the consideration of diverse data availability scenario. The hardware also appears to be reported more often w.r.t. Table~\ref{tab:sota}. However, we believe that the most significant improvement (which is not captured in these tables) is an \textit{increased release of source-code}. Indeed, out of the 30 papers in Table~\ref{tab:sota}, only 10 publicly disclosed their source code (i.e.,~\cite{araujo2019improving, barradas2021flowlens, bortolameotti2017decanter, erba2020constrained, fu2021realtime, han2020unicorn, han2021deepaid, liang2021fare, Mirsky:Kitsune, yang2021cade}), which is a mere 33\%. Such a low percentage dramatically increased to 75\% in 2022: 12 out of 16 papers in Table~\ref{tab:sota2022} published their code (i.e.,~\cite{apruzzese2022sok, dhooge2022establishing, dodia2022exposing,erba2022assessing,feng2022cj,jacobs2022ai,king2022euler,sharma2022lumos,tekiner2022lightweight,van2022deepcase,wang2022maddc,wang2022enidrift}). Such a positive trend is encouraging for both research and practice, since it facilitates reproducibility and can also allow practitioners to directly assess research proposals in production environments.

\subsection{NID datasets: practitioners' opinion}
\label{sapp:pradata}
\vspace{-0.5em}
\noindent
The real-world utility of public NID datasets has been scrutinized by many works---the most relevant being the paper by Kenyon et al.~\cite{kenyon2020public}. In what follows, we extend the main takeaways of~\cite{kenyon2020public} by providing some original observations based on our interactions with practitioners.

\textbf{Context.}
The performance of any ML method depends on its training data (§\ref{ssec:research}). Due to the increasing popularity of ML, the research community on ML-NIDS can now benefit from dozens of publicly available datasets. We refer the reader to some surveys of recent datasets for diverse domains related to ML-NIDS:~\cite{dhooge2022establishing, sarhan2020netflow, ring2019survey,zipperle2022provenance,conti2021survey}. Despite the usefulness of such datasets in research, from the operational perspective the sheer concept of a dataset has intrinsic limitations. Let us explain.

\textbf{Problem.}
We (informally) asked practitioners about the practical relevance of publicly available datasets for ML-NIDS. The general consensus is that all datasets they are aware of are inappropriate to derive sound conclusions on the applicability of a given ML method. The reasons are diverse, but can be summarized as:
\begin{itemize}
    \item \textit{Unrealistic assumptions.} Many datasets have samples generated via `simulations' (e.g., \dataset{NB15}~\cite{UNSWNB15:Dataset}), and the labelling may be done either too rigorously or too loosely (e.g.,~\cite{engelen2021troubleshooting}). For instance, assuming that the ground truth is known for \textit{each} sample is overly optimistic (practitioners use coarse labelling strategies~\cite{van2022deepcase,IBM:labelling,RobustIntelligence_SaTML}).
    
    \item \textit{Fixed point in time and space}. In order to serve as a ``benchmark'' for research purposes, a ML-NIDS dataset must be immutable. As a result, even if the data comes from real networks and corresponds to true attacks (e.g., \dataset{CTU13}~\cite{Garcia:CTU}), its practical value quickly deteriorates as the state-of-the-art advances (e.g., new network services may replace previous standards, and the threat landscape evolves). For instance, showing that a ML-NIDS can detect botnet samples that were  `problematic' 10 years before is not very relevant today (from a practical perspective).
\end{itemize}
Practitioners also remarked that these problems do not undermine the scientific contribution of research papers.

\textbf{Mitigation.}
We asked practitioners if they had any recommendations to mitigate the problems affecting public datasets for ML-NIDS. Accordingly, existing datasets could be \textit{enhanced by generating `new'} datasets that capture recent trends---e.g., by using CALDERA~\cite{MITRE:URL}. For instance, the \dataset{IDS17}~\cite{Sharafaldin:Toward} was updated with a more comprehensive version.\footnote{Unfortunately, even this version was found to be flawed~\cite{liu2022error}.} Doing this, however, may be tough for researchers: creating a new dataset makes the result of prior works not comparable, thereby requiring the researchers to assess previous methods on the new dataset (which is necessary for a meaningful comparison~\cite{arp2022dos}). Unfortunately, performing such re-assessments is not simple due to the lack of source code,\footnote{Among the 46 papers we analysed in this SoK, only 22 released their source-code at the time of acceptance (i.e., 48\%, see {\scriptsize \app{\ref{sapp:2022}}}).} preventing a simple (and bias-free) implementation of prior baselines~\cite{wolsing2022ipal, apruzzese2022position}. For this reason, practitioners endorse researchers to \textit{be as open as possible with their implementation}: alongside being helpful for future research, a `plug-and-play' artifact enables practitioners to assess the proposed ML-NIDS on their own environments---provided, of course, that the corresponding paper allows practitioners to estimate whether it would work in the first place.

\subsection{Comparison with a closely related work}
\label{sapp:comparison}
\vspace{-0.5em}
\noindent
We compare our SoK with the work by Arp et al.~\cite{arp2022dos}.

\textbf{Different goals.}
Arp et al.~\cite{arp2022dos} aim to provide recommendations that improve the soundness of ML assessments \textit{for future research}; in contrast, our recommendations aim to \textit{reduce the practitioners' skepticism on the practical value} of research papers. 
For this reason, some of our recommendations focus on aspects that are orthogonal to research, and contrast those of~\cite{arp2022dos}. For instance,~\cite{arp2022dos} claim that ``an evaluation of adversarial aspects is [...] a mandatory component in security research'', and suggest ``focusing on white-box attacks where possible''; in contrast, we argue that attackers with full-knowledge of the ML-NIDS is an extreme assumption in real environments (as also mentioned in~\cite{apruzzese2022position}), and our experiments focus on adaptive attackers with partial knowledge (which are more likely in reality). 

\textbf{Different focus.}
Arp et al.~\cite{arp2022dos} focus on generic applications of ML for security, and some recommendations have \textit{poor relevance in the specific NID context} (which is our focus). For example,~\cite{arp2022dos} emphasize the problem of ``temporal snooping'', which may be relevant when, e.g., analyzing malware samples, but not-so-much in when the analyses focuses on network activities over short timespans (as we explained in §\ref{sssec:dependency}; even our results show that there is barely any difference, performance-wise).

\textbf{Overlapping and Actionable recs.}
While some recommendations by~\cite{arp2022dos} can be applied for our cases (e.g., the base-rate fallacy §\ref{ssec:performance}), some of our recommendations are \textit{not elaborated in}~\cite{arp2022dos}. For instance, although~\cite{arp2022dos} recommend to ``move away from a laboratory setting [e.g., for runtime] and approximate a real-world setting [e.g., for open-world]'', there is no mentioning of how this could be done in NID: in contrast, we propose, e.g., `leaving-out' some malicious samples (§\ref{sssec:world}), and we perform an original experiment to showcase the importance of hardware on runtime (§\ref{ssec:cpu}).

\textbf{Literature Analysis and Validation.} The main theses of~\cite{arp2022dos} rely on an analysis of 30 papers over 10 years (from 2011 to 2021): in contrast, our SoK considers a higher number (46 in total) of more recent works (from 2017 until 2022).
Finally, the contributions by~\cite{arp2022dos} are exclusively based on prior literature and ``laboratory findings'', whereas our SoK has an additional validation phase supported by a user study with real practitioners.

\subsection{Pragmatic Assessments for other IDS}
\label{sapp:extensions}
\vspace{-0.5em}
\noindent
The focus of this SoK is on Machine Learning applications for Network Intrusion Detection Systems. Let us explain how our pragmatic assessment can be applied to other types of Intrusion Detection Systems (IDS)~\cite{khraisat2019survey}.

\textbf{Context.} What sets a (ML-based) NIDS apart from other IDS is the presence of a `network' element in its analysis (refer to §\ref{sec:background}).
By removing such an element, the ``Uniqueness of networks'' deployment challenge (§\ref{ssec:challenges}) disappears. From a practical perspective, this leads to \textit{a narrowing-down of the problem}: if the IDS does not have to account for the underlying network complexity, then it is easier to define the boundaries of what represents an `intrusion' or not. For example, detecting malware at the host level can be done a-priori, since ``malware is malicious everywhere, everytime'' (§\ref{sec:introduction}). Consequently, we argue that the results of similar researches are more directly applicable to reality. As a matter of fact, many commercial security products integrate state-of-the-art ML methods: e.g., \textit{deep learning} is used by Sophos to \textit{detect malware}~\cite{sophos_intercept}; and also by other companies to \textit{detect phishing webpages}~\cite{apruzzese2022position}. Therefore, we believe that there is a reduced necessity for pragmatic assessments in IDS that do not envision the underlying network complexity.

\textbf{Extension.} Research papers on other IDS can, however, still embrace our proposed pragmatic assessment notion: \textit{all our recommendations can be broadly applied to ML-IDS}. Nonetheless, in these cases, we argue that \textit{the role of hardware is even more important}. Indeed, while an organization (may) have the possibility of deploying the ML elements of a NIDS on diverse machines, in the case of host-IDS there is less room for doing this, since the analysis must be performed on the specific host\footnote{Of course, an organization can choose to deploy the ML element of a host-IDS on a powerful remote machine, but doing this for all machines of an organization may be impractical. Alternatively, if an organization used a centralized server that simultaneously analyzes all the low-level operations of the hosts, then this would resemble a NIDS.}. For example, consider our original experiment in §\ref{ssec:cpu}: the inference time can be substantially different (3x in our case) even for CPUs mounting ``an intel i5''. Hence, papers on host-IDS (including, e.g., commodity antiviruses) should put high emphasis on the size of \smabb{T} and \smabb{E}, and on the runtime for both training and testing---while clearly specifying the hardware specifications. We also endorse future researchers to consider different hardware configurations: this can be done, e.g., by downclocking the CPU, or running the experiments on a virtual machine and regulating the allocated computational resources (as we did in our experiments).

\section{Experiments: Configuration Settings}
\label{app:settings}
\noindent
We now provide an exhaustive description of our massive experimental campaign. 
Our experiments focus on supervised ML models for NID that analyze NetFlows. As explained in §\ref{ssec:setup}, such settings allow to witness the effects of all the `factors' described in our paper---while guaranteeing reproducibility. Indeed, using NetFlows showcases the role of data preprocessing, supervised settings highlight the importance of labelling, the generic `intrusion detection' epitomizes the distinction between open and closed worlds, and several public datasets are available.

We begin by presenting the considered datasets. Then, we thoroughly explain all the diverse configuration settings to meet all the conditions of a pragmatic assessment.

\subsection{Public Datasets}
\label{sapp:datasets}
\vspace{-0.5em}
\noindent
To provide meaningful results, for our evaluation we consider five datasets that include recent traffic and attack patterns, and which span across large and small network segments. We focus on datasets that are publicly available and validated by the state-of-the-art. In particular, we consider the following five datasets: \dataset{CTU13}, \dataset{NB15}, \dataset{UF-NB15}, \dataset{CICIDS17}, \dataset{GTCS}. Let us explain our choice.

\begin{itemize}
    \item \dataset{CTU13}~\cite{Garcia:CTU} is one of the largest publicly available datasets for NID. The data in \dataset{CTU13} is generated in a \textit{large network environment} (\smamath{\sim300} hosts), and contains attacks generated by diverse \textit{botnet} families.
    \item \dataset{NB15}~\cite{UNSWNB15:Dataset} is well-known~\cite{yuan2021recompose,singla2020preparing} and contains \textit{many attacks}, from DoS~\cite{pellegrino2015cashing} to shellcode injections.
    \item \dataset{UF-NB15}~\cite{sarhan2020netflow} is generated from the \textit{exact same} traffic of \dataset{NB15}, but the NetFlows derive from a different tool (i.e., \dataset{UF-NB15} has different \smacal{P} than \dataset{NB15}).
    \item \dataset{CICIDS17}~\cite{Sharafaldin:Toward} is among the most popular datasets (e.g.,~\cite{pontes2021new,leichtnam2020sec2graph}) for NID. Its original version was found to present labelling flaws~\cite{engelen2021troubleshooting}, so we perform our experiments on the \textit{fixed version} of \dataset{CICIDS17}.
    \item \dataset{GTCS}~\cite{mahfouz2020ensemble} is a very recent dataset. It includes similar attacks as those in \dataset{CICIDS17}, but the network is \textit{smaller} (i.e., it has less than a dozen hosts).
\end{itemize}

An in-depth view of such datasets is provided by Table~\ref{tab:samples}, showing the exact amount of samples per class. For \dataset{CICIDS17} we merge some underrepresented families into a single class (i.e., \textit{other}); whereas for \dataset{NB15}, \dataset{UF-NB15} we exclude some families because they had significantly mismatching numbers (in terms of available samples) which---we believe---could be due to labelling issues.

\begin{table}[!h]
    \centering
    \caption{Distribution of samples for each Dataset.}
    \resizebox{0.6\columnwidth}{!}{
        \begin{tabular}{c?c|c|c}
            \toprule
            \textbf{Dataset} & Class & \begin{tabular}{c} Attack \\ Family \end{tabular} & Samples \\
            \midrule
            
            \multirow{7}{*}{\dataset{CTU13}}
            & 0 & \textit{Benign} & 16\,748\,326 \\ \cline{2-4}
            & 1 & \textit{neris} & 205\,928 \\ \cline{2-4}
            & 2 & \textit{rbot} & 143\,918 \\ \cline{2-4}
            & 3 & \textit{nsis} & 2\,168 \\ \cline{2-4}
            & 4 & \textit{virut} & 40\,904 \\ \cline{2-4}
            & 5 & \textit{donbot} & 4\,630 \\ \cline{2-4}
            & 6 & \textit{murlo} & 6\,127  \\ 
            \midrule
            
            \multirow{5}{*}{\dataset{GTCS}}
            & 0 & \textit{Benign} & 139\,186 \\ \cline{2-4}
            & 1 & \textit{ddos} & 131\,211 \\ \cline{2-4}
            & 2 & \textit{bot} & 93\,021 \\ \cline{2-4}
            & 3 & \textit{brute} & 83\,857 \\ \cline{2-4}
            & 4 & \textit{inf} & 70\,202 \\ 
            \midrule

            \multirow{9}{*}{\dataset{NB15}}
& 0 & \textit{Benign} & 2\,218\,764 \\ \cline{2-4}
& 1 & \textit{expl} & 44\,525 \\ \cline{2-4}
& 2 & \textit{recon} & 13\,987 \\ \cline{2-4}
& 3 & \textit{dos} & 16\,353 \\ \cline{2-4}
& 4 & \textit{shell} & 1\,511 \\ \cline{2-4}
& 5 & \textit{fuzz} & 24\,246 \\ \cline{2-4}
& 6 & \textit{bdoor} & 2\,329 \\ \cline{2-4}
& 7 & \textit{ana} & 2\,677 \\ 
            \midrule
            
            \multirow{9}{*}{\dataset{UF-NB15}}
& 0 & \textit{Benign} & 2\,295\,222 \\ \cline{2-4}
& 1 & \textit{expl} & 31\,551 \\ \cline{2-4}
& 2 & \textit{recon} & 12\,779 \\ \cline{2-4}
& 3 & \textit{dos} & 5\,794 \\ \cline{2-4}
& 4 & \textit{shell} & 1\,427 \\ \cline{2-4}
& 5 & \textit{fuzz} & 22\,310 \\ \cline{2-4}
& 6 & \textit{bdoor} & 2\,169 \\ \cline{2-4}
& 7 & \textit{ana} & 2\,299 \\ 
            \midrule
            
            \multirow{10}{*}{\dataset{CICIDS17}}
            & 0 & \textit{Benign} & 1\,666\,837 \\ \cline{2-4}
            & 1 & \textit{ddos} & 95\,123 \\ \cline{2-4}
            & 2 & \textit{geye} & 7\,567 \\ \cline{2-4}
            & 3 & \textit{hulk} & 158\,469 \\ \cline{2-4}
            & 4 & \textit{http} & 1\,742 \\ \cline{2-4}
            & 5 & \textit{loris} & 4\,001 \\ \cline{2-4}
            & 6 & \textit{ftp} & 3\,973 \\ \cline{2-4}
            & 7 & \textit{pscan} & 159\,151 \\ \cline{2-4}
            & 8 & \textit{ssh} & 2\,980 \\ \cline{2-4}
            & 9 & \textit{other} & 971 \\

            \bottomrule
        \end{tabular}
    }
    \label{tab:samples}
\end{table}

In our experiments, we treat each dataset \smabb{D} as a separate environment, and we do not perform any mixing due to the intrinsic risks of such operations~\cite{apruzzese2022cross}.

\subsection{Hardware specifications}
\label{sapp:hardware}
\vspace{-0.5em}
\noindent
We carry out our evaluation on three different platforms each with different computational resources.
\begin{itemize}
    \item \textit{High-end} (default). A dedicated server for ML experiments, running an Intel Xeon W-2195@2.3GHz (36 cores), 256GB RAM. The OS is Ubuntu 20.04.
    
    \item \textit{Desktop}: Intel Core i5-4670@3.2GHz (4 cores) and 8GB of RAM. The OS is Windows 10.
    
    \item \textit{Laptop}: Intel Core i5-430M@2.5GHz (4 cores) and 8GB of RAM. The OS is Windows 10.
    
    \item \textit{Workstation}: Intel Core-i7 10750HQ@2.6GHz (12 cores) with 32GB RAM. The OS is Windows 10.
    
    \item \textit{Low-end}. A `downclocked' variant of the workstation, running on a Virtual Machine (using Ubuntu 20.04) that is set up to use only 4 cores (using at most 40\% of the frequency) and 8GB of RAM. 
    
    \item \textit{IoT}. A Raspberry Pi 4B with 2GB of RAM (4 cores).
\end{itemize}
We do not use GPU acceleration to ensure fairness.

We perform the majority of our experiments on the \textit{high-end} platform. The reason (as explained in~§\ref{ssec:factors}) is that hardware only affects\footnote{We verified this manually: all our ML models we develop across all our platforms achieve ultimately comparable detection performance---despite being trained/tested on different platforms} the runtime of an ML model. Hence, we use the other platforms to compare the \textit{training} and \textit{inference} runtime of each ML model. We do this only on the \dataset{GTCS} dataset, as runtime scales almost linearly with the size of the analyzed data.

\subsection{Data Availability}
\label{sapp:availability}
\vspace{-0.5em}
\noindent
Data, especially when labelled, is not cheap to obtain, and thus it is important to consider also cases in which the amount of labelled data is a hard constraint. To ensure fair and consistent comparisons, we always compose the evaluation partition \smabb{E} by choosing 20\% of the available samples for each class in \smabb{D}. 
Then we consider four `data availability' scenarios that regulate \smabb{T}:
\begin{itemize}
    \item \textit{Abundant}: we use all the remaining samples as training (i.e., \smabb{T} is 80\% of \smabb{D}).
    \item \textit{Modest}: we use half of the remaining data (i.e., \smabb{T} is 40\% of \smabb{D}).
    \item \textit{Scarce}: the training data is restricted to only a fifth of the remaining samples (i.e., \smabb{T} is 15\% of \smabb{D})
    \item \textit{Limited}: we use only 100 samples per class as training data \smabb{T}.
\end{itemize}
Nevertheless, we set a cap on the maximum amount of samples that are considered for any evaluation. Specifically, whenever we choose a dataset, the amount of benign samples that are put in \smabb{D} cannot exceed 500k, whereas the amount of malicious samples for each class cannot exceed 166k (i.e., one third of the benign samples). Indeed, some datasets (e.g., \dataset{CTU13}) contain millions of samples which are realistically difficult to manage (labelling issues are common in NID~\cite{engelen2021troubleshooting}). Moreover, we do not want the malicious samples to be more present than benign samples because this is not realistic: in reality, attacks are a ``needle in a haystack''~\cite{milajerdi2019holmes}. We remark, however, that we perform hundreds of trials---each drawing a different amount of samples from a source dataset to compose \smabb{D} (and hence \smabb{T} and \smabb{E}). Hence, we can reasonably assume that all samples in each dataset are analyzed by some ML model (either for training or testing such a model).

\subsection{Feature sets}
\label{sapp:features}
\vspace{-0.5em}
\noindent
NetFlows exporters can generate diverse features. In some cases, however, some features cannot be computed because the source PCAP data does not contain the necessary pieces of information (see~\cite{vormayr2020my}). Hence, we consider two cases of feature sets for each source dataset:
\begin{itemize}
    \item \textit{Complete}: we use \textit{all} the features provided with each dataset (and, hence, by the respective NetFlow exporter). To avoid classification bias~\cite{arp2022dos}, we omit the plain IP address and network ports (we replace the latter with their IANA categories, as done in~\cite{apruzzese2020deep}).
    \item \textit{Essential}: we use a subset of about half of the original features, which include the essential NetFlow fields (e.g., duration, packets, bytes~\cite{Cisco:Flow}). 
\end{itemize}
Such distinction is also used to shape the different operational scenarios (described in \app{\ref{app:scenarios}}).

\subsection{Design of the ML Pipelines}
\label{sapp:pipeline}
\begin{figure*}
    \centering
    
        \centering
    \begin{subfigure}[t]{0.20\textwidth}
        \centering
        \includegraphics[width=\columnwidth]{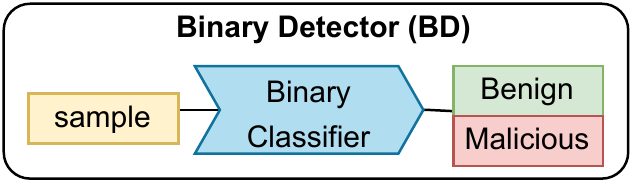}
        \caption{\footnotesize{BD Detector}}
         \label{sfig:bd}
    \end{subfigure}%
    ~ 
    \begin{subfigure}[t]{0.20\textwidth}
        \centering
        \includegraphics[width=\columnwidth]{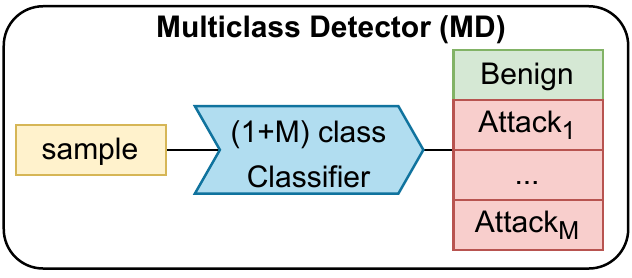}
        \caption{\footnotesize{MD Detector}}
         \label{sfig:md}
    \end{subfigure}
    ~ 
    \begin{subfigure}[t]{0.5\textwidth}
        \centering
        \includegraphics[width=\columnwidth]{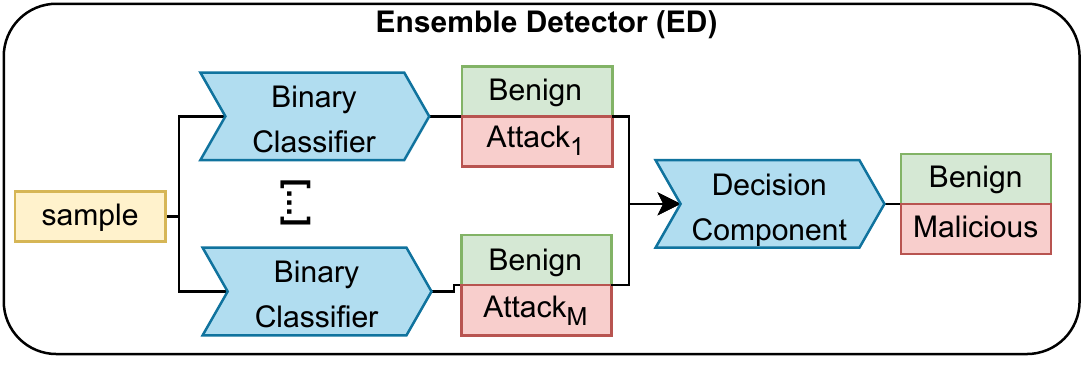}
        \caption{\footnotesize{Ensembles. The `decision component' determines ED-o, ED-s, ED-v.}}
         \label{sfig:ed}
    \end{subfigure}
    
    \caption{Design of the ML pipelines entailed in our considered ML-NIDS.}
    \label{fig:design}
\end{figure*}

\vspace{-0.5em}
\noindent
We always focus on ML components that operate as a \textit{detection} engine within the NIDS. 
We consider a wide array of such detectors, each with its own pipeline, which we now describe. A schematic is given in Fig.~\ref{fig:design}. 
\begin{itemize}
    \item Binary Detector (\variable{BD}, Fig.\ref{sfig:bd}). It consists of a single binary classifier: a sample is benign or malicious.
    \item Multiclass Detector (\variable{MD}, Fig.\ref{sfig:md}). It consists of a single 1+M-class classifier which infers whether a sample is benign, or belongs to one among M classes.
    \item Binary+Multiclass Detector (\variable{BMD}, Fig.\ref{fig:bmd}). This pipeline envisions a cascade of two ML models: the first is a binary classifier (i.e., the \variable{BD}), and the second is an M-class classifier which must determine the family of the malicious samples provided as output by the first classifier.
    \item Ensemble Detector (Fig.\ref{sfig:ed}). This pipeline consists of an ensemble of M binary classifiers, each specialized on a single type of attack---which is common in NID (e.g.,~\cite{Mirsky:Kitsune}). All such classifiers independently analyze each sample, and the output is produced by a final decision component. In particular, we consider 3 variants---each producing a binary output:
    \begin{itemize}
        \item Logical Or (\variable{ED-o}), where a sample is considered malicious if at least one classifier says so.
        \item Majority Voting (\variable{ED-v}), where a sample is considered malicious if at least M/2 classifiers say so.
        \item Stacked (\variable{ED-s}), where an additional ML model analyzes the predictions of all the M classifiers.
    \end{itemize}
\end{itemize}
We recall (cf. \app{\ref{sapp:flawed}}) that some past works propose ensemble detectors where each classifier receives \textit{only} the malicious samples that it can recognize (e.g.,~\cite{apruzzese2020deep}), and the results are taken by averaging the performance of each classifier. Despite the poor pragmatic value, we find instructive to consider also such `redundant' design, which we denote as \variable{ED} (cf. Fig.\ref{fig:edf}).

With respect to the NIDS architecture in Fig.\ref{fig:mlnids}, all our detectors can be placed in the exemplary ``ML pipeline'', where the preprocessing is done by the NetFlow tool.

\subsection{Selected ML Algorithms}
\label{sapp:algorithms}
\vspace{-0.5em}
\noindent
We create 4 variants of each detector, each using a specific ML algorithm. Of course, there exist dozens of ML algorithms, and benchmarking all of them is unfeasible and also outside the scope of our paper\footnote{In our source-code, changing the ML algorithm is a one-liner.}. Our focus is hence on a select subset of ML algorithms that have found use for ML-NIDS based on NetFlows (and that are known to be `easy to explain'). In particular, we use:
\begin{itemize}
    \item \textit{Decision Tree} (DT). One of the most popular classification algorithms for NID (e.g.,~\cite{pontes2021new, vinayakumar2019deep, kumar2021improving}). 
    \item \textit{Random Forest} (RF). A well-known (e.g.,~\cite{apruzzese2022cross}) \textit{ensemble} method, where each estimator is a single DT.
    \item \textit{Logistic Regression} (LR). Among the most common ML algorithms (also for NID~\cite{bapat2018identifying}), it relies on different decision mechanisms than tree-based algorithms (i.e., DT and RF).
    \item \textit{Histogram Gradient-boosting} (HGB). This algorithm leverages a novel boosting technique~\cite{ke2017lightgbm} that makes training significantly faster with respect to other gradient-based algorithms\footnote{We also assessed neural nets, but HGB always outmatched them---which is why we chose HGB as exemplary gradient-based algorithms.}.
\end{itemize}
From a resource utilization viewpoint, the learning phase for RF, LR and HGB use all cores available on a specific platform; whereas DT only uses a single core. On the other hand, the inference phase always uses a single core.

To the best of our knowledge, we are the first to evaluate detectors using HGB in our testbed, which is why we focus on this ML algorithm in our main paper. 
\section{Experiments: Operational Scenarios}
\label{app:scenarios}

\noindent
For each dataset, all the considered ML-NIDS are assessed in three different scenarios, which include both ``closed'' and ``open'' world settings. Specifically:
\begin{itemize}
    \item \textbf{Known Attacks (Baseline)}. This is the optimistic ``closed world'' setting: we use \textit{all} the available attack classes for both \smabb{T} and \smabb{E}.
    \item \textbf{Unknown Attacks}. To assess the performance against unknown attacks, we use the exclusion technique presented in §\ref{sssec:world}: we re-train each detector on \textit{all but one} of the available attack classes in \smabb{T}, and then test it on the leftout class (by using 20\% of its samples in \smabb{E}). For simplicity, we do such retraining by considering the `complete' feature set.
    \item \textbf{Adversarial Attacks}. We carry out adversarial \textit{evasion} attacks based on a well-known prior work~\cite{Apruzzese:Evading}, which envisions attacks that are both realistically feasible (i.e., the attacker treats the ML-NIDS as a black-box and cannot observe the output because it is accessible only from security administrators~\cite{apruzzese2021modeling}) and physically realizable (by extending the communications with junk bytes of data). To comply with the settings in~\cite{Apruzzese:Evading}, we consider the Essential feature set. More details are in \app{\ref{sapp:adversarial}}.
\end{itemize}
We re-assess every scenario through many trials, each involving a different \smabb{T} and \smabb{E}. Let us explain.

\subsection{Dependencies and Repetitions}
\label{sapp:dependencies}
\vspace{-0.5em}
\noindent
All the three scenarios are assessed by assuming both `static' and `temporal' data dependencies (§\ref{sssec:dependency}). 
\begin{itemize}
    \item \textbf{Static dependency.} Under this assumption, we always compose \smabb{T} and \smabb{E} by random sampling from \smabb{D}. To ensure statistically significant results that account for randomness, we perform a massive amount of trials for each setting. Specifically, we repeat the training/testing: 100 times for the \textit{Abundant}, \textit{Moderate}, and \textit{Scarce} labelling budgets; and 1000 times for the \textit{Limited} labelling budget. 
    
    \item \textbf{Temporal dependency.} To take into account the potential temporal dependency between samples, we repeat the same 3 operational scenarios, but by changing the way we compose \smabb{T} and \smabb{E}. Specifically---instead of randomly sampling from \smabb{D}---we compose \smabb{E} by selecting the \textit{most recent} samples, whereas we compose \smabb{T} by selecting the \textit{first appearing} samples. Of course, the composition of \smabb{T} is done according to the considered data availability setting---hence allowing a fair comparison with the `static' scenarios. As explained in §\ref{sssec:dependency}, the temporal scenarios are assessed only once per dataset (because they assume a deterministic `appearance' of samples).    
\end{itemize}
As an example, consider the \textit{Moderate} data availability setting. 
For the \textit{static} assumption, we first create \smabb{E} by randomly choosing 20\% of the samples for each class in a given \smabb{D}, and then create \smabb{T} by choosing 40\% of the remaining samples per class. For the \textit{temporal} assumption, we select the last 20\% samples for each class for \smabb{E}, and the first 40\% samples for \smabb{T}---resulting in a time-gap of 40\% samples between \smabb{T} and \smabb{E}.

\subsection{Adaptive ``adversarial'' Attacks}
\label{sapp:adversarial}
\vspace{-0.5em}
\noindent
We describe our adaptive attacks, which resemble the well-known paper by Apruzzese and Colajanni~\cite{Apruzzese:Evading}.

\textbf{Threat Model.} 
The \textit{defender} is an organization that adopts a ML-NIDS to detect malicious activities occurring in their network---a deployment scenario similar to the one depicted in Fig.~\ref{fig:network}. In particular, the NIDS includes an ML component that analyzes NetFlows, outputting whether a NetFlow is benign or malicious according to the data seen during its training stage. 
The \textit{attacker} is assumed to have already infiltrated the network (by, e.g., exploiting some zero day vulnerability, or by successfully `phishing' some employees). As such, the attacker is \textit{capable} of controlling some hosts, e.g., by manipulating the network communications. The attacker \textit{knows} that the organization adopts a ML-NIDS, but is agnostic of the exact functionalities of such ML-NIDS; moreover, the attacker cannot observe the output of the ML-NIDS because the attacker has no access to the admin console of the ML-NIDS. The attacker \textit{wants} to maintain access to the network: hence, the attacker is aware that they must operate stealthily and continuously change their activities to avoid being detected, especially if the ML-NIDS is retrained with data pertaining to more recent attacks. The \textit{strategy} adopted by attacker is to  modify the network communications (by, e.g., adding junk payloads) of the controlled machines, which results in `adversarial perturbations' that will affect the data analyzed by the ML-NIDS. Such an erratic behavior can confuse the ML-NIDS, potentially bypassing its detection. 
Such a threat model denotes attacks that are \textit{feasible} to stage~\cite{apruzzese2021modeling}, and hence likely to occur in reality.

\textbf{Implementation.}
We create the adversarial perturbations by manipulating the NetFlows samples. Such perturbations can be considered to be applied in the `feature space'~\cite{Pierazzi:Intriguing}. To ensure that the resulting adversarial samples are physically realizable~\cite{tong2019improving}, we follow strict rules.
\begin{itemize}
    \item We only perturb NetFlows whose source host is within the internal network. Indeed, our threat model assumes that the attacker has access and can control some machines within the target network.
    \item We only perturb UDP NetFlows. This is because other protocols may not allow the introduction of perturbations at the network-level.\footnote{For instance, some protocols (e.g., ICMP) have payload restrictions, whereas others perform additional communications (e.g., TCP's three-way handshake) that could result in unreliable adversarial samples.}
    \item The perturbations increase (by tiny amounts) the duration or the exchanged bytes---which are both common NetFlow features (and `controllable' by our attacker). We do not decrease such features because it may result in corrupted packets. Moreover, we ensure that the resulting `adversarial NetFlow' does not violate physical constraints (e.g., exceeding the MTU, or the maximum NetFlow duration).
    \item After applying the perturbation, we re-create the sample by taking into account inter-features dependencies (e.g., we recalculate the `bytes per second').
\end{itemize}
Finally, to replicate the scenario considered by Apruzzese and Colajanni~\cite{Apruzzese:Evading}, we only assess such attacks against the detectors that use the Essential feature set. This choice is also motivated by the fact that it is impossible to predict the effects of our perturbations to some features of the Complete feature set. As such, we consider the Essential setting to ensure that all of our adversarial samples are physically realizable. Nonetheless, a recent work~\cite{apruzzese2022spacephish} also suggests that our perturbations can be considered as a ``worst-case'' scenario, wherein an attacker has compromised the NetFlow exporter and is able to manipulate the preprocessing operations of the ML-NIDS.

\subsection{Performance Evaluation}
\label{sapp:metrics}
\vspace{-0.5em}
\noindent
We consider several performance metrics. Specifically:
\begin{itemize}
    \item \textit{False Positive Rate} (\smamath{fpr}), because any security system must exhibit low rates of false alarms.
    \item \textit{True Positive Rate} (\smamath{tpr}), because our primary focus is on intrusion detection (a positive is a malicious sample, irrespective of its class). Furthermore, our adversarial attacks focus on \textit{evasion}, and hence will affect the \smamath{tpr}.
    \item \textit{Accuracy} (\smamath{Acc}). Due to the base rate fallacy~\cite{arp2022dos}, we only use \smamath{Acc} to assess the multi-classification capabilities for the malicious classes (i.e., only for \variable{MD} and \variable{BMD}).
    \item \textit{Training time}, which is the time (in seconds) to train a given ML model on \smabb{T}. 
    \item \textit{Inference time}, which is the time that a (trained) ML model needs to analyze all samples in \smabb{E}.
\end{itemize}
For each dataset and data availability setting, we choose an ML algorithm (i.e., either RF, DT, LR, HGB). Then, we proceed by adopting the following workflow. 

\subsubsection{Training}
First, we develop all our detectors, i.e.: \variable{BD}, \variable{MD}, \variable{BMD}, as well as all the `specialized' detectors of \variable{ED} (i.e., \variable{ED-v}, \variable{ED-o}, \variable{ED-s}). All such detectors come in three variants: one using the Complete feature set; one using the Essential feature set; and one using the Complete feature set but trained without considering a specific malicious class---i.e., the unknown attacks (The latter yields M-1 sub-variants of each detector.) All such detectors are trained on the same \smabb{T} (with the appropriate changes of features or classes), and we measure their training time. 

\subsubsection{Inference}
Then, we test each detector on \smabb{E} by computing the \smamath{fpr} and \smamath{tpr} while measuring the time required to analyze \smabb{E}. For \variable{MD}, we consider a sample to be detected if it is classified as any attack class. Moreover, because the \variable{BMD} uses \variable{BD} as first detection layer, it follows that the \smamath{fpr} and \smamath{tpr} of \variable{BMD} are always the same as \variable{BD}. Indeed, to measure the benefit of \variable{BMD} we measure its \smamath{Acc} on the malicious samples predicted by \variable{BD}; to allow a fair comparison, we also measure \smamath{Acc} for \variable{MD}---but only on the malicious samples, otherwise the results would be skewed in favor of the benign samples which are analyzed by \variable{MD} but not by the multi-class classifier of \variable{BMD}.

\subsubsection{Adversarial Robustness}
Next, we craft the adversarial samples (as explained in \app{\ref{sapp:adversarial}}). We isolate from \smabb{E} the NetFlows that meet our criteria (must be UDP and start from an internal host, and of course be malicious), and re-compute the \smamath{tpr} of all our detectors on such `clean' samples (which should be different from the initial \smamath{tpr} computed on the whole \smabb{E}). Then, we apply the perturbations and analyze the resulting adversarial samples with all our detectors: if such `adversarial' \smamath{tpr} is \textit{lower} than the one on the `clean' malicious samples, then the attack is successful. 

\subsubsection{Reiterate and finalize}
All the procedures above are then repeated 100 times for the Abundant, Moderate and Scarce data availability settings, and 1000 times for the Limited data availability setting. Finally, we repeat all such experiments one last time by considering the temporal dependency (and hence choosing \smabb{T} and \smabb{E} accordingly).
\section{Experiments: Benchmark Results}
\label{app:benchmark}
\noindent
All our results are provided in a series of tables, each reporting the results achieved by all combinations of ML pipelines and ML algorithms for the increasing settings of data availability on a given dataset. The values reported in each table vary depending on the purpose of each table.

\subsection{Detection Performance (binary)}
\label{sapp:detection}
\vspace{-0.5em}
\noindent
We report our results for \textit{binary classification} by distinguishing the ``closed'' and ``open'' world settings.

\begin{itemize}
    \item \textbf{``Closed World''}: \dataset{CTU13} in Tables~\ref{tab:ctu_baseline},
\dataset{GTCS} in Tables~\ref{tab:gtcs_baseline},
\dataset{NB15} in Tables~\ref{tab:nb15_baseline},
\dataset{UF-NB15} in Tables~\ref{tab:ufnb15_baseline},
\dataset{CICIDS17} in Tables~\ref{tab:ids17_baseline}. Each of these tables reports the the \smamath{tpr} and \smamath{fpr} achieved by all our considered ML models, where we also differentiate the Essential from the Complete feature set. In particular, every table contains two subtables: the former (e.g., Table~\ref{tab:ctu_baseline_static}) reports the results in the absence of temporal dependencies, and hence the values denote the average metric (and standard deviation) across the many trials we performed. Whereas the latter (e.g., Table~\ref{tab:ctu_baseline_temporal}) reports the results of the single trial in which the samples are assumed to have temporal dependencies.

    \item \textbf{``Open World''}: \dataset{CTU13} in Tables~\ref{tab:ctu_open},
\dataset{GTCS} in Tables~\ref{tab:gtcs_open},
\dataset{NB15} in Tables~\ref{tab:nb15_open},
\dataset{UF-NB15} in Tables~\ref{tab:ufnb15_open},
\dataset{CICIDS17} in Tables~\ref{tab:ids17_open}. Each of these tables considers the twofold perspective of `unknown' attacks and `adversarial' attacks. For the former, which assume detectors using the Complete feature set, we report the \smamath{tpr} (on the `unknown' samples), but also the \smamath{fpr} (indeed, by excluding one class from \smabb{T}, the performance on the benign samples can also change). For the latter, which assume the Essential feature set, we report the \smamath{tpr} on the `original' NetFlows (which can vary from the one in the `open world' scenario because such samples are a subset of \smabb{E}) and on the `adversarial' NetFlows. Each of these tables contains two subtables, one for the `static' (e.g., Table~\ref{tab:ctu_open_static}) and one for the `temporal' (e.g., Table~\ref{tab:ctu_open_temporal}) dependency case.
\end{itemize}

We do not report the results of \variable{BMD} in any of these tables because they are identical to \variable{BD}: if a malicious sample `evades' the \variable{BD}, then it will logically also evade \variable{BMD} (whereas benign samples are not analyzed by the multi-class classifier of \variable{BMD}).

During our experiments, we observed that (especially when the size of \smabb{T} is huge) detectors based on LR tend to \textit{classify every sample as benign}: in these cases we report a 0 for both \smamath{tpr} and \smamath{fpr} (the detector is clearly unusable).

\subsection{Attack Identification (multiclass)}
\label{sapp:classification}
\vspace{-0.5em}
\noindent
Next, we focus on the classification performance on the \textit{malicious} samples. Such performance is measured via the \smamath{Acc}, which is computed for the \variable{MD} and \variable{BMD} detectors and only by taking into account the malicious samples (for the case of \variable{MD}, because these detectors also analyze benign samples). These values are computed only for the ``closed world'' settings, because any `unknown' attack is---by definition---misclassified (and the same can be said for the adversarial attacks).
All such results are reported in five tables (one per dataset): \dataset{CTU13} in Table~\ref{tab:ctu_multi},
\dataset{GTCS} in Table~\ref{tab:gtcs_multi},
\dataset{NB15} in Table~\ref{tab:ufnb15_multi},
\dataset{UF-NB15} in Table~\ref{tab:ufnb15_multi},
\dataset{CICIDS17} in Table~\ref{tab:ids17_multi}. All such tables include the \smamath{Acc} for both the Essential and Complete feature set, in both the static and temporal dependency scenarios.

\subsection{Runtime Performance (high-end platform)}
\label{sapp:highend}
\vspace{-0.5em}
\noindent
We report the runtime of all our ML models on the \textit{high-end} platform. The runtime for the temporal and static dependency scenarios is always the same.

\begin{itemize}
    \item \textit{Training.} We provide five tables, one per dataset: \dataset{CTU13} in Table~\ref{tab:ctu_training},
\dataset{GTCS} in Table~\ref{tab:gtcs_training},
\dataset{NB15} in Table~\ref{tab:nb15_training},
\dataset{UF-NB15} in Table~\ref{tab:ufnb15_training},
\dataset{CICIDS17} in Table~\ref{tab:ids17_training}. In these tables we report both the actual time (in seconds) and the standard deviation across all our trials. The training time of \variable{ED} is the sum of the training times for all the classifiers that compose the ensemble---which is the same for both \variable{ED-v} and \variable{ED-o}. The training time of \variable{ED-s} is always superior because it also requires training the stacked classifier.

    \item \textit{Testing.} We provide five tables, one per dataset: \dataset{CTU13} in Table~\ref{tab:ctu_testing},
\dataset{GTCS} in Table~\ref{tab:gtcs_testing},
\dataset{NB15} in Table~\ref{tab:nb15_testing},
\dataset{UF-NB15} in Table~\ref{tab:ufnb15_testing},
\dataset{CICIDS17} in Table~\ref{tab:ids17_testing}. In these tables we report only the actual time (in seconds), because variations were almost imperceptible. The testing time of \variable{ED} is the sum of the testing times for all the classifiers that compose the ensemble---which is the same for both \variable{ED-v} and \variable{ED-o}. The testing time of \variable{ED-s} is not necessarily superior than those of \variable{ED} because the stacked component can take a decision immediately.
\end{itemize}

In all cases, the runtime for \variable{BMD} is (almost) equivalent to the sum of \variable{BD} and \variable{MD}.

\subsection{Runtime Performance (other platforms)}
\label{sapp:runtime}
\vspace{-0.5em}
\noindent
We report the computational runtime of all our ML models as measured on the \textit{other hardware} platforms. We do this on a single dataset, \dataset{GTCS}, because it was the only one that could be processed by (most) of our machines.
Indeed, \textit{the Raspberry Pi4 was not able to run any} of our experiments (aside from those using the Limited data availability), due to a lack of available RAM memory. 
Such phenomenon motivated us to create a dedicated Virtual Machine (the \textit{low-end} platform) having a computational power similar to a Raspberry Pi4, but with significantly more RAM---enabling the development of ML models trained on \dataset{GTCS}.

We report all such results in Tables~\ref{tab:hardware}, which contains four subtables---each dedicated to a specific platform. These experiments are repeated 10 times and we report the average training and testing time.


\begin{table*}
  \centering
  \caption{\dataset{CTU13} binary classification results (\scmath{fpr} and \scmath{tpr}) against `known' attacks seen during the training stage (closed world).}

    \begin{subtable}[htbp]{1.99\columnwidth}
        \resizebox{1.0\columnwidth}{!}{
        \begin{tabular}{c|c ? cc|cc? cc|cc? cc|cc? cc|cc}
             \multicolumn{2}{c?}{Available Data} & \multicolumn{4}{c?}{Limited (100 per class) [N=1000]} & \multicolumn{4}{c?}{Scarce (15\% of \scbb{D}) [N=100]} &  \multicolumn{4}{c?}{Moderate (40\% of \scbb{D}) [N=100]} &  \multicolumn{4}{c}{Abundant (80\% of \scbb{D}) [N=100]} \\ \hline
             \multicolumn{2}{c?}{Features} &
             \multicolumn{2}{c|}{Complete} & \multicolumn{2}{c?}{Essential} & \multicolumn{2}{c|}{Complete} & \multicolumn{2}{c?}{Essential} & \multicolumn{2}{c|}{Complete} & \multicolumn{2}{c?}{Essential} &  
             \multicolumn{2}{c|}{Complete} & \multicolumn{2}{c}{Essential} \\ \hline
            Alg. & Design & \smamath{fpr} & \smamath{tpr} & \smamath{fpr} & \smamath{tpr} & \smamath{fpr} & \smamath{tpr} & \smamath{fpr} & \smamath{tpr} & \smamath{fpr} & \smamath{tpr} & \smamath{fpr} & \smamath{tpr} & \smamath{fpr} & \smamath{tpr} & \smamath{fpr} & \smamath{tpr} \\
             \toprule
             
             \multirow{6}{*}{\rotatebox{90}{\parbox{40pt}\centering\textbf{RF}}} 
             & \variable{BD} & \tabvalue{0.090}{0.016} & \tabvalue{0.998}{0.002} & \tabvalue{0.253}{0.027} & \tabvalue{0.987}{0.005}& \tabvalue{0.001}{0.000} & \tabvalue{0.999}{0.000} & \tabvalue{0.016}{0.000} & \tabvalue{0.981}{0.001}& \tabvalue{0.001}{0.000} & \tabvalue{0.999}{0.000} & \tabvalue{0.014}{0.000} & \tabvalue{0.984}{0.000}& \tabvalue{0.001}{0.000} & \tabvalue{0.999}{0.000} & \tabvalue{0.013}{0.000} & \tabvalue{0.985}{0.000}\\

            & \variable{MD} & \tabvalue{0.081}{0.017} & \tabvalue{0.996}{0.003} & \tabvalue{0.212}{0.027} & \tabvalue{0.972}{0.008}& \tabvalue{0.001}{0.000} & \tabvalue{0.998}{0.000} & \tabvalue{0.015}{0.000} & \tabvalue{0.979}{0.001}& \tabvalue{0.001}{0.000} & \tabvalue{0.999}{0.000} & \tabvalue{0.014}{0.000} & \tabvalue{0.983}{0.001}& \tabvalue{0.001}{0.000} & \tabvalue{0.999}{0.000} & \tabvalue{0.013}{0.000} & \tabvalue{0.984}{0.000}\\
            
            & \variable{ED-v} & \tabvalue{0.013}{0.009} & \tabvalue{0.693}{0.156} & \tabvalue{0.039}{0.016} & \tabvalue{0.470}{0.100}& \tabvalue{0.000}{0.000} & \tabvalue{0.348}{0.008} & \tabvalue{0.001}{0.000} & \tabvalue{0.302}{0.005}& \tabvalue{0.000}{0.000} & \tabvalue{0.359}{0.007} & \tabvalue{0.001}{0.000} & \tabvalue{0.302}{0.004}& \tabvalue{0.000}{0.000} & \tabvalue{0.363}{0.007} & \tabvalue{0.001}{0.000} & \tabvalue{0.303}{0.004}\\
            
            & \variable{ED-s} & \tabvalue{0.087}{0.016} & \tabvalue{0.997}{0.003} & \tabvalue{0.218}{0.027} & \tabvalue{0.975}{0.008}& \tabvalue{0.001}{0.000} & \tabvalue{0.998}{0.000} & \tabvalue{0.016}{0.000} & \tabvalue{0.979}{0.001}& \tabvalue{0.001}{0.000} & \tabvalue{0.999}{0.000} & \tabvalue{0.014}{0.000} & \tabvalue{0.983}{0.001}& \tabvalue{0.001}{0.000} & \tabvalue{0.999}{0.000} & \tabvalue{0.013}{0.000} & \tabvalue{0.984}{0.000}\\
            
            & \variable{ED-o} & \tabvalue{0.094}{0.017} & \tabvalue{0.997}{0.003} & \tabvalue{0.220}{0.027} & \tabvalue{0.975}{0.008}& \tabvalue{0.001}{0.000} & \tabvalue{0.998}{0.000} & \tabvalue{0.016}{0.000} & \tabvalue{0.979}{0.001}& \tabvalue{0.001}{0.000} & \tabvalue{0.999}{0.000} & \tabvalue{0.014}{0.000} & \tabvalue{0.983}{0.001}& \tabvalue{0.001}{0.000} & \tabvalue{0.999}{0.000} & \tabvalue{0.013}{0.000} & \tabvalue{0.984}{0.000}\\
            
            & \variable{ED} & \tabvalue{0.024}{0.007} & \tabvalue{0.986}{0.007} & \tabvalue{0.060}{0.011} & \tabvalue{0.942}{0.014}& \tabvalue{0.000}{0.000} & \tabvalue{0.997}{0.000} & \tabvalue{0.003}{0.000} & \tabvalue{0.971}{0.001}& \tabvalue{0.000}{0.000} & \tabvalue{0.998}{0.000} & \tabvalue{0.003}{0.000} & \tabvalue{0.976}{0.001}& \tabvalue{0.000}{0.000} & \tabvalue{0.998}{0.000} & \tabvalue{0.003}{0.000} & \tabvalue{0.979}{0.001}\\
             
            \midrule
            \multirow{6}{*}{\rotatebox{90}{\parbox{40pt}\centering\textbf{DT}}} 
             & \variable{BD} & \tabvalue{0.090}{0.024} & \tabvalue{0.991}{0.005} & \tabvalue{0.243}{0.031} & \tabvalue{0.963}{0.010}& \tabvalue{0.002}{0.000} & \tabvalue{0.998}{0.000} & \tabvalue{0.023}{0.001} & \tabvalue{0.972}{0.001}& \tabvalue{0.002}{0.000} & \tabvalue{0.998}{0.000} & \tabvalue{0.019}{0.001} & \tabvalue{0.976}{0.001}& \tabvalue{0.001}{0.000} & \tabvalue{0.999}{0.000} & \tabvalue{0.017}{0.000} & \tabvalue{0.979}{0.001}\\

            & \variable{MD} & \tabvalue{0.093}{0.023} & \tabvalue{0.994}{0.004} & \tabvalue{0.256}{0.035} & \tabvalue{0.966}{0.011}& \tabvalue{0.002}{0.000} & \tabvalue{0.998}{0.000} & \tabvalue{0.023}{0.001} & \tabvalue{0.972}{0.001}& \tabvalue{0.002}{0.000} & \tabvalue{0.998}{0.000} & \tabvalue{0.019}{0.001} & \tabvalue{0.976}{0.001}& \tabvalue{0.002}{0.000} & \tabvalue{0.998}{0.000} & \tabvalue{0.017}{0.000} & \tabvalue{0.978}{0.001}\\
            
            & \variable{ED-v} & \tabvalue{0.026}{0.016} & \tabvalue{0.820}{0.154} & \tabvalue{0.051}{0.018} & \tabvalue{0.506}{0.128}& \tabvalue{0.000}{0.000} & \tabvalue{0.501}{0.101} & \tabvalue{0.001}{0.000} & \tabvalue{0.310}{0.005}& \tabvalue{0.000}{0.000} & \tabvalue{0.544}{0.127} & \tabvalue{0.001}{0.000} & \tabvalue{0.312}{0.006}& \tabvalue{0.000}{0.000} & \tabvalue{0.508}{0.101} & \tabvalue{0.001}{0.000} & \tabvalue{0.313}{0.005}\\
            
            & \variable{ED-s} & \tabvalue{0.104}{0.026} & \tabvalue{0.997}{0.003} & \tabvalue{0.308}{0.037} & \tabvalue{0.979}{0.008}& \tabvalue{0.003}{0.000} & \tabvalue{0.998}{0.000} & \tabvalue{0.025}{0.001} & \tabvalue{0.975}{0.001}& \tabvalue{0.002}{0.000} & \tabvalue{0.999}{0.000} & \tabvalue{0.021}{0.001} & \tabvalue{0.978}{0.001}& \tabvalue{0.002}{0.000} & \tabvalue{0.999}{0.000} & \tabvalue{0.018}{0.000} & \tabvalue{0.981}{0.001}\\
            
            & \variable{ED-o} & \tabvalue{0.137}{0.029} & \tabvalue{0.998}{0.002} & \tabvalue{0.325}{0.036} & \tabvalue{0.980}{0.008}& \tabvalue{0.003}{0.000} & \tabvalue{0.998}{0.000} & \tabvalue{0.025}{0.001} & \tabvalue{0.975}{0.001}& \tabvalue{0.002}{0.000} & \tabvalue{0.999}{0.000} & \tabvalue{0.021}{0.001} & \tabvalue{0.978}{0.001}& \tabvalue{0.002}{0.000} & \tabvalue{0.999}{0.000} & \tabvalue{0.018}{0.000} & \tabvalue{0.981}{0.001}\\
            
            & \variable{ED} & \tabvalue{0.039}{0.010} & \tabvalue{0.980}{0.009} & \tabvalue{0.088}{0.013} & \tabvalue{0.933}{0.015}& \tabvalue{0.001}{0.000} & \tabvalue{0.996}{0.000} & \tabvalue{0.005}{0.000} & \tabvalue{0.963}{0.001}& \tabvalue{0.001}{0.000} & \tabvalue{0.997}{0.000} & \tabvalue{0.004}{0.000} & \tabvalue{0.969}{0.001}& \tabvalue{0.000}{0.000} & \tabvalue{0.997}{0.000} & \tabvalue{0.004}{0.000} & \tabvalue{0.972}{0.001}\\
             \midrule
            \multirow{6}{*}{\rotatebox{90}{\parbox{40pt}\centering\textbf{LR}}} 
            & \variable{BD} & \tabvalue{0.524}{0.061} & \tabvalue{0.954}{0.021} & \tabvalue{0.909}{0.100} & \tabvalue{0.992}{0.046}& \tabvalue{0.195}{0.041} & \tabvalue{0.785}{0.091} & \tabvalue{0.159}{0.208} & \tabvalue{0.546}{0.169}& \tabvalue{0.194}{0.031} & \tabvalue{0.797}{0.090} & \tabvalue{0.149}{0.184} & \tabvalue{0.547}{0.176}& \tabvalue{0.187}{0.035} & \tabvalue{0.788}{0.096} & \tabvalue{0.123}{0.183} & \tabvalue{0.501}{0.177}\\
            
            & \variable{MD} & \tabvalue{0.437}{0.049} & \tabvalue{0.853}{0.073} & \tabvalue{0.795}{0.255} & \tabvalue{0.900}{0.158}& \tabvalue{0.007}{0.011} & \tabvalue{0.066}{0.136} & \tabvalue{0.069}{0.014} & \tabvalue{0.510}{0.081}& \tabvalue{0.003}{0.005} & \tabvalue{0.024}{0.060} & \tabvalue{0.069}{0.011} & \tabvalue{0.517}{0.074}& \tabvalue{0.003}{0.003} & \tabvalue{0.020}{0.035} & \tabvalue{0.065}{0.015} & \tabvalue{0.503}{0.086}\\
            
            & \variable{ED-v} & \tabvalue{0.132}{0.040} & \tabvalue{0.759}{0.107} & \tabvalue{0.130}{0.072} & \tabvalue{0.660}{0.091}& \tabvalue{0.001}{0.001} & \tabvalue{0.005}{0.017} & \tabvalue{0.001}{0.004} & \tabvalue{0.148}{0.182}& \tabvalue{0.000}{0.001} & \tabvalue{0.004}{0.013} & \tabvalue{0.002}{0.004} & \tabvalue{0.159}{0.179}& \tabvalue{0.000}{0.000} & \tabvalue{0.001}{0.006} & \tabvalue{0.001}{0.002} & \tabvalue{0.116}{0.156}\\
            
            & \variable{ED-s} & \tabvalue{0.364}{0.068} & \tabvalue{0.928}{0.034} & \tabvalue{0.509}{0.334} & \tabvalue{0.749}{0.422}& \tabvalue{0.032}{0.014} & \tabvalue{0.606}{0.057} & \tabvalue{0.027}{0.020} & \tabvalue{0.479}{0.097}& \tabvalue{0.027}{0.009} & \tabvalue{0.584}{0.047} & \tabvalue{0.027}{0.020} & \tabvalue{0.474}{0.108}& \tabvalue{0.029}{0.011} & \tabvalue{0.575}{0.050} & \tabvalue{0.034}{0.022} & \tabvalue{0.477}{0.103}\\
            
            & \variable{ED-o} & \tabvalue{0.416}{0.043} & \tabvalue{0.934}{0.035} & \tabvalue{0.615}{0.228} & \tabvalue{0.922}{0.136}& \tabvalue{0.034}{0.014} & \tabvalue{0.606}{0.057} & \tabvalue{0.037}{0.020} & \tabvalue{0.486}{0.098}& \tabvalue{0.029}{0.009} & \tabvalue{0.586}{0.047} & \tabvalue{0.036}{0.021} & \tabvalue{0.479}{0.107}& \tabvalue{0.030}{0.011} & \tabvalue{0.577}{0.050} & \tabvalue{0.042}{0.021} & \tabvalue{0.484}{0.108}\\
            
            & \variable{ED} & \tabvalue{0.148}{0.020} & \tabvalue{0.883}{0.046} & \tabvalue{0.190}{0.070} & \tabvalue{0.836}{0.123}& \tabvalue{0.007}{0.003} & \tabvalue{0.444}{0.073} & \tabvalue{0.008}{0.005} & \tabvalue{0.366}{0.060}& \tabvalue{0.006}{0.002} & \tabvalue{0.413}{0.045} & \tabvalue{0.008}{0.005} & \tabvalue{0.366}{0.058}& \tabvalue{0.006}{0.002} & \tabvalue{0.400}{0.041} & \tabvalue{0.008}{0.004} & \tabvalue{0.361}{0.052}\\
             \midrule
            \multirow{6}{*}{\rotatebox{90}{\parbox{40pt}\centering\textbf{HGB}}} 
             & \variable{BD} & \tabvalue{0.088}{0.018} & \tabvalue{0.999}{0.002} & \tabvalue{0.253}{0.025} & \tabvalue{0.983}{0.006}& \tabvalue{0.002}{0.000} & \tabvalue{0.999}{0.000} & \tabvalue{0.027}{0.001} & \tabvalue{0.972}{0.001}& \tabvalue{0.002}{0.000} & \tabvalue{0.999}{0.000} & \tabvalue{0.026}{0.001} & \tabvalue{0.972}{0.001}& \tabvalue{0.002}{0.000} & \tabvalue{0.999}{0.000} & \tabvalue{0.026}{0.001} & \tabvalue{0.972}{0.001}\\

            & \variable{MD} & \tabvalue{0.075}{0.017} & \tabvalue{0.998}{0.002} & \tabvalue{0.213}{0.025} & \tabvalue{0.973}{0.008}& \tabvalue{0.005}{0.001} & \tabvalue{0.997}{0.001} & \tabvalue{0.031}{0.002} & \tabvalue{0.960}{0.006}& \tabvalue{0.004}{0.001} & \tabvalue{0.998}{0.000} & \tabvalue{0.030}{0.001} & \tabvalue{0.958}{0.006}& \tabvalue{0.003}{0.000} & \tabvalue{0.998}{0.001} & \tabvalue{0.029}{0.001} & \tabvalue{0.956}{0.005}\\
            
            & \variable{ED-v} & \tabvalue{0.018}{0.013} & \tabvalue{0.775}{0.149} & \tabvalue{0.056}{0.021} & \tabvalue{0.505}{0.098}& \tabvalue{0.000}{0.000} & \tabvalue{0.466}{0.038} & \tabvalue{0.001}{0.000} & \tabvalue{0.317}{0.004}& \tabvalue{0.000}{0.000} & \tabvalue{0.472}{0.028} & \tabvalue{0.001}{0.000} & \tabvalue{0.321}{0.005}& \tabvalue{0.000}{0.000} & \tabvalue{0.463}{0.028} & \tabvalue{0.001}{0.000} & \tabvalue{0.322}{0.004}\\
            
            & \variable{ED-s} & \tabvalue{0.079}{0.018} & \tabvalue{0.998}{0.003} & \tabvalue{0.238}{0.027} & \tabvalue{0.977}{0.008}& \tabvalue{0.002}{0.001} & \tabvalue{0.999}{0.000} & \tabvalue{0.025}{0.001} & \tabvalue{0.974}{0.001}& \tabvalue{0.002}{0.000} & \tabvalue{0.999}{0.000} & \tabvalue{0.025}{0.001} & \tabvalue{0.974}{0.002}& \tabvalue{0.002}{0.000} & \tabvalue{0.999}{0.000} & \tabvalue{0.025}{0.001} & \tabvalue{0.974}{0.002}\\
            
            & \variable{ED-o} & \tabvalue{0.088}{0.019} & \tabvalue{0.998}{0.002} & \tabvalue{0.243}{0.028} & \tabvalue{0.978}{0.008}& \tabvalue{0.002}{0.001} & \tabvalue{0.999}{0.000} & \tabvalue{0.025}{0.001} & \tabvalue{0.974}{0.001}& \tabvalue{0.002}{0.000} & \tabvalue{0.999}{0.000} & \tabvalue{0.025}{0.001} & \tabvalue{0.974}{0.002}& \tabvalue{0.002}{0.000} & \tabvalue{0.999}{0.000} & \tabvalue{0.025}{0.001} & \tabvalue{0.974}{0.002}\\
            
            & \variable{ED} & \tabvalue{0.026}{0.009} & \tabvalue{0.991}{0.006} & \tabvalue{0.073}{0.013} & \tabvalue{0.944}{0.014}& \tabvalue{0.001}{0.000} & \tabvalue{0.997}{0.000} & \tabvalue{0.005}{0.000} & \tabvalue{0.965}{0.002}& \tabvalue{0.001}{0.000} & \tabvalue{0.998}{0.000} & \tabvalue{0.005}{0.000} & \tabvalue{0.965}{0.002}& \tabvalue{0.000}{0.000} & \tabvalue{0.998}{0.000} & \tabvalue{0.005}{0.000} & \tabvalue{0.966}{0.002}\\
            
            \bottomrule
        \end{tabular}
        }
                
        \caption{\textit{Static Dependency}: Results by assuming the absence of temporal dependencies among samples (\scbb{T} and \scbb{E} are randomly sampled from \scbb{D}).} 
    \label{tab:ctu_baseline_static}
    \end{subtable}

    \begin{subtable}[htbp]{1.8\columnwidth}
        \resizebox{1.0\columnwidth}{!}{
        \begin{tabular}{c|c ? cc|cc? cc|cc? cc|cc? cc|cc}
             \multicolumn{2}{c?}{Available Data} & \multicolumn{4}{c?}{Limited (100 per class) [N=1]} & \multicolumn{4}{c?}{Scarce (15\% of \scbb{D}) [N=1]} &  \multicolumn{4}{c?}{Moderate (40\% of \scbb{D}) [N=1]} &  \multicolumn{4}{c}{Abundant (80\% of \scbb{D}) [N=1]} \\ \hline
             \multicolumn{2}{c?}{Features} &
             \multicolumn{2}{c|}{Complete} & \multicolumn{2}{c?}{Essential} & \multicolumn{2}{c|}{Complete} & \multicolumn{2}{c?}{Essential} & \multicolumn{2}{c|}{Complete} & \multicolumn{2}{c?}{Essential} &  
             \multicolumn{2}{c|}{Complete} & \multicolumn{2}{c}{Essential} \\ \hline
            Alg. & Design & \smamath{fpr} & \smamath{tpr} & \smamath{fpr} & \smamath{tpr} & \smamath{fpr} & \smamath{tpr} & \smamath{fpr} & \smamath{tpr} & \smamath{fpr} & \smamath{tpr} & \smamath{fpr} & \smamath{tpr} & \smamath{fpr} & \smamath{tpr} & \smamath{fpr} & \smamath{tpr} \\
             \toprule
             
             \multirow{6}{*}{\rotatebox{90}{\parbox{40pt}\centering\textbf{RF}}} 
             & \variable{BD} & \tabvaluetemp{0.107} & \tabvaluetemp{0.995} & \tabvaluetemp{0.248} & \tabvaluetemp{0.986}& \tabvaluetemp{0.001} & \tabvaluetemp{0.999} & \tabvaluetemp{0.016} & \tabvaluetemp{0.980}& \tabvaluetemp{0.001} & \tabvaluetemp{0.999} & \tabvaluetemp{0.014} & \tabvaluetemp{0.983}& \tabvaluetemp{0.001} & \tabvaluetemp{0.999} & \tabvaluetemp{0.013} & \tabvaluetemp{0.986}\\

            & \variable{MD} & \tabvaluetemp{0.086} & \tabvaluetemp{0.995} & \tabvaluetemp{0.205} & \tabvaluetemp{0.972}& \tabvaluetemp{0.001} & \tabvaluetemp{0.999} & \tabvaluetemp{0.015} & \tabvaluetemp{0.979}& \tabvaluetemp{0.001} & \tabvaluetemp{0.999} & \tabvaluetemp{0.013} & \tabvaluetemp{0.982}& \tabvaluetemp{0.001} & \tabvaluetemp{0.999} & \tabvaluetemp{0.013} & \tabvaluetemp{0.985}\\
            
            & \variable{ED-v} & \tabvaluetemp{0.010} & \tabvaluetemp{0.902} & \tabvaluetemp{0.046} & \tabvaluetemp{0.811}& \tabvaluetemp{0.000} & \tabvaluetemp{0.351} & \tabvaluetemp{0.001} & \tabvaluetemp{0.303}& \tabvaluetemp{0.000} & \tabvaluetemp{0.361} & \tabvaluetemp{0.001} & \tabvaluetemp{0.305}& \tabvaluetemp{0.000} & \tabvaluetemp{0.366} & \tabvaluetemp{0.001} & \tabvaluetemp{0.308}\\
            
            & \variable{ED-s} & \tabvaluetemp{0.089} & \tabvaluetemp{0.994} & \tabvaluetemp{0.211} & \tabvaluetemp{0.978}& \tabvaluetemp{0.001} & \tabvaluetemp{0.999} & \tabvaluetemp{0.015} & \tabvaluetemp{0.979}& \tabvaluetemp{0.001} & \tabvaluetemp{0.999} & \tabvaluetemp{0.013} & \tabvaluetemp{0.982}& \tabvaluetemp{0.001} & \tabvaluetemp{0.999} & \tabvaluetemp{0.013} & \tabvaluetemp{0.985}\\
            
            & \variable{ED-o} & \tabvaluetemp{0.090} & \tabvaluetemp{0.995} & \tabvaluetemp{0.212} & \tabvaluetemp{0.978}& \tabvaluetemp{0.001} & \tabvaluetemp{0.999} & \tabvaluetemp{0.015} & \tabvaluetemp{0.979}& \tabvaluetemp{0.001} & \tabvaluetemp{0.999} & \tabvaluetemp{0.013} & \tabvaluetemp{0.982}& \tabvaluetemp{0.001} & \tabvaluetemp{0.999} & \tabvaluetemp{0.013} & \tabvaluetemp{0.985}\\
            
            & \variable{ED} & \tabvaluetemp{0.024} & \tabvaluetemp{0.990} & \tabvaluetemp{0.061} & \tabvaluetemp{0.948}& \tabvaluetemp{0.000} & \tabvaluetemp{0.996} & \tabvaluetemp{0.003} & \tabvaluetemp{0.971}& \tabvaluetemp{0.000} & \tabvaluetemp{0.998} & \tabvaluetemp{0.003} & \tabvaluetemp{0.976}& \tabvaluetemp{0.000} & \tabvaluetemp{0.999} & \tabvaluetemp{0.003} & \tabvaluetemp{0.979}\\
             
            \midrule
            \multirow{6}{*}{\rotatebox{90}{\parbox{40pt}\centering\textbf{DT}}} 
             & \variable{BD} & \tabvaluetemp{0.131} & \tabvaluetemp{0.982} & \tabvaluetemp{0.304} & \tabvaluetemp{0.976}& \tabvaluetemp{0.002} & \tabvaluetemp{0.998} & \tabvaluetemp{0.022} & \tabvaluetemp{0.971}& \tabvaluetemp{0.002} & \tabvaluetemp{0.998} & \tabvaluetemp{0.020} & \tabvaluetemp{0.975}& \tabvaluetemp{0.001} & \tabvaluetemp{0.999} & \tabvaluetemp{0.017} & \tabvaluetemp{0.979}\\

            & \variable{MD} & \tabvaluetemp{0.135} & \tabvaluetemp{0.995} & \tabvaluetemp{0.272} & \tabvaluetemp{0.977}& \tabvaluetemp{0.002} & \tabvaluetemp{0.997} & \tabvaluetemp{0.022} & \tabvaluetemp{0.973}& \tabvaluetemp{0.002} & \tabvaluetemp{0.998} & \tabvaluetemp{0.020} & \tabvaluetemp{0.977}& \tabvaluetemp{0.002} & \tabvaluetemp{0.998} & \tabvaluetemp{0.018} & \tabvaluetemp{0.979}\\
            
            & \variable{ED-v} & \tabvaluetemp{0.019} & \tabvaluetemp{0.939} & \tabvaluetemp{0.075} & \tabvaluetemp{0.860}& \tabvaluetemp{0.000} & \tabvaluetemp{0.768} & \tabvaluetemp{0.001} & \tabvaluetemp{0.319}& \tabvaluetemp{0.000} & \tabvaluetemp{0.498} & \tabvaluetemp{0.001} & \tabvaluetemp{0.316}& \tabvaluetemp{0.000} & \tabvaluetemp{0.484} & \tabvaluetemp{0.001} & \tabvaluetemp{0.306}\\
            
            & \variable{ED-s} & \tabvaluetemp{0.187} & \tabvaluetemp{0.997} & \tabvaluetemp{0.339} & \tabvaluetemp{0.979}& \tabvaluetemp{0.003} & \tabvaluetemp{0.998} & \tabvaluetemp{0.025} & \tabvaluetemp{0.975}& \tabvaluetemp{0.002} & \tabvaluetemp{0.999} & \tabvaluetemp{0.021} & \tabvaluetemp{0.979}& \tabvaluetemp{0.002} & \tabvaluetemp{0.999} & \tabvaluetemp{0.018} & \tabvaluetemp{0.981}\\
            
            & \variable{ED-o} & \tabvaluetemp{0.220} & \tabvaluetemp{0.997} & \tabvaluetemp{0.341} & \tabvaluetemp{0.979}& \tabvaluetemp{0.003} & \tabvaluetemp{0.998} & \tabvaluetemp{0.025} & \tabvaluetemp{0.975}& \tabvaluetemp{0.002} & \tabvaluetemp{0.999} & \tabvaluetemp{0.021} & \tabvaluetemp{0.979}& \tabvaluetemp{0.002} & \tabvaluetemp{0.999} & \tabvaluetemp{0.018} & \tabvaluetemp{0.981}\\
            
            & \variable{ED} & \tabvaluetemp{0.057} & \tabvaluetemp{0.991} & \tabvaluetemp{0.099} & \tabvaluetemp{0.940}& \tabvaluetemp{0.001} & \tabvaluetemp{0.995} & \tabvaluetemp{0.005} & \tabvaluetemp{0.962}& \tabvaluetemp{0.000} & \tabvaluetemp{0.997} & \tabvaluetemp{0.004} & \tabvaluetemp{0.968}& \tabvaluetemp{0.000} & \tabvaluetemp{0.997} & \tabvaluetemp{0.004} & \tabvaluetemp{0.973}\\
             \midrule
            \multirow{6}{*}{\rotatebox{90}{\parbox{40pt}\centering\textbf{LR}}} 
             & \variable{BD} & \tabvaluetemp{0.576} & \tabvaluetemp{0.983} & \tabvaluetemp{0.989} & \tabvaluetemp{0.991}& \tabvaluetemp{0.044} & \tabvaluetemp{0.105} & \tabvaluetemp{0.108} & \tabvaluetemp{0.663}& \tabvaluetemp{0.224} & \tabvaluetemp{0.842} & \tabvaluetemp{0.032} & \tabvaluetemp{0.359}& \tabvaluetemp{0.208} & \tabvaluetemp{0.845} & \tabvaluetemp{0.154} & \tabvaluetemp{0.645}\\

            & \variable{MD} & \tabvaluetemp{0.419} & \tabvaluetemp{0.888} & \tabvaluetemp{0.931} & \tabvaluetemp{0.981}& \tabvaluetemp{0.000} & \tabvaluetemp{0.001} & \tabvaluetemp{0.073} & \tabvaluetemp{0.554}& \tabvaluetemp{0.000} & \tabvaluetemp{0.001} & \tabvaluetemp{0.045} & \tabvaluetemp{0.360}& \tabvaluetemp{0.000} & \tabvaluetemp{0.001} & \tabvaluetemp{0.067} & \tabvaluetemp{0.554}\\
            
            & \variable{ED-v} & \tabvaluetemp{0.141} & \tabvaluetemp{0.875} & \tabvaluetemp{0.105} & \tabvaluetemp{0.600}& \tabvaluetemp{0.000} & \tabvaluetemp{0.000} & \tabvaluetemp{0.002} & \tabvaluetemp{0.296}& \tabvaluetemp{0.001} & \tabvaluetemp{0.000} & \tabvaluetemp{0.000} & \tabvaluetemp{0.000}& \tabvaluetemp{0.000} & \tabvaluetemp{0.000} & \tabvaluetemp{0.000} & \tabvaluetemp{0.000}\\
            
            & \variable{ED-s} & \tabvaluetemp{0.448} & \tabvaluetemp{0.950} & \tabvaluetemp{0.564} & \tabvaluetemp{0.977}& \tabvaluetemp{0.021} & \tabvaluetemp{0.549} & \tabvaluetemp{0.022} & \tabvaluetemp{0.561}& \tabvaluetemp{0.021} & \tabvaluetemp{0.554} & \tabvaluetemp{0.068} & \tabvaluetemp{0.568}& \tabvaluetemp{0.023} & \tabvaluetemp{0.552} & \tabvaluetemp{0.032} & \tabvaluetemp{0.414}\\
            
            & \variable{ED-o} & \tabvaluetemp{0.465} & \tabvaluetemp{0.951} & \tabvaluetemp{0.606} & \tabvaluetemp{0.983}& \tabvaluetemp{0.023} & \tabvaluetemp{0.550} & \tabvaluetemp{0.031} & \tabvaluetemp{0.568}& \tabvaluetemp{0.022} & \tabvaluetemp{0.554} & \tabvaluetemp{0.068} & \tabvaluetemp{0.568}& \tabvaluetemp{0.023} & \tabvaluetemp{0.552} & \tabvaluetemp{0.035} & \tabvaluetemp{0.414}\\
            
            & \variable{ED} & \tabvaluetemp{0.148} & \tabvaluetemp{0.923} & \tabvaluetemp{0.168} & \tabvaluetemp{0.930}& \tabvaluetemp{0.004} & \tabvaluetemp{0.369} & \tabvaluetemp{0.008} & \tabvaluetemp{0.383}& \tabvaluetemp{0.004} & \tabvaluetemp{0.370} & \tabvaluetemp{0.012} & \tabvaluetemp{0.388}& \tabvaluetemp{0.006} & \tabvaluetemp{0.421} & \tabvaluetemp{0.007} & \tabvaluetemp{0.316}\\
             \midrule
            \multirow{6}{*}{\rotatebox{90}{\parbox{40pt}\centering\textbf{HGB}}} 
            & \variable{BD} & \tabvaluetemp{0.082} & \tabvaluetemp{1.000} & \tabvaluetemp{0.251} & \tabvaluetemp{0.983}& \tabvaluetemp{0.003} & \tabvaluetemp{0.998} & \tabvaluetemp{0.027} & \tabvaluetemp{0.971}& \tabvaluetemp{0.001} & \tabvaluetemp{0.999} & \tabvaluetemp{0.026} & \tabvaluetemp{0.972}& \tabvaluetemp{0.002} & \tabvaluetemp{0.999} & \tabvaluetemp{0.027} & \tabvaluetemp{0.970}\\

            & \variable{MD} & \tabvaluetemp{0.073} & \tabvaluetemp{0.999} & \tabvaluetemp{0.202} & \tabvaluetemp{0.966}& \tabvaluetemp{0.005} & \tabvaluetemp{0.998} & \tabvaluetemp{0.036} & \tabvaluetemp{0.960}& \tabvaluetemp{0.005} & \tabvaluetemp{0.999} & \tabvaluetemp{0.030} & \tabvaluetemp{0.957}& \tabvaluetemp{0.003} & \tabvaluetemp{0.997} & \tabvaluetemp{0.030} & \tabvaluetemp{0.959}\\
            
            & \variable{ED-v} & \tabvaluetemp{0.025} & \tabvaluetemp{0.938} & \tabvaluetemp{0.065} & \tabvaluetemp{0.843}& \tabvaluetemp{0.000} & \tabvaluetemp{0.492} & \tabvaluetemp{0.001} & \tabvaluetemp{0.316}& \tabvaluetemp{0.000} & \tabvaluetemp{0.440} & \tabvaluetemp{0.001} & \tabvaluetemp{0.317}& \tabvaluetemp{0.000} & \tabvaluetemp{0.455} & \tabvaluetemp{0.001} & \tabvaluetemp{0.339}\\
            
            & \variable{ED-s} & \tabvaluetemp{0.097} & \tabvaluetemp{0.999} & \tabvaluetemp{0.271} & \tabvaluetemp{0.976}& \tabvaluetemp{0.002} & \tabvaluetemp{0.999} & \tabvaluetemp{0.025} & \tabvaluetemp{0.974}& \tabvaluetemp{0.002} & \tabvaluetemp{0.999} & \tabvaluetemp{0.026} & \tabvaluetemp{0.973}& \tabvaluetemp{0.002} & \tabvaluetemp{0.999} & \tabvaluetemp{0.025} & \tabvaluetemp{0.971}\\
            
            & \variable{ED-o} & \tabvaluetemp{0.111} & \tabvaluetemp{1.000} & \tabvaluetemp{0.277} & \tabvaluetemp{0.976}& \tabvaluetemp{0.002} & \tabvaluetemp{0.999} & \tabvaluetemp{0.025} & \tabvaluetemp{0.974}& \tabvaluetemp{0.002} & \tabvaluetemp{0.999} & \tabvaluetemp{0.026} & \tabvaluetemp{0.973}& \tabvaluetemp{0.002} & \tabvaluetemp{0.999} & \tabvaluetemp{0.025} & \tabvaluetemp{0.971}\\
            
            & \variable{ED} & \tabvaluetemp{0.033} & \tabvaluetemp{0.962} & \tabvaluetemp{0.087} & \tabvaluetemp{0.938}& \tabvaluetemp{0.000} & \tabvaluetemp{0.997} & \tabvaluetemp{0.005} & \tabvaluetemp{0.966}& \tabvaluetemp{0.000} & \tabvaluetemp{0.998} & \tabvaluetemp{0.005} & \tabvaluetemp{0.963}& \tabvaluetemp{0.000} & \tabvaluetemp{0.998} & \tabvaluetemp{0.005} & \tabvaluetemp{0.962}\\
            
            \bottomrule
        \end{tabular}
        }
                
        \caption{\textit{Temporal Dependency}: Results by assuming the presence of temporal dependencies among samples (the `first' samples of \scbb{D} are put in \scbb{T}, while the last 20\% represent \scbb{E}).} 
    \label{tab:ctu_baseline_temporal}
    \end{subtable}

    \label{tab:ctu_baseline}
\end{table*}

\begin{table*}
  \centering
  \caption{\dataset{CTU13}. Results against adversarial (original \scmath{tpr} and adversarial \scmath{tpr}) and unknown attacks (the \scmath{tpr} is the average on the `unknown' attacks, while the \scmath{fpr} is due to training on a new \scbb{T} that does not have the `unknown' class.).}

    \begin{subtable}[htbp]{1.99\columnwidth}
        \resizebox{1.0\columnwidth}{!}{
        \begin{tabular}{c|c ? cc|cc? cc|cc? cc|cc? cc|cc}
\multicolumn{2}{c?}{Available Data} & \multicolumn{4}{c?}{Limited (100 per class) [N=1000]} & \multicolumn{4}{c?}{Scarce (15\% of \scbb{D}) [N=100]} &  \multicolumn{4}{c?}{Moderate (40\% of \scbb{D}) [N=100]} &  \multicolumn{4}{c}{Abundant (80\% of \scbb{D}) [N=100]} \\ \hline
             \multicolumn{2}{c?}{Scenario} &
             \multicolumn{2}{c|}{Adversarial Attacks} & \multicolumn{2}{c?}{Unknown Attacks} & \multicolumn{2}{c|}{Adversarial Attacks} & \multicolumn{2}{c?}{Unknown Attacks} & \multicolumn{2}{c|}{Adversarial Attacks} & \multicolumn{2}{c?}{Unknown Attacks} & \multicolumn{2}{c|}{Adversarial Attacks} & \multicolumn{2}{c?}{Unknown Attacks}  \\ \hline
            Alg. & Design & 
            \footnotesize{$tpr$} \tiny{(org)} & \footnotesize{$tpr$} \tiny{(adv)} & \footnotesize{$fpr$} & \footnotesize{$tpr$} & \footnotesize{$tpr$} \tiny{(org)} & \footnotesize{$tpr$} \tiny{(adv)} & \footnotesize{$fpr$} & \footnotesize{$tpr$} & \footnotesize{$tpr$} \tiny{(org)} & \footnotesize{$tpr$} \tiny{(adv)} & \footnotesize{$fpr$} & \footnotesize{$tpr$} & \footnotesize{$tpr$} \tiny{(org)} & \footnotesize{$tpr$} \tiny{(adv)} & \footnotesize{$fpr$} & \footnotesize{$tpr$} \\
             \toprule
             
             \multirow{5}{*}{\rotatebox{90}{\parbox{40pt}\centering\textbf{RF}}} 
            & \variable{BD} & \tabvalue{0.966}{0.015} & \tabvalue{0.924}{0.191} & \tabvalue{0.082}{0.015} & \tabvalue{0.922}{0.065}& \tabvalue{0.964}{0.002} & \tabvalue{0.003}{0.000} & \tabvalue{0.001}{0.000} & \tabvalue{0.741}{0.003}& \tabvalue{0.968}{0.001} & \tabvalue{0.003}{0.000} & \tabvalue{0.001}{0.000} & \tabvalue{0.741}{0.014}& \tabvalue{0.970}{0.001} & \tabvalue{0.003}{0.000} & \tabvalue{0.001}{0.000} & \tabvalue{0.734}{0.029}\\
            
            & \variable{MD} & \tabvalue{0.942}{0.022} & \tabvalue{0.716}{0.322} & \tabvalue{0.074}{0.016} & \tabvalue{0.858}{0.068}& \tabvalue{0.961}{0.002} & \tabvalue{0.003}{0.000} & \tabvalue{0.001}{0.000} & \tabvalue{0.738}{0.005}& \tabvalue{0.966}{0.001} & \tabvalue{0.003}{0.000} & \tabvalue{0.001}{0.000} & \tabvalue{0.729}{0.022}& \tabvalue{0.968}{0.001} & \tabvalue{0.003}{0.000} & \tabvalue{0.001}{0.000} & \tabvalue{0.717}{0.030}\\
            
            & \variable{ED-v} & \tabvalue{0.378}{0.163} & \tabvalue{0.077}{0.202} & \tabvalue{0.008}{0.006} & \tabvalue{0.540}{0.081}& \tabvalue{0.004}{0.002} & \tabvalue{0.000}{0.000} & \tabvalue{0.000}{0.000} & \tabvalue{0.397}{0.007}& \tabvalue{0.011}{0.003} & \tabvalue{0.000}{0.000} & \tabvalue{0.000}{0.000} & \tabvalue{0.394}{0.021}& \tabvalue{0.017}{0.003} & \tabvalue{0.000}{0.000} & \tabvalue{0.000}{0.000} & \tabvalue{0.386}{0.036}\\
            
            & \variable{ED-s} & \tabvalue{0.948}{0.021} & \tabvalue{0.728}{0.336} & \tabvalue{0.078}{0.015} & \tabvalue{0.853}{0.074}& \tabvalue{0.962}{0.002} & \tabvalue{0.002}{0.000} & \tabvalue{0.001}{0.000} & \tabvalue{0.736}{0.004}& \tabvalue{0.967}{0.001} & \tabvalue{0.003}{0.000} & \tabvalue{0.001}{0.000} & \tabvalue{0.737}{0.014}& \tabvalue{0.969}{0.001} & \tabvalue{0.003}{0.000} & \tabvalue{0.001}{0.000} & \tabvalue{0.731}{0.026}\\
            
            & \variable{ED-o} & \tabvalue{0.948}{0.021} & \tabvalue{0.728}{0.336} & \tabvalue{0.084}{0.017} & \tabvalue{0.863}{0.073}& \tabvalue{0.962}{0.002} & \tabvalue{0.002}{0.000} & \tabvalue{0.001}{0.000} & \tabvalue{0.736}{0.004}& \tabvalue{0.967}{0.001} & \tabvalue{0.003}{0.000} & \tabvalue{0.001}{0.000} & \tabvalue{0.737}{0.014}& \tabvalue{0.969}{0.001} & \tabvalue{0.003}{0.000} & \tabvalue{0.001}{0.000} & \tabvalue{0.731}{0.026}\\
             
            \midrule
            \multirow{5}{*}{\rotatebox{90}{\parbox{40pt}\centering\textbf{DT}}} 
            & \variable{BD} & \tabvalue{0.920}{0.029} & \tabvalue{0.729}{0.368} & \tabvalue{0.082}{0.019} & \tabvalue{0.902}{0.068}& \tabvalue{0.935}{0.002} & \tabvalue{0.139}{0.227} & \tabvalue{0.002}{0.000} & \tabvalue{0.804}{0.064}& \tabvalue{0.943}{0.002} & \tabvalue{0.186}{0.280} & \tabvalue{0.002}{0.000} & \tabvalue{0.790}{0.062}& \tabvalue{0.948}{0.002} & \tabvalue{0.243}{0.313} & \tabvalue{0.001}{0.000} & \tabvalue{0.791}{0.070}\\
            
            & \variable{MD} & \tabvalue{0.927}{0.030} & \tabvalue{0.795}{0.357} & \tabvalue{0.086}{0.019} & \tabvalue{0.910}{0.061}& \tabvalue{0.935}{0.003} & \tabvalue{0.224}{0.256} & \tabvalue{0.002}{0.000} & \tabvalue{0.807}{0.065}& \tabvalue{0.943}{0.002} & \tabvalue{0.238}{0.268} & \tabvalue{0.002}{0.000} & \tabvalue{0.799}{0.063}& \tabvalue{0.948}{0.002} & \tabvalue{0.244}{0.260} & \tabvalue{0.001}{0.000} & \tabvalue{0.784}{0.066}\\
            
            & \variable{ED-v} & \tabvalue{0.373}{0.159} & \tabvalue{0.340}{0.401} & \tabvalue{0.018}{0.012} & \tabvalue{0.646}{0.104}& \tabvalue{0.034}{0.012} & \tabvalue{0.002}{0.008} & \tabvalue{0.000}{0.000} & \tabvalue{0.468}{0.054}& \tabvalue{0.044}{0.013} & \tabvalue{0.003}{0.015} & \tabvalue{0.000}{0.000} & \tabvalue{0.477}{0.068}& \tabvalue{0.048}{0.010} & \tabvalue{0.009}{0.053} & \tabvalue{0.000}{0.000} & \tabvalue{0.456}{0.064}\\
            
            & \variable{ED-s} & \tabvalue{0.952}{0.022} & \tabvalue{0.907}{0.242} & \tabvalue{0.096}{0.022} & \tabvalue{0.904}{0.068}& \tabvalue{0.943}{0.002} & \tabvalue{0.357}{0.318} & \tabvalue{0.003}{0.000} & \tabvalue{0.794}{0.062}& \tabvalue{0.950}{0.002} & \tabvalue{0.428}{0.315} & \tabvalue{0.002}{0.000} & \tabvalue{0.816}{0.063}& \tabvalue{0.954}{0.002} & \tabvalue{0.504}{0.290} & \tabvalue{0.001}{0.000} & \tabvalue{0.792}{0.069}\\
            
            & \variable{ED-o} & \tabvalue{0.954}{0.022} & \tabvalue{0.923}{0.220} & \tabvalue{0.124}{0.026} & \tabvalue{0.932}{0.056}& \tabvalue{0.943}{0.002} & \tabvalue{0.357}{0.318} & \tabvalue{0.003}{0.000} & \tabvalue{0.811}{0.061}& \tabvalue{0.950}{0.002} & \tabvalue{0.428}{0.315} & \tabvalue{0.002}{0.000} & \tabvalue{0.825}{0.061}& \tabvalue{0.954}{0.002} & \tabvalue{0.504}{0.290} & \tabvalue{0.001}{0.000} & \tabvalue{0.801}{0.070}\\
             \midrule
            \multirow{5}{*}{\rotatebox{90}{\parbox{40pt}\centering\textbf{LR}}} 
            & \variable{BD} & \tabvalue{0.993}{0.048} & \tabvalue{0.830}{0.366} & \tabvalue{0.479}{0.045} & \tabvalue{0.888}{0.050}& \tabvalue{0.152}{0.331} & \tabvalue{0.001}{0.001} & \tabvalue{0.139}{0.021} & \tabvalue{0.604}{0.066}& \tabvalue{0.150}{0.312} & \tabvalue{0.001}{0.000} & \tabvalue{0.136}{0.018} & \tabvalue{0.600}{0.064}& \tabvalue{0.120}{0.292} & \tabvalue{0.001}{0.001} & \tabvalue{0.133}{0.016} & \tabvalue{0.584}{0.065}\\
            
            & \variable{MD} & \tabvalue{0.788}{0.359} & \tabvalue{0.611}{0.455} & \tabvalue{0.422}{0.041} & \tabvalue{0.863}{0.080}& \tabvalue{0.014}{0.003} & \tabvalue{0.007}{0.006} & \tabvalue{0.008}{0.010} & \tabvalue{0.099}{0.113}& \tabvalue{0.014}{0.002} & \tabvalue{0.007}{0.005} & \tabvalue{0.004}{0.003} & \tabvalue{0.060}{0.051}& \tabvalue{0.013}{0.003} & \tabvalue{0.006}{0.004} & \tabvalue{0.003}{0.002} & \tabvalue{0.052}{0.036}\\
            
            & \variable{ED-v} & \tabvalue{0.067}{0.216} & \tabvalue{0.417}{0.468} & \tabvalue{0.094}{0.028} & \tabvalue{0.639}{0.088}& \tabvalue{0.000}{0.000} & \tabvalue{0.000}{0.000} & \tabvalue{0.000}{0.000} & \tabvalue{0.007}{0.025}& \tabvalue{0.000}{0.000} & \tabvalue{0.000}{0.000} & \tabvalue{0.000}{0.000} & \tabvalue{0.006}{0.022}& \tabvalue{0.000}{0.000} & \tabvalue{0.000}{0.000} & \tabvalue{0.000}{0.000} & \tabvalue{0.001}{0.007}\\
            
            & \variable{ED-s} & \tabvalue{0.984}{0.073} & \tabvalue{0.914}{0.260} & \tabvalue{0.342}{0.054} & \tabvalue{0.897}{0.038}& \tabvalue{0.005}{0.005} & \tabvalue{0.027}{0.145} & \tabvalue{0.028}{0.012} & \tabvalue{0.400}{0.071}& \tabvalue{0.005}{0.005} & \tabvalue{0.036}{0.170} & \tabvalue{0.024}{0.008} & \tabvalue{0.366}{0.050}& \tabvalue{0.006}{0.006} & \tabvalue{0.002}{0.004} & \tabvalue{0.024}{0.009} & \tabvalue{0.368}{0.053}\\
            
            & \variable{ED-o} & \tabvalue{0.770}{0.393} & \tabvalue{0.916}{0.255} & \tabvalue{0.388}{0.039} & \tabvalue{0.915}{0.026}& \tabvalue{0.006}{0.005} & \tabvalue{0.027}{0.145} & \tabvalue{0.029}{0.012} & \tabvalue{0.407}{0.080}& \tabvalue{0.007}{0.006} & \tabvalue{0.047}{0.182} & \tabvalue{0.025}{0.007} & \tabvalue{0.371}{0.050}& \tabvalue{0.008}{0.005} & \tabvalue{0.003}{0.005} & \tabvalue{0.026}{0.009} & \tabvalue{0.370}{0.053}\\
             \midrule
            \multirow{5}{*}{\rotatebox{90}{\parbox{40pt}\centering\textbf{HGB}}} 
            & \variable{BD} & \tabvalue{0.960}{0.018} & \tabvalue{0.897}{0.204} & \tabvalue{0.079}{0.017} & \tabvalue{0.898}{0.068}& \tabvalue{0.939}{0.003} & \tabvalue{0.015}{0.067} & \tabvalue{0.002}{0.000} & \tabvalue{0.747}{0.006}& \tabvalue{0.940}{0.004} & \tabvalue{0.010}{0.020} & \tabvalue{0.002}{0.000} & \tabvalue{0.751}{0.013}& \tabvalue{0.941}{0.003} & \tabvalue{0.010}{0.037} & \tabvalue{0.001}{0.000} & \tabvalue{0.755}{0.022}\\
            
            & \variable{MD} & \tabvalue{0.947}{0.022} & \tabvalue{0.794}{0.293} & \tabvalue{0.069}{0.016} & \tabvalue{0.841}{0.072}& \tabvalue{0.925}{0.008} & \tabvalue{0.109}{0.195} & \tabvalue{0.005}{0.001} & \tabvalue{0.753}{0.033}& \tabvalue{0.921}{0.009} & \tabvalue{0.110}{0.203} & \tabvalue{0.003}{0.001} & \tabvalue{0.765}{0.044}& \tabvalue{0.920}{0.007} & \tabvalue{0.148}{0.213} & \tabvalue{0.003}{0.000} & \tabvalue{0.785}{0.058}\\
            
            & \variable{ED-v} & \tabvalue{0.427}{0.163} & \tabvalue{0.649}{0.330} & \tabvalue{0.013}{0.010} & \tabvalue{0.651}{0.080}& \tabvalue{0.020}{0.014} & \tabvalue{0.000}{0.000} & \tabvalue{0.000}{0.000} & \tabvalue{0.456}{0.045}& \tabvalue{0.034}{0.015} & \tabvalue{0.000}{0.000} & \tabvalue{0.000}{0.000} & \tabvalue{0.467}{0.037}& \tabvalue{0.039}{0.014} & \tabvalue{0.000}{0.000} & \tabvalue{0.000}{0.000} & \tabvalue{0.476}{0.051}\\
            
            & \variable{ED-s} & \tabvalue{0.951}{0.021} & \tabvalue{0.962}{0.091} & \tabvalue{0.072}{0.016} & \tabvalue{0.880}{0.071}& \tabvalue{0.946}{0.004} & \tabvalue{0.037}{0.123} & \tabvalue{0.002}{0.000} & \tabvalue{0.757}{0.031}& \tabvalue{0.947}{0.004} & \tabvalue{0.046}{0.110} & \tabvalue{0.002}{0.000} & \tabvalue{0.755}{0.026}& \tabvalue{0.948}{0.004} & \tabvalue{0.046}{0.080} & \tabvalue{0.002}{0.000} & \tabvalue{0.760}{0.032}\\
            
            & \variable{ED-o} & \tabvalue{0.953}{0.021} & \tabvalue{0.966}{0.083} & \tabvalue{0.080}{0.018} & \tabvalue{0.898}{0.066}& \tabvalue{0.946}{0.004} & \tabvalue{0.037}{0.123} & \tabvalue{0.002}{0.000} & \tabvalue{0.766}{0.042}& \tabvalue{0.947}{0.004} & \tabvalue{0.046}{0.110} & \tabvalue{0.002}{0.000} & \tabvalue{0.758}{0.029}& \tabvalue{0.948}{0.004} & \tabvalue{0.053}{0.104} & \tabvalue{0.002}{0.000} & \tabvalue{0.760}{0.032}\\
            
            \bottomrule
        \end{tabular}
        }
                
        \caption{\textit{Static Dependency}: Results by assuming the absence of temporal dependencies among samples (\scbb{T} and \scbb{E} are randomly sampled from \scbb{D}).} 
    \label{tab:ctu_open_static}
    \end{subtable}

    \begin{subtable}[htbp]{1.8\columnwidth}
        \resizebox{1.0\columnwidth}{!}{
        \begin{tabular}{c|c ? cc|cc? cc|cc? cc|cc? cc|cc}
\multicolumn{2}{c?}{Available Data} & \multicolumn{4}{c?}{Limited (100 per class) [N=1]} & \multicolumn{4}{c?}{Scarce (15\% of \scbb{D}) [N=1]} &  \multicolumn{4}{c?}{Moderate (40\% of \scbb{D}) [N=1]} &  \multicolumn{4}{c}{Abundant (80\% of \scbb{D}) [N=1]} \\ \hline
             \multicolumn{2}{c?}{Scenario} &
             \multicolumn{2}{c|}{Adversarial Attacks} & \multicolumn{2}{c?}{Unknown Attacks} & \multicolumn{2}{c|}{Adversarial Attacks} & \multicolumn{2}{c?}{Unknown Attacks} & \multicolumn{2}{c|}{Adversarial Attacks} & \multicolumn{2}{c?}{Unknown Attacks} & \multicolumn{2}{c|}{Adversarial Attacks} & \multicolumn{2}{c?}{Unknown Attacks}  \\ \hline
            Alg. & Design & 
            \footnotesize{$tpr$} \tiny{(org)} & \footnotesize{$tpr$} \tiny{(adv)} & \footnotesize{$fpr$} & \footnotesize{$tpr$} & \footnotesize{$tpr$} \tiny{(org)} & \footnotesize{$tpr$} \tiny{(adv)} & \footnotesize{$fpr$} & \footnotesize{$tpr$} & \footnotesize{$tpr$} \tiny{(org)} & \footnotesize{$tpr$} \tiny{(adv)} & \footnotesize{$fpr$} & \footnotesize{$tpr$} & \footnotesize{$tpr$} \tiny{(org)} & \footnotesize{$tpr$} \tiny{(adv)} & \footnotesize{$fpr$} & \footnotesize{$tpr$} \\
             \toprule
             
             \multirow{5}{*}{\rotatebox{90}{\parbox{40pt}\centering\textbf{RF}}} 
              & \variable{BD} & \tabvaluetemp{0.954} & \tabvaluetemp{0.999} & \tabvaluetemp{0.095} & \tabvaluetemp{0.972}& \tabvaluetemp{0.965} & \tabvaluetemp{0.003} & \tabvaluetemp{0.001} & \tabvaluetemp{0.743}& \tabvaluetemp{0.968} & \tabvaluetemp{0.003} & \tabvaluetemp{0.001} & \tabvaluetemp{0.746}& \tabvaluetemp{0.970} & \tabvaluetemp{0.003} & \tabvaluetemp{0.001} & \tabvaluetemp{0.746}\\

                & \variable{MD} & \tabvaluetemp{0.933} & \tabvaluetemp{0.857} & \tabvaluetemp{0.074} & \tabvaluetemp{0.937}& \tabvaluetemp{0.962} & \tabvaluetemp{0.003} & \tabvaluetemp{0.001} & \tabvaluetemp{0.739}& \tabvaluetemp{0.966} & \tabvaluetemp{0.003} & \tabvaluetemp{0.001} & \tabvaluetemp{0.743}& \tabvaluetemp{0.968} & \tabvaluetemp{0.003} & \tabvaluetemp{0.001} & \tabvaluetemp{0.739}\\
                
                & \variable{ED-v} & \tabvaluetemp{0.522} & \tabvaluetemp{0.000} & \tabvaluetemp{0.007} & \tabvaluetemp{0.704}& \tabvaluetemp{0.002} & \tabvaluetemp{0.000} & \tabvaluetemp{0.000} & \tabvaluetemp{0.396}& \tabvaluetemp{0.010} & \tabvaluetemp{0.000} & \tabvaluetemp{0.000} & \tabvaluetemp{0.402}& \tabvaluetemp{0.012} & \tabvaluetemp{0.000} & \tabvaluetemp{0.000} & \tabvaluetemp{0.393}\\
                
                & \variable{ED-s} & \tabvaluetemp{0.947} & \tabvaluetemp{0.234} & \tabvaluetemp{0.079} & \tabvaluetemp{0.967}& \tabvaluetemp{0.963} & \tabvaluetemp{0.003} & \tabvaluetemp{0.001} & \tabvaluetemp{0.737}& \tabvaluetemp{0.968} & \tabvaluetemp{0.003} & \tabvaluetemp{0.001} & \tabvaluetemp{0.746}& \tabvaluetemp{0.968} & \tabvaluetemp{0.003} & \tabvaluetemp{0.001} & \tabvaluetemp{0.742}\\
                
                & \variable{ED-o} & \tabvaluetemp{0.947} & \tabvaluetemp{0.234} & \tabvaluetemp{0.081} & \tabvaluetemp{0.969}& \tabvaluetemp{0.963} & \tabvaluetemp{0.003} & \tabvaluetemp{0.001} & \tabvaluetemp{0.737}& \tabvaluetemp{0.968} & \tabvaluetemp{0.003} & \tabvaluetemp{0.001} & \tabvaluetemp{0.746}& \tabvaluetemp{0.968} & \tabvaluetemp{0.003} & \tabvaluetemp{0.001} & \tabvaluetemp{0.742}\\
             
            \midrule
            \multirow{5}{*}{\rotatebox{90}{\parbox{40pt}\centering\textbf{DT}}} 
             & \variable{BD} & \tabvaluetemp{0.941} & \tabvaluetemp{0.988} & \tabvaluetemp{0.105} & \tabvaluetemp{0.908}& \tabvaluetemp{0.933} & \tabvaluetemp{0.005} & \tabvaluetemp{0.002} & \tabvaluetemp{0.742}& \tabvaluetemp{0.942} & \tabvaluetemp{0.096} & \tabvaluetemp{0.001} & \tabvaluetemp{0.876}& \tabvaluetemp{0.950} & \tabvaluetemp{0.013} & \tabvaluetemp{0.001} & \tabvaluetemp{0.752}\\

            & \variable{MD} & \tabvaluetemp{0.947} & \tabvaluetemp{0.331} & \tabvaluetemp{0.109} & \tabvaluetemp{0.905}& \tabvaluetemp{0.939} & \tabvaluetemp{0.558} & \tabvaluetemp{0.002} & \tabvaluetemp{0.745}& \tabvaluetemp{0.947} & \tabvaluetemp{0.160} & \tabvaluetemp{0.002} & \tabvaluetemp{0.732}& \tabvaluetemp{0.949} & \tabvaluetemp{0.005} & \tabvaluetemp{0.001} & \tabvaluetemp{0.865}\\
            
            & \variable{ED-v} & \tabvaluetemp{0.661} & \tabvaluetemp{0.001} & \tabvaluetemp{0.012} & \tabvaluetemp{0.754}& \tabvaluetemp{0.050} & \tabvaluetemp{0.000} & \tabvaluetemp{0.000} & \tabvaluetemp{0.447}& \tabvaluetemp{0.051} & \tabvaluetemp{0.000} & \tabvaluetemp{0.000} & \tabvaluetemp{0.563}& \tabvaluetemp{0.039} & \tabvaluetemp{0.000} & \tabvaluetemp{0.000} & \tabvaluetemp{0.476}\\
            
            & \variable{ED-s} & \tabvaluetemp{0.949} & \tabvaluetemp{0.982} & \tabvaluetemp{0.168} & \tabvaluetemp{0.893}& \tabvaluetemp{0.944} & \tabvaluetemp{0.164} & \tabvaluetemp{0.002} & \tabvaluetemp{0.866}& \tabvaluetemp{0.952} & \tabvaluetemp{0.025} & \tabvaluetemp{0.002} & \tabvaluetemp{0.754}& \tabvaluetemp{0.957} & \tabvaluetemp{0.026} & \tabvaluetemp{0.001} & \tabvaluetemp{0.752}\\
            
            & \variable{ED-o} & \tabvaluetemp{0.949} & \tabvaluetemp{0.982} & \tabvaluetemp{0.200} & \tabvaluetemp{0.906}& \tabvaluetemp{0.944} & \tabvaluetemp{0.164} & \tabvaluetemp{0.002} & \tabvaluetemp{0.866}& \tabvaluetemp{0.952} & \tabvaluetemp{0.025} & \tabvaluetemp{0.002} & \tabvaluetemp{0.754}& \tabvaluetemp{0.957} & \tabvaluetemp{0.026} & \tabvaluetemp{0.001} & \tabvaluetemp{0.757}\\
             \midrule
            \multirow{5}{*}{\rotatebox{90}{\parbox{40pt}\centering\textbf{LR}}} 
             & \variable{BD} & \tabvaluetemp{0.995} & \tabvaluetemp{0.998} & \tabvaluetemp{0.516} & \tabvaluetemp{0.926}& \tabvaluetemp{0.013} & \tabvaluetemp{0.001} & \tabvaluetemp{0.065} & \tabvaluetemp{0.409}& \tabvaluetemp{0.009} & \tabvaluetemp{0.001} & \tabvaluetemp{0.134} & \tabvaluetemp{0.630}& \tabvaluetemp{0.339} & \tabvaluetemp{0.001} & \tabvaluetemp{0.150} & \tabvaluetemp{0.648}\\

            & \variable{MD} & \tabvaluetemp{0.994} & \tabvaluetemp{0.033} & \tabvaluetemp{0.403} & \tabvaluetemp{0.818}& \tabvaluetemp{0.013} & \tabvaluetemp{0.006} & \tabvaluetemp{0.004} & \tabvaluetemp{0.019}& \tabvaluetemp{0.009} & \tabvaluetemp{0.001} & \tabvaluetemp{0.004} & \tabvaluetemp{0.018}& \tabvaluetemp{0.014} & \tabvaluetemp{0.006} & \tabvaluetemp{0.004} & \tabvaluetemp{0.068}\\
            
            & \variable{ED-v} & \tabvaluetemp{0.014} & \tabvaluetemp{0.008} & \tabvaluetemp{0.093} & \tabvaluetemp{0.662}& \tabvaluetemp{0.000} & \tabvaluetemp{0.000} & \tabvaluetemp{0.000} & \tabvaluetemp{0.000}& \tabvaluetemp{0.000} & \tabvaluetemp{0.000} & \tabvaluetemp{0.000} & \tabvaluetemp{0.000}& \tabvaluetemp{0.000} & \tabvaluetemp{0.000} & \tabvaluetemp{0.000} & \tabvaluetemp{0.000}\\
            
            & \variable{ED-s} & \tabvaluetemp{0.987} & \tabvaluetemp{0.020} & \tabvaluetemp{0.410} & \tabvaluetemp{0.912}& \tabvaluetemp{0.001} & \tabvaluetemp{0.001} & \tabvaluetemp{0.018} & \tabvaluetemp{0.346}& \tabvaluetemp{0.014} & \tabvaluetemp{0.014} & \tabvaluetemp{0.018} & \tabvaluetemp{0.366}& \tabvaluetemp{0.012} & \tabvaluetemp{0.001} & \tabvaluetemp{0.021} & \tabvaluetemp{0.369}\\
            
            & \variable{ED-o} & \tabvaluetemp{0.987} & \tabvaluetemp{0.020} & \tabvaluetemp{0.421} & \tabvaluetemp{0.915}& \tabvaluetemp{0.001} & \tabvaluetemp{0.001} & \tabvaluetemp{0.019} & \tabvaluetemp{0.347}& \tabvaluetemp{0.014} & \tabvaluetemp{0.014} & \tabvaluetemp{0.019} & \tabvaluetemp{0.366}& \tabvaluetemp{0.012} & \tabvaluetemp{0.001} & \tabvaluetemp{0.021} & \tabvaluetemp{0.369}\\
             \midrule
            \multirow{5}{*}{\rotatebox{90}{\parbox{40pt}\centering\textbf{HGB}}} 
            & \variable{BD} & \tabvaluetemp{0.955} & \tabvaluetemp{0.998} & \tabvaluetemp{0.082} & \tabvaluetemp{0.921}& \tabvaluetemp{0.939} & \tabvaluetemp{0.002} & \tabvaluetemp{0.002} & \tabvaluetemp{0.752}& \tabvaluetemp{0.943} & \tabvaluetemp{0.044} & \tabvaluetemp{0.002} & \tabvaluetemp{0.749}& \tabvaluetemp{0.935} & \tabvaluetemp{0.003} & \tabvaluetemp{0.001} & \tabvaluetemp{0.748}\\

            & \variable{MD} & \tabvaluetemp{0.916} & \tabvaluetemp{0.996} & \tabvaluetemp{0.070} & \tabvaluetemp{0.844}& \tabvaluetemp{0.919} & \tabvaluetemp{0.419} & \tabvaluetemp{0.004} & \tabvaluetemp{0.749}& \tabvaluetemp{0.922} & \tabvaluetemp{0.122} & \tabvaluetemp{0.003} & \tabvaluetemp{0.747}& \tabvaluetemp{0.921} & \tabvaluetemp{0.004} & \tabvaluetemp{0.002} & \tabvaluetemp{0.744}\\
            
            & \variable{ED-v} & \tabvaluetemp{0.612} & \tabvaluetemp{0.006} & \tabvaluetemp{0.020} & \tabvaluetemp{0.669}& \tabvaluetemp{0.008} & \tabvaluetemp{0.000} & \tabvaluetemp{0.000} & \tabvaluetemp{0.462}& \tabvaluetemp{0.023} & \tabvaluetemp{0.000} & \tabvaluetemp{0.000} & \tabvaluetemp{0.559}& \tabvaluetemp{0.099} & \tabvaluetemp{0.000} & \tabvaluetemp{0.000} & \tabvaluetemp{0.464}\\
            
            & \variable{ED-s} & \tabvaluetemp{0.940} & \tabvaluetemp{0.992} & \tabvaluetemp{0.088} & \tabvaluetemp{0.937}& \tabvaluetemp{0.947} & \tabvaluetemp{0.005} & \tabvaluetemp{0.002} & \tabvaluetemp{0.749}& \tabvaluetemp{0.949} & \tabvaluetemp{0.021} & \tabvaluetemp{0.002} & \tabvaluetemp{0.753}& \tabvaluetemp{0.943} & \tabvaluetemp{0.010} & \tabvaluetemp{0.002} & \tabvaluetemp{0.752}\\
            
            & \variable{ED-o} & \tabvaluetemp{0.940} & \tabvaluetemp{0.992} & \tabvaluetemp{0.100} & \tabvaluetemp{0.939}& \tabvaluetemp{0.947} & \tabvaluetemp{0.005} & \tabvaluetemp{0.002} & \tabvaluetemp{0.749}& \tabvaluetemp{0.949} & \tabvaluetemp{0.021} & \tabvaluetemp{0.002} & \tabvaluetemp{0.753}& \tabvaluetemp{0.943} & \tabvaluetemp{0.010} & \tabvaluetemp{0.002} & \tabvaluetemp{0.752}\\
            
            \bottomrule
        \end{tabular}
        }
                
        \caption{\textit{Temporal Dependency}: Results by assuming the presence of temporal dependencies among samples (the `first' samples of \scbb{D} are put in \scbb{T}, while the last 20\% represent \scbb{E}).} 
    \label{tab:ctu_open_temporal}
    \end{subtable}

    \label{tab:ctu_open}
\end{table*}

\begin{table*}
  \centering
  \caption{\dataset{GTCS} binary classification results (\scmath{fpr} and \scmath{tpr}) against `known' attacks seen during the training stage (closed world).}

    \begin{subtable}[htbp]{1.99\columnwidth}
        \resizebox{1.0\columnwidth}{!}{
        \begin{tabular}{c|c ? cc|cc? cc|cc? cc|cc? cc|cc}
             \multicolumn{2}{c?}{Available Data} & \multicolumn{4}{c?}{Limited (100 per class) [N=1000]} & \multicolumn{4}{c?}{Scarce (15\% of \scbb{D}) [N=100]} &  \multicolumn{4}{c?}{Moderate (40\% of \scbb{D}) [N=100]} &  \multicolumn{4}{c}{Abundant (80\% of \scbb{D}) [N=100]} \\ \hline
             \multicolumn{2}{c?}{Features} &
             \multicolumn{2}{c|}{Complete} & \multicolumn{2}{c?}{Essential} & \multicolumn{2}{c|}{Complete} & \multicolumn{2}{c?}{Essential} & \multicolumn{2}{c|}{Complete} & \multicolumn{2}{c?}{Essential} &  
             \multicolumn{2}{c|}{Complete} & \multicolumn{2}{c}{Essential} \\ \hline
            Alg. & Design & \smamath{fpr} & \smamath{tpr} & \smamath{fpr} & \smamath{tpr} & \smamath{fpr} & \smamath{tpr} & \smamath{fpr} & \smamath{tpr} & \smamath{fpr} & \smamath{tpr} & \smamath{fpr} & \smamath{tpr} & \smamath{fpr} & \smamath{tpr} & \smamath{fpr} & \smamath{tpr} \\
             \toprule
             
             \multirow{6}{*}{\rotatebox{90}{\parbox{40pt}\centering\textbf{RF}}} 
             & \variable{BD} & \tabvalue{0.130}{0.021} & \tabvalue{0.977}{0.009} & \tabvalue{0.264}{0.039} & \tabvalue{0.971}{0.008}& \tabvalue{0.033}{0.001} & \tabvalue{0.992}{0.001} & \tabvalue{0.116}{0.012} & \tabvalue{0.974}{0.005}& \tabvalue{0.030}{0.001} & \tabvalue{0.993}{0.000} & \tabvalue{0.119}{0.016} & \tabvalue{0.977}{0.004}& \tabvalue{0.027}{0.001} & \tabvalue{0.994}{0.000} & \tabvalue{0.123}{0.007} & \tabvalue{0.977}{0.002}\\

& \variable{MD} & \tabvalue{0.135}{0.025} & \tabvalue{0.974}{0.009} & \tabvalue{0.233}{0.037} & \tabvalue{0.970}{0.008}& \tabvalue{0.032}{0.001} & \tabvalue{0.992}{0.001} & \tabvalue{0.115}{0.015} & \tabvalue{0.975}{0.006}& \tabvalue{0.029}{0.001} & \tabvalue{0.993}{0.000} & \tabvalue{0.121}{0.018} & \tabvalue{0.979}{0.004}& \tabvalue{0.027}{0.001} & \tabvalue{0.994}{0.000} & \tabvalue{0.126}{0.009} & \tabvalue{0.980}{0.003}\\

& \variable{ED-v} & \tabvalue{0.005}{0.003} & \tabvalue{0.632}{0.160} & \tabvalue{0.011}{0.006} & \tabvalue{0.587}{0.109}& \tabvalue{0.000}{0.000} & \tabvalue{0.596}{0.081} & \tabvalue{0.000}{0.000} & \tabvalue{0.506}{0.021}& \tabvalue{0.000}{0.000} & \tabvalue{0.574}{0.066} & \tabvalue{0.000}{0.000} & \tabvalue{0.509}{0.026}& \tabvalue{0.000}{0.000} & \tabvalue{0.542}{0.055} & \tabvalue{0.000}{0.000} & \tabvalue{0.506}{0.002}\\

& \variable{ED-s} & \tabvalue{0.123}{0.022} & \tabvalue{0.976}{0.009} & \tabvalue{0.231}{0.036} & \tabvalue{0.969}{0.008}& \tabvalue{0.032}{0.001} & \tabvalue{0.992}{0.001} & \tabvalue{0.117}{0.013} & \tabvalue{0.976}{0.005}& \tabvalue{0.029}{0.001} & \tabvalue{0.993}{0.000} & \tabvalue{0.120}{0.018} & \tabvalue{0.979}{0.004}& \tabvalue{0.027}{0.001} & \tabvalue{0.994}{0.000} & \tabvalue{0.124}{0.011} & \tabvalue{0.980}{0.004}\\

& \variable{ED-o} & \tabvalue{0.124}{0.022} & \tabvalue{0.977}{0.009} & \tabvalue{0.232}{0.036} & \tabvalue{0.969}{0.008}& \tabvalue{0.032}{0.001} & \tabvalue{0.992}{0.001} & \tabvalue{0.117}{0.013} & \tabvalue{0.976}{0.005}& \tabvalue{0.029}{0.001} & \tabvalue{0.993}{0.000} & \tabvalue{0.120}{0.018} & \tabvalue{0.979}{0.004}& \tabvalue{0.027}{0.001} & \tabvalue{0.994}{0.000} & \tabvalue{0.124}{0.011} & \tabvalue{0.980}{0.004}\\

& \variable{ED} & \tabvalue{0.032}{0.006} & \tabvalue{0.974}{0.009} & \tabvalue{0.061}{0.010} & \tabvalue{0.967}{0.008}& \tabvalue{0.008}{0.000} & \tabvalue{0.992}{0.001} & \tabvalue{0.029}{0.003} & \tabvalue{0.976}{0.005}& \tabvalue{0.007}{0.000} & \tabvalue{0.993}{0.000} & \tabvalue{0.030}{0.005} & \tabvalue{0.979}{0.004}& \tabvalue{0.007}{0.000} & \tabvalue{0.994}{0.000} & \tabvalue{0.031}{0.003} & \tabvalue{0.979}{0.004}\\
             
            \midrule
            \multirow{6}{*}{\rotatebox{90}{\parbox{40pt}\centering\textbf{DT}}} 
            & \variable{BD} & \tabvalue{0.131}{0.018} & \tabvalue{0.976}{0.007} & \tabvalue{0.261}{0.037} & \tabvalue{0.955}{0.008}& \tabvalue{0.033}{0.001} & \tabvalue{0.988}{0.001} & \tabvalue{0.100}{0.002} & \tabvalue{0.963}{0.001}& \tabvalue{0.029}{0.001} & \tabvalue{0.990}{0.000} & \tabvalue{0.097}{0.009} & \tabvalue{0.966}{0.003}& \tabvalue{0.026}{0.001} & \tabvalue{0.990}{0.000} & \tabvalue{0.097}{0.002} & \tabvalue{0.966}{0.001}\\

& \variable{MD} & \tabvalue{0.143}{0.021} & \tabvalue{0.979}{0.008} & \tabvalue{0.278}{0.039} & \tabvalue{0.955}{0.009}& \tabvalue{0.033}{0.001} & \tabvalue{0.988}{0.001} & \tabvalue{0.100}{0.002} & \tabvalue{0.963}{0.001}& \tabvalue{0.028}{0.001} & \tabvalue{0.990}{0.000} & \tabvalue{0.097}{0.009} & \tabvalue{0.966}{0.003}& \tabvalue{0.026}{0.001} & \tabvalue{0.991}{0.000} & \tabvalue{0.097}{0.002} & \tabvalue{0.966}{0.001}\\

& \variable{ED-v} & \tabvalue{0.019}{0.013} & \tabvalue{0.692}{0.151} & \tabvalue{0.024}{0.012} & \tabvalue{0.592}{0.129}& \tabvalue{0.000}{0.000} & \tabvalue{0.488}{0.120} & \tabvalue{0.001}{0.000} & \tabvalue{0.469}{0.058}& \tabvalue{0.000}{0.000} & \tabvalue{0.480}{0.106} & \tabvalue{0.001}{0.000} & \tabvalue{0.449}{0.045}& \tabvalue{0.000}{0.000} & \tabvalue{0.462}{0.098} & \tabvalue{0.001}{0.000} & \tabvalue{0.437}{0.015}\\

& \variable{ED-s} & \tabvalue{0.140}{0.020} & \tabvalue{0.978}{0.007} & \tabvalue{0.271}{0.037} & \tabvalue{0.958}{0.008}& \tabvalue{0.033}{0.001} & \tabvalue{0.988}{0.001} & \tabvalue{0.100}{0.002} & \tabvalue{0.963}{0.001}& \tabvalue{0.028}{0.001} & \tabvalue{0.990}{0.000} & \tabvalue{0.096}{0.009} & \tabvalue{0.966}{0.003}& \tabvalue{0.026}{0.001} & \tabvalue{0.991}{0.000} & \tabvalue{0.097}{0.002} & \tabvalue{0.966}{0.001}\\

& \variable{ED-o} & \tabvalue{0.149}{0.022} & \tabvalue{0.980}{0.007} & \tabvalue{0.275}{0.038} & \tabvalue{0.959}{0.008}& \tabvalue{0.033}{0.001} & \tabvalue{0.988}{0.001} & \tabvalue{0.100}{0.002} & \tabvalue{0.963}{0.001}& \tabvalue{0.028}{0.001} & \tabvalue{0.990}{0.000} & \tabvalue{0.096}{0.009} & \tabvalue{0.966}{0.003}& \tabvalue{0.026}{0.001} & \tabvalue{0.991}{0.000} & \tabvalue{0.097}{0.002} & \tabvalue{0.966}{0.001}\\

& \variable{ED} & \tabvalue{0.042}{0.008} & \tabvalue{0.975}{0.007} & \tabvalue{0.075}{0.011} & \tabvalue{0.949}{0.009}& \tabvalue{0.008}{0.000} & \tabvalue{0.988}{0.001} & \tabvalue{0.025}{0.001} & \tabvalue{0.963}{0.001}& \tabvalue{0.007}{0.000} & \tabvalue{0.990}{0.000} & \tabvalue{0.024}{0.002} & \tabvalue{0.966}{0.003}& \tabvalue{0.006}{0.000} & \tabvalue{0.991}{0.000} & \tabvalue{0.025}{0.000} & \tabvalue{0.966}{0.001}\\
             \midrule
            \multirow{6}{*}{\rotatebox{90}{\parbox{40pt}\centering\textbf{LR}}} 
            & \variable{BD} & \tabvalue{0.067}{0.250} & \tabvalue{0.067}{0.250} & \tabvalue{0.610}{0.109} & \tabvalue{0.983}{0.045}& \tabvalue{0.000}{0.000} & \tabvalue{0.000}{0.000} & \tabvalue{0.440}{0.046} & \tabvalue{0.953}{0.029}& \tabvalue{0.000}{0.000} & \tabvalue{0.000}{0.000} & \tabvalue{0.431}{0.077} & \tabvalue{0.932}{0.136}& \tabvalue{0.000}{0.000} & \tabvalue{0.000}{0.000} & \tabvalue{0.439}{0.048} & \tabvalue{0.952}{0.029}\\

& \variable{MD} & \tabvalue{0.106}{0.279} & \tabvalue{0.195}{0.297} & \tabvalue{0.532}{0.085} & \tabvalue{0.968}{0.040}& \tabvalue{0.000}{0.000} & \tabvalue{0.000}{0.000} & \tabvalue{0.249}{0.040} & \tabvalue{0.881}{0.016}& \tabvalue{0.000}{0.000} & \tabvalue{0.000}{0.000} & \tabvalue{0.245}{0.052} & \tabvalue{0.863}{0.125}& \tabvalue{0.000}{0.000} & \tabvalue{0.000}{0.000} & \tabvalue{0.245}{0.035} & \tabvalue{0.879}{0.016}\\

& \variable{ED-v} & \tabvalue{0.009}{0.019} & \tabvalue{0.199}{0.244} & \tabvalue{0.135}{0.051} & \tabvalue{0.627}{0.166}& \tabvalue{0.000}{0.000} & \tabvalue{0.025}{0.072} & \tabvalue{0.022}{0.011} & \tabvalue{0.353}{0.090}& \tabvalue{0.000}{0.000} & \tabvalue{0.013}{0.053} & \tabvalue{0.022}{0.011} & \tabvalue{0.358}{0.109}& \tabvalue{0.000}{0.000} & \tabvalue{0.000}{0.000} & \tabvalue{0.023}{0.010} & \tabvalue{0.371}{0.102}\\

& \variable{ED-s} & \tabvalue{0.028}{0.074} & \tabvalue{0.230}{0.384} & \tabvalue{0.523}{0.113} & \tabvalue{0.958}{0.121}& \tabvalue{0.007}{0.043} & \tabvalue{0.076}{0.229} & \tabvalue{0.207}{0.047} & \tabvalue{0.870}{0.018}& \tabvalue{0.013}{0.083} & \tabvalue{0.062}{0.212} & \tabvalue{0.201}{0.069} & \tabvalue{0.865}{0.018}& \tabvalue{0.000}{0.000} & \tabvalue{0.000}{0.000} & \tabvalue{0.187}{0.050} & \tabvalue{0.861}{0.013}\\

& \variable{ED-o} & \tabvalue{0.052}{0.066} & \tabvalue{0.540}{0.311} & \tabvalue{0.543}{0.085} & \tabvalue{0.972}{0.036}& \tabvalue{0.008}{0.043} & \tabvalue{0.306}{0.171} & \tabvalue{0.237}{0.028} & \tabvalue{0.878}{0.012}& \tabvalue{0.002}{0.005} & \tabvalue{0.277}{0.131} & \tabvalue{0.230}{0.039} & \tabvalue{0.867}{0.069}& \tabvalue{0.000}{0.000} & \tabvalue{0.233}{0.003} & \tabvalue{0.234}{0.022} & \tabvalue{0.876}{0.012}\\

& \variable{ED} & \tabvalue{0.015}{0.017} & \tabvalue{0.423}{0.166} & \tabvalue{0.173}{0.029} & \tabvalue{0.956}{0.043}& \tabvalue{0.002}{0.011} & \tabvalue{0.274}{0.123} & \tabvalue{0.065}{0.008} & \tabvalue{0.858}{0.010}& \tabvalue{0.000}{0.001} & \tabvalue{0.256}{0.104} & \tabvalue{0.063}{0.011} & \tabvalue{0.848}{0.070}& \tabvalue{0.000}{0.000} & \tabvalue{0.222}{0.001} & \tabvalue{0.064}{0.007} & \tabvalue{0.858}{0.012}\\
             \midrule
            \multirow{6}{*}{\rotatebox{90}{\parbox{40pt}\centering\textbf{HGB}}} 
            & \variable{BD} & \tabvalue{0.145}{0.023} & \tabvalue{0.989}{0.005} & \tabvalue{0.259}{0.042} & \tabvalue{0.964}{0.007}& \tabvalue{0.042}{0.001} & \tabvalue{0.997}{0.000} & \tabvalue{0.077}{0.004} & \tabvalue{0.977}{0.002}& \tabvalue{0.041}{0.002} & \tabvalue{0.997}{0.000} & \tabvalue{0.077}{0.006} & \tabvalue{0.978}{0.003}& \tabvalue{0.040}{0.001} & \tabvalue{0.997}{0.000} & \tabvalue{0.078}{0.004} & \tabvalue{0.979}{0.002}\\

& \variable{MD} & \tabvalue{0.135}{0.024} & \tabvalue{0.987}{0.006} & \tabvalue{0.233}{0.039} & \tabvalue{0.960}{0.008}& \tabvalue{0.037}{0.001} & \tabvalue{0.997}{0.000} & \tabvalue{0.072}{0.003} & \tabvalue{0.978}{0.001}& \tabvalue{0.035}{0.001} & \tabvalue{0.997}{0.000} & \tabvalue{0.072}{0.004} & \tabvalue{0.980}{0.002}& \tabvalue{0.034}{0.001} & \tabvalue{0.998}{0.000} & \tabvalue{0.073}{0.003} & \tabvalue{0.981}{0.001}\\

& \variable{ED-v} & \tabvalue{0.010}{0.008} & \tabvalue{0.732}{0.115} & \tabvalue{0.019}{0.011} & \tabvalue{0.605}{0.130}& \tabvalue{0.000}{0.000} & \tabvalue{0.682}{0.048} & \tabvalue{0.001}{0.000} & \tabvalue{0.516}{0.013}& \tabvalue{0.000}{0.000} & \tabvalue{0.681}{0.033} & \tabvalue{0.000}{0.000} & \tabvalue{0.518}{0.022}& \tabvalue{0.000}{0.000} & \tabvalue{0.699}{0.029} & \tabvalue{0.000}{0.000} & \tabvalue{0.515}{0.003}\\

& \variable{ED-s} & \tabvalue{0.147}{0.024} & \tabvalue{0.987}{0.006} & \tabvalue{0.265}{0.046} & \tabvalue{0.961}{0.010}& \tabvalue{0.040}{0.001} & \tabvalue{0.997}{0.000} & \tabvalue{0.073}{0.003} & \tabvalue{0.978}{0.001}& \tabvalue{0.038}{0.002} & \tabvalue{0.997}{0.000} & \tabvalue{0.074}{0.005} & \tabvalue{0.980}{0.002}& \tabvalue{0.038}{0.001} & \tabvalue{0.998}{0.000} & \tabvalue{0.076}{0.002} & \tabvalue{0.981}{0.001}\\

& \variable{ED-o} & \tabvalue{0.157}{0.026} & \tabvalue{0.988}{0.005} & \tabvalue{0.266}{0.046} & \tabvalue{0.961}{0.010}& \tabvalue{0.040}{0.001} & \tabvalue{0.997}{0.000} & \tabvalue{0.073}{0.003} & \tabvalue{0.978}{0.001}& \tabvalue{0.038}{0.001} & \tabvalue{0.997}{0.000} & \tabvalue{0.074}{0.005} & \tabvalue{0.980}{0.002}& \tabvalue{0.038}{0.001} & \tabvalue{0.998}{0.000} & \tabvalue{0.076}{0.002} & \tabvalue{0.981}{0.001}\\

& \variable{ED} & \tabvalue{0.042}{0.008} & \tabvalue{0.985}{0.006} & \tabvalue{0.072}{0.013} & \tabvalue{0.952}{0.011}& \tabvalue{0.010}{0.000} & \tabvalue{0.997}{0.000} & \tabvalue{0.018}{0.001} & \tabvalue{0.978}{0.001}& \tabvalue{0.010}{0.000} & \tabvalue{0.997}{0.000} & \tabvalue{0.019}{0.001} & \tabvalue{0.980}{0.002}& \tabvalue{0.009}{0.000} & \tabvalue{0.998}{0.000} & \tabvalue{0.019}{0.001} & \tabvalue{0.981}{0.001}\\
            
            \bottomrule
        \end{tabular}
        }
                
        \caption{\textit{Static Dependency}: Results by assuming the absence of temporal dependencies among samples (\scbb{T} and \scbb{E} are randomly sampled from \scbb{D}).} 
    \label{tab:gtcs_baseline_static}
    \end{subtable}

    \begin{subtable}[htbp]{1.8\columnwidth}
        \resizebox{1.0\columnwidth}{!}{
        \begin{tabular}{c|c ? cc|cc? cc|cc? cc|cc? cc|cc}
             \multicolumn{2}{c?}{Available Data} & \multicolumn{4}{c?}{Limited (100 per class) [N=1]} & \multicolumn{4}{c?}{Scarce (15\% of \scbb{D}) [N=1]} &  \multicolumn{4}{c?}{Moderate (40\% of \scbb{D}) [N=1]} &  \multicolumn{4}{c}{Abundant (80\% of \scbb{D}) [N=1]} \\ \hline
             \multicolumn{2}{c?}{Features} &
             \multicolumn{2}{c|}{Complete} & \multicolumn{2}{c?}{Essential} & \multicolumn{2}{c|}{Complete} & \multicolumn{2}{c?}{Essential} & \multicolumn{2}{c|}{Complete} & \multicolumn{2}{c?}{Essential} &  
             \multicolumn{2}{c|}{Complete} & \multicolumn{2}{c}{Essential} \\ \hline
            Alg. & Design & \smamath{fpr} & \smamath{tpr} & \smamath{fpr} & \smamath{tpr} & \smamath{fpr} & \smamath{tpr} & \smamath{fpr} & \smamath{tpr} & \smamath{fpr} & \smamath{tpr} & \smamath{fpr} & \smamath{tpr} & \smamath{fpr} & \smamath{tpr} & \smamath{fpr} & \smamath{tpr} \\
             \toprule
             
             \multirow{6}{*}{\rotatebox{90}{\parbox{40pt}\centering\textbf{RF}}} 
             & \variable{BD} & \tabvaluetemp{0.109} & \tabvaluetemp{0.970} & \tabvaluetemp{0.272} & \tabvaluetemp{0.972}& \tabvaluetemp{0.024} & \tabvaluetemp{0.983} & \tabvaluetemp{0.040} & \tabvaluetemp{0.928}& \tabvaluetemp{0.030} & \tabvaluetemp{0.993} & \tabvaluetemp{0.124} & \tabvaluetemp{0.978}& \tabvaluetemp{0.026} & \tabvaluetemp{0.994} & \tabvaluetemp{0.121} & \tabvaluetemp{0.975}\\

& \variable{MD} & \tabvaluetemp{0.098} & \tabvaluetemp{0.938} & \tabvaluetemp{0.244} & \tabvaluetemp{0.958}& \tabvaluetemp{0.024} & \tabvaluetemp{0.984} & \tabvaluetemp{0.039} & \tabvaluetemp{0.927}& \tabvaluetemp{0.030} & \tabvaluetemp{0.993} & \tabvaluetemp{0.112} & \tabvaluetemp{0.974}& \tabvaluetemp{0.026} & \tabvaluetemp{0.994} & \tabvaluetemp{0.120} & \tabvaluetemp{0.977}\\

& \variable{ED-v} & \tabvaluetemp{0.002} & \tabvaluetemp{0.496} & \tabvaluetemp{0.004} & \tabvaluetemp{0.473}& \tabvaluetemp{0.000} & \tabvaluetemp{0.524} & \tabvaluetemp{0.000} & \tabvaluetemp{0.505}& \tabvaluetemp{0.000} & \tabvaluetemp{0.609} & \tabvaluetemp{0.000} & \tabvaluetemp{0.508}& \tabvaluetemp{0.000} & \tabvaluetemp{0.484} & \tabvaluetemp{0.000} & \tabvaluetemp{0.506}\\

& \variable{ED-s} & \tabvaluetemp{0.083} & \tabvaluetemp{0.954} & \tabvaluetemp{0.231} & \tabvaluetemp{0.958}& \tabvaluetemp{0.024} & \tabvaluetemp{0.985} & \tabvaluetemp{0.034} & \tabvaluetemp{0.926}& \tabvaluetemp{0.030} & \tabvaluetemp{0.993} & \tabvaluetemp{0.127} & \tabvaluetemp{0.978}& \tabvaluetemp{0.026} & \tabvaluetemp{0.994} & \tabvaluetemp{0.108} & \tabvaluetemp{0.974}\\

& \variable{ED-o} & \tabvaluetemp{0.083} & \tabvaluetemp{0.954} & \tabvaluetemp{0.231} & \tabvaluetemp{0.958}& \tabvaluetemp{0.024} & \tabvaluetemp{0.985} & \tabvaluetemp{0.034} & \tabvaluetemp{0.926}& \tabvaluetemp{0.030} & \tabvaluetemp{0.993} & \tabvaluetemp{0.127} & \tabvaluetemp{0.978}& \tabvaluetemp{0.026} & \tabvaluetemp{0.994} & \tabvaluetemp{0.108} & \tabvaluetemp{0.974}\\

& \variable{ED} & \tabvaluetemp{0.021} & \tabvaluetemp{0.951} & \tabvaluetemp{0.059} & \tabvaluetemp{0.956}& \tabvaluetemp{0.006} & \tabvaluetemp{0.985} & \tabvaluetemp{0.009} & \tabvaluetemp{0.926}& \tabvaluetemp{0.007} & \tabvaluetemp{0.993} & \tabvaluetemp{0.032} & \tabvaluetemp{0.978}& \tabvaluetemp{0.006} & \tabvaluetemp{0.994} & \tabvaluetemp{0.027} & \tabvaluetemp{0.974}\\
             
            \midrule
            \multirow{6}{*}{\rotatebox{90}{\parbox{40pt}\centering\textbf{DT}}} 
            & \variable{BD} & \tabvaluetemp{0.123} & \tabvaluetemp{0.971} & \tabvaluetemp{0.273} & \tabvaluetemp{0.960}& \tabvaluetemp{0.029} & \tabvaluetemp{0.981} & \tabvaluetemp{0.084} & \tabvaluetemp{0.951}& \tabvaluetemp{0.028} & \tabvaluetemp{0.988} & \tabvaluetemp{0.099} & \tabvaluetemp{0.964}& \tabvaluetemp{0.026} & \tabvaluetemp{0.990} & \tabvaluetemp{0.097} & \tabvaluetemp{0.964}\\

& \variable{MD} & \tabvaluetemp{0.116} & \tabvaluetemp{0.949} & \tabvaluetemp{0.260} & \tabvaluetemp{0.942}& \tabvaluetemp{0.029} & \tabvaluetemp{0.982} & \tabvaluetemp{0.083} & \tabvaluetemp{0.951}& \tabvaluetemp{0.027} & \tabvaluetemp{0.989} & \tabvaluetemp{0.099} & \tabvaluetemp{0.964}& \tabvaluetemp{0.025} & \tabvaluetemp{0.990} & \tabvaluetemp{0.099} & \tabvaluetemp{0.964}\\

& \variable{ED-v} & \tabvaluetemp{0.012} & \tabvaluetemp{0.584} & \tabvaluetemp{0.009} & \tabvaluetemp{0.571}& \tabvaluetemp{0.000} & \tabvaluetemp{0.544} & \tabvaluetemp{0.001} & \tabvaluetemp{0.412}& \tabvaluetemp{0.000} & \tabvaluetemp{0.534} & \tabvaluetemp{0.001} & \tabvaluetemp{0.426}& \tabvaluetemp{0.000} & \tabvaluetemp{0.658} & \tabvaluetemp{0.001} & \tabvaluetemp{0.441}\\

& \variable{ED-s} & \tabvaluetemp{0.146} & \tabvaluetemp{0.958} & \tabvaluetemp{0.250} & \tabvaluetemp{0.942}& \tabvaluetemp{0.027} & \tabvaluetemp{0.981} & \tabvaluetemp{0.084} & \tabvaluetemp{0.951}& \tabvaluetemp{0.027} & \tabvaluetemp{0.989} & \tabvaluetemp{0.099} & \tabvaluetemp{0.964}& \tabvaluetemp{0.025} & \tabvaluetemp{0.990} & \tabvaluetemp{0.098} & \tabvaluetemp{0.964}\\

& \variable{ED-o} & \tabvaluetemp{0.148} & \tabvaluetemp{0.958} & \tabvaluetemp{0.250} & \tabvaluetemp{0.942}& \tabvaluetemp{0.027} & \tabvaluetemp{0.981} & \tabvaluetemp{0.084} & \tabvaluetemp{0.951}& \tabvaluetemp{0.027} & \tabvaluetemp{0.989} & \tabvaluetemp{0.099} & \tabvaluetemp{0.964}& \tabvaluetemp{0.025} & \tabvaluetemp{0.990} & \tabvaluetemp{0.098} & \tabvaluetemp{0.964}\\

& \variable{ED} & \tabvaluetemp{0.040} & \tabvaluetemp{0.957} & \tabvaluetemp{0.065} & \tabvaluetemp{0.923}& \tabvaluetemp{0.007} & \tabvaluetemp{0.981} & \tabvaluetemp{0.021} & \tabvaluetemp{0.950}& \tabvaluetemp{0.007} & \tabvaluetemp{0.989} & \tabvaluetemp{0.025} & \tabvaluetemp{0.964}& \tabvaluetemp{0.006} & \tabvaluetemp{0.990} & \tabvaluetemp{0.025} & \tabvaluetemp{0.964}\\
             \midrule
            \multirow{6}{*}{\rotatebox{90}{\parbox{40pt}\centering\textbf{LR}}} 
            & \variable{BD} & \tabvaluetemp{0.000} & \tabvaluetemp{0.000} & \tabvaluetemp{0.498} & \tabvaluetemp{0.989}& \tabvaluetemp{0.000} & \tabvaluetemp{0.000} & \tabvaluetemp{0.340} & \tabvaluetemp{0.915}& \tabvaluetemp{0.000} & \tabvaluetemp{0.000} & \tabvaluetemp{0.488} & \tabvaluetemp{0.913}& \tabvaluetemp{0.000} & \tabvaluetemp{0.000} & \tabvaluetemp{0.441} & \tabvaluetemp{0.973}\\

& \variable{MD} & \tabvaluetemp{0.001} & \tabvaluetemp{0.197} & \tabvaluetemp{0.430} & \tabvaluetemp{0.982}& \tabvaluetemp{0.000} & \tabvaluetemp{0.000} & \tabvaluetemp{0.263} & \tabvaluetemp{0.907}& \tabvaluetemp{0.000} & \tabvaluetemp{0.000} & \tabvaluetemp{0.271} & \tabvaluetemp{0.875}& \tabvaluetemp{0.000} & \tabvaluetemp{0.000} & \tabvaluetemp{0.273} & \tabvaluetemp{0.883}\\

& \variable{ED-v} & \tabvaluetemp{0.034} & \tabvaluetemp{0.772} & \tabvaluetemp{0.159} & \tabvaluetemp{0.741}& \tabvaluetemp{0.000} & \tabvaluetemp{0.000} & \tabvaluetemp{0.044} & \tabvaluetemp{0.535}& \tabvaluetemp{0.000} & \tabvaluetemp{0.000} & \tabvaluetemp{0.019} & \tabvaluetemp{0.309}& \tabvaluetemp{0.000} & \tabvaluetemp{0.000} & \tabvaluetemp{0.016} & \tabvaluetemp{0.308}\\

& \variable{ED-s} & \tabvaluetemp{0.061} & \tabvaluetemp{0.877} & \tabvaluetemp{0.430} & \tabvaluetemp{0.983}& \tabvaluetemp{0.000} & \tabvaluetemp{0.000} & \tabvaluetemp{0.266} & \tabvaluetemp{0.896}& \tabvaluetemp{0.000} & \tabvaluetemp{0.000} & \tabvaluetemp{0.248} & \tabvaluetemp{0.877}& \tabvaluetemp{0.000} & \tabvaluetemp{0.000} & \tabvaluetemp{0.189} & \tabvaluetemp{0.876}\\

& \variable{ED-o} & \tabvaluetemp{0.061} & \tabvaluetemp{0.877} & \tabvaluetemp{0.449} & \tabvaluetemp{0.983}& \tabvaluetemp{0.001} & \tabvaluetemp{0.239} & \tabvaluetemp{0.266} & \tabvaluetemp{0.896}& \tabvaluetemp{0.001} & \tabvaluetemp{0.237} & \tabvaluetemp{0.248} & \tabvaluetemp{0.877}& \tabvaluetemp{0.000} & \tabvaluetemp{0.232} & \tabvaluetemp{0.189} & \tabvaluetemp{0.876}\\

& \variable{ED} & \tabvaluetemp{0.024} & \tabvaluetemp{0.813} & \tabvaluetemp{0.155} & \tabvaluetemp{0.961}& \tabvaluetemp{0.000} & \tabvaluetemp{0.222} & \tabvaluetemp{0.078} & \tabvaluetemp{0.879}& \tabvaluetemp{0.000} & \tabvaluetemp{0.222} & \tabvaluetemp{0.067} & \tabvaluetemp{0.853}& \tabvaluetemp{0.000} & \tabvaluetemp{0.222} & \tabvaluetemp{0.051} & \tabvaluetemp{0.856}\\
             \midrule
            \multirow{6}{*}{\rotatebox{90}{\parbox{40pt}\centering\textbf{HGB}}} 
            & \variable{BD} & \tabvaluetemp{0.116} & \tabvaluetemp{0.990} & \tabvaluetemp{0.259} & \tabvaluetemp{0.960}& \tabvaluetemp{0.036} & \tabvaluetemp{0.994} & \tabvaluetemp{0.050} & \tabvaluetemp{0.964}& \tabvaluetemp{0.038} & \tabvaluetemp{0.997} & \tabvaluetemp{0.086} & \tabvaluetemp{0.983}& \tabvaluetemp{0.038} & \tabvaluetemp{0.998} & \tabvaluetemp{0.072} & \tabvaluetemp{0.978}\\

& \variable{MD} & \tabvaluetemp{0.106} & \tabvaluetemp{0.987} & \tabvaluetemp{0.215} & \tabvaluetemp{0.959}& \tabvaluetemp{0.029} & \tabvaluetemp{0.993} & \tabvaluetemp{0.049} & \tabvaluetemp{0.967}& \tabvaluetemp{0.033} & \tabvaluetemp{0.997} & \tabvaluetemp{0.077} & \tabvaluetemp{0.983}& \tabvaluetemp{0.033} & \tabvaluetemp{0.998} & \tabvaluetemp{0.071} & \tabvaluetemp{0.982}\\

& \variable{ED-v} & \tabvaluetemp{0.005} & \tabvaluetemp{0.695} & \tabvaluetemp{0.013} & \tabvaluetemp{0.290}& \tabvaluetemp{0.000} & \tabvaluetemp{0.598} & \tabvaluetemp{0.000} & \tabvaluetemp{0.521}& \tabvaluetemp{0.000} & \tabvaluetemp{0.729} & \tabvaluetemp{0.000} & \tabvaluetemp{0.515}& \tabvaluetemp{0.000} & \tabvaluetemp{0.687} & \tabvaluetemp{0.000} & \tabvaluetemp{0.516}\\

& \variable{ED-s} & \tabvaluetemp{0.111} & \tabvaluetemp{0.965} & \tabvaluetemp{0.236} & \tabvaluetemp{0.942}& \tabvaluetemp{0.032} & \tabvaluetemp{0.994} & \tabvaluetemp{0.044} & \tabvaluetemp{0.962}& \tabvaluetemp{0.035} & \tabvaluetemp{0.997} & \tabvaluetemp{0.079} & \tabvaluetemp{0.982}& \tabvaluetemp{0.036} & \tabvaluetemp{0.998} & \tabvaluetemp{0.075} & \tabvaluetemp{0.981}\\

& \variable{ED-o} & \tabvaluetemp{0.113} & \tabvaluetemp{0.967} & \tabvaluetemp{0.236} & \tabvaluetemp{0.942}& \tabvaluetemp{0.032} & \tabvaluetemp{0.994} & \tabvaluetemp{0.044} & \tabvaluetemp{0.962}& \tabvaluetemp{0.035} & \tabvaluetemp{0.997} & \tabvaluetemp{0.079} & \tabvaluetemp{0.982}& \tabvaluetemp{0.036} & \tabvaluetemp{0.998} & \tabvaluetemp{0.075} & \tabvaluetemp{0.981}\\

& \variable{ED} & \tabvaluetemp{0.030} & \tabvaluetemp{0.965} & \tabvaluetemp{0.062} & \tabvaluetemp{0.926}& \tabvaluetemp{0.008} & \tabvaluetemp{0.994} & \tabvaluetemp{0.011} & \tabvaluetemp{0.962}& \tabvaluetemp{0.009} & \tabvaluetemp{0.997} & \tabvaluetemp{0.020} & \tabvaluetemp{0.982}& \tabvaluetemp{0.009} & \tabvaluetemp{0.998} & \tabvaluetemp{0.019} & \tabvaluetemp{0.981}\\
            
            \bottomrule
        \end{tabular}
        }
                
        \caption{\textit{Temporal Dependency}: Results by assuming the presence of temporal dependencies among samples (the `first' samples of \scbb{D} are put in \scbb{T}, while the last 20\% represent \scbb{E}).} 
    \label{tab:gtcs_baseline_temporal}
    \end{subtable}

    \label{tab:gtcs_baseline}
\end{table*}

\begin{table*}
  \centering
  \caption{\dataset{GTCS}. Results against adversarial (original \scmath{tpr} and adversarial \scmath{tpr}) and unknown attacks (the \scmath{tpr} is the average on the `unknown' attacks, while the \scmath{fpr} is due to training on a new \scbb{T} that does not have the `unknown' class.).}

    \begin{subtable}[htbp]{1.99\columnwidth}
        \resizebox{1.0\columnwidth}{!}{
        \begin{tabular}{c|c ? cc|cc? cc|cc? cc|cc? cc|cc}
\multicolumn{2}{c?}{Available Data} & \multicolumn{4}{c?}{Limited (100 per class) [N=1000]} & \multicolumn{4}{c?}{Scarce (15\% of \scbb{D}) [N=100]} &  \multicolumn{4}{c?}{Moderate (40\% of \scbb{D}) [N=100]} &  \multicolumn{4}{c}{Abundant (80\% of \scbb{D}) [N=100]} \\ \hline
             \multicolumn{2}{c?}{Scenario} &
             \multicolumn{2}{c|}{Adversarial Attacks} & \multicolumn{2}{c?}{Unknown Attacks} & \multicolumn{2}{c|}{Adversarial Attacks} & \multicolumn{2}{c?}{Unknown Attacks} & \multicolumn{2}{c|}{Adversarial Attacks} & \multicolumn{2}{c?}{Unknown Attacks} & \multicolumn{2}{c|}{Adversarial Attacks} & \multicolumn{2}{c?}{Unknown Attacks}  \\ \hline
            Alg. & Design & 
            \footnotesize{$tpr$} \tiny{(org)} & \footnotesize{$tpr$} \tiny{(adv)} & \footnotesize{$fpr$} & \footnotesize{$tpr$} & \footnotesize{$tpr$} \tiny{(org)} & \footnotesize{$tpr$} \tiny{(adv)} & \footnotesize{$fpr$} & \footnotesize{$tpr$} & \footnotesize{$tpr$} \tiny{(org)} & \footnotesize{$tpr$} \tiny{(adv)} & \footnotesize{$fpr$} & \footnotesize{$tpr$} & \footnotesize{$tpr$} \tiny{(org)} & \footnotesize{$tpr$} \tiny{(adv)} & \footnotesize{$fpr$} & \footnotesize{$tpr$} \\
             \toprule
             
             \multirow{5}{*}{\rotatebox{90}{\parbox{40pt}\centering\textbf{RF}}} 
             & \variable{BD} & \tabvalue{0.881}{0.082} & \tabvalue{0.211}{0.378} & \tabvalue{0.098}{0.016} & \tabvalue{0.686}{0.083}& \tabvalue{0.817}{0.076} & \tabvalue{0.490}{0.480} & \tabvalue{0.025}{0.001} & \tabvalue{0.574}{0.075}& \tabvalue{0.851}{0.052} & \tabvalue{0.594}{0.465} & \tabvalue{0.022}{0.001} & \tabvalue{0.539}{0.066}& \tabvalue{0.858}{0.034} & \tabvalue{0.674}{0.450} & \tabvalue{0.021}{0.001} & \tabvalue{0.510}{0.061}\\

& \variable{MD} & \tabvalue{0.877}{0.081} & \tabvalue{0.221}{0.391} & \tabvalue{0.101}{0.018} & \tabvalue{0.518}{0.139}& \tabvalue{0.856}{0.091} & \tabvalue{0.147}{0.337} & \tabvalue{0.024}{0.001} & \tabvalue{0.525}{0.079}& \tabvalue{0.906}{0.063} & \tabvalue{0.254}{0.413} & \tabvalue{0.022}{0.001} & \tabvalue{0.488}{0.069}& \tabvalue{0.916}{0.052} & \tabvalue{0.266}{0.432} & \tabvalue{0.020}{0.001} & \tabvalue{0.466}{0.056}\\

& \variable{ED-v} & \tabvalue{0.000}{0.000} & \tabvalue{0.000}{0.000} & \tabvalue{0.002}{0.002} & \tabvalue{0.022}{0.035}& \tabvalue{0.000}{0.000} & \tabvalue{0.000}{0.000} & \tabvalue{0.000}{0.000} & \tabvalue{0.000}{0.000}& \tabvalue{0.000}{0.000} & \tabvalue{0.000}{0.000} & \tabvalue{0.000}{0.000} & \tabvalue{0.000}{0.000}& \tabvalue{0.000}{0.000} & \tabvalue{0.000}{0.000} & \tabvalue{0.000}{0.000} & \tabvalue{0.000}{0.000}\\

& \variable{ED-s} & \tabvalue{0.867}{0.088} & \tabvalue{0.193}{0.379} & \tabvalue{0.094}{0.017} & \tabvalue{0.603}{0.135}& \tabvalue{0.884}{0.079} & \tabvalue{0.170}{0.365} & \tabvalue{0.024}{0.001} & \tabvalue{0.550}{0.076}& \tabvalue{0.913}{0.061} & \tabvalue{0.365}{0.463} & \tabvalue{0.022}{0.001} & \tabvalue{0.533}{0.059}& \tabvalue{0.917}{0.059} & \tabvalue{0.494}{0.483} & \tabvalue{0.020}{0.001} & \tabvalue{0.502}{0.055}\\

& \variable{ED-o} & \tabvalue{0.867}{0.088} & \tabvalue{0.193}{0.379} & \tabvalue{0.094}{0.017} & \tabvalue{0.602}{0.136}& \tabvalue{0.884}{0.079} & \tabvalue{0.170}{0.365} & \tabvalue{0.024}{0.001} & \tabvalue{0.550}{0.076}& \tabvalue{0.913}{0.061} & \tabvalue{0.367}{0.464} & \tabvalue{0.022}{0.001} & \tabvalue{0.532}{0.060}& \tabvalue{0.917}{0.059} & \tabvalue{0.494}{0.483} & \tabvalue{0.020}{0.001} & \tabvalue{0.502}{0.055}\\
             
            \midrule
            \multirow{5}{*}{\rotatebox{90}{\parbox{40pt}\centering\textbf{DT}}} 
             & \variable{BD} & \tabvalue{0.762}{0.078} & \tabvalue{0.319}{0.457} & \tabvalue{0.104}{0.014} & \tabvalue{0.658}{0.134}& \tabvalue{0.669}{0.011} & \tabvalue{0.816}{0.382} & \tabvalue{0.025}{0.001} & \tabvalue{0.541}{0.107}& \tabvalue{0.698}{0.025} & \tabvalue{0.944}{0.213} & \tabvalue{0.022}{0.001} & \tabvalue{0.525}{0.091}& \tabvalue{0.703}{0.008} & \tabvalue{0.896}{0.298} & \tabvalue{0.020}{0.001} & \tabvalue{0.500}{0.093}\\

& \variable{MD} & \tabvalue{0.766}{0.080} & \tabvalue{0.347}{0.462} & \tabvalue{0.113}{0.015} & \tabvalue{0.606}{0.160}& \tabvalue{0.669}{0.011} & \tabvalue{0.323}{0.458} & \tabvalue{0.025}{0.001} & \tabvalue{0.480}{0.121}& \tabvalue{0.698}{0.025} & \tabvalue{0.427}{0.483} & \tabvalue{0.021}{0.001} & \tabvalue{0.491}{0.079}& \tabvalue{0.702}{0.008} & \tabvalue{0.443}{0.487} & \tabvalue{0.019}{0.001} & \tabvalue{0.491}{0.076}\\

& \variable{ED-v} & \tabvalue{0.020}{0.038} & \tabvalue{0.001}{0.031} & \tabvalue{0.010}{0.007} & \tabvalue{0.117}{0.137}& \tabvalue{0.000}{0.000} & \tabvalue{0.000}{0.000} & \tabvalue{0.000}{0.000} & \tabvalue{0.017}{0.042}& \tabvalue{0.000}{0.000} & \tabvalue{0.000}{0.000} & \tabvalue{0.000}{0.000} & \tabvalue{0.008}{0.030}& \tabvalue{0.000}{0.000} & \tabvalue{0.000}{0.000} & \tabvalue{0.000}{0.000} & \tabvalue{0.000}{0.000}\\

& \variable{ED-s} & \tabvalue{0.766}{0.079} & \tabvalue{0.322}{0.458} & \tabvalue{0.112}{0.016} & \tabvalue{0.651}{0.137}& \tabvalue{0.669}{0.011} & \tabvalue{0.554}{0.477} & \tabvalue{0.024}{0.001} & \tabvalue{0.454}{0.121}& \tabvalue{0.698}{0.026} & \tabvalue{0.442}{0.479} & \tabvalue{0.021}{0.001} & \tabvalue{0.453}{0.106}& \tabvalue{0.702}{0.008} & \tabvalue{0.353}{0.468} & \tabvalue{0.019}{0.001} & \tabvalue{0.440}{0.097}\\

& \variable{ED-o} & \tabvalue{0.766}{0.079} & \tabvalue{0.322}{0.458} & \tabvalue{0.116}{0.018} & \tabvalue{0.657}{0.137}& \tabvalue{0.669}{0.011} & \tabvalue{0.554}{0.477} & \tabvalue{0.024}{0.001} & \tabvalue{0.454}{0.121}& \tabvalue{0.698}{0.026} & \tabvalue{0.442}{0.479} & \tabvalue{0.021}{0.001} & \tabvalue{0.453}{0.106}& \tabvalue{0.702}{0.008} & \tabvalue{0.353}{0.468} & \tabvalue{0.019}{0.001} & \tabvalue{0.440}{0.097}\\
             \midrule
            \multirow{5}{*}{\rotatebox{90}{\parbox{40pt}\centering\textbf{LR}}} 
             & \variable{BD} & \tabvalue{0.949}{0.188} & \tabvalue{0.003}{0.054} & \tabvalue{0.027}{0.068} & \tabvalue{0.042}{0.080}& \tabvalue{0.585}{0.478} & \tabvalue{0.000}{0.000} & \tabvalue{0.002}{0.016} & \tabvalue{0.002}{0.018}& \tabvalue{0.532}{0.482} & \tabvalue{0.000}{0.000} & \tabvalue{0.000}{0.000} & \tabvalue{0.000}{0.000}& \tabvalue{0.551}{0.480} & \tabvalue{0.000}{0.000} & \tabvalue{0.000}{0.000} & \tabvalue{0.000}{0.000}\\

& \variable{MD} & \tabvalue{0.847}{0.345} & \tabvalue{0.173}{0.374} & \tabvalue{0.107}{0.218} & \tabvalue{0.165}{0.246}& \tabvalue{0.041}{0.191} & \tabvalue{0.000}{0.000} & \tabvalue{0.002}{0.014} & \tabvalue{0.004}{0.025}& \tabvalue{0.031}{0.166} & \tabvalue{0.000}{0.000} & \tabvalue{0.009}{0.032} & \tabvalue{0.018}{0.062}& \tabvalue{0.040}{0.190} & \tabvalue{0.000}{0.000} & \tabvalue{0.003}{0.019} & \tabvalue{0.008}{0.041}\\

& \variable{ED-v} & \tabvalue{0.000}{0.000} & \tabvalue{0.000}{0.000} & \tabvalue{0.005}{0.010} & \tabvalue{0.057}{0.122}& \tabvalue{0.000}{0.000} & \tabvalue{0.000}{0.000} & \tabvalue{0.000}{0.000} & \tabvalue{0.001}{0.004}& \tabvalue{0.000}{0.000} & \tabvalue{0.000}{0.000} & \tabvalue{0.000}{0.000} & \tabvalue{0.000}{0.001}& \tabvalue{0.000}{0.000} & \tabvalue{0.000}{0.000} & \tabvalue{0.000}{0.000} & \tabvalue{0.000}{0.000}\\

& \variable{ED-s} & \tabvalue{0.858}{0.333} & \tabvalue{0.171}{0.373} & \tabvalue{0.032}{0.053} & \tabvalue{0.230}{0.234}& \tabvalue{0.001}{0.003} & \tabvalue{0.000}{0.000} & \tabvalue{0.005}{0.033} & \tabvalue{0.047}{0.141}& \tabvalue{0.019}{0.132} & \tabvalue{0.000}{0.000} & \tabvalue{0.013}{0.082} & \tabvalue{0.044}{0.158}& \tabvalue{0.000}{0.001} & \tabvalue{0.000}{0.000} & \tabvalue{0.000}{0.000} & \tabvalue{0.000}{0.000}\\

& \variable{ED-o} & \tabvalue{0.836}{0.354} & \tabvalue{0.150}{0.353} & \tabvalue{0.041}{0.050} & \tabvalue{0.282}{0.241}& \tabvalue{0.002}{0.003} & \tabvalue{0.000}{0.000} & \tabvalue{0.006}{0.033} & \tabvalue{0.066}{0.137}& \tabvalue{0.002}{0.003} & \tabvalue{0.000}{0.000} & \tabvalue{0.001}{0.004} & \tabvalue{0.041}{0.095}& \tabvalue{0.001}{0.003} & \tabvalue{0.000}{0.000} & \tabvalue{0.000}{0.000} & \tabvalue{0.015}{0.004}\\
             \midrule
            \multirow{5}{*}{\rotatebox{90}{\parbox{40pt}\centering\textbf{HGB}}} 
             & \variable{BD} & \tabvalue{0.814}{0.084} & \tabvalue{0.211}{0.394} & \tabvalue{0.111}{0.018} & \tabvalue{0.725}{0.082}& \tabvalue{0.848}{0.023} & \tabvalue{0.000}{0.000} & \tabvalue{0.031}{0.001} & \tabvalue{0.606}{0.044}& \tabvalue{0.864}{0.029} & \tabvalue{0.006}{0.041} & \tabvalue{0.030}{0.001} & \tabvalue{0.605}{0.035}& \tabvalue{0.868}{0.025} & \tabvalue{0.000}{0.000} & \tabvalue{0.029}{0.001} & \tabvalue{0.618}{0.024}\\

& \variable{MD} & \tabvalue{0.804}{0.083} & \tabvalue{0.180}{0.369} & \tabvalue{0.105}{0.018} & \tabvalue{0.644}{0.117}& \tabvalue{0.858}{0.011} & \tabvalue{0.001}{0.002} & \tabvalue{0.027}{0.001} & \tabvalue{0.581}{0.053}& \tabvalue{0.882}{0.017} & \tabvalue{0.020}{0.137} & \tabvalue{0.026}{0.001} & \tabvalue{0.584}{0.046}& \tabvalue{0.890}{0.007} & \tabvalue{0.001}{0.002} & \tabvalue{0.025}{0.001} & \tabvalue{0.590}{0.049}\\

& \variable{ED-v} & \tabvalue{0.008}{0.020} & \tabvalue{0.002}{0.044} & \tabvalue{0.005}{0.005} & \tabvalue{0.166}{0.146}& \tabvalue{0.000}{0.000} & \tabvalue{0.000}{0.000} & \tabvalue{0.000}{0.000} & \tabvalue{0.006}{0.025}& \tabvalue{0.000}{0.000} & \tabvalue{0.000}{0.000} & \tabvalue{0.000}{0.000} & \tabvalue{0.000}{0.000}& \tabvalue{0.000}{0.000} & \tabvalue{0.000}{0.000} & \tabvalue{0.000}{0.000} & \tabvalue{0.000}{0.000}\\

& \variable{ED-s} & \tabvalue{0.814}{0.087} & \tabvalue{0.154}{0.334} & \tabvalue{0.114}{0.019} & \tabvalue{0.680}{0.104}& \tabvalue{0.870}{0.018} & \tabvalue{0.001}{0.002} & \tabvalue{0.030}{0.001} & \tabvalue{0.620}{0.042}& \tabvalue{0.894}{0.017} & \tabvalue{0.028}{0.153} & \tabvalue{0.029}{0.001} & \tabvalue{0.615}{0.037}& \tabvalue{0.905}{0.009} & \tabvalue{0.003}{0.020} & \tabvalue{0.028}{0.001} & \tabvalue{0.634}{0.028}\\

& \variable{ED-o} & \tabvalue{0.814}{0.087} & \tabvalue{0.154}{0.334} & \tabvalue{0.120}{0.021} & \tabvalue{0.687}{0.103}& \tabvalue{0.870}{0.018} & \tabvalue{0.001}{0.002} & \tabvalue{0.030}{0.001} & \tabvalue{0.620}{0.042}& \tabvalue{0.895}{0.017} & \tabvalue{0.029}{0.159} & \tabvalue{0.029}{0.001} & \tabvalue{0.616}{0.033}& \tabvalue{0.905}{0.009} & \tabvalue{0.003}{0.020} & \tabvalue{0.028}{0.001} & \tabvalue{0.634}{0.028}\\
            
            \bottomrule
        \end{tabular}
        }
                
        \caption{\textit{Static Dependency}: Results by assuming the absence of temporal dependencies among samples (\scbb{T} and \scbb{E} are randomly sampled from \scbb{D}).} 
    \label{tab:gtcs_open_static}
    \end{subtable}

    \begin{subtable}[htbp]{1.8\columnwidth}
        \resizebox{1.0\columnwidth}{!}{
        \begin{tabular}{c|c ? cc|cc? cc|cc? cc|cc? cc|cc}
\multicolumn{2}{c?}{Available Data} & \multicolumn{4}{c?}{Limited (100 per class) [N=1]} & \multicolumn{4}{c?}{Scarce (15\% of \scbb{D}) [N=1]} &  \multicolumn{4}{c?}{Moderate (40\% of \scbb{D}) [N=1]} &  \multicolumn{4}{c}{Abundant (80\% of \scbb{D}) [N=1]} \\ \hline
             \multicolumn{2}{c?}{Scenario} &
             \multicolumn{2}{c|}{Adversarial Attacks} & \multicolumn{2}{c?}{Unknown Attacks} & \multicolumn{2}{c|}{Adversarial Attacks} & \multicolumn{2}{c?}{Unknown Attacks} & \multicolumn{2}{c|}{Adversarial Attacks} & \multicolumn{2}{c?}{Unknown Attacks} & \multicolumn{2}{c|}{Adversarial Attacks} & \multicolumn{2}{c?}{Unknown Attacks}  \\ \hline
            Alg. & Design & 
            \footnotesize{$tpr$} \tiny{(org)} & \footnotesize{$tpr$} \tiny{(adv)} & \footnotesize{$fpr$} & \footnotesize{$tpr$} & \footnotesize{$tpr$} \tiny{(org)} & \footnotesize{$tpr$} \tiny{(adv)} & \footnotesize{$fpr$} & \footnotesize{$tpr$} & \footnotesize{$tpr$} \tiny{(org)} & \footnotesize{$tpr$} \tiny{(adv)} & \footnotesize{$fpr$} & \footnotesize{$tpr$} & \footnotesize{$tpr$} \tiny{(org)} & \footnotesize{$tpr$} \tiny{(adv)} & \footnotesize{$fpr$} & \footnotesize{$tpr$} \\
             \toprule
             
             \multirow{5}{*}{\rotatebox{90}{\parbox{40pt}\centering\textbf{RF}}} 
             & \variable{BD} & \tabvaluetemp{0.967} & \tabvaluetemp{0.977} & \tabvaluetemp{0.073} & \tabvaluetemp{0.377}& \tabvaluetemp{0.234} & \tabvaluetemp{0.014} & \tabvaluetemp{0.018} & \tabvaluetemp{0.621}& \tabvaluetemp{0.909} & \tabvaluetemp{0.978} & \tabvaluetemp{0.022} & \tabvaluetemp{0.457}& \tabvaluetemp{0.851} & \tabvaluetemp{0.978} & \tabvaluetemp{0.020} & \tabvaluetemp{0.571}\\

& \variable{MD} & \tabvaluetemp{0.968} & \tabvaluetemp{0.149} & \tabvaluetemp{0.070} & \tabvaluetemp{0.132}& \tabvaluetemp{0.258} & \tabvaluetemp{0.001} & \tabvaluetemp{0.017} & \tabvaluetemp{0.507}& \tabvaluetemp{0.864} & \tabvaluetemp{0.978} & \tabvaluetemp{0.022} & \tabvaluetemp{0.566}& \tabvaluetemp{0.881} & \tabvaluetemp{0.366} & \tabvaluetemp{0.019} & \tabvaluetemp{0.476}\\

& \variable{ED-v} & \tabvaluetemp{0.000} & \tabvaluetemp{0.000} & \tabvaluetemp{0.001} & \tabvaluetemp{0.000}& \tabvaluetemp{0.000} & \tabvaluetemp{0.000} & \tabvaluetemp{0.000} & \tabvaluetemp{0.000}& \tabvaluetemp{0.000} & \tabvaluetemp{0.000} & \tabvaluetemp{0.000} & \tabvaluetemp{0.000}& \tabvaluetemp{0.000} & \tabvaluetemp{0.000} & \tabvaluetemp{0.000} & \tabvaluetemp{0.000}\\

& \variable{ED-s} & \tabvaluetemp{0.972} & \tabvaluetemp{0.975} & \tabvaluetemp{0.063} & \tabvaluetemp{0.478}& \tabvaluetemp{0.243} & \tabvaluetemp{0.001} & \tabvaluetemp{0.018} & \tabvaluetemp{0.506}& \tabvaluetemp{0.915} & \tabvaluetemp{0.978} & \tabvaluetemp{0.022} & \tabvaluetemp{0.565}& \tabvaluetemp{0.847} & \tabvaluetemp{0.001} & \tabvaluetemp{0.019} & \tabvaluetemp{0.441}\\

& \variable{ED-o} & \tabvaluetemp{0.972} & \tabvaluetemp{0.975} & \tabvaluetemp{0.063} & \tabvaluetemp{0.478}& \tabvaluetemp{0.243} & \tabvaluetemp{0.001} & \tabvaluetemp{0.018} & \tabvaluetemp{0.506}& \tabvaluetemp{0.915} & \tabvaluetemp{0.978} & \tabvaluetemp{0.022} & \tabvaluetemp{0.565}& \tabvaluetemp{0.847} & \tabvaluetemp{0.001} & \tabvaluetemp{0.019} & \tabvaluetemp{0.441}\\
             
            \midrule
            \multirow{5}{*}{\rotatebox{90}{\parbox{40pt}\centering\textbf{DT}}} 
             & \variable{BD} & \tabvaluetemp{0.742} & \tabvaluetemp{0.982} & \tabvaluetemp{0.094} & \tabvaluetemp{0.560}& \tabvaluetemp{0.599} & \tabvaluetemp{0.992} & \tabvaluetemp{0.021} & \tabvaluetemp{0.476}& \tabvaluetemp{0.714} & \tabvaluetemp{0.993} & \tabvaluetemp{0.021} & \tabvaluetemp{0.580}& \tabvaluetemp{0.698} & \tabvaluetemp{0.993} & \tabvaluetemp{0.019} & \tabvaluetemp{0.377}\\

& \variable{MD} & \tabvaluetemp{0.787} & \tabvaluetemp{0.000} & \tabvaluetemp{0.098} & \tabvaluetemp{0.268}& \tabvaluetemp{0.598} & \tabvaluetemp{0.988} & \tabvaluetemp{0.021} & \tabvaluetemp{0.545}& \tabvaluetemp{0.713} & \tabvaluetemp{0.982} & \tabvaluetemp{0.021} & \tabvaluetemp{0.367}& \tabvaluetemp{0.699} & \tabvaluetemp{0.989} & \tabvaluetemp{0.019} & \tabvaluetemp{0.514}\\

& \variable{ED-v} & \tabvaluetemp{0.000} & \tabvaluetemp{0.000} & \tabvaluetemp{0.006} & \tabvaluetemp{0.249}& \tabvaluetemp{0.000} & \tabvaluetemp{0.000} & \tabvaluetemp{0.000} & \tabvaluetemp{0.125}& \tabvaluetemp{0.000} & \tabvaluetemp{0.000} & \tabvaluetemp{0.000} & \tabvaluetemp{0.000}& \tabvaluetemp{0.000} & \tabvaluetemp{0.000} & \tabvaluetemp{0.000} & \tabvaluetemp{0.000}\\

& \variable{ED-s} & \tabvaluetemp{0.816} & \tabvaluetemp{0.011} & \tabvaluetemp{0.112} & \tabvaluetemp{0.517}& \tabvaluetemp{0.598} & \tabvaluetemp{0.000} & \tabvaluetemp{0.020} & \tabvaluetemp{0.519}& \tabvaluetemp{0.713} & \tabvaluetemp{0.978} & \tabvaluetemp{0.020} & \tabvaluetemp{0.475}& \tabvaluetemp{0.698} & \tabvaluetemp{0.000} & \tabvaluetemp{0.019} & \tabvaluetemp{0.604}\\

& \variable{ED-o} & \tabvaluetemp{0.816} & \tabvaluetemp{0.011} & \tabvaluetemp{0.114} & \tabvaluetemp{0.522}& \tabvaluetemp{0.598} & \tabvaluetemp{0.000} & \tabvaluetemp{0.020} & \tabvaluetemp{0.519}& \tabvaluetemp{0.713} & \tabvaluetemp{0.978} & \tabvaluetemp{0.020} & \tabvaluetemp{0.475}& \tabvaluetemp{0.698} & \tabvaluetemp{0.000} & \tabvaluetemp{0.019} & \tabvaluetemp{0.604}\\
             \midrule
            \multirow{5}{*}{\rotatebox{90}{\parbox{40pt}\centering\textbf{LR}}} 
             & \variable{BD} & \tabvaluetemp{0.979} & \tabvaluetemp{0.000} & \tabvaluetemp{0.000} & \tabvaluetemp{0.000}& \tabvaluetemp{0.000} & \tabvaluetemp{0.000} & \tabvaluetemp{0.000} & \tabvaluetemp{0.000}& \tabvaluetemp{0.000} & \tabvaluetemp{0.000} & \tabvaluetemp{0.000} & \tabvaluetemp{0.000}& \tabvaluetemp{0.977} & \tabvaluetemp{0.000} & \tabvaluetemp{0.000} & \tabvaluetemp{0.000}\\

& \variable{MD} & \tabvaluetemp{0.981} & \tabvaluetemp{0.002} & \tabvaluetemp{0.025} & \tabvaluetemp{0.087}& \tabvaluetemp{0.000} & \tabvaluetemp{0.000} & \tabvaluetemp{0.000} & \tabvaluetemp{0.000}& \tabvaluetemp{0.000} & \tabvaluetemp{0.000} & \tabvaluetemp{0.000} & \tabvaluetemp{0.000}& \tabvaluetemp{0.010} & \tabvaluetemp{0.000} & \tabvaluetemp{0.000} & \tabvaluetemp{0.000}\\

& \variable{ED-v} & \tabvaluetemp{0.000} & \tabvaluetemp{0.000} & \tabvaluetemp{0.018} & \tabvaluetemp{0.313}& \tabvaluetemp{0.000} & \tabvaluetemp{0.000} & \tabvaluetemp{0.000} & \tabvaluetemp{0.000}& \tabvaluetemp{0.000} & \tabvaluetemp{0.000} & \tabvaluetemp{0.000} & \tabvaluetemp{0.000}& \tabvaluetemp{0.000} & \tabvaluetemp{0.000} & \tabvaluetemp{0.000} & \tabvaluetemp{0.000}\\

& \variable{ED-s} & \tabvaluetemp{0.981} & \tabvaluetemp{0.002} & \tabvaluetemp{0.051} & \tabvaluetemp{0.771}& \tabvaluetemp{0.000} & \tabvaluetemp{0.000} & \tabvaluetemp{0.000} & \tabvaluetemp{0.000}& \tabvaluetemp{0.000} & \tabvaluetemp{0.000} & \tabvaluetemp{0.000} & \tabvaluetemp{0.000}& \tabvaluetemp{0.000} & \tabvaluetemp{0.000} & \tabvaluetemp{0.000} & \tabvaluetemp{0.000}\\

& \variable{ED-o} & \tabvaluetemp{0.981} & \tabvaluetemp{0.002} & \tabvaluetemp{0.054} & \tabvaluetemp{0.772}& \tabvaluetemp{0.000} & \tabvaluetemp{0.000} & \tabvaluetemp{0.001} & \tabvaluetemp{0.024}& \tabvaluetemp{0.000} & \tabvaluetemp{0.000} & \tabvaluetemp{0.001} & \tabvaluetemp{0.020}& \tabvaluetemp{0.000} & \tabvaluetemp{0.000} & \tabvaluetemp{0.000} & \tabvaluetemp{0.014}\\
             \midrule
            \multirow{5}{*}{\rotatebox{90}{\parbox{40pt}\centering\textbf{HGB}}} 
             & \variable{BD} & \tabvaluetemp{0.868} & \tabvaluetemp{0.683} & \tabvaluetemp{0.090} & \tabvaluetemp{0.636}& \tabvaluetemp{0.732} & \tabvaluetemp{0.000} & \tabvaluetemp{0.025} & \tabvaluetemp{0.498}& \tabvaluetemp{0.933} & \tabvaluetemp{0.000} & \tabvaluetemp{0.028} & \tabvaluetemp{0.605}& \tabvaluetemp{0.857} & \tabvaluetemp{0.000} & \tabvaluetemp{0.028} & \tabvaluetemp{0.620}\\

& \variable{MD} & \tabvaluetemp{0.867} & \tabvaluetemp{0.612} & \tabvaluetemp{0.082} & \tabvaluetemp{0.385}& \tabvaluetemp{0.764} & \tabvaluetemp{0.000} & \tabvaluetemp{0.022} & \tabvaluetemp{0.498}& \tabvaluetemp{0.930} & \tabvaluetemp{0.000} & \tabvaluetemp{0.025} & \tabvaluetemp{0.513}& \tabvaluetemp{0.911} & \tabvaluetemp{0.001} & \tabvaluetemp{0.024} & \tabvaluetemp{0.623}\\

& \variable{ED-v} & \tabvaluetemp{0.020} & \tabvaluetemp{0.000} & \tabvaluetemp{0.002} & \tabvaluetemp{0.026}& \tabvaluetemp{0.002} & \tabvaluetemp{0.000} & \tabvaluetemp{0.000} & \tabvaluetemp{0.000}& \tabvaluetemp{0.000} & \tabvaluetemp{0.000} & \tabvaluetemp{0.000} & \tabvaluetemp{0.000}& \tabvaluetemp{0.000} & \tabvaluetemp{0.000} & \tabvaluetemp{0.000} & \tabvaluetemp{0.000}\\

& \variable{ED-s} & \tabvaluetemp{0.859} & \tabvaluetemp{0.992} & \tabvaluetemp{0.085} & \tabvaluetemp{0.634}& \tabvaluetemp{0.702} & \tabvaluetemp{0.000} & \tabvaluetemp{0.024} & \tabvaluetemp{0.581}& \tabvaluetemp{0.938} & \tabvaluetemp{0.002} & \tabvaluetemp{0.026} & \tabvaluetemp{0.665}& \tabvaluetemp{0.923} & \tabvaluetemp{0.002} & \tabvaluetemp{0.027} & \tabvaluetemp{0.622}\\

& \variable{ED-o} & \tabvaluetemp{0.859} & \tabvaluetemp{0.992} & \tabvaluetemp{0.086} & \tabvaluetemp{0.634}& \tabvaluetemp{0.702} & \tabvaluetemp{0.000} & \tabvaluetemp{0.024} & \tabvaluetemp{0.581}& \tabvaluetemp{0.938} & \tabvaluetemp{0.002} & \tabvaluetemp{0.026} & \tabvaluetemp{0.665}& \tabvaluetemp{0.923} & \tabvaluetemp{0.002} & \tabvaluetemp{0.027} & \tabvaluetemp{0.622}\\
            
            \bottomrule
        \end{tabular}
        }
                
        \caption{\textit{Temporal Dependency}: Results by assuming the presence of temporal dependencies among samples (the `first' samples of \scbb{D} are put in \scbb{T}, while the last 20\% represent \scbb{E}).} 
    \label{tab:gtcs_open_temporal}
    \end{subtable}

    \label{tab:gtcs_open}
\end{table*}

\begin{table*}
  \centering
  \caption{\dataset{NB15} binary classification results (\scmath{fpr} and \scmath{tpr}) against `known' attacks seen during the training stage (closed world).}

    \begin{subtable}[htbp]{1.99\columnwidth}
        \resizebox{1.0\columnwidth}{!}{
        \begin{tabular}{c|c ? cc|cc? cc|cc? cc|cc? cc|cc}
             \multicolumn{2}{c?}{Available Data} & \multicolumn{4}{c?}{Limited (100 per class) [N=1000]} & \multicolumn{4}{c?}{Scarce (15\% of \scbb{D}) [N=100]} &  \multicolumn{4}{c?}{Moderate (40\% of \scbb{D}) [N=100]} &  \multicolumn{4}{c}{Abundant (80\% of \scbb{D}) [N=100]} \\ \hline
             \multicolumn{2}{c?}{Features} &
             \multicolumn{2}{c|}{Complete} & \multicolumn{2}{c?}{Essential} & \multicolumn{2}{c|}{Complete} & \multicolumn{2}{c?}{Essential} & \multicolumn{2}{c|}{Complete} & \multicolumn{2}{c?}{Essential} &  
             \multicolumn{2}{c|}{Complete} & \multicolumn{2}{c}{Essential} \\ \hline
            Alg. & Design & \smamath{fpr} & \smamath{tpr} & \smamath{fpr} & \smamath{tpr} & \smamath{fpr} & \smamath{tpr} & \smamath{fpr} & \smamath{tpr} & \smamath{fpr} & \smamath{tpr} & \smamath{fpr} & \smamath{tpr} & \smamath{fpr} & \smamath{tpr} & \smamath{fpr} & \smamath{tpr} \\
             \toprule
             
             \multirow{6}{*}{\rotatebox{90}{\parbox{40pt}\centering\textbf{RF}}} 
             & \variable{BD} & \tabvalue{0.021}{0.002} & \tabvalue{1.000}{0.000} & \tabvalue{0.022}{0.002} & \tabvalue{0.997}{0.001}& \tabvalue{0.010}{0.000} & \tabvalue{0.992}{0.001} & \tabvalue{0.010}{0.000} & \tabvalue{0.989}{0.001}& \tabvalue{0.010}{0.000} & \tabvalue{0.991}{0.001} & \tabvalue{0.009}{0.000} & \tabvalue{0.989}{0.001}& \tabvalue{0.009}{0.000} & \tabvalue{0.991}{0.001} & \tabvalue{0.009}{0.000} & \tabvalue{0.989}{0.001}\\

& \variable{MD} & \tabvalue{0.020}{0.001} & \tabvalue{0.998}{0.001} & \tabvalue{0.020}{0.002} & \tabvalue{0.995}{0.001}& \tabvalue{0.010}{0.000} & \tabvalue{0.987}{0.001} & \tabvalue{0.009}{0.000} & \tabvalue{0.984}{0.001}& \tabvalue{0.009}{0.000} & \tabvalue{0.988}{0.001} & \tabvalue{0.009}{0.000} & \tabvalue{0.985}{0.001}& \tabvalue{0.008}{0.000} & \tabvalue{0.988}{0.001} & \tabvalue{0.008}{0.000} & \tabvalue{0.986}{0.001}\\

& \variable{ED-v} & \tabvalue{0.019}{0.002} & \tabvalue{0.970}{0.035} & \tabvalue{0.018}{0.002} & \tabvalue{0.959}{0.040}& \tabvalue{0.000}{0.000} & \tabvalue{0.531}{0.023} & \tabvalue{0.000}{0.000} & \tabvalue{0.554}{0.021}& \tabvalue{0.000}{0.000} & \tabvalue{0.564}{0.019} & \tabvalue{0.000}{0.000} & \tabvalue{0.575}{0.015}& \tabvalue{0.000}{0.000} & \tabvalue{0.579}{0.015} & \tabvalue{0.000}{0.000} & \tabvalue{0.586}{0.013}\\

& \variable{ED-s} & \tabvalue{0.021}{0.002} & \tabvalue{0.999}{0.001} & \tabvalue{0.023}{0.002} & \tabvalue{0.996}{0.001}& \tabvalue{0.010}{0.000} & \tabvalue{0.988}{0.001} & \tabvalue{0.009}{0.000} & \tabvalue{0.985}{0.001}& \tabvalue{0.009}{0.000} & \tabvalue{0.988}{0.001} & \tabvalue{0.008}{0.000} & \tabvalue{0.986}{0.001}& \tabvalue{0.008}{0.000} & \tabvalue{0.988}{0.001} & \tabvalue{0.008}{0.000} & \tabvalue{0.987}{0.001}\\

& \variable{ED-o} & \tabvalue{0.022}{0.002} & \tabvalue{1.000}{0.001} & \tabvalue{0.023}{0.002} & \tabvalue{0.997}{0.001}& \tabvalue{0.010}{0.000} & \tabvalue{0.988}{0.001} & \tabvalue{0.009}{0.000} & \tabvalue{0.985}{0.001}& \tabvalue{0.009}{0.000} & \tabvalue{0.988}{0.001} & \tabvalue{0.008}{0.000} & \tabvalue{0.986}{0.001}& \tabvalue{0.008}{0.000} & \tabvalue{0.988}{0.001} & \tabvalue{0.008}{0.000} & \tabvalue{0.987}{0.001}\\

& \variable{ED} & \tabvalue{0.018}{0.002} & \tabvalue{0.997}{0.002} & \tabvalue{0.017}{0.002} & \tabvalue{0.991}{0.003}& \tabvalue{0.002}{0.000} & \tabvalue{0.978}{0.002} & \tabvalue{0.002}{0.000} & \tabvalue{0.975}{0.002}& \tabvalue{0.001}{0.000} & \tabvalue{0.981}{0.001} & \tabvalue{0.001}{0.000} & \tabvalue{0.979}{0.001}& \tabvalue{0.001}{0.000} & \tabvalue{0.983}{0.001} & \tabvalue{0.001}{0.000} & \tabvalue{0.982}{0.001}\\
             
            \midrule
            \multirow{6}{*}{\rotatebox{90}{\parbox{40pt}\centering\textbf{DT}}} 
            & \variable{BD} & \tabvalue{0.022}{0.004} & \tabvalue{0.996}{0.004} & \tabvalue{0.028}{0.012} & \tabvalue{0.994}{0.004}& \tabvalue{0.008}{0.000} & \tabvalue{0.958}{0.002} & \tabvalue{0.008}{0.000} & \tabvalue{0.958}{0.002}& \tabvalue{0.008}{0.000} & \tabvalue{0.962}{0.002} & \tabvalue{0.008}{0.000} & \tabvalue{0.962}{0.002}& \tabvalue{0.007}{0.000} & \tabvalue{0.965}{0.001} & \tabvalue{0.007}{0.000} & \tabvalue{0.966}{0.001}\\

& \variable{MD} & \tabvalue{0.022}{0.004} & \tabvalue{0.996}{0.004} & \tabvalue{0.028}{0.012} & \tabvalue{0.994}{0.004}& \tabvalue{0.008}{0.000} & \tabvalue{0.959}{0.002} & \tabvalue{0.008}{0.000} & \tabvalue{0.959}{0.002}& \tabvalue{0.008}{0.000} & \tabvalue{0.962}{0.002} & \tabvalue{0.008}{0.000} & \tabvalue{0.963}{0.002}& \tabvalue{0.007}{0.000} & \tabvalue{0.966}{0.001} & \tabvalue{0.007}{0.000} & \tabvalue{0.966}{0.001}\\

& \variable{ED-v} & \tabvalue{0.017}{0.002} & \tabvalue{0.932}{0.068} & \tabvalue{0.016}{0.002} & \tabvalue{0.955}{0.045}& \tabvalue{0.000}{0.000} & \tabvalue{0.613}{0.028} & \tabvalue{0.000}{0.000} & \tabvalue{0.641}{0.031}& \tabvalue{0.000}{0.000} & \tabvalue{0.616}{0.026} & \tabvalue{0.000}{0.000} & \tabvalue{0.637}{0.027}& \tabvalue{0.000}{0.000} & \tabvalue{0.610}{0.022} & \tabvalue{0.000}{0.000} & \tabvalue{0.628}{0.022}\\

& \variable{ED-s} & \tabvalue{0.022}{0.004} & \tabvalue{0.997}{0.003} & \tabvalue{0.029}{0.012} & \tabvalue{0.995}{0.002}& \tabvalue{0.009}{0.000} & \tabvalue{0.967}{0.002} & \tabvalue{0.008}{0.000} & \tabvalue{0.966}{0.002}& \tabvalue{0.008}{0.000} & \tabvalue{0.969}{0.002} & \tabvalue{0.008}{0.000} & \tabvalue{0.968}{0.001}& \tabvalue{0.007}{0.000} & \tabvalue{0.970}{0.001} & \tabvalue{0.007}{0.000} & \tabvalue{0.971}{0.001}\\

& \variable{ED-o} & \tabvalue{0.023}{0.005} & \tabvalue{0.999}{0.002} & \tabvalue{0.030}{0.013} & \tabvalue{0.996}{0.002}& \tabvalue{0.009}{0.000} & \tabvalue{0.967}{0.002} & \tabvalue{0.008}{0.000} & \tabvalue{0.967}{0.002}& \tabvalue{0.008}{0.000} & \tabvalue{0.969}{0.002} & \tabvalue{0.008}{0.000} & \tabvalue{0.968}{0.001}& \tabvalue{0.007}{0.000} & \tabvalue{0.970}{0.001} & \tabvalue{0.007}{0.000} & \tabvalue{0.971}{0.001}\\

& \variable{ED} & \tabvalue{0.016}{0.002} & \tabvalue{0.982}{0.012} & \tabvalue{0.017}{0.003} & \tabvalue{0.981}{0.011}& \tabvalue{0.002}{0.000} & \tabvalue{0.946}{0.002} & \tabvalue{0.002}{0.000} & \tabvalue{0.946}{0.002}& \tabvalue{0.001}{0.000} & \tabvalue{0.953}{0.002} & \tabvalue{0.001}{0.000} & \tabvalue{0.953}{0.002}& \tabvalue{0.001}{0.000} & \tabvalue{0.959}{0.002} & \tabvalue{0.001}{0.000} & \tabvalue{0.959}{0.002}\\

             \midrule
            \multirow{6}{*}{\rotatebox{90}{\parbox{40pt}\centering\textbf{LR}}} 
             & \variable{BD} & \tabvalue{0.361}{0.016} & \tabvalue{0.973}{0.006} & \tabvalue{0.462}{0.199} & \tabvalue{0.963}{0.088}& \tabvalue{0.002}{0.001} & \tabvalue{0.047}{0.007} & \tabvalue{0.008}{0.004} & \tabvalue{0.450}{0.153}& \tabvalue{0.003}{0.001} & \tabvalue{0.048}{0.007} & \tabvalue{0.008}{0.004} & \tabvalue{0.458}{0.157}& \tabvalue{0.003}{0.001} & \tabvalue{0.046}{0.006} & \tabvalue{0.008}{0.005} & \tabvalue{0.456}{0.148}\\

& \variable{MD} & \tabvalue{0.499}{0.201} & \tabvalue{0.873}{0.140} & \tabvalue{0.636}{0.295} & \tabvalue{0.869}{0.142}& \tabvalue{0.001}{0.000} & \tabvalue{0.005}{0.002} & \tabvalue{0.001}{0.006} & \tabvalue{0.022}{0.024}& \tabvalue{0.001}{0.000} & \tabvalue{0.006}{0.003} & \tabvalue{0.001}{0.006} & \tabvalue{0.021}{0.031}& \tabvalue{0.001}{0.000} & \tabvalue{0.006}{0.004} & \tabvalue{0.000}{0.000} & \tabvalue{0.020}{0.019}\\

& \variable{ED-v} & \tabvalue{0.147}{0.008} & \tabvalue{0.853}{0.021} & \tabvalue{0.019}{0.010} & \tabvalue{0.673}{0.056}& \tabvalue{0.000}{0.000} & \tabvalue{0.000}{0.000} & \tabvalue{0.001}{0.001} & \tabvalue{0.327}{0.024}& \tabvalue{0.000}{0.000} & \tabvalue{0.000}{0.000} & \tabvalue{0.001}{0.001} & \tabvalue{0.338}{0.023}& \tabvalue{0.000}{0.000} & \tabvalue{0.000}{0.000} & \tabvalue{0.001}{0.000} & \tabvalue{0.335}{0.021}\\

& \variable{ED-s} & \tabvalue{0.305}{0.053} & \tabvalue{0.951}{0.021} & \tabvalue{0.169}{0.054} & \tabvalue{0.961}{0.024}& \tabvalue{0.013}{0.001} & \tabvalue{0.201}{0.012} & \tabvalue{0.009}{0.003} & \tabvalue{0.598}{0.060}& \tabvalue{0.013}{0.000} & \tabvalue{0.201}{0.007} & \tabvalue{0.008}{0.002} & \tabvalue{0.578}{0.049}& \tabvalue{0.013}{0.000} & \tabvalue{0.203}{0.005} & \tabvalue{0.007}{0.002} & \tabvalue{0.562}{0.045}\\

& \variable{ED-o} & \tabvalue{0.329}{0.021} & \tabvalue{0.960}{0.010} & \tabvalue{0.194}{0.057} & \tabvalue{0.967}{0.037}& \tabvalue{0.013}{0.001} & \tabvalue{0.202}{0.012} & \tabvalue{0.010}{0.003} & \tabvalue{0.605}{0.061}& \tabvalue{0.013}{0.000} & \tabvalue{0.202}{0.007} & \tabvalue{0.010}{0.002} & \tabvalue{0.587}{0.049}& \tabvalue{0.013}{0.000} & \tabvalue{0.203}{0.005} & \tabvalue{0.009}{0.002} & \tabvalue{0.571}{0.046}\\

& \variable{ED} & \tabvalue{0.175}{0.013} & \tabvalue{0.924}{0.034} & \tabvalue{0.049}{0.015} & \tabvalue{0.894}{0.055}& \tabvalue{0.002}{0.000} & \tabvalue{0.101}{0.007} & \tabvalue{0.002}{0.001} & \tabvalue{0.436}{0.052}& \tabvalue{0.002}{0.000} & \tabvalue{0.101}{0.004} & \tabvalue{0.003}{0.001} & \tabvalue{0.438}{0.048}& \tabvalue{0.002}{0.000} & \tabvalue{0.101}{0.003} & \tabvalue{0.002}{0.001} & \tabvalue{0.422}{0.038}\\

             \midrule
            \multirow{6}{*}{\rotatebox{90}{\parbox{40pt}\centering\textbf{HGB}}} 
            & \variable{BD} & \tabvalue{0.022}{0.003} & \tabvalue{0.999}{0.001} & \tabvalue{0.024}{0.006} & \tabvalue{0.997}{0.001}& \tabvalue{0.010}{0.000} & \tabvalue{0.991}{0.002} & \tabvalue{0.010}{0.000} & \tabvalue{0.991}{0.002}& \tabvalue{0.010}{0.000} & \tabvalue{0.993}{0.001} & \tabvalue{0.010}{0.000} & \tabvalue{0.993}{0.001}& \tabvalue{0.010}{0.000} & \tabvalue{0.994}{0.001} & \tabvalue{0.010}{0.000} & \tabvalue{0.994}{0.001}\\

& \variable{MD} & \tabvalue{0.022}{0.003} & \tabvalue{0.999}{0.001} & \tabvalue{0.023}{0.003} & \tabvalue{0.996}{0.002}& \tabvalue{0.012}{0.001} & \tabvalue{0.986}{0.003} & \tabvalue{0.011}{0.001} & \tabvalue{0.983}{0.003}& \tabvalue{0.011}{0.001} & \tabvalue{0.989}{0.002} & \tabvalue{0.011}{0.001} & \tabvalue{0.986}{0.002}& \tabvalue{0.011}{0.000} & \tabvalue{0.990}{0.001} & \tabvalue{0.011}{0.001} & \tabvalue{0.986}{0.002}\\

& \variable{ED-v} & \tabvalue{0.016}{0.003} & \tabvalue{0.921}{0.074} & \tabvalue{0.016}{0.003} & \tabvalue{0.936}{0.060}& \tabvalue{0.000}{0.000} & \tabvalue{0.600}{0.023} & \tabvalue{0.000}{0.000} & \tabvalue{0.635}{0.027}& \tabvalue{0.000}{0.000} & \tabvalue{0.618}{0.019} & \tabvalue{0.000}{0.000} & \tabvalue{0.643}{0.022}& \tabvalue{0.000}{0.000} & \tabvalue{0.628}{0.017} & \tabvalue{0.000}{0.000} & \tabvalue{0.653}{0.021}\\

& \variable{ED-s} & \tabvalue{0.021}{0.003} & \tabvalue{0.998}{0.002} & \tabvalue{0.023}{0.005} & \tabvalue{0.996}{0.002}& \tabvalue{0.010}{0.000} & \tabvalue{0.986}{0.002} & \tabvalue{0.010}{0.000} & \tabvalue{0.986}{0.002}& \tabvalue{0.010}{0.000} & \tabvalue{0.990}{0.001} & \tabvalue{0.010}{0.000} & \tabvalue{0.990}{0.001}& \tabvalue{0.010}{0.000} & \tabvalue{0.991}{0.001} & \tabvalue{0.010}{0.000} & \tabvalue{0.991}{0.001}\\

& \variable{ED-o} & \tabvalue{0.023}{0.005} & \tabvalue{0.999}{0.001} & \tabvalue{0.026}{0.008} & \tabvalue{0.997}{0.001}& \tabvalue{0.010}{0.001} & \tabvalue{0.987}{0.002} & \tabvalue{0.010}{0.001} & \tabvalue{0.987}{0.002}& \tabvalue{0.010}{0.000} & \tabvalue{0.990}{0.001} & \tabvalue{0.010}{0.001} & \tabvalue{0.991}{0.001}& \tabvalue{0.010}{0.000} & \tabvalue{0.992}{0.001} & \tabvalue{0.010}{0.001} & \tabvalue{0.992}{0.001}\\

& \variable{ED} & \tabvalue{0.015}{0.003} & \tabvalue{0.992}{0.005} & \tabvalue{0.016}{0.003} & \tabvalue{0.989}{0.005}& \tabvalue{0.002}{0.000} & \tabvalue{0.976}{0.002} & \tabvalue{0.002}{0.000} & \tabvalue{0.975}{0.002}& \tabvalue{0.002}{0.000} & \tabvalue{0.983}{0.002} & \tabvalue{0.002}{0.000} & \tabvalue{0.982}{0.002}& \tabvalue{0.002}{0.000} & \tabvalue{0.986}{0.001} & \tabvalue{0.002}{0.000} & \tabvalue{0.985}{0.001}\\

            \bottomrule
        \end{tabular}
        }
                
        \caption{\textit{Static Dependency}: Results by assuming the absence of temporal dependencies among samples (\scbb{T} and \scbb{E} are randomly sampled from \scbb{D}).} 
    \label{tab:nb15_baseline_static}
    \end{subtable}

    \begin{subtable}[htbp]{1.8\columnwidth}
        \resizebox{1.0\columnwidth}{!}{
        \begin{tabular}{c|c ? cc|cc? cc|cc? cc|cc? cc|cc}
             \multicolumn{2}{c?}{Available Data} & \multicolumn{4}{c?}{Limited (100 per class) [N=1]} & \multicolumn{4}{c?}{Scarce (15\% of \scbb{D}) [N=1]} &  \multicolumn{4}{c?}{Moderate (40\% of \scbb{D}) [N=1]} &  \multicolumn{4}{c}{Abundant (80\% of \scbb{D}) [N=1]} \\ \hline
             \multicolumn{2}{c?}{Features} &
             \multicolumn{2}{c|}{Complete} & \multicolumn{2}{c?}{Essential} & \multicolumn{2}{c|}{Complete} & \multicolumn{2}{c?}{Essential} & \multicolumn{2}{c|}{Complete} & \multicolumn{2}{c?}{Essential} &  
             \multicolumn{2}{c|}{Complete} & \multicolumn{2}{c}{Essential} \\ \hline
            Alg. & Design & \smamath{fpr} & \smamath{tpr} & \smamath{fpr} & \smamath{tpr} & \smamath{fpr} & \smamath{tpr} & \smamath{fpr} & \smamath{tpr} & \smamath{fpr} & \smamath{tpr} & \smamath{fpr} & \smamath{tpr} & \smamath{fpr} & \smamath{tpr} & \smamath{fpr} & \smamath{tpr} \\
             \toprule
             
             \multirow{6}{*}{\rotatebox{90}{\parbox{40pt}\centering\textbf{RF}}} 
            & \variable{BD} & \tabvaluetemp{0.019} & \tabvaluetemp{1.000} & \tabvaluetemp{0.023} & \tabvaluetemp{0.997}& \tabvaluetemp{0.010} & \tabvaluetemp{0.994} & \tabvaluetemp{0.010} & \tabvaluetemp{0.990}& \tabvaluetemp{0.010} & \tabvaluetemp{0.991} & \tabvaluetemp{0.010} & \tabvaluetemp{0.988}& \tabvaluetemp{0.009} & \tabvaluetemp{0.990} & \tabvaluetemp{0.008} & \tabvaluetemp{0.988}\\

& \variable{MD} & \tabvaluetemp{0.019} & \tabvaluetemp{0.997} & \tabvaluetemp{0.019} & \tabvaluetemp{0.996}& \tabvaluetemp{0.009} & \tabvaluetemp{0.989} & \tabvaluetemp{0.009} & \tabvaluetemp{0.984}& \tabvaluetemp{0.010} & \tabvaluetemp{0.986} & \tabvaluetemp{0.009} & \tabvaluetemp{0.984}& \tabvaluetemp{0.008} & \tabvaluetemp{0.988} & \tabvaluetemp{0.008} & \tabvaluetemp{0.985}\\

& \variable{ED-v} & \tabvaluetemp{0.020} & \tabvaluetemp{0.974} & \tabvaluetemp{0.019} & \tabvaluetemp{0.982}& \tabvaluetemp{0.000} & \tabvaluetemp{0.535} & \tabvaluetemp{0.000} & \tabvaluetemp{0.566}& \tabvaluetemp{0.000} & \tabvaluetemp{0.563} & \tabvaluetemp{0.000} & \tabvaluetemp{0.573}& \tabvaluetemp{0.000} & \tabvaluetemp{0.578} & \tabvaluetemp{0.000} & \tabvaluetemp{0.582}\\

& \variable{ED-s} & \tabvaluetemp{0.020} & \tabvaluetemp{1.000} & \tabvaluetemp{0.024} & \tabvaluetemp{0.998}& \tabvaluetemp{0.009} & \tabvaluetemp{0.987} & \tabvaluetemp{0.009} & \tabvaluetemp{0.985}& \tabvaluetemp{0.009} & \tabvaluetemp{0.986} & \tabvaluetemp{0.009} & \tabvaluetemp{0.985}& \tabvaluetemp{0.008} & \tabvaluetemp{0.987} & \tabvaluetemp{0.008} & \tabvaluetemp{0.985}\\

& \variable{ED-o} & \tabvaluetemp{0.021} & \tabvaluetemp{1.000} & \tabvaluetemp{0.024} & \tabvaluetemp{0.998}& \tabvaluetemp{0.009} & \tabvaluetemp{0.987} & \tabvaluetemp{0.009} & \tabvaluetemp{0.985}& \tabvaluetemp{0.009} & \tabvaluetemp{0.986} & \tabvaluetemp{0.009} & \tabvaluetemp{0.985}& \tabvaluetemp{0.008} & \tabvaluetemp{0.987} & \tabvaluetemp{0.008} & \tabvaluetemp{0.985}\\

& \variable{ED} & \tabvaluetemp{0.019} & \tabvaluetemp{0.996} & \tabvaluetemp{0.019} & \tabvaluetemp{0.992}& \tabvaluetemp{0.002} & \tabvaluetemp{0.977} & \tabvaluetemp{0.001} & \tabvaluetemp{0.974}& \tabvaluetemp{0.001} & \tabvaluetemp{0.981} & \tabvaluetemp{0.001} & \tabvaluetemp{0.980}& \tabvaluetemp{0.001} & \tabvaluetemp{0.983} & \tabvaluetemp{0.001} & \tabvaluetemp{0.980}\\

            \midrule
            \multirow{6}{*}{\rotatebox{90}{\parbox{40pt}\centering\textbf{DT}}} 
            & \variable{BD} & \tabvaluetemp{0.020} & \tabvaluetemp{1.000} & \tabvaluetemp{0.032} & \tabvaluetemp{0.997}& \tabvaluetemp{0.008} & \tabvaluetemp{0.959} & \tabvaluetemp{0.008} & \tabvaluetemp{0.959}& \tabvaluetemp{0.007} & \tabvaluetemp{0.962} & \tabvaluetemp{0.007} & \tabvaluetemp{0.961}& \tabvaluetemp{0.007} & \tabvaluetemp{0.966} & \tabvaluetemp{0.007} & \tabvaluetemp{0.968}\\

& \variable{MD} & \tabvaluetemp{0.020} & \tabvaluetemp{1.000} & \tabvaluetemp{0.029} & \tabvaluetemp{0.997}& \tabvaluetemp{0.008} & \tabvaluetemp{0.962} & \tabvaluetemp{0.008} & \tabvaluetemp{0.960}& \tabvaluetemp{0.007} & \tabvaluetemp{0.961} & \tabvaluetemp{0.008} & \tabvaluetemp{0.962}& \tabvaluetemp{0.007} & \tabvaluetemp{0.967} & \tabvaluetemp{0.007} & \tabvaluetemp{0.969}\\

& \variable{ED-v} & \tabvaluetemp{0.019} & \tabvaluetemp{0.990} & \tabvaluetemp{0.019} & \tabvaluetemp{0.990}& \tabvaluetemp{0.000} & \tabvaluetemp{0.585} & \tabvaluetemp{0.000} & \tabvaluetemp{0.585}& \tabvaluetemp{0.000} & \tabvaluetemp{0.640} & \tabvaluetemp{0.000} & \tabvaluetemp{0.671}& \tabvaluetemp{0.000} & \tabvaluetemp{0.571} & \tabvaluetemp{0.000} & \tabvaluetemp{0.622}\\

& \variable{ED-s} & \tabvaluetemp{0.020} & \tabvaluetemp{1.000} & \tabvaluetemp{0.030} & \tabvaluetemp{0.997}& \tabvaluetemp{0.009} & \tabvaluetemp{0.971} & \tabvaluetemp{0.008} & \tabvaluetemp{0.965}& \tabvaluetemp{0.008} & \tabvaluetemp{0.969} & \tabvaluetemp{0.008} & \tabvaluetemp{0.966}& \tabvaluetemp{0.007} & \tabvaluetemp{0.970} & \tabvaluetemp{0.007} & \tabvaluetemp{0.971}\\

& \variable{ED-o} & \tabvaluetemp{0.025} & \tabvaluetemp{1.000} & \tabvaluetemp{0.030} & \tabvaluetemp{0.997}& \tabvaluetemp{0.009} & \tabvaluetemp{0.971} & \tabvaluetemp{0.008} & \tabvaluetemp{0.965}& \tabvaluetemp{0.008} & \tabvaluetemp{0.969} & \tabvaluetemp{0.008} & \tabvaluetemp{0.966}& \tabvaluetemp{0.007} & \tabvaluetemp{0.970} & \tabvaluetemp{0.007} & \tabvaluetemp{0.971}\\

& \variable{ED} & \tabvaluetemp{0.019} & \tabvaluetemp{0.993} & \tabvaluetemp{0.020} & \tabvaluetemp{0.992}& \tabvaluetemp{0.002} & \tabvaluetemp{0.949} & \tabvaluetemp{0.002} & \tabvaluetemp{0.947}& \tabvaluetemp{0.001} & \tabvaluetemp{0.954} & \tabvaluetemp{0.001} & \tabvaluetemp{0.953}& \tabvaluetemp{0.001} & \tabvaluetemp{0.958} & \tabvaluetemp{0.001} & \tabvaluetemp{0.959}\\

             \midrule
            \multirow{6}{*}{\rotatebox{90}{\parbox{40pt}\centering\textbf{LR}}} 
             & \variable{BD} & \tabvaluetemp{0.356} & \tabvaluetemp{0.971} & \tabvaluetemp{0.341} & \tabvaluetemp{0.983}& \tabvaluetemp{0.004} & \tabvaluetemp{0.045} & \tabvaluetemp{0.008} & \tabvaluetemp{0.548}& \tabvaluetemp{0.004} & \tabvaluetemp{0.045} & \tabvaluetemp{0.004} & \tabvaluetemp{0.306}& \tabvaluetemp{0.002} & \tabvaluetemp{0.044} & \tabvaluetemp{0.008} & \tabvaluetemp{0.533}\\

& \variable{MD} & \tabvaluetemp{0.397} & \tabvaluetemp{0.547} & \tabvaluetemp{0.937} & \tabvaluetemp{0.963}& \tabvaluetemp{0.000} & \tabvaluetemp{0.005} & \tabvaluetemp{0.000} & \tabvaluetemp{0.006}& \tabvaluetemp{0.000} & \tabvaluetemp{0.005} & \tabvaluetemp{0.000} & \tabvaluetemp{0.009}& \tabvaluetemp{0.000} & \tabvaluetemp{0.005} & \tabvaluetemp{0.001} & \tabvaluetemp{0.035}\\

& \variable{ED-v} & \tabvaluetemp{0.144} & \tabvaluetemp{0.861} & \tabvaluetemp{0.017} & \tabvaluetemp{0.689}& \tabvaluetemp{0.000} & \tabvaluetemp{0.000} & \tabvaluetemp{0.001} & \tabvaluetemp{0.329}& \tabvaluetemp{0.000} & \tabvaluetemp{0.000} & \tabvaluetemp{0.000} & \tabvaluetemp{0.314}& \tabvaluetemp{0.000} & \tabvaluetemp{0.000} & \tabvaluetemp{0.001} & \tabvaluetemp{0.326}\\

& \variable{ED-s} & \tabvaluetemp{0.321} & \tabvaluetemp{0.957} & \tabvaluetemp{0.065} & \tabvaluetemp{0.951}& \tabvaluetemp{0.012} & \tabvaluetemp{0.195} & \tabvaluetemp{0.010} & \tabvaluetemp{0.630}& \tabvaluetemp{0.013} & \tabvaluetemp{0.213} & \tabvaluetemp{0.004} & \tabvaluetemp{0.479}& \tabvaluetemp{0.013} & \tabvaluetemp{0.198} & \tabvaluetemp{0.006} & \tabvaluetemp{0.538}\\

& \variable{ED-o} & \tabvaluetemp{0.321} & \tabvaluetemp{0.958} & \tabvaluetemp{0.149} & \tabvaluetemp{0.980}& \tabvaluetemp{0.012} & \tabvaluetemp{0.197} & \tabvaluetemp{0.010} & \tabvaluetemp{0.630}& \tabvaluetemp{0.013} & \tabvaluetemp{0.213} & \tabvaluetemp{0.007} & \tabvaluetemp{0.488}& \tabvaluetemp{0.013} & \tabvaluetemp{0.198} & \tabvaluetemp{0.011} & \tabvaluetemp{0.558}\\

& \variable{ED} & \tabvaluetemp{0.166} & \tabvaluetemp{0.914} & \tabvaluetemp{0.037} & \tabvaluetemp{0.922}& \tabvaluetemp{0.002} & \tabvaluetemp{0.102} & \tabvaluetemp{0.002} & \tabvaluetemp{0.474}& \tabvaluetemp{0.002} & \tabvaluetemp{0.108} & \tabvaluetemp{0.002} & \tabvaluetemp{0.375}& \tabvaluetemp{0.002} & \tabvaluetemp{0.097} & \tabvaluetemp{0.003} & \tabvaluetemp{0.411}\\
             \midrule
            \multirow{6}{*}{\rotatebox{90}{\parbox{40pt}\centering\textbf{HGB}}} 
             & \variable{BD} & \tabvaluetemp{0.023} & \tabvaluetemp{1.000} & \tabvaluetemp{0.028} & \tabvaluetemp{0.997}& \tabvaluetemp{0.010} & \tabvaluetemp{0.990} & \tabvaluetemp{0.010} & \tabvaluetemp{0.990}& \tabvaluetemp{0.010} & \tabvaluetemp{0.993} & \tabvaluetemp{0.010} & \tabvaluetemp{0.992}& \tabvaluetemp{0.011} & \tabvaluetemp{0.995} & \tabvaluetemp{0.011} & \tabvaluetemp{0.995}\\

& \variable{MD} & \tabvaluetemp{0.024} & \tabvaluetemp{1.000} & \tabvaluetemp{0.028} & \tabvaluetemp{0.997}& \tabvaluetemp{0.012} & \tabvaluetemp{0.987} & \tabvaluetemp{0.011} & \tabvaluetemp{0.991}& \tabvaluetemp{0.011} & \tabvaluetemp{0.989} & \tabvaluetemp{0.011} & \tabvaluetemp{0.985}& \tabvaluetemp{0.012} & \tabvaluetemp{0.990} & \tabvaluetemp{0.011} & \tabvaluetemp{0.987}\\

& \variable{ED-v} & \tabvaluetemp{0.021} & \tabvaluetemp{0.947} & \tabvaluetemp{0.020} & \tabvaluetemp{0.867}& \tabvaluetemp{0.000} & \tabvaluetemp{0.581} & \tabvaluetemp{0.000} & \tabvaluetemp{0.646}& \tabvaluetemp{0.000} & \tabvaluetemp{0.640} & \tabvaluetemp{0.000} & \tabvaluetemp{0.682}& \tabvaluetemp{0.000} & \tabvaluetemp{0.622} & \tabvaluetemp{0.000} & \tabvaluetemp{0.657}\\

& \variable{ED-s} & \tabvaluetemp{0.023} & \tabvaluetemp{1.000} & \tabvaluetemp{0.027} & \tabvaluetemp{0.996}& \tabvaluetemp{0.010} & \tabvaluetemp{0.988} & \tabvaluetemp{0.010} & \tabvaluetemp{0.986}& \tabvaluetemp{0.009} & \tabvaluetemp{0.989} & \tabvaluetemp{0.010} & \tabvaluetemp{0.990}& \tabvaluetemp{0.010} & \tabvaluetemp{0.992} & \tabvaluetemp{0.011} & \tabvaluetemp{0.992}\\

& \variable{ED-o} & \tabvaluetemp{0.023} & \tabvaluetemp{1.000} & \tabvaluetemp{0.028} & \tabvaluetemp{0.997}& \tabvaluetemp{0.010} & \tabvaluetemp{0.989} & \tabvaluetemp{0.010} & \tabvaluetemp{0.987}& \tabvaluetemp{0.010} & \tabvaluetemp{0.989} & \tabvaluetemp{0.010} & \tabvaluetemp{0.990}& \tabvaluetemp{0.010} & \tabvaluetemp{0.993} & \tabvaluetemp{0.011} & \tabvaluetemp{0.992}\\

& \variable{ED} & \tabvaluetemp{0.020} & \tabvaluetemp{0.996} & \tabvaluetemp{0.020} & \tabvaluetemp{0.990}& \tabvaluetemp{0.002} & \tabvaluetemp{0.977} & \tabvaluetemp{0.002} & \tabvaluetemp{0.976}& \tabvaluetemp{0.002} & \tabvaluetemp{0.982} & \tabvaluetemp{0.002} & \tabvaluetemp{0.983}& \tabvaluetemp{0.002} & \tabvaluetemp{0.986} & \tabvaluetemp{0.002} & \tabvaluetemp{0.984}\\
            
            \bottomrule
        \end{tabular}
        }
                
        \caption{\textit{Temporal Dependency}: Results by assuming the presence of temporal dependencies among samples (the `first' samples of \scbb{D} are put in \scbb{T}, while the last 20\% represent \scbb{E}).} 
    \label{tab:nb15_baseline_temporal}
    \end{subtable}

    \label{tab:nb15_baseline}
\end{table*}

\begin{table*}
  \centering
  \caption{\dataset{NB15}. Results against adversarial (original \scmath{tpr} and adversarial \scmath{tpr}) and unknown attacks (the \scmath{tpr} is the average on the `unknown' attacks, while the \scmath{fpr} is due to training on a new \scbb{T} that does not have the `unknown' class.).}

    \begin{subtable}[htbp]{1.99\columnwidth}
        \resizebox{1.0\columnwidth}{!}{
        \begin{tabular}{c|c ? cc|cc? cc|cc? cc|cc? cc|cc}
\multicolumn{2}{c?}{Available Data} & \multicolumn{4}{c?}{Limited (100 per class) [N=1000]} & \multicolumn{4}{c?}{Scarce (15\% of \scbb{D}) [N=100]} &  \multicolumn{4}{c?}{Moderate (40\% of \scbb{D}) [N=100]} &  \multicolumn{4}{c}{Abundant (80\% of \scbb{D}) [N=100]} \\ \hline
             \multicolumn{2}{c?}{Scenario} &
             \multicolumn{2}{c|}{Adversarial Attacks} & \multicolumn{2}{c?}{Unknown Attacks} & \multicolumn{2}{c|}{Adversarial Attacks} & \multicolumn{2}{c?}{Unknown Attacks} & \multicolumn{2}{c|}{Adversarial Attacks} & \multicolumn{2}{c?}{Unknown Attacks} & \multicolumn{2}{c|}{Adversarial Attacks} & \multicolumn{2}{c?}{Unknown Attacks}  \\ \hline
            Alg. & Design & 
            \footnotesize{$tpr$} \tiny{(org)} & \footnotesize{$tpr$} \tiny{(adv)} & \footnotesize{$fpr$} & \footnotesize{$tpr$} & \footnotesize{$tpr$} \tiny{(org)} & \footnotesize{$tpr$} \tiny{(adv)} & \footnotesize{$fpr$} & \footnotesize{$tpr$} & \footnotesize{$tpr$} \tiny{(org)} & \footnotesize{$tpr$} \tiny{(adv)} & \footnotesize{$fpr$} & \footnotesize{$tpr$} & \footnotesize{$tpr$} \tiny{(org)} & \footnotesize{$tpr$} \tiny{(adv)} & \footnotesize{$fpr$} & \footnotesize{$tpr$} \\
             \toprule
             
             \multirow{5}{*}{\rotatebox{90}{\parbox{40pt}\centering\textbf{RF}}} 
             & \variable{BD} & \tabvalue{0.984}{0.007} & \tabvalue{0.976}{0.010} & \tabvalue{0.021}{0.002} & \tabvalue{1.000}{0.001}& \tabvalue{0.972}{0.005} & \tabvalue{0.998}{0.003} & \tabvalue{0.009}{0.000} & \tabvalue{0.890}{0.003}& \tabvalue{0.974}{0.004} & \tabvalue{1.000}{0.001} & \tabvalue{0.008}{0.000} & \tabvalue{0.886}{0.002}& \tabvalue{0.975}{0.003} & \tabvalue{1.000}{0.000} & \tabvalue{0.008}{0.000} & \tabvalue{0.883}{0.003}\\

& \variable{MD} & \tabvalue{0.971}{0.006} & \tabvalue{0.552}{0.263} & \tabvalue{0.020}{0.001} & \tabvalue{0.995}{0.005}& \tabvalue{0.967}{0.006} & \tabvalue{0.989}{0.008} & \tabvalue{0.008}{0.000} & \tabvalue{0.861}{0.006}& \tabvalue{0.972}{0.004} & \tabvalue{0.996}{0.003} & \tabvalue{0.008}{0.000} & \tabvalue{0.863}{0.004}& \tabvalue{0.973}{0.003} & \tabvalue{0.998}{0.002} & \tabvalue{0.007}{0.000} & \tabvalue{0.863}{0.003}\\

& \variable{ED-v} & \tabvalue{0.896}{0.098} & \tabvalue{0.171}{0.229} & \tabvalue{0.018}{0.002} & \tabvalue{0.958}{0.035}& \tabvalue{0.349}{0.082} & \tabvalue{0.062}{0.054} & \tabvalue{0.000}{0.000} & \tabvalue{0.547}{0.028}& \tabvalue{0.384}{0.086} & \tabvalue{0.072}{0.053} & \tabvalue{0.000}{0.000} & \tabvalue{0.559}{0.027}& \tabvalue{0.400}{0.062} & \tabvalue{0.064}{0.040} & \tabvalue{0.000}{0.000} & \tabvalue{0.559}{0.024}\\

& \variable{ED-s} & \tabvalue{0.975}{0.008} & \tabvalue{0.861}{0.154} & \tabvalue{0.021}{0.002} & \tabvalue{0.998}{0.003}& \tabvalue{0.968}{0.005} & \tabvalue{0.982}{0.020} & \tabvalue{0.008}{0.000} & \tabvalue{0.879}{0.004}& \tabvalue{0.972}{0.003} & \tabvalue{0.993}{0.014} & \tabvalue{0.008}{0.000} & \tabvalue{0.879}{0.003}& \tabvalue{0.974}{0.003} & \tabvalue{0.993}{0.016} & \tabvalue{0.007}{0.000} & \tabvalue{0.879}{0.003}\\

& \variable{ED-o} & \tabvalue{0.977}{0.006} & \tabvalue{0.886}{0.128} & \tabvalue{0.022}{0.002} & \tabvalue{0.999}{0.002}& \tabvalue{0.968}{0.005} & \tabvalue{0.982}{0.020} & \tabvalue{0.008}{0.000} & \tabvalue{0.879}{0.004}& \tabvalue{0.972}{0.003} & \tabvalue{0.993}{0.014} & \tabvalue{0.008}{0.000} & \tabvalue{0.879}{0.003}& \tabvalue{0.974}{0.003} & \tabvalue{0.993}{0.016} & \tabvalue{0.007}{0.000} & \tabvalue{0.879}{0.003}\\
             
            \midrule
            \multirow{5}{*}{\rotatebox{90}{\parbox{40pt}\centering\textbf{DT}}} 
             & \variable{BD} & \tabvalue{0.975}{0.015} & \tabvalue{0.973}{0.056} & \tabvalue{0.021}{0.003} & \tabvalue{0.992}{0.008}& \tabvalue{0.916}{0.009} & \tabvalue{0.865}{0.127} & \tabvalue{0.007}{0.000} & \tabvalue{0.866}{0.010}& \tabvalue{0.925}{0.006} & \tabvalue{0.882}{0.126} & \tabvalue{0.007}{0.000} & \tabvalue{0.864}{0.007}& \tabvalue{0.932}{0.006} & \tabvalue{0.877}{0.121} & \tabvalue{0.006}{0.000} & \tabvalue{0.862}{0.006}\\

& \variable{MD} & \tabvalue{0.974}{0.017} & \tabvalue{0.982}{0.015} & \tabvalue{0.022}{0.003} & \tabvalue{0.985}{0.014}& \tabvalue{0.918}{0.008} & \tabvalue{0.866}{0.271} & \tabvalue{0.007}{0.000} & \tabvalue{0.863}{0.009}& \tabvalue{0.925}{0.006} & \tabvalue{0.767}{0.367} & \tabvalue{0.007}{0.000} & \tabvalue{0.863}{0.009}& \tabvalue{0.932}{0.006} & \tabvalue{0.716}{0.337} & \tabvalue{0.006}{0.000} & \tabvalue{0.862}{0.008}\\

& \variable{ED-v} & \tabvalue{0.914}{0.141} & \tabvalue{0.880}{0.223} & \tabvalue{0.015}{0.003} & \tabvalue{0.905}{0.067}& \tabvalue{0.367}{0.079} & \tabvalue{0.395}{0.252} & \tabvalue{0.000}{0.000} & \tabvalue{0.563}{0.027}& \tabvalue{0.357}{0.072} & \tabvalue{0.378}{0.225} & \tabvalue{0.000}{0.000} & \tabvalue{0.563}{0.021}& \tabvalue{0.336}{0.077} & \tabvalue{0.341}{0.186} & \tabvalue{0.000}{0.000} & \tabvalue{0.560}{0.015}\\

& \variable{ED-s} & \tabvalue{0.978}{0.012} & \tabvalue{0.977}{0.047} & \tabvalue{0.021}{0.003} & \tabvalue{0.994}{0.006}& \tabvalue{0.923}{0.007} & \tabvalue{0.972}{0.064} & \tabvalue{0.008}{0.000} & \tabvalue{0.890}{0.005}& \tabvalue{0.930}{0.006} & \tabvalue{0.969}{0.058} & \tabvalue{0.007}{0.000} & \tabvalue{0.886}{0.005}& \tabvalue{0.936}{0.005} & \tabvalue{0.951}{0.064} & \tabvalue{0.007}{0.000} & \tabvalue{0.881}{0.004}\\

& \variable{ED-o} & \tabvalue{0.980}{0.010} & \tabvalue{0.982}{0.036} & \tabvalue{0.023}{0.004} & \tabvalue{0.997}{0.004}& \tabvalue{0.923}{0.007} & \tabvalue{0.972}{0.064} & \tabvalue{0.008}{0.000} & \tabvalue{0.890}{0.005}& \tabvalue{0.930}{0.006} & \tabvalue{0.969}{0.058} & \tabvalue{0.007}{0.000} & \tabvalue{0.886}{0.005}& \tabvalue{0.936}{0.005} & \tabvalue{0.951}{0.064} & \tabvalue{0.007}{0.000} & \tabvalue{0.881}{0.004}\\
             \midrule
            \multirow{5}{*}{\rotatebox{90}{\parbox{40pt}\centering\textbf{LR}}} 
             & \variable{BD} & \tabvalue{0.958}{0.027} & \tabvalue{0.807}{0.355} & \tabvalue{0.349}{0.017} & \tabvalue{0.975}{0.006}& \tabvalue{0.295}{0.256} & \tabvalue{0.500}{0.473} & \tabvalue{0.003}{0.001} & \tabvalue{0.055}{0.005}& \tabvalue{0.321}{0.264} & \tabvalue{0.545}{0.474} & \tabvalue{0.004}{0.001} & \tabvalue{0.055}{0.004}& \tabvalue{0.322}{0.266} & \tabvalue{0.558}{0.469} & \tabvalue{0.004}{0.000} & \tabvalue{0.055}{0.002}\\

& \variable{MD} & \tabvalue{0.984}{0.051} & \tabvalue{0.160}{0.254} & \tabvalue{0.446}{0.143} & \tabvalue{0.910}{0.087}& \tabvalue{0.075}{0.097} & \tabvalue{0.000}{0.000} & \tabvalue{0.001}{0.000} & \tabvalue{0.010}{0.003}& \tabvalue{0.073}{0.115} & \tabvalue{0.000}{0.000} & \tabvalue{0.001}{0.000} & \tabvalue{0.010}{0.003}& \tabvalue{0.075}{0.084} & \tabvalue{0.000}{0.000} & \tabvalue{0.001}{0.000} & \tabvalue{0.010}{0.003}\\

& \variable{ED-v} & \tabvalue{0.660}{0.120} & \tabvalue{0.032}{0.133} & \tabvalue{0.145}{0.007} & \tabvalue{0.896}{0.020}& \tabvalue{0.064}{0.040} & \tabvalue{0.000}{0.000} & \tabvalue{0.000}{0.000} & \tabvalue{0.000}{0.000}& \tabvalue{0.080}{0.061} & \tabvalue{0.000}{0.000} & \tabvalue{0.000}{0.000} & \tabvalue{0.000}{0.000}& \tabvalue{0.072}{0.050} & \tabvalue{0.000}{0.000} & \tabvalue{0.000}{0.000} & \tabvalue{0.000}{0.000}\\

& \variable{ED-s} & \tabvalue{0.956}{0.022} & \tabvalue{0.863}{0.266} & \tabvalue{0.283}{0.044} & \tabvalue{0.957}{0.007}& \tabvalue{0.567}{0.129} & \tabvalue{0.239}{0.373} & \tabvalue{0.011}{0.001} & \tabvalue{0.143}{0.008}& \tabvalue{0.585}{0.108} & \tabvalue{0.380}{0.414} & \tabvalue{0.011}{0.000} & \tabvalue{0.143}{0.005}& \tabvalue{0.565}{0.111} & \tabvalue{0.313}{0.407} & \tabvalue{0.011}{0.000} & \tabvalue{0.143}{0.005}\\

& \variable{ED-o} & \tabvalue{0.961}{0.019} & \tabvalue{0.917}{0.189} & \tabvalue{0.309}{0.020} & \tabvalue{0.963}{0.006}& \tabvalue{0.574}{0.127} & \tabvalue{0.300}{0.390} & \tabvalue{0.011}{0.001} & \tabvalue{0.146}{0.008}& \tabvalue{0.596}{0.104} & \tabvalue{0.455}{0.410} & \tabvalue{0.011}{0.000} & \tabvalue{0.146}{0.005}& \tabvalue{0.577}{0.109} & \tabvalue{0.350}{0.411} & \tabvalue{0.011}{0.000} & \tabvalue{0.146}{0.005}\\
             \midrule
            \multirow{5}{*}{\rotatebox{90}{\parbox{40pt}\centering\textbf{HGB}}} 
             & \variable{BD} & \tabvalue{0.983}{0.008} & \tabvalue{0.972}{0.060} & \tabvalue{0.021}{0.002} & \tabvalue{0.997}{0.004}& \tabvalue{0.983}{0.005} & \tabvalue{0.992}{0.019} & \tabvalue{0.009}{0.000} & \tabvalue{0.890}{0.003}& \tabvalue{0.989}{0.003} & \tabvalue{0.992}{0.033} & \tabvalue{0.009}{0.000} & \tabvalue{0.887}{0.002}& \tabvalue{0.990}{0.003} & \tabvalue{0.991}{0.024} & \tabvalue{0.009}{0.000} & \tabvalue{0.886}{0.003}\\

& \variable{MD} & \tabvalue{0.979}{0.010} & \tabvalue{0.941}{0.111} & \tabvalue{0.021}{0.002} & \tabvalue{0.993}{0.008}& \tabvalue{0.957}{0.011} & \tabvalue{0.956}{0.085} & \tabvalue{0.010}{0.001} & \tabvalue{0.858}{0.012}& \tabvalue{0.966}{0.007} & \tabvalue{0.962}{0.073} & \tabvalue{0.010}{0.000} & \tabvalue{0.850}{0.016}& \tabvalue{0.967}{0.005} & \tabvalue{0.974}{0.027} & \tabvalue{0.010}{0.000} & \tabvalue{0.846}{0.018}\\

& \variable{ED-v} & \tabvalue{0.817}{0.217} & \tabvalue{0.763}{0.260} & \tabvalue{0.014}{0.003} & \tabvalue{0.888}{0.064}& \tabvalue{0.429}{0.090} & \tabvalue{0.458}{0.203} & \tabvalue{0.000}{0.000} & \tabvalue{0.589}{0.030}& \tabvalue{0.457}{0.076} & \tabvalue{0.477}{0.196} & \tabvalue{0.000}{0.000} & \tabvalue{0.596}{0.024}& \tabvalue{0.469}{0.065} & \tabvalue{0.500}{0.188} & \tabvalue{0.000}{0.000} & \tabvalue{0.588}{0.018}\\

& \variable{ED-s} & \tabvalue{0.980}{0.009} & \tabvalue{0.963}{0.072} & \tabvalue{0.021}{0.002} & \tabvalue{0.990}{0.011}& \tabvalue{0.978}{0.006} & \tabvalue{0.980}{0.030} & \tabvalue{0.008}{0.000} & \tabvalue{0.881}{0.004}& \tabvalue{0.984}{0.004} & \tabvalue{0.982}{0.034} & \tabvalue{0.008}{0.000} & \tabvalue{0.880}{0.003}& \tabvalue{0.986}{0.004} & \tabvalue{0.983}{0.026} & \tabvalue{0.008}{0.000} & \tabvalue{0.879}{0.003}\\

& \variable{ED-o} & \tabvalue{0.982}{0.008} & \tabvalue{0.975}{0.032} & \tabvalue{0.022}{0.004} & \tabvalue{0.995}{0.007}& \tabvalue{0.979}{0.005} & \tabvalue{0.981}{0.030} & \tabvalue{0.009}{0.001} & \tabvalue{0.887}{0.005}& \tabvalue{0.986}{0.003} & \tabvalue{0.984}{0.031} & \tabvalue{0.009}{0.000} & \tabvalue{0.884}{0.004}& \tabvalue{0.987}{0.003} & \tabvalue{0.986}{0.023} & \tabvalue{0.009}{0.000} & \tabvalue{0.882}{0.003}\\
            
            \bottomrule
        \end{tabular}
        }
                
        \caption{\textit{Static Dependency}: Results by assuming the absence of temporal dependencies among samples (\scbb{T} and \scbb{E} are randomly sampled from \scbb{D}).} 
    \label{tab:nb15_open_static}
    \end{subtable}

    \begin{subtable}[htbp]{1.8\columnwidth}
        \resizebox{1.0\columnwidth}{!}{
        \begin{tabular}{c|c ? cc|cc? cc|cc? cc|cc? cc|cc}
\multicolumn{2}{c?}{Available Data} & \multicolumn{4}{c?}{Limited (100 per class) [N=1]} & \multicolumn{4}{c?}{Scarce (15\% of \scbb{D}) [N=1]} &  \multicolumn{4}{c?}{Moderate (40\% of \scbb{D}) [N=1]} &  \multicolumn{4}{c}{Abundant (80\% of \scbb{D}) [N=1]} \\ \hline
             \multicolumn{2}{c?}{Scenario} &
             \multicolumn{2}{c|}{Adversarial Attacks} & \multicolumn{2}{c?}{Unknown Attacks} & \multicolumn{2}{c|}{Adversarial Attacks} & \multicolumn{2}{c?}{Unknown Attacks} & \multicolumn{2}{c|}{Adversarial Attacks} & \multicolumn{2}{c?}{Unknown Attacks} & \multicolumn{2}{c|}{Adversarial Attacks} & \multicolumn{2}{c?}{Unknown Attacks}  \\ \hline
            Alg. & Design & 
            \footnotesize{$tpr$} \tiny{(org)} & \footnotesize{$tpr$} \tiny{(adv)} & \footnotesize{$fpr$} & \footnotesize{$tpr$} & \footnotesize{$tpr$} \tiny{(org)} & \footnotesize{$tpr$} \tiny{(adv)} & \footnotesize{$fpr$} & \footnotesize{$tpr$} & \footnotesize{$tpr$} \tiny{(org)} & \footnotesize{$tpr$} \tiny{(adv)} & \footnotesize{$fpr$} & \footnotesize{$tpr$} & \footnotesize{$tpr$} \tiny{(org)} & \footnotesize{$tpr$} \tiny{(adv)} & \footnotesize{$fpr$} & \footnotesize{$tpr$} \\
             \toprule
             
             \multirow{5}{*}{\rotatebox{90}{\parbox{40pt}\centering\textbf{RF}}} 
              & \variable{BD} & \tabvaluetemp{0.985} & \tabvaluetemp{0.983} & \tabvaluetemp{0.020} & \tabvaluetemp{1.000}& \tabvaluetemp{0.967} & \tabvaluetemp{0.998} & \tabvaluetemp{0.009} & \tabvaluetemp{0.890}& \tabvaluetemp{0.972} & \tabvaluetemp{1.000} & \tabvaluetemp{0.009} & \tabvaluetemp{0.886}& \tabvaluetemp{0.973} & \tabvaluetemp{1.000} & \tabvaluetemp{0.008} & \tabvaluetemp{0.882}\\

& \variable{MD} & \tabvaluetemp{0.974} & \tabvaluetemp{0.529} & \tabvaluetemp{0.019} & \tabvaluetemp{0.996}& \tabvaluetemp{0.957} & \tabvaluetemp{0.960} & \tabvaluetemp{0.008} & \tabvaluetemp{0.859}& \tabvaluetemp{0.969} & \tabvaluetemp{0.998} & \tabvaluetemp{0.008} & \tabvaluetemp{0.865}& \tabvaluetemp{0.974} & \tabvaluetemp{1.000} & \tabvaluetemp{0.007} & \tabvaluetemp{0.865}\\

& \variable{ED-v} & \tabvaluetemp{0.928} & \tabvaluetemp{0.407} & \tabvaluetemp{0.020} & \tabvaluetemp{0.970}& \tabvaluetemp{0.416} & \tabvaluetemp{0.151} & \tabvaluetemp{0.000} & \tabvaluetemp{0.583}& \tabvaluetemp{0.398} & \tabvaluetemp{0.098} & \tabvaluetemp{0.000} & \tabvaluetemp{0.531}& \tabvaluetemp{0.452} & \tabvaluetemp{0.161} & \tabvaluetemp{0.000} & \tabvaluetemp{0.550}\\

& \variable{ED-s} & \tabvaluetemp{0.982} & \tabvaluetemp{0.645} & \tabvaluetemp{0.020} & \tabvaluetemp{1.000}& \tabvaluetemp{0.956} & \tabvaluetemp{0.975} & \tabvaluetemp{0.008} & \tabvaluetemp{0.879}& \tabvaluetemp{0.970} & \tabvaluetemp{1.000} & \tabvaluetemp{0.008} & \tabvaluetemp{0.877}& \tabvaluetemp{0.972} & \tabvaluetemp{0.994} & \tabvaluetemp{0.007} & \tabvaluetemp{0.877}\\

& \variable{ED-o} & \tabvaluetemp{0.983} & \tabvaluetemp{0.925} & \tabvaluetemp{0.021} & \tabvaluetemp{1.000}& \tabvaluetemp{0.956} & \tabvaluetemp{0.975} & \tabvaluetemp{0.008} & \tabvaluetemp{0.879}& \tabvaluetemp{0.970} & \tabvaluetemp{1.000} & \tabvaluetemp{0.008} & \tabvaluetemp{0.877}& \tabvaluetemp{0.972} & \tabvaluetemp{0.994} & \tabvaluetemp{0.007} & \tabvaluetemp{0.877}\\
             
            \midrule
            \multirow{5}{*}{\rotatebox{90}{\parbox{40pt}\centering\textbf{DT}}} 
            & \variable{BD} & \tabvaluetemp{0.982} & \tabvaluetemp{0.982} & \tabvaluetemp{0.020} & \tabvaluetemp{1.000}& \tabvaluetemp{0.914} & \tabvaluetemp{0.915} & \tabvaluetemp{0.007} & \tabvaluetemp{0.872}& \tabvaluetemp{0.927} & \tabvaluetemp{0.660} & \tabvaluetemp{0.007} & \tabvaluetemp{0.870}& \tabvaluetemp{0.934} & \tabvaluetemp{0.955} & \tabvaluetemp{0.006} & \tabvaluetemp{0.858}\\

& \variable{MD} & \tabvaluetemp{0.982} & \tabvaluetemp{0.983} & \tabvaluetemp{0.021} & \tabvaluetemp{1.000}& \tabvaluetemp{0.929} & \tabvaluetemp{0.991} & \tabvaluetemp{0.007} & \tabvaluetemp{0.877}& \tabvaluetemp{0.935} & \tabvaluetemp{0.783} & \tabvaluetemp{0.007} & \tabvaluetemp{0.834}& \tabvaluetemp{0.928} & \tabvaluetemp{0.982} & \tabvaluetemp{0.006} & \tabvaluetemp{0.864}\\

& \variable{ED-v} & \tabvaluetemp{0.978} & \tabvaluetemp{0.978} & \tabvaluetemp{0.019} & \tabvaluetemp{0.989}& \tabvaluetemp{0.306} & \tabvaluetemp{0.168} & \tabvaluetemp{0.000} & \tabvaluetemp{0.541}& \tabvaluetemp{0.383} & \tabvaluetemp{0.293} & \tabvaluetemp{0.000} & \tabvaluetemp{0.558}& \tabvaluetemp{0.368} & \tabvaluetemp{0.013} & \tabvaluetemp{0.000} & \tabvaluetemp{0.545}\\

& \variable{ED-s} & \tabvaluetemp{0.982} & \tabvaluetemp{0.983} & \tabvaluetemp{0.020} & \tabvaluetemp{0.998}& \tabvaluetemp{0.928} & \tabvaluetemp{0.993} & \tabvaluetemp{0.008} & \tabvaluetemp{0.896}& \tabvaluetemp{0.930} & \tabvaluetemp{0.987} & \tabvaluetemp{0.007} & \tabvaluetemp{0.881}& \tabvaluetemp{0.937} & \tabvaluetemp{0.839} & \tabvaluetemp{0.006} & \tabvaluetemp{0.885}\\

& \variable{ED-o} & \tabvaluetemp{0.982} & \tabvaluetemp{0.983} & \tabvaluetemp{0.024} & \tabvaluetemp{1.000}& \tabvaluetemp{0.928} & \tabvaluetemp{0.993} & \tabvaluetemp{0.008} & \tabvaluetemp{0.896}& \tabvaluetemp{0.930} & \tabvaluetemp{0.987} & \tabvaluetemp{0.007} & \tabvaluetemp{0.881}& \tabvaluetemp{0.937} & \tabvaluetemp{0.839} & \tabvaluetemp{0.006} & \tabvaluetemp{0.885}\\
             \midrule
            \multirow{5}{*}{\rotatebox{90}{\parbox{40pt}\centering\textbf{LR}}} 
             & \variable{BD} & \tabvaluetemp{0.974} & \tabvaluetemp{0.993} & \tabvaluetemp{0.346} & \tabvaluetemp{0.974}& \tabvaluetemp{0.563} & \tabvaluetemp{0.971} & \tabvaluetemp{0.004} & \tabvaluetemp{0.058}& \tabvaluetemp{0.015} & \tabvaluetemp{0.000} & \tabvaluetemp{0.004} & \tabvaluetemp{0.055}& \tabvaluetemp{0.578} & \tabvaluetemp{0.999} & \tabvaluetemp{0.003} & \tabvaluetemp{0.052}\\

& \variable{MD} & \tabvaluetemp{0.999} & \tabvaluetemp{0.249} & \tabvaluetemp{0.219} & \tabvaluetemp{0.594}& \tabvaluetemp{0.001} & \tabvaluetemp{0.000} & \tabvaluetemp{0.001} & \tabvaluetemp{0.008}& \tabvaluetemp{0.007} & \tabvaluetemp{0.000} & \tabvaluetemp{0.002} & \tabvaluetemp{0.012}& \tabvaluetemp{0.167} & \tabvaluetemp{0.000} & \tabvaluetemp{0.001} & \tabvaluetemp{0.008}\\

& \variable{ED-v} & \tabvaluetemp{0.476} & \tabvaluetemp{0.005} & \tabvaluetemp{0.144} & \tabvaluetemp{0.908}& \tabvaluetemp{0.062} & \tabvaluetemp{0.000} & \tabvaluetemp{0.000} & \tabvaluetemp{0.000}& \tabvaluetemp{0.013} & \tabvaluetemp{0.000} & \tabvaluetemp{0.000} & \tabvaluetemp{0.000}& \tabvaluetemp{0.077} & \tabvaluetemp{0.000} & \tabvaluetemp{0.000} & \tabvaluetemp{0.000}\\

& \variable{ED-s} & \tabvaluetemp{0.937} & \tabvaluetemp{0.999} & \tabvaluetemp{0.274} & \tabvaluetemp{0.951}& \tabvaluetemp{0.739} & \tabvaluetemp{0.002} & \tabvaluetemp{0.011} & \tabvaluetemp{0.136}& \tabvaluetemp{0.192} & \tabvaluetemp{0.883} & \tabvaluetemp{0.012} & \tabvaluetemp{0.152}& \tabvaluetemp{0.555} & \tabvaluetemp{0.001} & \tabvaluetemp{0.011} & \tabvaluetemp{0.139}\\

& \variable{ED-o} & \tabvaluetemp{0.941} & \tabvaluetemp{1.000} & \tabvaluetemp{0.298} & \tabvaluetemp{0.960}& \tabvaluetemp{0.739} & \tabvaluetemp{0.002} & \tabvaluetemp{0.011} & \tabvaluetemp{0.144}& \tabvaluetemp{0.210} & \tabvaluetemp{0.883} & \tabvaluetemp{0.012} & \tabvaluetemp{0.154}& \tabvaluetemp{0.587} & \tabvaluetemp{0.836} & \tabvaluetemp{0.011} & \tabvaluetemp{0.143}\\
             \midrule
            \multirow{5}{*}{\rotatebox{90}{\parbox{40pt}\centering\textbf{HGB}}} 
             & \variable{BD} & \tabvaluetemp{0.979} & \tabvaluetemp{0.975} & \tabvaluetemp{0.023} & \tabvaluetemp{0.999}& \tabvaluetemp{0.984} & \tabvaluetemp{0.989} & \tabvaluetemp{0.009} & \tabvaluetemp{0.891}& \tabvaluetemp{0.991} & \tabvaluetemp{0.999} & \tabvaluetemp{0.008} & \tabvaluetemp{0.889}& \tabvaluetemp{0.989} & \tabvaluetemp{0.999} & \tabvaluetemp{0.009} & \tabvaluetemp{0.887}\\

& \variable{MD} & \tabvaluetemp{0.979} & \tabvaluetemp{0.975} & \tabvaluetemp{0.024} & \tabvaluetemp{0.999}& \tabvaluetemp{0.988} & \tabvaluetemp{0.989} & \tabvaluetemp{0.010} & \tabvaluetemp{0.825}& \tabvaluetemp{0.969} & \tabvaluetemp{0.980} & \tabvaluetemp{0.010} & \tabvaluetemp{0.853}& \tabvaluetemp{0.972} & \tabvaluetemp{0.979} & \tabvaluetemp{0.010} & \tabvaluetemp{0.826}\\

& \variable{ED-v} & \tabvaluetemp{0.972} & \tabvaluetemp{0.000} & \tabvaluetemp{0.021} & \tabvaluetemp{0.942}& \tabvaluetemp{0.421} & \tabvaluetemp{0.380} & \tabvaluetemp{0.000} & \tabvaluetemp{0.599}& \tabvaluetemp{0.527} & \tabvaluetemp{0.244} & \tabvaluetemp{0.000} & \tabvaluetemp{0.593}& \tabvaluetemp{0.507} & \tabvaluetemp{0.207} & \tabvaluetemp{0.000} & \tabvaluetemp{0.603}\\

& \variable{ED-s} & \tabvaluetemp{0.980} & \tabvaluetemp{0.975} & \tabvaluetemp{0.023} & \tabvaluetemp{0.998}& \tabvaluetemp{0.974} & \tabvaluetemp{0.965} & \tabvaluetemp{0.008} & \tabvaluetemp{0.882}& \tabvaluetemp{0.988} & \tabvaluetemp{0.998} & \tabvaluetemp{0.008} & \tabvaluetemp{0.883}& \tabvaluetemp{0.983} & \tabvaluetemp{0.993} & \tabvaluetemp{0.009} & \tabvaluetemp{0.878}\\

& \variable{ED-o} & \tabvaluetemp{0.980} & \tabvaluetemp{0.975} & \tabvaluetemp{0.023} & \tabvaluetemp{0.999}& \tabvaluetemp{0.976} & \tabvaluetemp{0.965} & \tabvaluetemp{0.009} & \tabvaluetemp{0.891}& \tabvaluetemp{0.988} & \tabvaluetemp{0.998} & \tabvaluetemp{0.008} & \tabvaluetemp{0.886}& \tabvaluetemp{0.984} & \tabvaluetemp{0.993} & \tabvaluetemp{0.009} & \tabvaluetemp{0.884}\\
            
            \bottomrule
        \end{tabular}
        }
                
        \caption{\textit{Temporal Dependency}: Results by assuming the presence of temporal dependencies among samples (the `first' samples of \scbb{D} are put in \scbb{T}, while the last 20\% represent \scbb{E}).} 
    \label{tab:nb15_open_temporal}
    \end{subtable}

    \label{tab:nb15_open}
\end{table*}

\begin{table*}
  \centering
  \caption{\dataset{UF-NB15} binary classification results (\scmath{fpr} and \scmath{tpr}) against `known' attacks seen during the training stage (closed world).}

    \begin{subtable}[htbp]{1.99\columnwidth}
        \resizebox{1.0\columnwidth}{!}{
        \begin{tabular}{c|c ? cc|cc? cc|cc? cc|cc? cc|cc}
             \multicolumn{2}{c?}{Available Data} & \multicolumn{4}{c?}{Limited (100 per class) [N=1000]} & \multicolumn{4}{c?}{Scarce (15\% of \scbb{D}) [N=100]} &  \multicolumn{4}{c?}{Moderate (40\% of \scbb{D}) [N=100]} &  \multicolumn{4}{c}{Abundant (80\% of \scbb{D}) [N=100]} \\ \hline
             \multicolumn{2}{c?}{Features} &
             \multicolumn{2}{c|}{Complete} & \multicolumn{2}{c?}{Essential} & \multicolumn{2}{c|}{Complete} & \multicolumn{2}{c?}{Essential} & \multicolumn{2}{c|}{Complete} & \multicolumn{2}{c?}{Essential} &  
             \multicolumn{2}{c|}{Complete} & \multicolumn{2}{c}{Essential} \\ \hline
            Alg. & Design & \smamath{fpr} & \smamath{tpr} & \smamath{fpr} & \smamath{tpr} & \smamath{fpr} & \smamath{tpr} & \smamath{fpr} & \smamath{tpr} & \smamath{fpr} & \smamath{tpr} & \smamath{fpr} & \smamath{tpr} & \smamath{fpr} & \smamath{tpr} & \smamath{fpr} & \smamath{tpr} \\
             \toprule
             
             \multirow{6}{*}{\rotatebox{90}{\parbox{40pt}\centering\textbf{RF}}} 
             & \variable{BD} & \tabvalue{0.011}{0.003} & \tabvalue{0.999}{0.001} & \tabvalue{0.060}{0.027} & \tabvalue{0.991}{0.006}& \tabvalue{0.004}{0.000} & \tabvalue{0.994}{0.001} & \tabvalue{0.004}{0.000} & \tabvalue{0.989}{0.001}& \tabvalue{0.003}{0.000} & \tabvalue{0.994}{0.001} & \tabvalue{0.003}{0.000} & \tabvalue{0.990}{0.001}& \tabvalue{0.003}{0.000} & \tabvalue{0.994}{0.001} & \tabvalue{0.003}{0.000} & \tabvalue{0.991}{0.001}\\

& \variable{MD} & \tabvalue{0.009}{0.002} & \tabvalue{0.998}{0.001} & \tabvalue{0.050}{0.025} & \tabvalue{0.982}{0.009}& \tabvalue{0.003}{0.000} & \tabvalue{0.990}{0.001} & \tabvalue{0.004}{0.000} & \tabvalue{0.985}{0.001}& \tabvalue{0.003}{0.000} & \tabvalue{0.990}{0.001} & \tabvalue{0.003}{0.000} & \tabvalue{0.986}{0.001}& \tabvalue{0.003}{0.000} & \tabvalue{0.991}{0.001} & \tabvalue{0.003}{0.000} & \tabvalue{0.988}{0.001}\\

& \variable{ED-v} & \tabvalue{0.008}{0.002} & \tabvalue{0.913}{0.094} & \tabvalue{0.008}{0.003} & \tabvalue{0.699}{0.031}& \tabvalue{0.000}{0.000} & \tabvalue{0.423}{0.022} & \tabvalue{0.000}{0.000} & \tabvalue{0.389}{0.035}& \tabvalue{0.000}{0.000} & \tabvalue{0.421}{0.015} & \tabvalue{0.000}{0.000} & \tabvalue{0.383}{0.025}& \tabvalue{0.000}{0.000} & \tabvalue{0.420}{0.014} & \tabvalue{0.000}{0.000} & \tabvalue{0.383}{0.015}\\

& \variable{ED-s} & \tabvalue{0.011}{0.003} & \tabvalue{0.999}{0.001} & \tabvalue{0.057}{0.026} & \tabvalue{0.984}{0.009}& \tabvalue{0.003}{0.000} & \tabvalue{0.991}{0.001} & \tabvalue{0.004}{0.000} & \tabvalue{0.986}{0.001}& \tabvalue{0.003}{0.000} & \tabvalue{0.991}{0.001} & \tabvalue{0.003}{0.000} & \tabvalue{0.987}{0.001}& \tabvalue{0.003}{0.000} & \tabvalue{0.991}{0.001} & \tabvalue{0.003}{0.000} & \tabvalue{0.988}{0.001}\\

& \variable{ED-o} & \tabvalue{0.012}{0.003} & \tabvalue{0.999}{0.001} & \tabvalue{0.058}{0.027} & \tabvalue{0.985}{0.009}& \tabvalue{0.003}{0.000} & \tabvalue{0.991}{0.001} & \tabvalue{0.004}{0.000} & \tabvalue{0.986}{0.001}& \tabvalue{0.003}{0.000} & \tabvalue{0.991}{0.001} & \tabvalue{0.003}{0.000} & \tabvalue{0.987}{0.001}& \tabvalue{0.003}{0.000} & \tabvalue{0.991}{0.001} & \tabvalue{0.003}{0.000} & \tabvalue{0.988}{0.001}\\

& \variable{ED} & \tabvalue{0.008}{0.001} & \tabvalue{0.996}{0.002} & \tabvalue{0.018}{0.007} & \tabvalue{0.969}{0.011}& \tabvalue{0.001}{0.000} & \tabvalue{0.979}{0.002} & \tabvalue{0.001}{0.000} & \tabvalue{0.967}{0.002}& \tabvalue{0.001}{0.000} & \tabvalue{0.980}{0.001} & \tabvalue{0.001}{0.000} & \tabvalue{0.972}{0.002}& \tabvalue{0.001}{0.000} & \tabvalue{0.982}{0.001} & \tabvalue{0.001}{0.000} & \tabvalue{0.975}{0.001}\\

            \midrule
            \multirow{6}{*}{\rotatebox{90}{\parbox{40pt}\centering\textbf{DT}}} 
             & \variable{BD} & \tabvalue{0.011}{0.004} & \tabvalue{0.997}{0.003} & \tabvalue{0.065}{0.027} & \tabvalue{0.975}{0.010}& \tabvalue{0.003}{0.000} & \tabvalue{0.979}{0.002} & \tabvalue{0.004}{0.000} & \tabvalue{0.974}{0.002}& \tabvalue{0.003}{0.000} & \tabvalue{0.981}{0.001} & \tabvalue{0.003}{0.000} & \tabvalue{0.979}{0.001}& \tabvalue{0.003}{0.000} & \tabvalue{0.983}{0.001} & \tabvalue{0.003}{0.000} & \tabvalue{0.981}{0.001}\\

& \variable{MD} & \tabvalue{0.011}{0.004} & \tabvalue{0.997}{0.003} & \tabvalue{0.072}{0.031} & \tabvalue{0.976}{0.010}& \tabvalue{0.003}{0.000} & \tabvalue{0.980}{0.002} & \tabvalue{0.004}{0.000} & \tabvalue{0.975}{0.002}& \tabvalue{0.003}{0.000} & \tabvalue{0.982}{0.001} & \tabvalue{0.003}{0.000} & \tabvalue{0.979}{0.001}& \tabvalue{0.003}{0.000} & \tabvalue{0.983}{0.001} & \tabvalue{0.003}{0.000} & \tabvalue{0.981}{0.001}\\

& \variable{ED-v} & \tabvalue{0.008}{0.003} & \tabvalue{0.934}{0.085} & \tabvalue{0.008}{0.007} & \tabvalue{0.698}{0.046}& \tabvalue{0.000}{0.000} & \tabvalue{0.419}{0.032} & \tabvalue{0.000}{0.000} & \tabvalue{0.362}{0.042}& \tabvalue{0.000}{0.000} & \tabvalue{0.413}{0.029} & \tabvalue{0.000}{0.000} & \tabvalue{0.377}{0.032}& \tabvalue{0.000}{0.000} & \tabvalue{0.400}{0.028} & \tabvalue{0.000}{0.000} & \tabvalue{0.379}{0.022}\\

& \variable{ED-s} & \tabvalue{0.012}{0.004} & \tabvalue{0.998}{0.002} & \tabvalue{0.094}{0.034} & \tabvalue{0.983}{0.009}& \tabvalue{0.003}{0.000} & \tabvalue{0.985}{0.001} & \tabvalue{0.005}{0.000} & \tabvalue{0.980}{0.002}& \tabvalue{0.003}{0.000} & \tabvalue{0.986}{0.001} & \tabvalue{0.004}{0.000} & \tabvalue{0.983}{0.001}& \tabvalue{0.003}{0.000} & \tabvalue{0.987}{0.001} & \tabvalue{0.004}{0.000} & \tabvalue{0.984}{0.001}\\

& \variable{ED-o} & \tabvalue{0.013}{0.004} & \tabvalue{0.999}{0.001} & \tabvalue{0.103}{0.034} & \tabvalue{0.985}{0.008}& \tabvalue{0.003}{0.000} & \tabvalue{0.985}{0.001} & \tabvalue{0.005}{0.000} & \tabvalue{0.980}{0.002}& \tabvalue{0.003}{0.000} & \tabvalue{0.986}{0.001} & \tabvalue{0.004}{0.000} & \tabvalue{0.983}{0.001}& \tabvalue{0.003}{0.000} & \tabvalue{0.987}{0.001} & \tabvalue{0.004}{0.000} & \tabvalue{0.984}{0.001}\\

& \variable{ED} & \tabvalue{0.008}{0.002} & \tabvalue{0.993}{0.007} & \tabvalue{0.025}{0.009} & \tabvalue{0.948}{0.015}& \tabvalue{0.001}{0.000} & \tabvalue{0.964}{0.002} & \tabvalue{0.001}{0.000} & \tabvalue{0.950}{0.003}& \tabvalue{0.001}{0.000} & \tabvalue{0.969}{0.002} & \tabvalue{0.001}{0.000} & \tabvalue{0.959}{0.002}& \tabvalue{0.001}{0.000} & \tabvalue{0.971}{0.002} & \tabvalue{0.001}{0.000} & \tabvalue{0.964}{0.002}\\

             \midrule
            \multirow{6}{*}{\rotatebox{90}{\parbox{40pt}\centering\textbf{LR}}} 
             & \variable{BD} & \tabvalue{0.701}{0.111} & \tabvalue{0.966}{0.052} & \tabvalue{0.469}{0.144} & \tabvalue{0.890}{0.054}& \tabvalue{0.009}{0.018} & \tabvalue{0.241}{0.096} & \tabvalue{0.007}{0.005} & \tabvalue{0.220}{0.048}& \tabvalue{0.005}{0.011} & \tabvalue{0.268}{0.074} & \tabvalue{0.007}{0.005} & \tabvalue{0.242}{0.050}& \tabvalue{0.005}{0.007} & \tabvalue{0.268}{0.080} & \tabvalue{0.006}{0.003} & \tabvalue{0.259}{0.049}\\

& \variable{MD} & \tabvalue{0.579}{0.320} & \tabvalue{0.832}{0.167} & \tabvalue{0.651}{0.184} & \tabvalue{0.922}{0.103}& \tabvalue{0.004}{0.001} & \tabvalue{0.284}{0.003} & \tabvalue{0.014}{0.025} & \tabvalue{0.279}{0.065}& \tabvalue{0.005}{0.001} & \tabvalue{0.284}{0.003} & \tabvalue{0.023}{0.033} & \tabvalue{0.298}{0.015}& \tabvalue{0.005}{0.001} & \tabvalue{0.284}{0.003} & \tabvalue{0.019}{0.029} & \tabvalue{0.297}{0.031}\\

& \variable{ED-v} & \tabvalue{0.039}{0.019} & \tabvalue{0.683}{0.061} & \tabvalue{0.029}{0.018} & \tabvalue{0.588}{0.059}& \tabvalue{0.000}{0.000} & \tabvalue{0.081}{0.100} & \tabvalue{0.000}{0.000} & \tabvalue{0.116}{0.023}& \tabvalue{0.000}{0.000} & \tabvalue{0.053}{0.084} & \tabvalue{0.000}{0.000} & \tabvalue{0.123}{0.018}& \tabvalue{0.000}{0.000} & \tabvalue{0.072}{0.095} & \tabvalue{0.000}{0.000} & \tabvalue{0.122}{0.017}\\

& \variable{ED-s} & \tabvalue{0.298}{0.111} & \tabvalue{0.922}{0.031} & \tabvalue{0.228}{0.064} & \tabvalue{0.906}{0.029}& \tabvalue{0.008}{0.002} & \tabvalue{0.469}{0.051} & \tabvalue{0.005}{0.001} & \tabvalue{0.419}{0.043}& \tabvalue{0.009}{0.002} & \tabvalue{0.464}{0.037} & \tabvalue{0.005}{0.001} & \tabvalue{0.414}{0.046}& \tabvalue{0.009}{0.002} & \tabvalue{0.463}{0.032} & \tabvalue{0.005}{0.001} & \tabvalue{0.415}{0.034}\\

& \variable{ED-o} & \tabvalue{0.460}{0.118} & \tabvalue{0.955}{0.048} & \tabvalue{0.283}{0.087} & \tabvalue{0.914}{0.030}& \tabvalue{0.045}{0.035} & \tabvalue{0.500}{0.046} & \tabvalue{0.007}{0.003} & \tabvalue{0.432}{0.042}& \tabvalue{0.030}{0.031} & \tabvalue{0.491}{0.029} & \tabvalue{0.007}{0.003} & \tabvalue{0.426}{0.045}& \tabvalue{0.019}{0.022} & \tabvalue{0.487}{0.024} & \tabvalue{0.007}{0.001} & \tabvalue{0.426}{0.032}\\

& \variable{ED} & \tabvalue{0.129}{0.030} & \tabvalue{0.809}{0.071} & \tabvalue{0.079}{0.020} & \tabvalue{0.807}{0.048}& \tabvalue{0.007}{0.005} & \tabvalue{0.235}{0.039} & \tabvalue{0.002}{0.000} & \tabvalue{0.230}{0.033}& \tabvalue{0.005}{0.004} & \tabvalue{0.233}{0.040} & \tabvalue{0.002}{0.000} & \tabvalue{0.232}{0.035}& \tabvalue{0.003}{0.003} & \tabvalue{0.233}{0.036} & \tabvalue{0.001}{0.000} & \tabvalue{0.227}{0.030}\\

             \midrule
            \multirow{6}{*}{\rotatebox{90}{\parbox{40pt}\centering\textbf{HGB}}} 
             & \variable{BD} & \tabvalue{0.011}{0.003} & \tabvalue{0.999}{0.001} & \tabvalue{0.058}{0.025} & \tabvalue{0.988}{0.006}& \tabvalue{0.004}{0.000} & \tabvalue{0.997}{0.001} & \tabvalue{0.005}{0.000} & \tabvalue{0.994}{0.001}& \tabvalue{0.004}{0.000} & \tabvalue{0.998}{0.000} & \tabvalue{0.005}{0.000} & \tabvalue{0.996}{0.001}& \tabvalue{0.004}{0.000} & \tabvalue{0.999}{0.000} & \tabvalue{0.005}{0.000} & \tabvalue{0.997}{0.001}\\

& \variable{MD} & \tabvalue{0.011}{0.003} & \tabvalue{0.998}{0.001} & \tabvalue{0.054}{0.025} & \tabvalue{0.984}{0.007}& \tabvalue{0.004}{0.000} & \tabvalue{0.990}{0.002} & \tabvalue{0.006}{0.002} & \tabvalue{0.976}{0.004}& \tabvalue{0.004}{0.000} & \tabvalue{0.992}{0.002} & \tabvalue{0.005}{0.001} & \tabvalue{0.979}{0.004}& \tabvalue{0.004}{0.000} & \tabvalue{0.992}{0.001} & \tabvalue{0.005}{0.001} & \tabvalue{0.980}{0.004}\\

& \variable{ED-v} & \tabvalue{0.008}{0.002} & \tabvalue{0.913}{0.101} & \tabvalue{0.012}{0.008} & \tabvalue{0.690}{0.057}& \tabvalue{0.000}{0.000} & \tabvalue{0.462}{0.028} & \tabvalue{0.000}{0.000} & \tabvalue{0.434}{0.031}& \tabvalue{0.000}{0.000} & \tabvalue{0.452}{0.028} & \tabvalue{0.000}{0.000} & \tabvalue{0.426}{0.032}& \tabvalue{0.000}{0.000} & \tabvalue{0.438}{0.027} & \tabvalue{0.000}{0.000} & \tabvalue{0.427}{0.027}\\

& \variable{ED-s} & \tabvalue{0.011}{0.003} & \tabvalue{0.999}{0.002} & \tabvalue{0.065}{0.031} & \tabvalue{0.984}{0.008}& \tabvalue{0.004}{0.000} & \tabvalue{0.993}{0.001} & \tabvalue{0.004}{0.000} & \tabvalue{0.989}{0.002}& \tabvalue{0.003}{0.000} & \tabvalue{0.994}{0.001} & \tabvalue{0.004}{0.000} & \tabvalue{0.992}{0.002}& \tabvalue{0.003}{0.000} & \tabvalue{0.995}{0.001} & \tabvalue{0.004}{0.000} & \tabvalue{0.994}{0.001}\\

& \variable{ED-o} & \tabvalue{0.011}{0.004} & \tabvalue{0.999}{0.001} & \tabvalue{0.068}{0.032} & \tabvalue{0.985}{0.007}& \tabvalue{0.004}{0.001} & \tabvalue{0.994}{0.001} & \tabvalue{0.005}{0.001} & \tabvalue{0.990}{0.002}& \tabvalue{0.004}{0.000} & \tabvalue{0.995}{0.001} & \tabvalue{0.005}{0.001} & \tabvalue{0.993}{0.001}& \tabvalue{0.004}{0.000} & \tabvalue{0.996}{0.001} & \tabvalue{0.005}{0.002} & \tabvalue{0.995}{0.001}\\

& \variable{ED} & \tabvalue{0.007}{0.002} & \tabvalue{0.995}{0.003} & \tabvalue{0.021}{0.008} & \tabvalue{0.960}{0.015}& \tabvalue{0.001}{0.000} & \tabvalue{0.978}{0.002} & \tabvalue{0.001}{0.000} & \tabvalue{0.966}{0.005}& \tabvalue{0.001}{0.000} & \tabvalue{0.983}{0.002} & \tabvalue{0.001}{0.000} & \tabvalue{0.974}{0.005}& \tabvalue{0.001}{0.000} & \tabvalue{0.985}{0.004} & \tabvalue{0.001}{0.000} & \tabvalue{0.978}{0.005}\\

            \bottomrule
        \end{tabular}
        }
                
        \caption{\textit{Static Dependency}: Results by assuming the absence of temporal dependencies among samples (\scbb{T} and \scbb{E} are randomly sampled from \scbb{D}).} 
    \label{tab:ufnb15_baseline_static}
    \end{subtable}

    \begin{subtable}[htbp]{1.8\columnwidth}
        \resizebox{1.0\columnwidth}{!}{
        \begin{tabular}{c|c ? cc|cc? cc|cc? cc|cc? cc|cc}
             \multicolumn{2}{c?}{Available Data} & \multicolumn{4}{c?}{Limited (100 per class) [N=1]} & \multicolumn{4}{c?}{Scarce (15\% of \scbb{D}) [N=1]} &  \multicolumn{4}{c?}{Moderate (40\% of \scbb{D}) [N=1]} &  \multicolumn{4}{c}{Abundant (80\% of \scbb{D}) [N=1]} \\ \hline
             \multicolumn{2}{c?}{Features} &
             \multicolumn{2}{c|}{Complete} & \multicolumn{2}{c?}{Essential} & \multicolumn{2}{c|}{Complete} & \multicolumn{2}{c?}{Essential} & \multicolumn{2}{c|}{Complete} & \multicolumn{2}{c?}{Essential} &  
             \multicolumn{2}{c|}{Complete} & \multicolumn{2}{c}{Essential} \\ \hline
            Alg. & Design & \smamath{fpr} & \smamath{tpr} & \smamath{fpr} & \smamath{tpr} & \smamath{fpr} & \smamath{tpr} & \smamath{fpr} & \smamath{tpr} & \smamath{fpr} & \smamath{tpr} & \smamath{fpr} & \smamath{tpr} & \smamath{fpr} & \smamath{tpr} & \smamath{fpr} & \smamath{tpr} \\
             \toprule
             
             \multirow{6}{*}{\rotatebox{90}{\parbox{40pt}\centering\textbf{RF}}} 
             & \variable{BD} & \tabvaluetemp{0.009} & \tabvaluetemp{0.999} & \tabvaluetemp{0.050} & \tabvaluetemp{0.996}& \tabvaluetemp{0.004} & \tabvaluetemp{0.995} & \tabvaluetemp{0.004} & \tabvaluetemp{0.992}& \tabvaluetemp{0.003} & \tabvaluetemp{0.995} & \tabvaluetemp{0.003} & \tabvaluetemp{0.992}& \tabvaluetemp{0.003} & \tabvaluetemp{0.994} & \tabvaluetemp{0.003} & \tabvaluetemp{0.991}\\

& \variable{MD} & \tabvaluetemp{0.008} & \tabvaluetemp{0.998} & \tabvaluetemp{0.052} & \tabvaluetemp{0.994}& \tabvaluetemp{0.003} & \tabvaluetemp{0.992} & \tabvaluetemp{0.004} & \tabvaluetemp{0.988}& \tabvaluetemp{0.003} & \tabvaluetemp{0.993} & \tabvaluetemp{0.003} & \tabvaluetemp{0.988}& \tabvaluetemp{0.003} & \tabvaluetemp{0.991} & \tabvaluetemp{0.003} & \tabvaluetemp{0.988}\\

& \variable{ED-v} & \tabvaluetemp{0.008} & \tabvaluetemp{0.997} & \tabvaluetemp{0.007} & \tabvaluetemp{0.706}& \tabvaluetemp{0.000} & \tabvaluetemp{0.424} & \tabvaluetemp{0.000} & \tabvaluetemp{0.341}& \tabvaluetemp{0.000} & \tabvaluetemp{0.425} & \tabvaluetemp{0.000} & \tabvaluetemp{0.367}& \tabvaluetemp{0.000} & \tabvaluetemp{0.427} & \tabvaluetemp{0.000} & \tabvaluetemp{0.383}\\

& \variable{ED-s} & \tabvaluetemp{0.009} & \tabvaluetemp{0.999} & \tabvaluetemp{0.050} & \tabvaluetemp{0.995}& \tabvaluetemp{0.004} & \tabvaluetemp{0.993} & \tabvaluetemp{0.004} & \tabvaluetemp{0.989}& \tabvaluetemp{0.003} & \tabvaluetemp{0.993} & \tabvaluetemp{0.003} & \tabvaluetemp{0.990}& \tabvaluetemp{0.003} & \tabvaluetemp{0.992} & \tabvaluetemp{0.003} & \tabvaluetemp{0.988}\\

& \variable{ED-o} & \tabvaluetemp{0.009} & \tabvaluetemp{0.999} & \tabvaluetemp{0.050} & \tabvaluetemp{0.995}& \tabvaluetemp{0.004} & \tabvaluetemp{0.993} & \tabvaluetemp{0.004} & \tabvaluetemp{0.989}& \tabvaluetemp{0.003} & \tabvaluetemp{0.993} & \tabvaluetemp{0.003} & \tabvaluetemp{0.990}& \tabvaluetemp{0.003} & \tabvaluetemp{0.992} & \tabvaluetemp{0.003} & \tabvaluetemp{0.988}\\

& \variable{ED} & \tabvaluetemp{0.007} & \tabvaluetemp{0.997} & \tabvaluetemp{0.015} & \tabvaluetemp{0.985}& \tabvaluetemp{0.001} & \tabvaluetemp{0.980} & \tabvaluetemp{0.001} & \tabvaluetemp{0.971}& \tabvaluetemp{0.001} & \tabvaluetemp{0.983} & \tabvaluetemp{0.001} & \tabvaluetemp{0.974}& \tabvaluetemp{0.001} & \tabvaluetemp{0.982} & \tabvaluetemp{0.001} & \tabvaluetemp{0.975}\\

            \midrule
            \multirow{6}{*}{\rotatebox{90}{\parbox{40pt}\centering\textbf{DT}}} 
             & \variable{BD} & \tabvaluetemp{0.006} & \tabvaluetemp{0.998} & \tabvaluetemp{0.050} & \tabvaluetemp{0.954}& \tabvaluetemp{0.003} & \tabvaluetemp{0.978} & \tabvaluetemp{0.004} & \tabvaluetemp{0.975}& \tabvaluetemp{0.003} & \tabvaluetemp{0.980} & \tabvaluetemp{0.003} & \tabvaluetemp{0.980}& \tabvaluetemp{0.003} & \tabvaluetemp{0.984} & \tabvaluetemp{0.003} & \tabvaluetemp{0.982}\\

& \variable{MD} & \tabvaluetemp{0.010} & \tabvaluetemp{0.994} & \tabvaluetemp{0.088} & \tabvaluetemp{0.962}& \tabvaluetemp{0.003} & \tabvaluetemp{0.978} & \tabvaluetemp{0.004} & \tabvaluetemp{0.975}& \tabvaluetemp{0.003} & \tabvaluetemp{0.983} & \tabvaluetemp{0.003} & \tabvaluetemp{0.978}& \tabvaluetemp{0.003} & \tabvaluetemp{0.985} & \tabvaluetemp{0.003} & \tabvaluetemp{0.983}\\

& \variable{ED-v} & \tabvaluetemp{0.005} & \tabvaluetemp{0.808} & \tabvaluetemp{0.003} & \tabvaluetemp{0.647}& \tabvaluetemp{0.000} & \tabvaluetemp{0.492} & \tabvaluetemp{0.000} & \tabvaluetemp{0.385}& \tabvaluetemp{0.000} & \tabvaluetemp{0.431} & \tabvaluetemp{0.000} & \tabvaluetemp{0.398}& \tabvaluetemp{0.000} & \tabvaluetemp{0.428} & \tabvaluetemp{0.000} & \tabvaluetemp{0.321}\\

& \variable{ED-s} & \tabvaluetemp{0.011} & \tabvaluetemp{0.996} & \tabvaluetemp{0.073} & \tabvaluetemp{0.974}& \tabvaluetemp{0.003} & \tabvaluetemp{0.987} & \tabvaluetemp{0.005} & \tabvaluetemp{0.981}& \tabvaluetemp{0.003} & \tabvaluetemp{0.986} & \tabvaluetemp{0.004} & \tabvaluetemp{0.982}& \tabvaluetemp{0.003} & \tabvaluetemp{0.987} & \tabvaluetemp{0.003} & \tabvaluetemp{0.986}\\

& \variable{ED-o} & \tabvaluetemp{0.011} & \tabvaluetemp{1.000} & \tabvaluetemp{0.100} & \tabvaluetemp{0.991}& \tabvaluetemp{0.003} & \tabvaluetemp{0.987} & \tabvaluetemp{0.005} & \tabvaluetemp{0.981}& \tabvaluetemp{0.003} & \tabvaluetemp{0.986} & \tabvaluetemp{0.004} & \tabvaluetemp{0.982}& \tabvaluetemp{0.003} & \tabvaluetemp{0.987} & \tabvaluetemp{0.003} & \tabvaluetemp{0.986}\\

& \variable{ED} & \tabvaluetemp{0.006} & \tabvaluetemp{0.983} & \tabvaluetemp{0.021} & \tabvaluetemp{0.920}& \tabvaluetemp{0.001} & \tabvaluetemp{0.967} & \tabvaluetemp{0.001} & \tabvaluetemp{0.951}& \tabvaluetemp{0.001} & \tabvaluetemp{0.968} & \tabvaluetemp{0.001} & \tabvaluetemp{0.959}& \tabvaluetemp{0.001} & \tabvaluetemp{0.974} & \tabvaluetemp{0.001} & \tabvaluetemp{0.965}\\
             \midrule
            \multirow{6}{*}{\rotatebox{90}{\parbox{40pt}\centering\textbf{LR}}} 
             & \variable{BD} & \tabvaluetemp{0.603} & \tabvaluetemp{0.987} & \tabvaluetemp{0.362} & \tabvaluetemp{0.928}& \tabvaluetemp{0.001} & \tabvaluetemp{0.048} & \tabvaluetemp{0.013} & \tabvaluetemp{0.193}& \tabvaluetemp{0.002} & \tabvaluetemp{0.298} & \tabvaluetemp{0.008} & \tabvaluetemp{0.209}& \tabvaluetemp{0.003} & \tabvaluetemp{0.295} & \tabvaluetemp{0.004} & \tabvaluetemp{0.294}\\

& \variable{MD} & \tabvaluetemp{0.224} & \tabvaluetemp{0.737} & \tabvaluetemp{0.349} & \tabvaluetemp{0.870}& \tabvaluetemp{0.003} & \tabvaluetemp{0.280} & \tabvaluetemp{0.004} & \tabvaluetemp{0.285}& \tabvaluetemp{0.005} & \tabvaluetemp{0.284} & \tabvaluetemp{0.004} & \tabvaluetemp{0.290}& \tabvaluetemp{0.006} & \tabvaluetemp{0.286} & \tabvaluetemp{0.005} & \tabvaluetemp{0.297}\\

& \variable{ED-v} & \tabvaluetemp{0.062} & \tabvaluetemp{0.678} & \tabvaluetemp{0.088} & \tabvaluetemp{0.703}& \tabvaluetemp{0.000} & \tabvaluetemp{0.159} & \tabvaluetemp{0.000} & \tabvaluetemp{0.094}& \tabvaluetemp{0.000} & \tabvaluetemp{0.005} & \tabvaluetemp{0.000} & \tabvaluetemp{0.129}& \tabvaluetemp{0.000} & \tabvaluetemp{0.226} & \tabvaluetemp{0.000} & \tabvaluetemp{0.129}\\

& \variable{ED-s} & \tabvaluetemp{0.217} & \tabvaluetemp{0.952} & \tabvaluetemp{0.299} & \tabvaluetemp{0.961}& \tabvaluetemp{0.008} & \tabvaluetemp{0.469} & \tabvaluetemp{0.005} & \tabvaluetemp{0.409}& \tabvaluetemp{0.008} & \tabvaluetemp{0.456} & \tabvaluetemp{0.005} & \tabvaluetemp{0.409}& \tabvaluetemp{0.010} & \tabvaluetemp{0.472} & \tabvaluetemp{0.003} & \tabvaluetemp{0.454}\\

& \variable{ED-o} & \tabvaluetemp{0.433} & \tabvaluetemp{0.985} & \tabvaluetemp{0.300} & \tabvaluetemp{0.971}& \tabvaluetemp{0.080} & \tabvaluetemp{0.512} & \tabvaluetemp{0.007} & \tabvaluetemp{0.413}& \tabvaluetemp{0.081} & \tabvaluetemp{0.478} & \tabvaluetemp{0.006} & \tabvaluetemp{0.414}& \tabvaluetemp{0.011} & \tabvaluetemp{0.482} & \tabvaluetemp{0.007} & \tabvaluetemp{0.494}\\

& \variable{ED} & \tabvaluetemp{0.154} & \tabvaluetemp{0.795} & \tabvaluetemp{0.141} & \tabvaluetemp{0.916}& \tabvaluetemp{0.012} & \tabvaluetemp{0.241} & \tabvaluetemp{0.001} & \tabvaluetemp{0.201}& \tabvaluetemp{0.012} & \tabvaluetemp{0.198} & \tabvaluetemp{0.001} & \tabvaluetemp{0.231}& \tabvaluetemp{0.002} & \tabvaluetemp{0.244} & \tabvaluetemp{0.002} & \tabvaluetemp{0.238}\\
             \midrule
            \multirow{6}{*}{\rotatebox{90}{\parbox{40pt}\centering\textbf{HGB}}} 
            & \variable{BD} & \tabvaluetemp{0.012} & \tabvaluetemp{0.996} & \tabvaluetemp{0.062} & \tabvaluetemp{0.986}& \tabvaluetemp{0.004} & \tabvaluetemp{0.997} & \tabvaluetemp{0.005} & \tabvaluetemp{0.994}& \tabvaluetemp{0.004} & \tabvaluetemp{0.998} & \tabvaluetemp{0.005} & \tabvaluetemp{0.996}& \tabvaluetemp{0.004} & \tabvaluetemp{0.999} & \tabvaluetemp{0.004} & \tabvaluetemp{0.997}\\

& \variable{MD} & \tabvaluetemp{0.013} & \tabvaluetemp{0.996} & \tabvaluetemp{0.064} & \tabvaluetemp{0.979}& \tabvaluetemp{0.004} & \tabvaluetemp{0.987} & \tabvaluetemp{0.006} & \tabvaluetemp{0.973}& \tabvaluetemp{0.004} & \tabvaluetemp{0.992} & \tabvaluetemp{0.005} & \tabvaluetemp{0.977}& \tabvaluetemp{0.004} & \tabvaluetemp{0.993} & \tabvaluetemp{0.005} & \tabvaluetemp{0.985}\\

& \variable{ED-v} & \tabvaluetemp{0.005} & \tabvaluetemp{0.996} & \tabvaluetemp{0.040} & \tabvaluetemp{0.765}& \tabvaluetemp{0.000} & \tabvaluetemp{0.505} & \tabvaluetemp{0.001} & \tabvaluetemp{0.384}& \tabvaluetemp{0.000} & \tabvaluetemp{0.492} & \tabvaluetemp{0.000} & \tabvaluetemp{0.452}& \tabvaluetemp{0.000} & \tabvaluetemp{0.452} & \tabvaluetemp{0.000} & \tabvaluetemp{0.448}\\

& \variable{ED-s} & \tabvaluetemp{0.011} & \tabvaluetemp{0.996} & \tabvaluetemp{0.065} & \tabvaluetemp{0.980}& \tabvaluetemp{0.004} & \tabvaluetemp{0.993} & \tabvaluetemp{0.004} & \tabvaluetemp{0.989}& \tabvaluetemp{0.004} & \tabvaluetemp{0.993} & \tabvaluetemp{0.004} & \tabvaluetemp{0.991}& \tabvaluetemp{0.004} & \tabvaluetemp{0.997} & \tabvaluetemp{0.004} & \tabvaluetemp{0.994}\\

& \variable{ED-o} & \tabvaluetemp{0.031} & \tabvaluetemp{1.000} & \tabvaluetemp{0.065} & \tabvaluetemp{0.982}& \tabvaluetemp{0.004} & \tabvaluetemp{0.994} & \tabvaluetemp{0.005} & \tabvaluetemp{0.989}& \tabvaluetemp{0.004} & \tabvaluetemp{0.994} & \tabvaluetemp{0.005} & \tabvaluetemp{0.992}& \tabvaluetemp{0.004} & \tabvaluetemp{0.997} & \tabvaluetemp{0.005} & \tabvaluetemp{0.995}\\

& \variable{ED} & \tabvaluetemp{0.010} & \tabvaluetemp{0.997} & \tabvaluetemp{0.032} & \tabvaluetemp{0.959}& \tabvaluetemp{0.001} & \tabvaluetemp{0.979} & \tabvaluetemp{0.001} & \tabvaluetemp{0.963}& \tabvaluetemp{0.001} & \tabvaluetemp{0.983} & \tabvaluetemp{0.001} & \tabvaluetemp{0.972}& \tabvaluetemp{0.001} & \tabvaluetemp{0.988} & \tabvaluetemp{0.001} & \tabvaluetemp{0.977}\\
            
            \bottomrule
        \end{tabular}
        }
                
        \caption{\textit{Temporal Dependency}: Results by assuming the presence of temporal dependencies among samples (the `first' samples of \scbb{D} are put in \scbb{T}, while the last 20\% represent \scbb{E}).} 
    \label{tab:ufnb15_baseline_temporal}
    \end{subtable}

    \label{tab:ufnb15_baseline}
\end{table*}

\begin{table*}
  \centering
  \caption{\dataset{UF-NB15}. Results against adversarial (original \scmath{tpr} and adversarial \scmath{tpr}) and unknown attacks (the \scmath{tpr} is the average on the `unknown' attacks, while the \scmath{fpr} is due to training on a new \scbb{T} that does not have the `unknown' class.).}

    \begin{subtable}[htbp]{1.99\columnwidth}
        \resizebox{1.0\columnwidth}{!}{
        \begin{tabular}{c|c ? cc|cc? cc|cc? cc|cc? cc|cc}
\multicolumn{2}{c?}{Available Data} & \multicolumn{4}{c?}{Limited (100 per class) [N=1000]} & \multicolumn{4}{c?}{Scarce (15\% of \scbb{D}) [N=100]} &  \multicolumn{4}{c?}{Moderate (40\% of \scbb{D}) [N=100]} &  \multicolumn{4}{c}{Abundant (80\% of \scbb{D}) [N=100]} \\ \hline
             \multicolumn{2}{c?}{Scenario} &
             \multicolumn{2}{c|}{Adversarial Attacks} & \multicolumn{2}{c?}{Unknown Attacks} & \multicolumn{2}{c|}{Adversarial Attacks} & \multicolumn{2}{c?}{Unknown Attacks} & \multicolumn{2}{c|}{Adversarial Attacks} & \multicolumn{2}{c?}{Unknown Attacks} & \multicolumn{2}{c|}{Adversarial Attacks} & \multicolumn{2}{c?}{Unknown Attacks}  \\ \hline
            Alg. & Design & 
            \footnotesize{$tpr$} \tiny{(org)} & \footnotesize{$tpr$} \tiny{(adv)} & \footnotesize{$fpr$} & \footnotesize{$tpr$} & \footnotesize{$tpr$} \tiny{(org)} & \footnotesize{$tpr$} \tiny{(adv)} & \footnotesize{$fpr$} & \footnotesize{$tpr$} & \footnotesize{$tpr$} \tiny{(org)} & \footnotesize{$tpr$} \tiny{(adv)} & \footnotesize{$fpr$} & \footnotesize{$tpr$} & \footnotesize{$tpr$} \tiny{(org)} & \footnotesize{$tpr$} \tiny{(adv)} & \footnotesize{$fpr$} & \footnotesize{$tpr$} \\
             \toprule
             
             \multirow{5}{*}{\rotatebox{90}{\parbox{40pt}\centering\textbf{RF}}} 
             & \variable{BD} & \tabvalue{0.987}{0.006} & \tabvalue{0.460}{0.395} & \tabvalue{0.011}{0.002} & \tabvalue{0.999}{0.001}& \tabvalue{0.992}{0.003} & \tabvalue{0.956}{0.148} & \tabvalue{0.003}{0.000} & \tabvalue{0.937}{0.007}& \tabvalue{0.994}{0.002} & \tabvalue{0.996}{0.011} & \tabvalue{0.003}{0.000} & \tabvalue{0.921}{0.005}& \tabvalue{0.995}{0.002} & \tabvalue{0.992}{0.051} & \tabvalue{0.003}{0.000} & \tabvalue{0.911}{0.004}\\

& \variable{MD} & \tabvalue{0.982}{0.007} & \tabvalue{0.295}{0.293} & \tabvalue{0.009}{0.002} & \tabvalue{0.997}{0.002}& \tabvalue{0.990}{0.003} & \tabvalue{0.899}{0.222} & \tabvalue{0.003}{0.000} & \tabvalue{0.920}{0.008}& \tabvalue{0.993}{0.002} & \tabvalue{0.972}{0.067} & \tabvalue{0.003}{0.000} & \tabvalue{0.906}{0.007}& \tabvalue{0.995}{0.002} & \tabvalue{0.947}{0.123} & \tabvalue{0.003}{0.000} & \tabvalue{0.897}{0.005}\\

& \variable{ED-v} & \tabvalue{0.965}{0.025} & \tabvalue{0.002}{0.021} & \tabvalue{0.007}{0.001} & \tabvalue{0.864}{0.057}& \tabvalue{0.583}{0.134} & \tabvalue{0.000}{0.000} & \tabvalue{0.000}{0.000} & \tabvalue{0.478}{0.049}& \tabvalue{0.576}{0.108} & \tabvalue{0.000}{0.000} & \tabvalue{0.000}{0.000} & \tabvalue{0.474}{0.049}& \tabvalue{0.550}{0.077} & \tabvalue{0.000}{0.000} & \tabvalue{0.000}{0.000} & \tabvalue{0.473}{0.045}\\

& \variable{ED-s} & \tabvalue{0.983}{0.006} & \tabvalue{0.143}{0.211} & \tabvalue{0.011}{0.002} & \tabvalue{0.998}{0.003}& \tabvalue{0.990}{0.003} & \tabvalue{0.889}{0.222} & \tabvalue{0.003}{0.000} & \tabvalue{0.928}{0.006}& \tabvalue{0.993}{0.002} & \tabvalue{0.958}{0.101} & \tabvalue{0.003}{0.000} & \tabvalue{0.912}{0.005}& \tabvalue{0.995}{0.002} & \tabvalue{0.922}{0.149} & \tabvalue{0.003}{0.000} & \tabvalue{0.902}{0.004}\\

& \variable{ED-o} & \tabvalue{0.985}{0.005} & \tabvalue{0.149}{0.214} & \tabvalue{0.012}{0.003} & \tabvalue{0.999}{0.001}& \tabvalue{0.990}{0.003} & \tabvalue{0.889}{0.222} & \tabvalue{0.003}{0.000} & \tabvalue{0.928}{0.006}& \tabvalue{0.993}{0.002} & \tabvalue{0.958}{0.101} & \tabvalue{0.003}{0.000} & \tabvalue{0.912}{0.005}& \tabvalue{0.995}{0.002} & \tabvalue{0.922}{0.149} & \tabvalue{0.003}{0.000} & \tabvalue{0.902}{0.004}\\
             
            \midrule
            \multirow{5}{*}{\rotatebox{90}{\parbox{40pt}\centering\textbf{DT}}} 
             & \variable{BD} & \tabvalue{0.987}{0.010} & \tabvalue{0.605}{0.460} & \tabvalue{0.011}{0.003} & \tabvalue{0.996}{0.005}& \tabvalue{0.988}{0.004} & \tabvalue{0.727}{0.364} & \tabvalue{0.003}{0.000} & \tabvalue{0.892}{0.008}& \tabvalue{0.992}{0.002} & \tabvalue{0.803}{0.315} & \tabvalue{0.003}{0.000} & \tabvalue{0.883}{0.008}& \tabvalue{0.994}{0.002} & \tabvalue{0.834}{0.277} & \tabvalue{0.002}{0.000} & \tabvalue{0.878}{0.006}\\

& \variable{MD} & \tabvalue{0.987}{0.012} & \tabvalue{0.974}{0.091} & \tabvalue{0.011}{0.003} & \tabvalue{0.994}{0.007}& \tabvalue{0.988}{0.004} & \tabvalue{0.948}{0.137} & \tabvalue{0.003}{0.000} & \tabvalue{0.893}{0.011}& \tabvalue{0.992}{0.002} & \tabvalue{0.944}{0.127} & \tabvalue{0.003}{0.000} & \tabvalue{0.878}{0.010}& \tabvalue{0.994}{0.002} & \tabvalue{0.892}{0.217} & \tabvalue{0.002}{0.000} & \tabvalue{0.870}{0.009}\\

& \variable{ED-v} & \tabvalue{0.861}{0.198} & \tabvalue{0.076}{0.214} & \tabvalue{0.007}{0.002} & \tabvalue{0.902}{0.075}& \tabvalue{0.383}{0.210} & \tabvalue{0.025}{0.110} & \tabvalue{0.000}{0.000} & \tabvalue{0.442}{0.051}& \tabvalue{0.466}{0.153} & \tabvalue{0.002}{0.007} & \tabvalue{0.000}{0.000} & \tabvalue{0.443}{0.042}& \tabvalue{0.451}{0.117} & \tabvalue{0.006}{0.030} & \tabvalue{0.000}{0.000} & \tabvalue{0.429}{0.041}\\

& \variable{ED-s} & \tabvalue{0.989}{0.007} & \tabvalue{0.654}{0.391} & \tabvalue{0.011}{0.003} & \tabvalue{0.997}{0.004}& \tabvalue{0.990}{0.003} & \tabvalue{0.817}{0.280} & \tabvalue{0.003}{0.000} & \tabvalue{0.915}{0.008}& \tabvalue{0.993}{0.002} & \tabvalue{0.740}{0.301} & \tabvalue{0.003}{0.000} & \tabvalue{0.903}{0.006}& \tabvalue{0.994}{0.002} & \tabvalue{0.717}{0.292} & \tabvalue{0.003}{0.000} & \tabvalue{0.896}{0.005}\\

& \variable{ED-o} & \tabvalue{0.993}{0.005} & \tabvalue{0.930}{0.138} & \tabvalue{0.012}{0.004} & \tabvalue{0.998}{0.003}& \tabvalue{0.990}{0.003} & \tabvalue{0.818}{0.279} & \tabvalue{0.003}{0.000} & \tabvalue{0.915}{0.008}& \tabvalue{0.993}{0.002} & \tabvalue{0.740}{0.301} & \tabvalue{0.003}{0.000} & \tabvalue{0.903}{0.006}& \tabvalue{0.994}{0.002} & \tabvalue{0.717}{0.292} & \tabvalue{0.003}{0.000} & \tabvalue{0.896}{0.005}\\
             \midrule
            \multirow{5}{*}{\rotatebox{90}{\parbox{40pt}\centering\textbf{LR}}} 
             & \variable{BD} & \tabvalue{0.808}{0.311} & \tabvalue{0.324}{0.386} & \tabvalue{0.660}{0.084} & \tabvalue{0.950}{0.074}& \tabvalue{0.300}{0.375} & \tabvalue{0.018}{0.008} & \tabvalue{0.012}{0.013} & \tabvalue{0.409}{0.085}& \tabvalue{0.478}{0.431} & \tabvalue{0.019}{0.008} & \tabvalue{0.008}{0.008} & \tabvalue{0.429}{0.063}& \tabvalue{0.637}{0.411} & \tabvalue{0.020}{0.005} & \tabvalue{0.006}{0.005} & \tabvalue{0.440}{0.055}\\

& \variable{MD} & \tabvalue{0.950}{0.135} & \tabvalue{0.961}{0.092} & \tabvalue{0.513}{0.282} & \tabvalue{0.841}{0.116}& \tabvalue{0.897}{0.221} & \tabvalue{0.117}{0.236} & \tabvalue{0.003}{0.000} & \tabvalue{0.451}{0.020}& \tabvalue{0.955}{0.007} & \tabvalue{0.104}{0.165} & \tabvalue{0.003}{0.000} & \tabvalue{0.446}{0.014}& \tabvalue{0.944}{0.093} & \tabvalue{0.103}{0.143} & \tabvalue{0.003}{0.000} & \tabvalue{0.442}{0.011}\\

& \variable{ED-v} & \tabvalue{0.776}{0.281} & \tabvalue{0.020}{0.085} & \tabvalue{0.029}{0.014} & \tabvalue{0.730}{0.073}& \tabvalue{0.007}{0.022} & \tabvalue{0.000}{0.000} & \tabvalue{0.000}{0.000} & \tabvalue{0.084}{0.118}& \tabvalue{0.006}{0.020} & \tabvalue{0.000}{0.000} & \tabvalue{0.000}{0.000} & \tabvalue{0.057}{0.114}& \tabvalue{0.002}{0.005} & \tabvalue{0.000}{0.000} & \tabvalue{0.000}{0.000} & \tabvalue{0.079}{0.119}\\

& \variable{ED-s} & \tabvalue{0.968}{0.011} & \tabvalue{0.084}{0.201} & \tabvalue{0.265}{0.084} & \tabvalue{0.922}{0.020}& \tabvalue{0.863}{0.227} & \tabvalue{0.002}{0.004} & \tabvalue{0.007}{0.001} & \tabvalue{0.593}{0.043}& \tabvalue{0.874}{0.217} & \tabvalue{0.004}{0.007} & \tabvalue{0.008}{0.001} & \tabvalue{0.593}{0.034}& \tabvalue{0.921}{0.133} & \tabvalue{0.004}{0.009} & \tabvalue{0.008}{0.001} & \tabvalue{0.597}{0.026}\\

& \variable{ED-o} & \tabvalue{0.971}{0.011} & \tabvalue{0.212}{0.345} & \tabvalue{0.432}{0.106} & \tabvalue{0.957}{0.041}& \tabvalue{0.871}{0.217} & \tabvalue{0.018}{0.005} & \tabvalue{0.040}{0.030} & \tabvalue{0.630}{0.037}& \tabvalue{0.879}{0.215} & \tabvalue{0.019}{0.009} & \tabvalue{0.026}{0.027} & \tabvalue{0.622}{0.026}& \tabvalue{0.926}{0.133} & \tabvalue{0.020}{0.004} & \tabvalue{0.017}{0.019} & \tabvalue{0.622}{0.022}\\
             \midrule
            \multirow{5}{*}{\rotatebox{90}{\parbox{40pt}\centering\textbf{HGB}}} 
             & \variable{BD} & \tabvalue{0.985}{0.008} & \tabvalue{0.566}{0.418} & \tabvalue{0.011}{0.003} & \tabvalue{0.998}{0.003}& \tabvalue{0.995}{0.002} & \tabvalue{0.968}{0.091} & \tabvalue{0.004}{0.000} & \tabvalue{0.930}{0.007}& \tabvalue{0.999}{0.001} & \tabvalue{0.980}{0.038} & \tabvalue{0.004}{0.000} & \tabvalue{0.922}{0.005}& \tabvalue{0.999}{0.001} & \tabvalue{0.984}{0.033} & \tabvalue{0.003}{0.000} & \tabvalue{0.917}{0.003}\\

& \variable{MD} & \tabvalue{0.982}{0.009} & \tabvalue{0.471}{0.393} & \tabvalue{0.011}{0.003} & \tabvalue{0.997}{0.003}& \tabvalue{0.958}{0.014} & \tabvalue{0.474}{0.363} & \tabvalue{0.004}{0.001} & \tabvalue{0.913}{0.011}& \tabvalue{0.965}{0.012} & \tabvalue{0.619}{0.365} & \tabvalue{0.004}{0.000} & \tabvalue{0.906}{0.010}& \tabvalue{0.964}{0.012} & \tabvalue{0.644}{0.332} & \tabvalue{0.004}{0.001} & \tabvalue{0.901}{0.008}\\

& \variable{ED-v} & \tabvalue{0.796}{0.264} & \tabvalue{0.009}{0.044} & \tabvalue{0.007}{0.002} & \tabvalue{0.877}{0.076}& \tabvalue{0.640}{0.089} & \tabvalue{0.009}{0.054} & \tabvalue{0.000}{0.000} & \tabvalue{0.489}{0.044}& \tabvalue{0.626}{0.059} & \tabvalue{0.023}{0.101} & \tabvalue{0.000}{0.000} & \tabvalue{0.476}{0.045}& \tabvalue{0.620}{0.046} & \tabvalue{0.021}{0.092} & \tabvalue{0.000}{0.000} & \tabvalue{0.478}{0.042}\\

& \variable{ED-s} & \tabvalue{0.984}{0.008} & \tabvalue{0.484}{0.393} & \tabvalue{0.011}{0.003} & \tabvalue{0.997}{0.004}& \tabvalue{0.987}{0.007} & \tabvalue{0.821}{0.163} & \tabvalue{0.003}{0.000} & \tabvalue{0.918}{0.008}& \tabvalue{0.992}{0.007} & \tabvalue{0.843}{0.199} & \tabvalue{0.003}{0.000} & \tabvalue{0.908}{0.006}& \tabvalue{0.994}{0.005} & \tabvalue{0.872}{0.197} & \tabvalue{0.003}{0.000} & \tabvalue{0.902}{0.005}\\

& \variable{ED-o} & \tabvalue{0.987}{0.006} & \tabvalue{0.536}{0.387} & \tabvalue{0.011}{0.004} & \tabvalue{0.998}{0.003}& \tabvalue{0.990}{0.005} & \tabvalue{0.851}{0.129} & \tabvalue{0.004}{0.001} & \tabvalue{0.929}{0.008}& \tabvalue{0.996}{0.004} & \tabvalue{0.859}{0.189} & \tabvalue{0.003}{0.000} & \tabvalue{0.920}{0.007}& \tabvalue{0.997}{0.003} & \tabvalue{0.887}{0.182} & \tabvalue{0.003}{0.000} & \tabvalue{0.910}{0.006}\\
            
            \bottomrule
        \end{tabular}
        }
                
        \caption{\textit{Static Dependency}: Results by assuming the absence of temporal dependencies among samples (\scbb{T} and \scbb{E} are randomly sampled from \scbb{D}).} 
    \label{tab:ufnb15_open_static}
    \end{subtable}

    \begin{subtable}[htbp]{1.8\columnwidth}
        \resizebox{1.0\columnwidth}{!}{
        \begin{tabular}{c|c ? cc|cc? cc|cc? cc|cc? cc|cc}
\multicolumn{2}{c?}{Available Data} & \multicolumn{4}{c?}{Limited (100 per class) [N=1]} & \multicolumn{4}{c?}{Scarce (15\% of \scbb{D}) [N=1]} &  \multicolumn{4}{c?}{Moderate (40\% of \scbb{D}) [N=1]} &  \multicolumn{4}{c}{Abundant (80\% of \scbb{D}) [N=1]} \\ \hline
             \multicolumn{2}{c?}{Scenario} &
             \multicolumn{2}{c|}{Adversarial Attacks} & \multicolumn{2}{c?}{Unknown Attacks} & \multicolumn{2}{c|}{Adversarial Attacks} & \multicolumn{2}{c?}{Unknown Attacks} & \multicolumn{2}{c|}{Adversarial Attacks} & \multicolumn{2}{c?}{Unknown Attacks} & \multicolumn{2}{c|}{Adversarial Attacks} & \multicolumn{2}{c?}{Unknown Attacks}  \\ \hline
            Alg. & Design & 
            \footnotesize{$tpr$} \tiny{(org)} & \footnotesize{$tpr$} \tiny{(adv)} & \footnotesize{$fpr$} & \footnotesize{$tpr$} & \footnotesize{$tpr$} \tiny{(org)} & \footnotesize{$tpr$} \tiny{(adv)} & \footnotesize{$fpr$} & \footnotesize{$tpr$} & \footnotesize{$tpr$} \tiny{(org)} & \footnotesize{$tpr$} \tiny{(adv)} & \footnotesize{$fpr$} & \footnotesize{$tpr$} & \footnotesize{$tpr$} \tiny{(org)} & \footnotesize{$tpr$} \tiny{(adv)} & \footnotesize{$fpr$} & \footnotesize{$tpr$} \\
             \toprule
             
             \multirow{5}{*}{\rotatebox{90}{\parbox{40pt}\centering\textbf{RF}}} 
              & \variable{BD} & \tabvaluetemp{0.988} & \tabvaluetemp{0.063} & \tabvaluetemp{0.009} & \tabvaluetemp{0.999}& \tabvaluetemp{0.991} & \tabvaluetemp{1.000} & \tabvaluetemp{0.003} & \tabvaluetemp{0.942}& \tabvaluetemp{0.993} & \tabvaluetemp{1.000} & \tabvaluetemp{0.003} & \tabvaluetemp{0.923}& \tabvaluetemp{0.994} & \tabvaluetemp{1.000} & \tabvaluetemp{0.003} & \tabvaluetemp{0.909}\\

& \variable{MD} & \tabvaluetemp{0.986} & \tabvaluetemp{0.241} & \tabvaluetemp{0.008} & \tabvaluetemp{0.998}& \tabvaluetemp{0.990} & \tabvaluetemp{1.000} & \tabvaluetemp{0.003} & \tabvaluetemp{0.928}& \tabvaluetemp{0.993} & \tabvaluetemp{1.000} & \tabvaluetemp{0.003} & \tabvaluetemp{0.907}& \tabvaluetemp{0.994} & \tabvaluetemp{1.000} & \tabvaluetemp{0.003} & \tabvaluetemp{0.897}\\

& \variable{ED-v} & \tabvaluetemp{0.979} & \tabvaluetemp{0.000} & \tabvaluetemp{0.008} & \tabvaluetemp{0.994}& \tabvaluetemp{0.238} & \tabvaluetemp{0.000} & \tabvaluetemp{0.000} & \tabvaluetemp{0.542}& \tabvaluetemp{0.460} & \tabvaluetemp{0.000} & \tabvaluetemp{0.000} & \tabvaluetemp{0.472}& \tabvaluetemp{0.434} & \tabvaluetemp{0.000} & \tabvaluetemp{0.000} & \tabvaluetemp{0.447}\\

& \variable{ED-s} & \tabvaluetemp{0.984} & \tabvaluetemp{0.032} & \tabvaluetemp{0.009} & \tabvaluetemp{0.999}& \tabvaluetemp{0.989} & \tabvaluetemp{1.000} & \tabvaluetemp{0.003} & \tabvaluetemp{0.934}& \tabvaluetemp{0.993} & \tabvaluetemp{1.000} & \tabvaluetemp{0.003} & \tabvaluetemp{0.915}& \tabvaluetemp{0.994} & \tabvaluetemp{0.993} & \tabvaluetemp{0.003} & \tabvaluetemp{0.900}\\

& \variable{ED-o} & \tabvaluetemp{0.984} & \tabvaluetemp{0.032} & \tabvaluetemp{0.009} & \tabvaluetemp{0.999}& \tabvaluetemp{0.989} & \tabvaluetemp{1.000} & \tabvaluetemp{0.003} & \tabvaluetemp{0.934}& \tabvaluetemp{0.993} & \tabvaluetemp{1.000} & \tabvaluetemp{0.003} & \tabvaluetemp{0.915}& \tabvaluetemp{0.994} & \tabvaluetemp{0.993} & \tabvaluetemp{0.003} & \tabvaluetemp{0.900}\\
             
            \midrule
            \multirow{5}{*}{\rotatebox{90}{\parbox{40pt}\centering\textbf{DT}}} 
            & \variable{BD} & \tabvaluetemp{0.976} & \tabvaluetemp{0.976} & \tabvaluetemp{0.008} & \tabvaluetemp{0.995}& \tabvaluetemp{0.994} & \tabvaluetemp{0.991} & \tabvaluetemp{0.003} & \tabvaluetemp{0.895}& \tabvaluetemp{0.989} & \tabvaluetemp{0.980} & \tabvaluetemp{0.003} & \tabvaluetemp{0.893}& \tabvaluetemp{0.994} & \tabvaluetemp{0.956} & \tabvaluetemp{0.002} & \tabvaluetemp{0.883}\\

& \variable{MD} & \tabvaluetemp{0.968} & \tabvaluetemp{0.986} & \tabvaluetemp{0.008} & \tabvaluetemp{0.993}& \tabvaluetemp{0.988} & \tabvaluetemp{0.877} & \tabvaluetemp{0.003} & \tabvaluetemp{0.904}& \tabvaluetemp{0.987} & \tabvaluetemp{0.959} & \tabvaluetemp{0.003} & \tabvaluetemp{0.899}& \tabvaluetemp{0.994} & \tabvaluetemp{0.999} & \tabvaluetemp{0.002} & \tabvaluetemp{0.874}\\

& \variable{ED-v} & \tabvaluetemp{0.694} & \tabvaluetemp{0.000} & \tabvaluetemp{0.004} & \tabvaluetemp{0.813}& \tabvaluetemp{0.458} & \tabvaluetemp{0.000} & \tabvaluetemp{0.000} & \tabvaluetemp{0.524}& \tabvaluetemp{0.604} & \tabvaluetemp{0.000} & \tabvaluetemp{0.000} & \tabvaluetemp{0.451}& \tabvaluetemp{0.043} & \tabvaluetemp{0.006} & \tabvaluetemp{0.000} & \tabvaluetemp{0.449}\\

& \variable{ED-s} & \tabvaluetemp{0.991} & \tabvaluetemp{1.000} & \tabvaluetemp{0.011} & \tabvaluetemp{0.997}& \tabvaluetemp{0.993} & \tabvaluetemp{1.000} & \tabvaluetemp{0.003} & \tabvaluetemp{0.923}& \tabvaluetemp{0.990} & \tabvaluetemp{0.998} & \tabvaluetemp{0.003} & \tabvaluetemp{0.914}& \tabvaluetemp{0.995} & \tabvaluetemp{0.983} & \tabvaluetemp{0.003} & \tabvaluetemp{0.902}\\

& \variable{ED-o} & \tabvaluetemp{0.995} & \tabvaluetemp{1.000} & \tabvaluetemp{0.011} & \tabvaluetemp{0.998}& \tabvaluetemp{0.993} & \tabvaluetemp{1.000} & \tabvaluetemp{0.003} & \tabvaluetemp{0.923}& \tabvaluetemp{0.990} & \tabvaluetemp{0.998} & \tabvaluetemp{0.003} & \tabvaluetemp{0.914}& \tabvaluetemp{0.995} & \tabvaluetemp{0.983} & \tabvaluetemp{0.003} & \tabvaluetemp{0.902}\\
             \midrule
            \multirow{5}{*}{\rotatebox{90}{\parbox{40pt}\centering\textbf{LR}}} 
             & \variable{BD} & \tabvaluetemp{0.987} & \tabvaluetemp{0.992} & \tabvaluetemp{0.580} & \tabvaluetemp{0.963}& \tabvaluetemp{0.031} & \tabvaluetemp{0.013} & \tabvaluetemp{0.012} & \tabvaluetemp{0.478}& \tabvaluetemp{0.092} & \tabvaluetemp{0.024} & \tabvaluetemp{0.012} & \tabvaluetemp{0.478}& \tabvaluetemp{0.953} & \tabvaluetemp{0.023} & \tabvaluetemp{0.003} & \tabvaluetemp{0.333}\\

& \variable{MD} & \tabvaluetemp{0.940} & \tabvaluetemp{0.998} & \tabvaluetemp{0.201} & \tabvaluetemp{0.745}& \tabvaluetemp{0.949} & \tabvaluetemp{0.018} & \tabvaluetemp{0.003} & \tabvaluetemp{0.479}& \tabvaluetemp{0.954} & \tabvaluetemp{0.027} & \tabvaluetemp{0.003} & \tabvaluetemp{0.439}& \tabvaluetemp{0.953} & \tabvaluetemp{0.022} & \tabvaluetemp{0.004} & \tabvaluetemp{0.445}\\

& \variable{ED-v} & \tabvaluetemp{0.945} & \tabvaluetemp{0.979} & \tabvaluetemp{0.061} & \tabvaluetemp{0.799}& \tabvaluetemp{0.004} & \tabvaluetemp{0.000} & \tabvaluetemp{0.000} & \tabvaluetemp{0.223}& \tabvaluetemp{0.008} & \tabvaluetemp{0.000} & \tabvaluetemp{0.000} & \tabvaluetemp{0.001}& \tabvaluetemp{0.000} & \tabvaluetemp{0.000} & \tabvaluetemp{0.000} & \tabvaluetemp{0.274}\\

& \variable{ED-s} & \tabvaluetemp{0.974} & \tabvaluetemp{0.993} & \tabvaluetemp{0.230} & \tabvaluetemp{0.942}& \tabvaluetemp{0.946} & \tabvaluetemp{0.000} & \tabvaluetemp{0.007} & \tabvaluetemp{0.604}& \tabvaluetemp{0.947} & \tabvaluetemp{0.000} & \tabvaluetemp{0.007} & \tabvaluetemp{0.545}& \tabvaluetemp{0.943} & \tabvaluetemp{0.000} & \tabvaluetemp{0.009} & \tabvaluetemp{0.614}\\

& \variable{ED-o} & \tabvaluetemp{0.985} & \tabvaluetemp{0.995} & \tabvaluetemp{0.414} & \tabvaluetemp{0.984}& \tabvaluetemp{0.949} & \tabvaluetemp{0.018} & \tabvaluetemp{0.069} & \tabvaluetemp{0.649}& \tabvaluetemp{0.955} & \tabvaluetemp{0.023} & \tabvaluetemp{0.070} & \tabvaluetemp{0.614}& \tabvaluetemp{0.953} & \tabvaluetemp{0.022} & \tabvaluetemp{0.010} & \tabvaluetemp{0.627}\\
             \midrule
            \multirow{5}{*}{\rotatebox{90}{\parbox{40pt}\centering\textbf{HGB}}} 
             & \variable{BD} & \tabvaluetemp{0.979} & \tabvaluetemp{0.012} & \tabvaluetemp{0.012} & \tabvaluetemp{0.997}& \tabvaluetemp{0.996} & \tabvaluetemp{0.979} & \tabvaluetemp{0.004} & \tabvaluetemp{0.933}& \tabvaluetemp{1.000} & \tabvaluetemp{0.989} & \tabvaluetemp{0.004} & \tabvaluetemp{0.926}& \tabvaluetemp{0.999} & \tabvaluetemp{1.000} & \tabvaluetemp{0.003} & \tabvaluetemp{0.911}\\

& \variable{MD} & \tabvaluetemp{0.976} & \tabvaluetemp{0.012} & \tabvaluetemp{0.013} & \tabvaluetemp{0.997}& \tabvaluetemp{0.951} & \tabvaluetemp{0.280} & \tabvaluetemp{0.004} & \tabvaluetemp{0.907}& \tabvaluetemp{0.966} & \tabvaluetemp{0.589} & \tabvaluetemp{0.004} & \tabvaluetemp{0.913}& \tabvaluetemp{0.975} & \tabvaluetemp{0.778} & \tabvaluetemp{0.003} & \tabvaluetemp{0.891}\\

& \variable{ED-v} & \tabvaluetemp{0.969} & \tabvaluetemp{0.005} & \tabvaluetemp{0.004} & \tabvaluetemp{0.845}& \tabvaluetemp{0.593} & \tabvaluetemp{0.000} & \tabvaluetemp{0.000} & \tabvaluetemp{0.563}& \tabvaluetemp{0.660} & \tabvaluetemp{0.038} & \tabvaluetemp{0.000} & \tabvaluetemp{0.558}& \tabvaluetemp{0.599} & \tabvaluetemp{0.000} & \tabvaluetemp{0.000} & \tabvaluetemp{0.487}\\

& \variable{ED-s} & \tabvaluetemp{0.977} & \tabvaluetemp{0.008} & \tabvaluetemp{0.011} & \tabvaluetemp{0.997}& \tabvaluetemp{0.985} & \tabvaluetemp{0.882} & \tabvaluetemp{0.003} & \tabvaluetemp{0.919}& \tabvaluetemp{0.986} & \tabvaluetemp{0.942} & \tabvaluetemp{0.003} & \tabvaluetemp{0.913}& \tabvaluetemp{0.995} & \tabvaluetemp{0.991} & \tabvaluetemp{0.003} & \tabvaluetemp{0.899}\\

& \variable{ED-o} & \tabvaluetemp{0.980} & \tabvaluetemp{0.196} & \tabvaluetemp{0.028} & \tabvaluetemp{1.000}& \tabvaluetemp{0.985} & \tabvaluetemp{0.882} & \tabvaluetemp{0.004} & \tabvaluetemp{0.929}& \tabvaluetemp{0.988} & \tabvaluetemp{0.942} & \tabvaluetemp{0.004} & \tabvaluetemp{0.930}& \tabvaluetemp{0.996} & \tabvaluetemp{0.991} & \tabvaluetemp{0.003} & \tabvaluetemp{0.906}\\
            
            \bottomrule
        \end{tabular}
        }
                
        \caption{\textit{Temporal Dependency}: Results by assuming the presence of temporal dependencies among samples (the `first' samples of \scbb{D} are put in \scbb{T}, while the last 20\% represent \scbb{E}).} 
    \label{tab:ufnb15_open_temporal}
    \end{subtable}

    \label{tab:ufnb15_open}
\end{table*}

\begin{table*}
  \centering
  \caption{\dataset{CICIDS17} binary classification results (\scmath{fpr} and \scmath{tpr}) against `known' attacks seen during the training stage (closed world).}

    \begin{subtable}[htbp]{1.99\columnwidth}
        \resizebox{1.0\columnwidth}{!}{
        \begin{tabular}{c|c ? cc|cc? cc|cc? cc|cc? cc|cc}
             \multicolumn{2}{c?}{Available Data} & \multicolumn{4}{c?}{Limited (100 per class) [N=1000]} & \multicolumn{4}{c?}{Scarce (15\% of \scbb{D}) [N=100]} &  \multicolumn{4}{c?}{Moderate (40\% of \scbb{D}) [N=100]} &  \multicolumn{4}{c}{Abundant (80\% of \scbb{D}) [N=100]} \\ \hline
             \multicolumn{2}{c?}{Features} &
             \multicolumn{2}{c|}{Complete} & \multicolumn{2}{c?}{Essential} & \multicolumn{2}{c|}{Complete} & \multicolumn{2}{c?}{Essential} & \multicolumn{2}{c|}{Complete} & \multicolumn{2}{c?}{Essential} &  
             \multicolumn{2}{c|}{Complete} & \multicolumn{2}{c}{Essential} \\ \hline
            Alg. & Design & \smamath{fpr} & \smamath{tpr} & \smamath{fpr} & \smamath{tpr} & \smamath{fpr} & \smamath{tpr} & \smamath{fpr} & \smamath{tpr} & \smamath{fpr} & \smamath{tpr} & \smamath{fpr} & \smamath{tpr} & \smamath{fpr} & \smamath{tpr} & \smamath{fpr} & \smamath{tpr} \\
             \toprule
             
             \multirow{6}{*}{\rotatebox{90}{\parbox{40pt}\centering\textbf{RF}}} 
             & \variable{BD} & \tabvalue{0.109}{0.029} & \tabvalue{0.999}{0.001} & \tabvalue{0.102}{0.026} & \tabvalue{0.999}{0.001}& \tabvalue{0.001}{0.000} & \tabvalue{1.000}{0.000} & \tabvalue{0.001}{0.000} & \tabvalue{0.999}{0.000}& \tabvalue{0.001}{0.000} & \tabvalue{1.000}{0.000} & \tabvalue{0.001}{0.000} & \tabvalue{1.000}{0.000}& \tabvalue{0.001}{0.000} & \tabvalue{1.000}{0.000} & \tabvalue{0.001}{0.000} & \tabvalue{1.000}{0.000}\\

            & \variable{MD} & \tabvalue{0.072}{0.021} & \tabvalue{0.998}{0.002} & \tabvalue{0.064}{0.018} & \tabvalue{0.999}{0.001}& \tabvalue{0.001}{0.000} & \tabvalue{0.999}{0.000} & \tabvalue{0.001}{0.000} & \tabvalue{0.999}{0.000}& \tabvalue{0.001}{0.000} & \tabvalue{1.000}{0.000} & \tabvalue{0.001}{0.000} & \tabvalue{1.000}{0.000}& \tabvalue{0.001}{0.000} & \tabvalue{1.000}{0.000} & \tabvalue{0.001}{0.000} & \tabvalue{1.000}{0.000}\\
            
            & \variable{ED-v} & \tabvalue{0.000}{0.000} & \tabvalue{0.002}{0.018} & \tabvalue{0.000}{0.000} & \tabvalue{0.024}{0.061}& \tabvalue{0.000}{0.000} & \tabvalue{0.000}{0.000} & \tabvalue{0.000}{0.000} & \tabvalue{0.000}{0.000}& \tabvalue{0.000}{0.000} & \tabvalue{0.000}{0.000} & \tabvalue{0.000}{0.000} & \tabvalue{0.000}{0.000}& \tabvalue{0.000}{0.000} & \tabvalue{0.000}{0.000} & \tabvalue{0.000}{0.000} & \tabvalue{0.000}{0.000}\\
            
            & \variable{ED-s} & \tabvalue{0.076}{0.023} & \tabvalue{0.998}{0.002} & \tabvalue{0.066}{0.018} & \tabvalue{0.998}{0.001}& \tabvalue{0.001}{0.000} & \tabvalue{0.999}{0.000} & \tabvalue{0.001}{0.000} & \tabvalue{0.999}{0.000}& \tabvalue{0.001}{0.000} & \tabvalue{1.000}{0.000} & \tabvalue{0.001}{0.000} & \tabvalue{1.000}{0.000}& \tabvalue{0.001}{0.000} & \tabvalue{1.000}{0.000} & \tabvalue{0.001}{0.000} & \tabvalue{1.000}{0.000}\\
            
            & \variable{ED-o} & \tabvalue{0.079}{0.024} & \tabvalue{0.998}{0.002} & \tabvalue{0.066}{0.019} & \tabvalue{0.998}{0.001}& \tabvalue{0.001}{0.000} & \tabvalue{0.999}{0.000} & \tabvalue{0.001}{0.000} & \tabvalue{0.999}{0.000}& \tabvalue{0.001}{0.000} & \tabvalue{1.000}{0.000} & \tabvalue{0.001}{0.000} & \tabvalue{1.000}{0.000}& \tabvalue{0.001}{0.000} & \tabvalue{1.000}{0.000} & \tabvalue{0.001}{0.000} & \tabvalue{1.000}{0.000}\\
            
            & \variable{ED} & \tabvalue{0.011}{0.004} & \tabvalue{0.997}{0.002} & \tabvalue{0.009}{0.003} & \tabvalue{0.998}{0.001}& \tabvalue{0.000}{0.000} & \tabvalue{0.999}{0.000} & \tabvalue{0.000}{0.000} & \tabvalue{0.999}{0.000}& \tabvalue{0.000}{0.000} & \tabvalue{1.000}{0.000} & \tabvalue{0.000}{0.000} & \tabvalue{1.000}{0.000}& \tabvalue{0.000}{0.000} & \tabvalue{1.000}{0.000} & \tabvalue{0.000}{0.000} & \tabvalue{1.000}{0.000}\\

            \midrule
            \multirow{6}{*}{\rotatebox{90}{\parbox{40pt}\centering\textbf{DT}}} 
            & \variable{BD} & \tabvalue{0.101}{0.027} & \tabvalue{0.995}{0.004} & \tabvalue{0.106}{0.026} & \tabvalue{0.997}{0.004}& \tabvalue{0.001}{0.000} & \tabvalue{0.999}{0.000} & \tabvalue{0.002}{0.000} & \tabvalue{0.999}{0.000}& \tabvalue{0.001}{0.000} & \tabvalue{1.000}{0.000} & \tabvalue{0.001}{0.000} & \tabvalue{1.000}{0.000}& \tabvalue{0.001}{0.000} & \tabvalue{1.000}{0.000} & \tabvalue{0.001}{0.000} & \tabvalue{1.000}{0.000}\\

            & \variable{MD} & \tabvalue{0.125}{0.030} & \tabvalue{0.997}{0.002} & \tabvalue{0.119}{0.025} & \tabvalue{0.996}{0.003}& \tabvalue{0.001}{0.000} & \tabvalue{0.999}{0.000} & \tabvalue{0.001}{0.000} & \tabvalue{0.999}{0.000}& \tabvalue{0.001}{0.000} & \tabvalue{1.000}{0.000} & \tabvalue{0.001}{0.000} & \tabvalue{1.000}{0.000}& \tabvalue{0.001}{0.000} & \tabvalue{1.000}{0.000} & \tabvalue{0.001}{0.000} & \tabvalue{1.000}{0.000}\\
            
            & \variable{ED-v} & \tabvalue{0.000}{0.001} & \tabvalue{0.157}{0.184} & \tabvalue{0.000}{0.002} & \tabvalue{0.113}{0.135}& \tabvalue{0.000}{0.000} & \tabvalue{0.008}{0.028} & \tabvalue{0.000}{0.000} & \tabvalue{0.000}{0.003}& \tabvalue{0.000}{0.000} & \tabvalue{0.004}{0.015} & \tabvalue{0.000}{0.000} & \tabvalue{0.001}{0.012}& \tabvalue{0.000}{0.000} & \tabvalue{0.001}{0.003} & \tabvalue{0.000}{0.000} & \tabvalue{0.000}{0.000}\\
            
            & \variable{ED-s} & \tabvalue{0.131}{0.031} & \tabvalue{0.997}{0.003} & \tabvalue{0.118}{0.029} & \tabvalue{0.997}{0.003}& \tabvalue{0.001}{0.000} & \tabvalue{0.999}{0.000} & \tabvalue{0.002}{0.000} & \tabvalue{0.999}{0.000}& \tabvalue{0.001}{0.000} & \tabvalue{1.000}{0.000} & \tabvalue{0.001}{0.000} & \tabvalue{1.000}{0.000}& \tabvalue{0.001}{0.000} & \tabvalue{1.000}{0.000} & \tabvalue{0.001}{0.000} & \tabvalue{1.000}{0.000}\\
            
            & \variable{ED-o} & \tabvalue{0.151}{0.030} & \tabvalue{0.997}{0.003} & \tabvalue{0.131}{0.030} & \tabvalue{0.997}{0.003}& \tabvalue{0.001}{0.000} & \tabvalue{0.999}{0.000} & \tabvalue{0.002}{0.000} & \tabvalue{0.999}{0.000}& \tabvalue{0.001}{0.000} & \tabvalue{1.000}{0.000} & \tabvalue{0.001}{0.000} & \tabvalue{1.000}{0.000}& \tabvalue{0.001}{0.000} & \tabvalue{1.000}{0.000} & \tabvalue{0.001}{0.000} & \tabvalue{1.000}{0.000}\\
            
            & \variable{ED} & \tabvalue{0.023}{0.006} & \tabvalue{0.994}{0.003} & \tabvalue{0.019}{0.006} & \tabvalue{0.996}{0.003}& \tabvalue{0.000}{0.000} & \tabvalue{0.999}{0.000} & \tabvalue{0.000}{0.000} & \tabvalue{0.999}{0.000}& \tabvalue{0.000}{0.000} & \tabvalue{1.000}{0.000} & \tabvalue{0.000}{0.000} & \tabvalue{0.999}{0.000}& \tabvalue{0.000}{0.000} & \tabvalue{1.000}{0.000} & \tabvalue{0.000}{0.000} & \tabvalue{1.000}{0.000}\\
             \midrule
            \multirow{6}{*}{\rotatebox{90}{\parbox{40pt}\centering\textbf{LR}}} 
            & \variable{BD} & \tabvalue{0.482}{0.141} & \tabvalue{0.999}{0.001} & \tabvalue{0.621}{0.317} & \tabvalue{0.995}{0.071}& \tabvalue{0.113}{0.002} & \tabvalue{0.970}{0.001} & \tabvalue{0.012}{0.036} & \tabvalue{0.097}{0.289}& \tabvalue{0.112}{0.002} & \tabvalue{0.970}{0.001} & \tabvalue{0.013}{0.037} & \tabvalue{0.108}{0.304}& \tabvalue{0.113}{0.002} & \tabvalue{0.970}{0.001} & \tabvalue{0.029}{0.056} & \tabvalue{0.221}{0.404}\\

            & \variable{MD} & \tabvalue{0.640}{0.057} & \tabvalue{0.997}{0.015} & \tabvalue{0.968}{0.073} & \tabvalue{0.980}{0.086}& \tabvalue{0.079}{0.003} & \tabvalue{0.962}{0.003} & \tabvalue{0.001}{0.002} & \tabvalue{0.000}{0.001}& \tabvalue{0.080}{0.003} & \tabvalue{0.962}{0.003} & \tabvalue{0.000}{0.001} & \tabvalue{0.000}{0.001}& \tabvalue{0.079}{0.003} & \tabvalue{0.962}{0.003} & \tabvalue{0.001}{0.008} & \tabvalue{0.004}{0.037}\\
            
            & \variable{ED-v} & \tabvalue{0.002}{0.004} & \tabvalue{0.051}{0.098} & \tabvalue{0.005}{0.006} & \tabvalue{0.094}{0.124}& \tabvalue{0.000}{0.000} & \tabvalue{0.000}{0.001} & \tabvalue{0.000}{0.000} & \tabvalue{0.000}{0.000}& \tabvalue{0.000}{0.000} & \tabvalue{0.000}{0.000} & \tabvalue{0.000}{0.000} & \tabvalue{0.000}{0.000}& \tabvalue{0.000}{0.000} & \tabvalue{0.000}{0.000} & \tabvalue{0.000}{0.000} & \tabvalue{0.000}{0.000}\\
            
            & \variable{ED-s} & \tabvalue{0.212}{0.106} & \tabvalue{0.883}{0.314} & \tabvalue{0.312}{0.259} & \tabvalue{0.886}{0.315}& \tabvalue{0.046}{0.003} & \tabvalue{0.986}{0.002} & \tabvalue{0.044}{0.027} & \tabvalue{0.634}{0.206}& \tabvalue{0.047}{0.002} & \tabvalue{0.986}{0.001} & \tabvalue{0.036}{0.025} & \tabvalue{0.563}{0.209}& \tabvalue{0.047}{0.002} & \tabvalue{0.986}{0.001} & \tabvalue{0.045}{0.029} & \tabvalue{0.620}{0.194}\\
            
            & \variable{ED-o} & \tabvalue{0.238}{0.033} & \tabvalue{0.995}{0.004} & \tabvalue{0.414}{0.285} & \tabvalue{0.984}{0.116}& \tabvalue{0.050}{0.003} & \tabvalue{0.987}{0.001} & \tabvalue{0.072}{0.030} & \tabvalue{0.642}{0.207}& \tabvalue{0.050}{0.002} & \tabvalue{0.987}{0.001} & \tabvalue{0.072}{0.033} & \tabvalue{0.571}{0.209}& \tabvalue{0.050}{0.002} & \tabvalue{0.987}{0.001} & \tabvalue{0.080}{0.030} & \tabvalue{0.629}{0.193}\\
            
            & \variable{ED} & \tabvalue{0.043}{0.008} & \tabvalue{0.978}{0.007} & \tabvalue{0.068}{0.038} & \tabvalue{0.983}{0.062}& \tabvalue{0.006}{0.000} & \tabvalue{0.983}{0.001} & \tabvalue{0.010}{0.005} & \tabvalue{0.611}{0.186}& \tabvalue{0.006}{0.000} & \tabvalue{0.983}{0.001} & \tabvalue{0.010}{0.005} & \tabvalue{0.534}{0.164}& \tabvalue{0.006}{0.000} & \tabvalue{0.983}{0.001} & \tabvalue{0.012}{0.005} & \tabvalue{0.605}{0.175}\\
             \midrule
            \multirow{6}{*}{\rotatebox{90}{\parbox{40pt}\centering\textbf{HGB}}} 
            & \variable{BD} & \tabvalue{0.074}{0.026} & \tabvalue{0.999}{0.001} & \tabvalue{0.095}{0.024} & \tabvalue{0.999}{0.001}& \tabvalue{0.001}{0.000} & \tabvalue{1.000}{0.000} & \tabvalue{0.001}{0.000} & \tabvalue{0.999}{0.000}& \tabvalue{0.001}{0.000} & \tabvalue{1.000}{0.000} & \tabvalue{0.001}{0.000} & \tabvalue{1.000}{0.000}& \tabvalue{0.001}{0.000} & \tabvalue{1.000}{0.000} & \tabvalue{0.001}{0.000} & \tabvalue{1.000}{0.000}\\
            
            & \variable{MD} & \tabvalue{0.050}{0.019} & \tabvalue{0.998}{0.002} & \tabvalue{0.066}{0.019} & \tabvalue{0.998}{0.002}& \tabvalue{0.006}{0.002} & \tabvalue{0.998}{0.001} & \tabvalue{0.006}{0.004} & \tabvalue{0.998}{0.001}& \tabvalue{0.003}{0.002} & \tabvalue{0.999}{0.001} & \tabvalue{0.003}{0.001} & \tabvalue{0.999}{0.000}& \tabvalue{0.002}{0.000} & \tabvalue{0.999}{0.000} & \tabvalue{0.002}{0.000} & \tabvalue{0.999}{0.000}\\
            
            & \variable{ED-v} & \tabvalue{0.000}{0.001} & \tabvalue{0.066}{0.119} & \tabvalue{0.000}{0.001} & \tabvalue{0.045}{0.060}& \tabvalue{0.000}{0.000} & \tabvalue{0.000}{0.000} & \tabvalue{0.000}{0.000} & \tabvalue{0.045}{0.075}& \tabvalue{0.000}{0.000} & \tabvalue{0.000}{0.000} & \tabvalue{0.000}{0.000} & \tabvalue{0.046}{0.070}& \tabvalue{0.000}{0.000} & \tabvalue{0.000}{0.001} & \tabvalue{0.000}{0.000} & \tabvalue{0.063}{0.076}\\
            
            & \variable{ED-s} & \tabvalue{0.085}{0.027} & \tabvalue{0.997}{0.002} & \tabvalue{0.093}{0.025} & \tabvalue{0.996}{0.003}& \tabvalue{0.002}{0.000} & \tabvalue{0.999}{0.000} & \tabvalue{0.002}{0.000} & \tabvalue{0.999}{0.001}& \tabvalue{0.001}{0.000} & \tabvalue{1.000}{0.000} & \tabvalue{0.002}{0.001} & \tabvalue{0.999}{0.001}& \tabvalue{0.001}{0.000} & \tabvalue{1.000}{0.000} & \tabvalue{0.001}{0.000} & \tabvalue{0.999}{0.001}\\
            
            & \variable{ED-o} & \tabvalue{0.096}{0.028} & \tabvalue{0.998}{0.002} & \tabvalue{0.097}{0.026} & \tabvalue{0.997}{0.002}& \tabvalue{0.002}{0.001} & \tabvalue{0.999}{0.000} & \tabvalue{0.002}{0.001} & \tabvalue{0.999}{0.001}& \tabvalue{0.001}{0.000} & \tabvalue{1.000}{0.000} & \tabvalue{0.002}{0.001} & \tabvalue{0.999}{0.001}& \tabvalue{0.001}{0.000} & \tabvalue{1.000}{0.000} & \tabvalue{0.002}{0.001} & \tabvalue{1.000}{0.000}\\
            
            & \variable{ED} & \tabvalue{0.014}{0.005} & \tabvalue{0.995}{0.003} & \tabvalue{0.015}{0.005} & \tabvalue{0.995}{0.004}& \tabvalue{0.000}{0.000} & \tabvalue{0.999}{0.000} & \tabvalue{0.000}{0.000} & \tabvalue{0.999}{0.001}& \tabvalue{0.000}{0.000} & \tabvalue{1.000}{0.000} & \tabvalue{0.000}{0.000} & \tabvalue{0.999}{0.001}& \tabvalue{0.000}{0.000} & \tabvalue{1.000}{0.000} & \tabvalue{0.000}{0.000} & \tabvalue{0.999}{0.000}\\
            
            \bottomrule
        \end{tabular}
        }
                
        \caption{\textit{Static Dependency}: Results by assuming the absence of temporal dependencies among samples (\scbb{T} and \scbb{E} are randomly sampled from \scbb{D}).} 
    \label{tab:ids17_baseline_static}
    \end{subtable}

    \begin{subtable}[htbp]{1.8\columnwidth}
        \resizebox{1.0\columnwidth}{!}{
        \begin{tabular}{c|c ? cc|cc? cc|cc? cc|cc? cc|cc}
             \multicolumn{2}{c?}{Available Data} & \multicolumn{4}{c?}{Limited (100 per class) [N=1]} & \multicolumn{4}{c?}{Scarce (15\% of \scbb{D}) [N=1]} &  \multicolumn{4}{c?}{Moderate (40\% of \scbb{D}) [N=1]} &  \multicolumn{4}{c}{Abundant (80\% of \scbb{D}) [N=1]} \\ \hline
             \multicolumn{2}{c?}{Features} &
             \multicolumn{2}{c|}{Complete} & \multicolumn{2}{c?}{Essential} & \multicolumn{2}{c|}{Complete} & \multicolumn{2}{c?}{Essential} & \multicolumn{2}{c|}{Complete} & \multicolumn{2}{c?}{Essential} &  
             \multicolumn{2}{c|}{Complete} & \multicolumn{2}{c}{Essential} \\ \hline
            Alg. & Design & \smamath{fpr} & \smamath{tpr} & \smamath{fpr} & \smamath{tpr} & \smamath{fpr} & \smamath{tpr} & \smamath{fpr} & \smamath{tpr} & \smamath{fpr} & \smamath{tpr} & \smamath{fpr} & \smamath{tpr} & \smamath{fpr} & \smamath{tpr} & \smamath{fpr} & \smamath{tpr} \\
             \toprule
             
             \multirow{6}{*}{\rotatebox{90}{\parbox{40pt}\centering\textbf{RF}}} 
            & \variable{BD} & \tabvaluetemp{0.327} & \tabvaluetemp{1.000} & \tabvaluetemp{0.335} & \tabvaluetemp{1.000}& \tabvaluetemp{0.010} & \tabvaluetemp{0.990} & \tabvaluetemp{0.010} & \tabvaluetemp{0.990}& \tabvaluetemp{0.010} & \tabvaluetemp{0.994} & \tabvaluetemp{0.010} & \tabvaluetemp{0.994}& \tabvaluetemp{0.007} & \tabvaluetemp{0.994} & \tabvaluetemp{0.009} & \tabvaluetemp{0.994}\\
            
            & \variable{MD} & \tabvaluetemp{0.174} & \tabvaluetemp{0.996} & \tabvaluetemp{0.317} & \tabvaluetemp{1.000}& \tabvaluetemp{0.010} & \tabvaluetemp{0.989} & \tabvaluetemp{0.010} & \tabvaluetemp{0.989}& \tabvaluetemp{0.010} & \tabvaluetemp{0.993} & \tabvaluetemp{0.010} & \tabvaluetemp{0.994}& \tabvaluetemp{0.007} & \tabvaluetemp{0.993} & \tabvaluetemp{0.009} & \tabvaluetemp{0.994}\\
            
            & \variable{ED-v} & \tabvaluetemp{0.000} & \tabvaluetemp{0.044} & \tabvaluetemp{0.008} & \tabvaluetemp{0.440}& \tabvaluetemp{0.000} & \tabvaluetemp{0.000} & \tabvaluetemp{0.000} & \tabvaluetemp{0.000}& \tabvaluetemp{0.000} & \tabvaluetemp{0.000} & \tabvaluetemp{0.000} & \tabvaluetemp{0.000}& \tabvaluetemp{0.000} & \tabvaluetemp{0.000} & \tabvaluetemp{0.000} & \tabvaluetemp{0.000}\\
            
            & \variable{ED-s} & \tabvaluetemp{0.166} & \tabvaluetemp{0.994} & \tabvaluetemp{0.331} & \tabvaluetemp{0.999}& \tabvaluetemp{0.010} & \tabvaluetemp{0.989} & \tabvaluetemp{0.010} & \tabvaluetemp{0.989}& \tabvaluetemp{0.010} & \tabvaluetemp{0.992} & \tabvaluetemp{0.010} & \tabvaluetemp{0.993}& \tabvaluetemp{0.007} & \tabvaluetemp{0.992} & \tabvaluetemp{0.009} & \tabvaluetemp{0.993}\\
            
            & \variable{ED-o} & \tabvaluetemp{0.180} & \tabvaluetemp{0.995} & \tabvaluetemp{0.335} & \tabvaluetemp{0.999}& \tabvaluetemp{0.010} & \tabvaluetemp{0.989} & \tabvaluetemp{0.010} & \tabvaluetemp{0.989}& \tabvaluetemp{0.010} & \tabvaluetemp{0.992} & \tabvaluetemp{0.010} & \tabvaluetemp{0.993}& \tabvaluetemp{0.007} & \tabvaluetemp{0.992} & \tabvaluetemp{0.009} & \tabvaluetemp{0.993}\\
            
            & \variable{ED} & \tabvaluetemp{0.028} & \tabvaluetemp{0.987} & \tabvaluetemp{0.082} & \tabvaluetemp{0.990}& \tabvaluetemp{0.001} & \tabvaluetemp{0.988} & \tabvaluetemp{0.001} & \tabvaluetemp{0.989}& \tabvaluetemp{0.001} & \tabvaluetemp{0.992} & \tabvaluetemp{0.001} & \tabvaluetemp{0.993}& \tabvaluetemp{0.001} & \tabvaluetemp{0.992} & \tabvaluetemp{0.001} & \tabvaluetemp{0.992}\\
             
            \midrule
            \multirow{6}{*}{\rotatebox{90}{\parbox{40pt}\centering\textbf{DT}}} 
             & \variable{BD} & \tabvaluetemp{0.334} & \tabvaluetemp{0.976} & \tabvaluetemp{0.336} & \tabvaluetemp{0.999}& \tabvaluetemp{0.011} & \tabvaluetemp{0.990} & \tabvaluetemp{0.012} & \tabvaluetemp{0.989}& \tabvaluetemp{0.009} & \tabvaluetemp{0.994} & \tabvaluetemp{0.012} & \tabvaluetemp{0.988}& \tabvaluetemp{0.008} & \tabvaluetemp{0.995} & \tabvaluetemp{0.010} & \tabvaluetemp{0.994}\\

            & \variable{MD} & \tabvaluetemp{0.144} & \tabvaluetemp{0.983} & \tabvaluetemp{0.296} & \tabvaluetemp{0.992}& \tabvaluetemp{0.010} & \tabvaluetemp{0.988} & \tabvaluetemp{0.010} & \tabvaluetemp{0.993}& \tabvaluetemp{0.011} & \tabvaluetemp{0.996} & \tabvaluetemp{0.012} & \tabvaluetemp{0.997}& \tabvaluetemp{0.009} & \tabvaluetemp{0.998} & \tabvaluetemp{0.010} & \tabvaluetemp{0.994}\\
            
            & \variable{ED-v} & \tabvaluetemp{0.003} & \tabvaluetemp{0.536} & \tabvaluetemp{0.034} & \tabvaluetemp{0.248}& \tabvaluetemp{0.000} & \tabvaluetemp{0.000} & \tabvaluetemp{0.000} & \tabvaluetemp{0.000}& \tabvaluetemp{0.000} & \tabvaluetemp{0.003} & \tabvaluetemp{0.000} & \tabvaluetemp{0.000}& \tabvaluetemp{0.000} & \tabvaluetemp{0.000} & \tabvaluetemp{0.000} & \tabvaluetemp{0.000}\\
            
            & \variable{ED-s} & \tabvaluetemp{0.178} & \tabvaluetemp{0.995} & \tabvaluetemp{0.344} & \tabvaluetemp{1.000}& \tabvaluetemp{0.011} & \tabvaluetemp{0.992} & \tabvaluetemp{0.011} & \tabvaluetemp{0.992}& \tabvaluetemp{0.012} & \tabvaluetemp{0.995} & \tabvaluetemp{0.013} & \tabvaluetemp{0.991}& \tabvaluetemp{0.008} & \tabvaluetemp{0.994} & \tabvaluetemp{0.010} & \tabvaluetemp{0.996}\\
            
            & \variable{ED-o} & \tabvaluetemp{0.191} & \tabvaluetemp{0.995} & \tabvaluetemp{0.348} & \tabvaluetemp{1.000}& \tabvaluetemp{0.011} & \tabvaluetemp{0.992} & \tabvaluetemp{0.011} & \tabvaluetemp{0.992}& \tabvaluetemp{0.012} & \tabvaluetemp{0.995} & \tabvaluetemp{0.013} & \tabvaluetemp{0.991}& \tabvaluetemp{0.008} & \tabvaluetemp{0.994} & \tabvaluetemp{0.010} & \tabvaluetemp{0.996}\\
            
            & \variable{ED} & \tabvaluetemp{0.040} & \tabvaluetemp{0.980} & \tabvaluetemp{0.124} & \tabvaluetemp{0.988}& \tabvaluetemp{0.001} & \tabvaluetemp{0.989} & \tabvaluetemp{0.001} & \tabvaluetemp{0.992}& \tabvaluetemp{0.001} & \tabvaluetemp{0.993} & \tabvaluetemp{0.001} & \tabvaluetemp{0.988}& \tabvaluetemp{0.001} & \tabvaluetemp{0.993} & \tabvaluetemp{0.001} & \tabvaluetemp{0.995}\\
             \midrule
            \multirow{6}{*}{\rotatebox{90}{\parbox{40pt}\centering\textbf{LR}}} 
             & \variable{BD} & \tabvaluetemp{0.289} & \tabvaluetemp{0.986} & \tabvaluetemp{0.358} & \tabvaluetemp{1.000}& \tabvaluetemp{0.070} & \tabvaluetemp{0.961} & \tabvaluetemp{0.072} & \tabvaluetemp{0.974}& \tabvaluetemp{0.057} & \tabvaluetemp{0.965} & \tabvaluetemp{0.003} & \tabvaluetemp{0.000}& \tabvaluetemp{0.059} & \tabvaluetemp{0.963} & \tabvaluetemp{0.083} & \tabvaluetemp{0.972}\\

            & \variable{MD} & \tabvaluetemp{0.334} & \tabvaluetemp{0.675} & \tabvaluetemp{0.931} & \tabvaluetemp{0.989}& \tabvaluetemp{0.037} & \tabvaluetemp{0.958} & \tabvaluetemp{0.000} & \tabvaluetemp{0.000}& \tabvaluetemp{0.037} & \tabvaluetemp{0.954} & \tabvaluetemp{0.000} & \tabvaluetemp{0.000}& \tabvaluetemp{0.035} & \tabvaluetemp{0.954} & \tabvaluetemp{0.000} & \tabvaluetemp{0.000}\\
            
            & \variable{ED-v} & \tabvaluetemp{0.062} & \tabvaluetemp{0.285} & \tabvaluetemp{0.059} & \tabvaluetemp{0.599}& \tabvaluetemp{0.000} & \tabvaluetemp{0.000} & \tabvaluetemp{0.000} & \tabvaluetemp{0.000}& \tabvaluetemp{0.000} & \tabvaluetemp{0.000} & \tabvaluetemp{0.001} & \tabvaluetemp{0.000}& \tabvaluetemp{0.000} & \tabvaluetemp{0.000} & \tabvaluetemp{0.000} & \tabvaluetemp{0.000}\\
            
            & \variable{ED-s} & \tabvaluetemp{0.310} & \tabvaluetemp{0.986} & \tabvaluetemp{0.354} & \tabvaluetemp{0.999}& \tabvaluetemp{0.030} & \tabvaluetemp{0.984} & \tabvaluetemp{0.029} & \tabvaluetemp{0.602}& \tabvaluetemp{0.028} & \tabvaluetemp{0.982} & \tabvaluetemp{0.049} & \tabvaluetemp{0.596}& \tabvaluetemp{0.026} & \tabvaluetemp{0.981} & \tabvaluetemp{0.014} & \tabvaluetemp{0.388}\\
            
            & \variable{ED-o} & \tabvaluetemp{0.321} & \tabvaluetemp{0.988} & \tabvaluetemp{0.354} & \tabvaluetemp{0.999}& \tabvaluetemp{0.035} & \tabvaluetemp{0.984} & \tabvaluetemp{0.036} & \tabvaluetemp{0.603}& \tabvaluetemp{0.035} & \tabvaluetemp{0.986} & \tabvaluetemp{0.052} & \tabvaluetemp{0.597}& \tabvaluetemp{0.031} & \tabvaluetemp{0.985} & \tabvaluetemp{0.057} & \tabvaluetemp{0.399}\\
            
            & \variable{ED} & \tabvaluetemp{0.119} & \tabvaluetemp{0.972} & \tabvaluetemp{0.115} & \tabvaluetemp{0.990}& \tabvaluetemp{0.005} & \tabvaluetemp{0.977} & \tabvaluetemp{0.006} & \tabvaluetemp{0.600}& \tabvaluetemp{0.005} & \tabvaluetemp{0.979} & \tabvaluetemp{0.008} & \tabvaluetemp{0.593}& \tabvaluetemp{0.004} & \tabvaluetemp{0.979} & \tabvaluetemp{0.008} & \tabvaluetemp{0.394}\\
                         \midrule
            \multirow{6}{*}{\rotatebox{90}{\parbox{40pt}\centering\textbf{HGB}}} 
            & \variable{BD} & \tabvaluetemp{0.336} & \tabvaluetemp{1.000} & \tabvaluetemp{0.338} & \tabvaluetemp{1.000}& \tabvaluetemp{0.003} & \tabvaluetemp{0.992} & \tabvaluetemp{0.012} & \tabvaluetemp{0.985}& \tabvaluetemp{0.010} & \tabvaluetemp{0.999} & \tabvaluetemp{0.011} & \tabvaluetemp{0.990}& \tabvaluetemp{0.007} & \tabvaluetemp{0.999} & \tabvaluetemp{0.010} & \tabvaluetemp{0.995}\\

            & \variable{MD} & \tabvaluetemp{0.283} & \tabvaluetemp{0.999} & \tabvaluetemp{0.335} & \tabvaluetemp{1.000}& \tabvaluetemp{0.007} & \tabvaluetemp{0.991} & \tabvaluetemp{0.013} & \tabvaluetemp{0.988}& \tabvaluetemp{0.010} & \tabvaluetemp{0.992} & \tabvaluetemp{0.013} & \tabvaluetemp{0.994}& \tabvaluetemp{0.011} & \tabvaluetemp{0.996} & \tabvaluetemp{0.012} & \tabvaluetemp{0.997}\\
            
            & \variable{ED-v} & \tabvaluetemp{0.022} & \tabvaluetemp{0.546} & \tabvaluetemp{0.027} & \tabvaluetemp{0.192}& \tabvaluetemp{0.000} & \tabvaluetemp{0.000} & \tabvaluetemp{0.000} & \tabvaluetemp{0.002}& \tabvaluetemp{0.000} & \tabvaluetemp{0.000} & \tabvaluetemp{0.000} & \tabvaluetemp{0.061}& \tabvaluetemp{0.000} & \tabvaluetemp{0.000} & \tabvaluetemp{0.000} & \tabvaluetemp{0.053}\\
            
            & \variable{ED-s} & \tabvaluetemp{0.300} & \tabvaluetemp{0.999} & \tabvaluetemp{0.335} & \tabvaluetemp{1.000}& \tabvaluetemp{0.010} & \tabvaluetemp{0.992} & \tabvaluetemp{0.012} & \tabvaluetemp{0.993}& \tabvaluetemp{0.011} & \tabvaluetemp{0.994} & \tabvaluetemp{0.011} & \tabvaluetemp{0.994}& \tabvaluetemp{0.008} & \tabvaluetemp{0.994} & \tabvaluetemp{0.010} & \tabvaluetemp{0.993}\\
            
            & \variable{ED-o} & \tabvaluetemp{0.307} & \tabvaluetemp{1.000} & \tabvaluetemp{0.365} & \tabvaluetemp{1.000}& \tabvaluetemp{0.010} & \tabvaluetemp{0.992} & \tabvaluetemp{0.012} & \tabvaluetemp{0.993}& \tabvaluetemp{0.011} & \tabvaluetemp{0.994} & \tabvaluetemp{0.011} & \tabvaluetemp{0.994}& \tabvaluetemp{0.008} & \tabvaluetemp{0.994} & \tabvaluetemp{0.010} & \tabvaluetemp{0.993}\\
            
            & \variable{ED} & \tabvaluetemp{0.077} & \tabvaluetemp{0.986} & \tabvaluetemp{0.096} & \tabvaluetemp{0.975}& \tabvaluetemp{0.001} & \tabvaluetemp{0.991} & \tabvaluetemp{0.001} & \tabvaluetemp{0.991}& \tabvaluetemp{0.001} & \tabvaluetemp{0.993} & \tabvaluetemp{0.001} & \tabvaluetemp{0.992}& \tabvaluetemp{0.001} & \tabvaluetemp{0.994} & \tabvaluetemp{0.001} & \tabvaluetemp{0.992}\\
            
            \bottomrule
        \end{tabular}
        }
                
        \caption{\textit{Temporal Dependency}: Results by assuming the presence of temporal dependencies among samples (the `first' samples of \scbb{D} are put in \scbb{T}, while the last 20\% represent \scbb{E}).} 
    \label{tab:ids17_baseline_temporal}
    \end{subtable}

    \label{tab:ids17_baseline}
\end{table*}

\begin{table*}
  \centering
  \caption{\dataset{CICIDS17}. Results against adversarial (original \scmath{tpr} and adversarial \scmath{tpr}) and unknown attacks (the \scmath{tpr} is the average on the `unknown' attacks, while the \scmath{fpr} is due to training on a new \scbb{T} that does not have the `unknown' class.).}

    \begin{subtable}[htbp]{1.99\columnwidth}
        \resizebox{1.0\columnwidth}{!}{
\begin{tabular}{c|c ? cc|cc? cc|cc? cc|cc? cc|cc}
\multicolumn{2}{c?}{Available Data} & \multicolumn{4}{c?}{Limited (100 per class) [N=1000]} & \multicolumn{4}{c?}{Scarce (15\% of \scbb{D}) [N=100]} &  \multicolumn{4}{c?}{Moderate (40\% of \scbb{D}) [N=100]} &  \multicolumn{4}{c}{Abundant (80\% of \scbb{D}) [N=100]} \\ \hline
             \multicolumn{2}{c?}{Scenario} &
             \multicolumn{2}{c|}{Adversarial Attacks} & \multicolumn{2}{c?}{Unknown Attacks} & \multicolumn{2}{c|}{Adversarial Attacks} & \multicolumn{2}{c?}{Unknown Attacks} & \multicolumn{2}{c|}{Adversarial Attacks} & \multicolumn{2}{c?}{Unknown Attacks} & \multicolumn{2}{c|}{Adversarial Attacks} & \multicolumn{2}{c?}{Unknown Attacks}  \\ \hline
            Alg. & Design & 
            \footnotesize{$tpr$} \tiny{(org)} & \footnotesize{$tpr$} \tiny{(adv)} & \footnotesize{$fpr$} & \footnotesize{$tpr$} & \footnotesize{$tpr$} \tiny{(org)} & \footnotesize{$tpr$} \tiny{(adv)} & \footnotesize{$fpr$} & \footnotesize{$tpr$} & \footnotesize{$tpr$} \tiny{(org)} & \footnotesize{$tpr$} \tiny{(adv)} & \footnotesize{$fpr$} & \footnotesize{$tpr$} & \footnotesize{$tpr$} \tiny{(org)} & \footnotesize{$tpr$} \tiny{(adv)} & \footnotesize{$fpr$} & \footnotesize{$tpr$} \\
             \toprule
             
             \multirow{5}{*}{\rotatebox{90}{\parbox{40pt}\centering\textbf{RF}}} 
             & \variable{BD} & \tabvalue{0.999}{0.001} & \tabvalue{0.742}{0.221} & \tabvalue{0.099}{0.027} & \tabvalue{0.787}{0.075}& \tabvalue{0.999}{0.000} & \tabvalue{0.004}{0.002} & \tabvalue{0.001}{0.000} & \tabvalue{0.391}{0.021}& \tabvalue{1.000}{0.000} & \tabvalue{0.004}{0.002} & \tabvalue{0.001}{0.000} & \tabvalue{0.366}{0.020}& \tabvalue{1.000}{0.000} & \tabvalue{0.004}{0.001} & \tabvalue{0.001}{0.000} & \tabvalue{0.338}{0.016}\\
\bottomrule
        \end{tabular}
        }
                
        \caption{\textit{Static Dependency}: Results by assuming the absence of temporal dependencies among samples (\scbb{T} and \scbb{E} are randomly sampled from \scbb{D}).} 
    \label{tab:ids17_open_static}
    \end{subtable}

    \begin{subtable}[htbp]{1.8\columnwidth}
        \resizebox{1.0\columnwidth}{!}{
        \begin{tabular}{c|c ? cc|cc? cc|cc? cc|cc? cc|cc}
\multicolumn{2}{c?}{Available Data} & \multicolumn{4}{c?}{Limited (100 per class) [N=1]} & \multicolumn{4}{c?}{Scarce (15\% of \scbb{D}) [N=1]} &  \multicolumn{4}{c?}{Moderate (40\% of \scbb{D}) [N=1]} &  \multicolumn{4}{c}{Abundant (80\% of \scbb{D}) [N=1]} \\ \hline
             \multicolumn{2}{c?}{Scenario} &
             \multicolumn{2}{c|}{Adversarial Attacks} & \multicolumn{2}{c?}{Unknown Attacks} & \multicolumn{2}{c|}{Adversarial Attacks} & \multicolumn{2}{c?}{Unknown Attacks} & \multicolumn{2}{c|}{Adversarial Attacks} & \multicolumn{2}{c?}{Unknown Attacks} & \multicolumn{2}{c|}{Adversarial Attacks} & \multicolumn{2}{c?}{Unknown Attacks}  \\ \hline
            Alg. & Design & 
            \footnotesize{$tpr$} \tiny{(org)} & \footnotesize{$tpr$} \tiny{(adv)} & \footnotesize{$fpr$} & \footnotesize{$tpr$} & \footnotesize{$tpr$} \tiny{(org)} & \footnotesize{$tpr$} \tiny{(adv)} & \footnotesize{$fpr$} & \footnotesize{$tpr$} & \footnotesize{$tpr$} \tiny{(org)} & \footnotesize{$tpr$} \tiny{(adv)} & \footnotesize{$fpr$} & \footnotesize{$tpr$} & \footnotesize{$tpr$} \tiny{(org)} & \footnotesize{$tpr$} \tiny{(adv)} & \footnotesize{$fpr$} & \footnotesize{$tpr$} \\
             \toprule
             
             \multirow{5}{*}{\rotatebox{90}{\parbox{40pt}\centering\textbf{RF}}} 
             & \variable{BD} & \tabvaluetemp{1.000} & \tabvaluetemp{1.000} & \tabvaluetemp{0.325} & \tabvaluetemp{0.799}& \tabvaluetemp{0.990} & \tabvaluetemp{0.000} & \tabvaluetemp{0.009} & \tabvaluetemp{0.363}& \tabvaluetemp{0.994} & \tabvaluetemp{0.005} & \tabvaluetemp{0.009} & \tabvaluetemp{0.318}& \tabvaluetemp{0.994} & \tabvaluetemp{0.005} & \tabvaluetemp{0.006} & \tabvaluetemp{0.260}\\
\bottomrule
        \end{tabular}
        }
                
        \caption{\textit{Temporal Dependency}: Results by assuming the presence of temporal dependencies among samples (the `first' samples of \scbb{D} are put in \scbb{T}, while the last 20\% represent \scbb{E}).} 
    \label{tab:ids17_open_temporal}
    \end{subtable}

    \label{tab:ids17_open}
\end{table*}

\begin{table*}[!htbp]
  \centering
  \caption{\dataset{CTU13} Multi-classification results. Cells report the (average) Accuracy (and std. dev.) computed exclusively on the malicious samples.}
        \resizebox{2.0\columnwidth}{!}{
        \begin{tabular}{c|c ? cc|cc? cc|cc? cc|cc? cc|cc}
             \multicolumn{2}{c?}{Available Data} & \multicolumn{4}{c?}{Limited (100 per class)} & \multicolumn{4}{c?}{Scarce (15\% of \scbb{D})} &  \multicolumn{4}{c?}{Moderate (40\% of \scbb{D})} &  \multicolumn{4}{c}{Abundant (80\% of \scbb{D})} \\ \hline
             \multicolumn{2}{c?}{Dependency} &
             \multicolumn{2}{c|}{Static [N=1000]} & \multicolumn{2}{c?}{Temporal [N=1]} & \multicolumn{2}{c|}{Static [N=1000]} & \multicolumn{2}{c?}{Temporal [N=1]} & \multicolumn{2}{c|}{Static [N=1000]} & \multicolumn{2}{c?}{Temporal [N=1]} & \multicolumn{2}{c|}{Static [N=1000]} & \multicolumn{2}{c?}{Temporal [N=1]}  \\ \hline
            Alg. & Design & 
            Complete & Essential & Complete & Essential & 
            Complete & Essential & Complete & Essential & 
            Complete & Essential & Complete & Essential & 
            Complete & Essential & Complete & Essential \\
             \toprule
             
             \multirow{2}{*}{\rotatebox{90}{\parbox{40pt}\centering\textbf{RF}}}
                 & \variable{BMD} & \tabvalue{0.742}{0.021} & \tabvalue{0.722}{0.017} & \tabvaluetemp{0.734} & \tabvaluetemp{0.697} & \tabvalue{0.893}{0.001} & \tabvalue{0.877}{0.001} & \tabvaluetemp{0.895} & \tabvaluetemp{0.878} & \tabvalue{0.897}{0.001} & \tabvalue{0.880}{0.001} & \tabvaluetemp{0.899} & \tabvaluetemp{0.882}  & \tabvalue{0.900}{0.001} & \tabvalue{0.884}{0.001} & \tabvaluetemp{0.900} & \tabvaluetemp{0.884} \\
    
                & \variable{MD} & \tabvalue{0.741}{0.021} & \tabvalue{0.706}{0.018} & \tabvaluetemp{0.740} & \tabvaluetemp{0.678} & \tabvalue{0.892}{0.001} & \tabvalue{0.858}{0.001} & \tabvaluetemp{0.893} & \tabvaluetemp{0.859} & \tabvalue{0.896}{0.001} & \tabvalue{0.865}{0.001} & \tabvaluetemp{0.898} & \tabvaluetemp{0.865} & \tabvalue{0.900}{0.001} & \tabvalue{0.869}{0.001} & \tabvaluetemp{0.900} & \tabvaluetemp{0.870} \\
\bottomrule
             \end{tabular} 
             }
             \label{tab:ctu_multi}
\end{table*}

\begin{table*}[!htbp]
  \centering
  \caption{\dataset{GTCS} Multi-classification results. Cells report the (average) Accuracy (and std. dev.) computed exclusively on the malicious samples.}
        \resizebox{2.0\columnwidth}{!}{
        \begin{tabular}{c|c ? cc|cc? cc|cc? cc|cc? cc|cc}
             \multicolumn{2}{c?}{Available Data} & \multicolumn{4}{c?}{Limited (100 per class)} & \multicolumn{4}{c?}{Scarce (15\% of \scbb{D})} &  \multicolumn{4}{c?}{Moderate (40\% of \scbb{D})} &  \multicolumn{4}{c}{Abundant (80\% of \scbb{D})} \\ \hline
             \multicolumn{2}{c?}{Dependency} &
             \multicolumn{2}{c|}{Static [N=1000]} & \multicolumn{2}{c?}{Temporal [N=1]} & \multicolumn{2}{c|}{Static [N=1000]} & \multicolumn{2}{c?}{Temporal [N=1]} & \multicolumn{2}{c|}{Static [N=1000]} & \multicolumn{2}{c?}{Temporal [N=1]} & \multicolumn{2}{c|}{Static [N=1000]} & \multicolumn{2}{c?}{Temporal [N=1]}  \\ \hline
            Alg. & Design & 
            Complete & Essential & Complete & Essential & 
            Complete & Essential & Complete & Essential & 
            Complete & Essential & Complete & Essential & 
            Complete & Essential & Complete & Essential \\
             \toprule
             
             \multirow{2}{*}{\rotatebox{90}{\parbox{40pt}\centering\textbf{RF}}}
                & \variable{BMD} & \tabvalue{0.995}{0.002} & \tabvalue{0.990}{0.003} & \tabvaluetemp{0.985} & \tabvaluetemp{0.972} & \tabvalue{1.000}{0.000} & \tabvalue{0.997}{0.000} & \tabvaluetemp{1.000} & \tabvaluetemp{0.997} & \tabvalue{1.000}{0.000} & \tabvalue{0.997}{0.000} & \tabvaluetemp{1.000} & \tabvaluetemp{0.997} & \tabvalue{1.000}{0.000} & \tabvalue{0.997}{0.000} & \tabvaluetemp{1.000} & \tabvaluetemp{0.997} \\
                
                & \variable{MD} & \tabvalue{0.970}{0.009} & \tabvalue{0.961}{0.008} & \tabvaluetemp{0.935} & \tabvaluetemp{0.940} & \tabvalue{0.992}{0.001} & \tabvalue{0.972}{0.006} & \tabvaluetemp{0.984} & \tabvaluetemp{0.924} & \tabvalue{0.993}{0.000} & \tabvalue{0.977}{0.004} & \tabvaluetemp{0.993} & \tabvaluetemp{0.971} & \tabvalue{0.994}{0.000} & \tabvalue{0.977}{0.003} & \tabvaluetemp{0.994} & \tabvaluetemp{0.974}\\
\bottomrule
             \end{tabular} 
             }
             \label{tab:gtcs_multi}
\end{table*}

\begin{table*}[!htbp]
  \centering
  \caption{\dataset{NB15} Multi-classification results. Cells report the (average) Accuracy (and std. dev.) computed exclusively on the malicious samples.}
        \resizebox{2.0\columnwidth}{!}{
        \begin{tabular}{c|c ? cc|cc? cc|cc? cc|cc? cc|cc}
             \multicolumn{2}{c?}{Available Data} & \multicolumn{4}{c?}{Limited (100 per class)} & \multicolumn{4}{c?}{Scarce (15\% of \scbb{D})} &  \multicolumn{4}{c?}{Moderate (40\% of \scbb{D})} &  \multicolumn{4}{c}{Abundant (80\% of \scbb{D})} \\ \hline
             \multicolumn{2}{c?}{Dependency} &
             \multicolumn{2}{c|}{Static [N=1000]} & \multicolumn{2}{c?}{Temporal [N=1]} & \multicolumn{2}{c|}{Static [N=1000]} & \multicolumn{2}{c?}{Temporal [N=1]} & \multicolumn{2}{c|}{Static [N=1000]} & \multicolumn{2}{c?}{Temporal [N=1]} & \multicolumn{2}{c|}{Static [N=1000]} & \multicolumn{2}{c?}{Temporal [N=1]}  \\ \hline
            Alg. & Design & 
            Complete & Essential & Complete & Essential & 
            Complete & Essential & Complete & Essential & 
            Complete & Essential & Complete & Essential & 
            Complete & Essential & Complete & Essential \\
             \toprule
             
             \multirow{2}{*}{\rotatebox{90}{\parbox{40pt}\centering\textbf{RF}}}
                & \variable{BMD} & \tabvalue{0.572}{0.014} & \tabvalue{0.572}{0.015} & \tabvaluetemp{0.571} & \tabvaluetemp{0.578} & \tabvalue{0.698}{0.003} & \tabvalue{0.696}{0.003} & \tabvaluetemp{0.571} & \tabvaluetemp{0.578} & \tabvalue{0.710}{0.003} & \tabvalue{0.708}{0.003} & \tabvaluetemp{0.571} & \tabvaluetemp{0.578} & \tabvalue{0.712}{0.003} & \tabvalue{0.710}{0.003} & \tabvaluetemp{0.712} & \tabvaluetemp{0.711} \\
                
                & \variable{MD} & \tabvalue{0.572}{0.015} & \tabvalue{0.571}{0.016} & \tabvaluetemp{0.570} & \tabvaluetemp{0.575} & \tabvalue{0.688}{0.004} & \tabvalue{0.685}{0.004} & \tabvaluetemp{0.685} & \tabvaluetemp{0.680} & \tabvalue{0.701}{0.003} & \tabvalue{0.697}{0.004} & \tabvaluetemp{0.701} & \tabvaluetemp{0.697}  & \tabvalue{0.703}{0.003} & \tabvalue{0.700}{0.003} & \tabvaluetemp{0.703} & \tabvaluetemp{0.699} \\
\bottomrule
             \end{tabular} 
             }
             \label{tab:nb_multi}
\end{table*}

\begin{table*}[!htbp]
  \centering
  \caption{\dataset{UF-NB15} Multi-classification results. Cells report the (average) Accuracy (and std. dev.) computed exclusively on the malicious samples.}
        \resizebox{2.0\columnwidth}{!}{
        \begin{tabular}{c|c ? cc|cc? cc|cc? cc|cc? cc|cc}
             \multicolumn{2}{c?}{Available Data} & \multicolumn{4}{c?}{Limited (100 per class)} & \multicolumn{4}{c?}{Scarce (15\% of \scbb{D})} &  \multicolumn{4}{c?}{Moderate (40\% of \scbb{D})} &  \multicolumn{4}{c}{Abundant (80\% of \scbb{D})} \\ \hline
             \multicolumn{2}{c?}{Dependency} &
             \multicolumn{2}{c|}{Static [N=1000]} & \multicolumn{2}{c?}{Temporal [N=1]} & \multicolumn{2}{c|}{Static [N=1000]} & \multicolumn{2}{c?}{Temporal [N=1]} & \multicolumn{2}{c|}{Static [N=1000]} & \multicolumn{2}{c?}{Temporal [N=1]} & \multicolumn{2}{c|}{Static [N=1000]} & \multicolumn{2}{c?}{Temporal [N=1]}  \\ \hline
            Alg. & Design & 
            Complete & Essential & Complete & Essential & 
            Complete & Essential & Complete & Essential & 
            Complete & Essential & Complete & Essential & 
            Complete & Essential & Complete & Essential \\
             \toprule
             
             \multirow{2}{*}{\rotatebox{90}{\parbox{40pt}\centering\textbf{RF}}}
                & \variable{BMD} & \tabvalue{0.676}{0.015} & \tabvalue{0.657}{0.015} & \tabvaluetemp{0.691} & \tabvaluetemp{0.667} & \tabvalue{0.784}{0.003} & \tabvalue{0.767}{0.003} & \tabvaluetemp{0.787} & \tabvaluetemp{0.769} & \tabvalue{0.791}{0.003} & \tabvalue{0.776}{0.003} & \tabvaluetemp{0.791} & \tabvaluetemp{0.777}& \tabvalue{0.788}{0.003} & \tabvalue{0.775}{0.003} & \tabvaluetemp{0.787} & \tabvaluetemp{0.772}\\
                
                & \variable{MD} & \tabvalue{0.675}{0.017} & \tabvalue{0.645}{0.019} & \tabvaluetemp{0.688} & \tabvaluetemp{0.662} & \tabvalue{0.776}{0.003} & \tabvalue{0.757}{0.003} & \tabvaluetemp{0.779} & \tabvaluetemp{0.760} & \tabvalue{0.784}{0.003} & \tabvalue{0.767}{0.004} & \tabvaluetemp{0.785} & \tabvaluetemp{0.766} & \tabvalue{0.781}{0.003} & \tabvalue{0.766}{0.003} & \tabvaluetemp{0.779} & \tabvaluetemp{0.763} \\
\bottomrule
             \end{tabular} 
             }
             \label{tab:ufnb15_multi}
\end{table*}

\begin{table*}[!htbp]
  \centering
  \caption{\dataset{CICIDS17} Multi-classification results. Cells report the (average) Accuracy (and std. dev.) computed exclusively on the malicious samples.}
        \resizebox{2.0\columnwidth}{!}{
        \begin{tabular}{c|c ? cc|cc? cc|cc? cc|cc? cc|cc}
             \multicolumn{2}{c?}{Available Data} & \multicolumn{4}{c?}{Limited (100 per class)} & \multicolumn{4}{c?}{Scarce (15\% of \scbb{D})} &  \multicolumn{4}{c?}{Moderate (40\% of \scbb{D})} &  \multicolumn{4}{c}{Abundant (80\% of \scbb{D})} \\ \hline
             \multicolumn{2}{c?}{Dependency} &
             \multicolumn{2}{c|}{Static [N=1000]} & \multicolumn{2}{c?}{Temporal [N=1]} & \multicolumn{2}{c|}{Static [N=1000]} & \multicolumn{2}{c?}{Temporal [N=1]} & \multicolumn{2}{c|}{Static [N=1000]} & \multicolumn{2}{c?}{Temporal [N=1]} & \multicolumn{2}{c|}{Static [N=1000]} & \multicolumn{2}{c?}{Temporal [N=1]}  \\ \hline
            Alg. & Design & 
            Complete & Essential & Complete & Essential & 
            Complete & Essential & Complete & Essential & 
            Complete & Essential & Complete & Essential & 
            Complete & Essential & Complete & Essential \\
             \toprule
             
             \multirow{2}{*}{\rotatebox{90}{\parbox{40pt}\centering\textbf{RF}}}
                & \variable{BMD} & \tabvalue{0.998}{0.003} & \tabvalue{0.998}{0.001} & \tabvaluetemp{0.986} & \tabvaluetemp{0.985} & \tabvalue{1.000}{0.000} & \tabvalue{1.000}{0.000} & \tabvaluetemp{0.999} & \tabvaluetemp{0.999} & \tabvalue{1.000}{0.000} & \tabvalue{1.000}{0.000} & \tabvaluetemp{0.999} & \tabvaluetemp{0.998} & \tabvalue{1.000}{0.000} & \tabvalue{1.000}{0.000} & \tabvaluetemp{0.999} & \tabvaluetemp{0.999}  \\
                
                & \variable{MD} & \tabvalue{0.996}{0.003} & \tabvalue{0.997}{0.002} & \tabvaluetemp{0.984} & \tabvaluetemp{0.988} & \tabvalue{0.999}{0.000} & \tabvalue{0.999}{0.000} & \tabvaluetemp{0.989} & \tabvaluetemp{0.989} & \tabvalue{1.000}{0.000} & \tabvalue{1.000}{0.000} & \tabvaluetemp{0.992} & \tabvaluetemp{0.993} & \tabvalue{1.000}{0.000} & \tabvalue{1.000}{0.000} & \tabvaluetemp{0.993} & \tabvaluetemp{0.993} \\
\bottomrule
             \end{tabular} 
             }
             \label{tab:ids17_multi}
\end{table*}

\clearpage

\begin{table*}[!htbp]
  \centering
  \caption{\dataset{CTU13} Runtime (in seconds) for training the ML-NIDS on the \textit{high-end} platform..}
        \resizebox{1.5\columnwidth}{!}{
        %
}}
                & \variable{BD} & \tabvalue{0.82}{0.014} & \tabvalue{0.33}{0.011}& \tabvalue{1.71}{0.036} & \tabvalue{0.59}{0.123}& \tabvalue{4.94}{0.082} & \tabvalue{1.94}{0.783}& \tabvalue{11.72}{0.330} & \tabvalue{5.67}{0.465}\\
                
                & \variable{MD} & \tabvalue{0.86}{0.017} & \tabvalue{0.86}{0.014}& \tabvalue{1.84}{0.048} & \tabvalue{1.70}{0.076}& \tabvalue{5.06}{0.075} & \tabvalue{4.00}{0.068}& \tabvalue{11.57}{0.324} & \tabvalue{9.11}{0.305}\\
                
                & \variable{ED} & \tabvalue{4.25}{0.047} & \tabvalue{4.31}{0.042}& \tabvalue{6.84}{0.080} & \tabvalue{6.99}{0.107}& \tabvalue{16.44}{0.163} & \tabvalue{12.83}{0.164}& \tabvalue{37.44}{0.957} & \tabvalue{29.01}{0.718}\\
                
                & \variable{ED-s} & \tabvalue{5.92}{0.070} & \tabvalue{5.98}{0.067}& \tabvalue{10.53}{0.141} & \tabvalue{9.72}{0.184}& \tabvalue{23.28}{0.187} & \tabvalue{17.71}{0.212}& \tabvalue{50.05}{1.112} & \tabvalue{37.57}{0.796}\\
             \midrule
             
             \multirow{4}{*}{\rotatebox{90}{\parbox{40pt}\centering
             \begin{tabular}{c}
                  \textbf{DT} \\
                  \tiny{one core}
             \end{tabular}}}
                & \variable{BD} & \tabvalue{0.00}{0.000} & \tabvalue{0.00}{0.000}& \tabvalue{0.89}{0.060} & \tabvalue{0.16}{0.001}& \tabvalue{2.39}{0.175} & \tabvalue{0.95}{0.201} & \tabvalue{5.58}{0.510} & \tabvalue{1.87}{0.653}\\
                
                & \variable{MD} & \tabvalue{0.01}{0.000} & \tabvalue{0.00}{0.000}& \tabvalue{0.85}{0.038} & \tabvalue{0.37}{0.023}& \tabvalue{2.25}{0.118} & \tabvalue{0.97}{0.072}& \tabvalue{5.15}{0.435} & \tabvalue{2.23}{0.246}\\
                
                & \variable{ED} & \tabvalue{0.02}{0.001} & \tabvalue{0.01}{0.001}& \tabvalue{2.80}{0.126} & \tabvalue{1.24}{0.072}& \tabvalue{6.94}{0.326} & \tabvalue{3.27}{0.223}& \tabvalue{15.46}{1.244} & \tabvalue{7.62}{0.778}\\
                
                & \variable{ED-s} & \tabvalue{0.03}{0.001} & \tabvalue{0.02}{0.001}& \tabvalue{3.85}{0.148} & \tabvalue{1.33}{0.075}& \tabvalue{9.15}{0.376} & \tabvalue{3.52}{0.230}& \tabvalue{20.05}{1.537} & \tabvalue{8.15}{0.800}\\
             \midrule
             
             \multirow{4}{*}{\rotatebox{90}{\parbox{40pt}\centering
             \begin{tabular}{c}
                  \textbf{LR} \\
                  \tiny{all cores}
             \end{tabular}}}
                & \variable{BD} & \tabvalue{0.08}{0.385} & \tabvalue{0.01}{0.001}& \tabvalue{7.08}{0.931} & \tabvalue{3.44}{0.873}& \tabvalue{23.76}{6.714} & \tabvalue{11.34}{4.351}& \tabvalue{47.84}{12.698} & \tabvalue{21.85}{14.564}\\
                
                & \variable{MD} & \tabvalue{0.11}{0.026} & \tabvalue{0.11}{0.030}& \tabvalue{10.04}{1.527} & \tabvalue{9.64}{1.729}& \tabvalue{25.77}{6.915} & \tabvalue{27.96}{8.821}& \tabvalue{52.06}{14.071} & \tabvalue{56.45}{15.916}\\
                
                & \variable{ED} & \tabvalue{0.22}{0.155} & \tabvalue{0.18}{0.133}& \tabvalue{30.75}{3.091} & \tabvalue{12.33}{1.232}& \tabvalue{117.02}{30.614} & \tabvalue{35.94}{9.819}& \tabvalue{243.77}{66.388} & \tabvalue{95.58}{26.226}\\
                
                & \variable{ED-s} & \tabvalue{0.26}{0.186} & \tabvalue{0.21}{0.133}& \tabvalue{32.45}{3.079} & \tabvalue{12.84}{1.231}& \tabvalue{120.31}{31.382} & \tabvalue{37.07}{10.086}& \tabvalue{249.68}{67.708} & \tabvalue{97.72}{26.740}\\
             \midrule
             
             \multirow{4}{*}{\rotatebox{90}{\parbox{40pt}\centering
             \begin{tabular}{c}
                  \textbf{HGB} \\
                  \tiny{all cores}
             \end{tabular}}}
                & \variable{BD} & \tabvalue{0.33}{0.041} & \tabvalue{0.01}{0.012}& \tabvalue{1.52}{0.136} & \tabvalue{0.64}{0.191}& \tabvalue{2.75}{0.176} & \tabvalue{0.99}{0.033}& \tabvalue{4.72}{0.240} & \tabvalue{1.95}{0.547}\\
                
                & \variable{MD} & \tabvalue{2.25}{0.699} & \tabvalue{1.80}{0.120}& \tabvalue{2.04}{0.470} & \tabvalue{1.93}{0.474}& \tabvalue{3.88}{0.679} & \tabvalue{3.14}{0.706}& \tabvalue{5.94}{1.166} & \tabvalue{4.81}{0.857}\\
                
                & \variable{ED} & \tabvalue{0.66}{0.063} & \tabvalue{0.58}{0.057}& \tabvalue{6.90}{0.360} & \tabvalue{5.20}{0.341}& \tabvalue{12.41}{0.320} & \tabvalue{9.23}{0.563}& \tabvalue{19.80}{0.458} & \tabvalue{14.44}{0.936}\\
                
                & \variable{ED-s} & \tabvalue{0.76}{0.071} & \tabvalue{0.72}{0.067}& \tabvalue{9.30}{0.379} & \tabvalue{6.32}{0.343}& \tabvalue{17.69}{0.364} & \tabvalue{11.73}{0.628}& \tabvalue{30.29}{0.622} & \tabvalue{19.20}{1.033}\\

             \bottomrule
             \end{tabular} 
             }
             \label{tab:ctu_training}
\end{table*}

\begin{table*}[!htbp]
  \centering
  \caption{\dataset{GTCS}: Runtime (in seconds) for training the ML-NIDS on the \textit{high-end} platform..}
        \resizebox{1.5\columnwidth}{!}{
        \begin{tabular}{c|c ? cc? cc? cc? cc}
             \multicolumn{2}{c?}{Available Data} & \multicolumn{2}{c?}{Limited (100 per class) [N=1000]} & \multicolumn{2}{c?}{Scarce (15\% of \smabb{D}) [N=100]} &  \multicolumn{2}{c?}{Moderate (40\% of \smabb{D}) [N=100]} &  \multicolumn{2}{c}{Abundant (80\% of \smabb{D}) [N=100]} \\ 
              \hline
            Alg. {\tiny CPU} & Design & 
            Complete & Essential &
            Complete & Essential &
            Complete & Essential &
            Complete & Essential \\
             \toprule
             
             \multirow{4}{*}{\rotatebox{90}{\parbox{40pt}\centering
             \begin{tabular}{c}
                  \textbf{RF} \\
                  \tiny{all cores}
             \end{tabular}}}
                & \variable{BD} & \tabvalue{0.79}{0.050} & \tabvalue{0.29}{0.043}& \tabvalue{1.46}{0.033} & \tabvalue{1.63}{0.047}& \tabvalue{4.35}{1.474} & \tabvalue{2.81}{0.784}& \tabvalue{9.69}{0.195} & \tabvalue{5.38}{0.334}\\
                
                & \variable{MD} & \tabvalue{0.81}{0.053} & \tabvalue{0.81}{0.051}& \tabvalue{1.41}{0.043} & \tabvalue{1.07}{0.043}& \tabvalue{3.92}{1.304} & \tabvalue{2.40}{0.706}& \tabvalue{8.57}{0.165} & \tabvalue{4.89}{0.096}\\
                
                & \variable{ED} & \tabvalue{2.81}{0.129} & \tabvalue{2.87}{0.131}& \tabvalue{4.17}{0.056} & \tabvalue{4.14}{0.085}& \tabvalue{6.41}{1.188} & \tabvalue{4.39}{0.636}& \tabvalue{12.60}{0.211} & \tabvalue{7.14}{0.110}\\
                
                & \variable{ED-s} & \tabvalue{4.10}{0.203} & \tabvalue{4.17}{0.205}& \tabvalue{7.40}{0.079} & \tabvalue{6.36}{0.092}& \tabvalue{12.53}{1.637} & \tabvalue{7.65}{1.181}& \tabvalue{23.11}{0.383} & \tabvalue{11.77}{0.121}\\
             \midrule
             
             \multirow{4}{*}{\rotatebox{90}{\parbox{40pt}\centering
             \begin{tabular}{c}
                  \textbf{DT} \\
                  \tiny{one core}
             \end{tabular}}}
                & \variable{BD} & \tabvalue{0.01}{0.001} & \tabvalue{0.01}{0.001}& \tabvalue{1.11}{0.056} & \tabvalue{0.33}{0.074}& \tabvalue{3.19}{0.209} & \tabvalue{0.94}{0.114}& \tabvalue{7.15}{0.576} & \tabvalue{2.11}{0.674}\\
                
                & \variable{MD} & \tabvalue{0.01}{0.000} & \tabvalue{0.00}{0.000}& \tabvalue{1.02}{0.048} & \tabvalue{0.42}{0.017}& \tabvalue{2.90}{0.181} & \tabvalue{1.15}{0.081}& \tabvalue{6.41}{0.476} & \tabvalue{2.52}{0.208}\\
                
                & \variable{ED} & \tabvalue{0.02}{0.001} & \tabvalue{0.01}{0.000}& \tabvalue{1.27}{0.028} & \tabvalue{0.54}{0.018}& \tabvalue{3.50}{0.230} & \tabvalue{1.48}{0.097}& \tabvalue{7.63}{0.393} & \tabvalue{3.29}{0.262}\\
                
                & \variable{ED-s} & \tabvalue{0.03}{0.001} & \tabvalue{0.01}{0.001}& \tabvalue{2.29}{0.049} & \tabvalue{0.58}{0.019}& \tabvalue{6.39}{0.242} & \tabvalue{1.61}{0.173}& \tabvalue{13.64}{0.545} & \tabvalue{3.55}{0.274}\\
             \midrule
             
             \multirow{4}{*}{\rotatebox{90}{\parbox{40pt}\centering
             \begin{tabular}{c}
                  \textbf{LR} \\
                  \tiny{all cores}
             \end{tabular}}}
                & \variable{BD} & \tabvalue{0.03}{0.176} & \tabvalue{0.01}{0.087}& \tabvalue{1.58}{0.221} & \tabvalue{2.05}{0.712}& \tabvalue{4.94}{0.406} & \tabvalue{9.04}{1.013}& \tabvalue{10.19}{0.683} & \tabvalue{17.06}{1.741}\\
                
                & \variable{MD} & \tabvalue{0.08}{0.034} & \tabvalue{0.09}{0.022}& \tabvalue{5.26}{0.581} & \tabvalue{3.34}{0.155}& \tabvalue{13.00}{1.305} & \tabvalue{8.87}{0.318}& \tabvalue{26.36}{2.640} & \tabvalue{17.19}{0.737}\\
                
                & \variable{ED} & \tabvalue{0.10}{0.128} & \tabvalue{0.11}{0.086}& \tabvalue{5.79}{0.811} & \tabvalue{2.45}{0.206}& \tabvalue{20.00}{2.969} & \tabvalue{7.43}{1.444}& \tabvalue{46.23}{1.968} & \tabvalue{20.92}{2.338}\\
                
                & \variable{ED-s} & \tabvalue{0.13}{0.152} & \tabvalue{0.13}{0.112}& \tabvalue{7.09}{0.818} & \tabvalue{2.71}{0.217}& \tabvalue{23.20}{3.036} & \tabvalue{7.86}{1.726}& \tabvalue{52.53}{1.981} & \tabvalue{21.62}{2.347}\\
             \midrule
             
             \multirow{4}{*}{\rotatebox{90}{\parbox{40pt}\centering
             \begin{tabular}{c}
                  \textbf{HGB} \\
                  \tiny{all cores}
             \end{tabular}}}
                & \variable{BD} & \tabvalue{0.32}{0.042} & \tabvalue{0.29}{0.026}& \tabvalue{1.76}{0.041} & \tabvalue{1.03}{0.047}& \tabvalue{3.56}{0.474} & \tabvalue{1.80}{0.347}& \tabvalue{5.80}{0.069} & \tabvalue{2.96}{0.143}\\
                
                & \variable{MD} & \tabvalue{1.21}{0.119} & \tabvalue{1.11}{0.101}& \tabvalue{5.81}{0.431} & \tabvalue{4.23}{0.210}& \tabvalue{10.51}{1.440} & \tabvalue{7.08}{0.879}& \tabvalue{16.52}{0.211} & \tabvalue{11.00}{0.868}\\
                
                & \variable{ED} & \tabvalue{0.45}{0.057} & \tabvalue{0.43}{0.047}& \tabvalue{4.55}{0.148} & \tabvalue{2.88}{0.124}& \tabvalue{8.46}{0.632} & \tabvalue{4.57}{0.464}& \tabvalue{13.98}{0.183} & \tabvalue{7.30}{0.126}\\
                
                & \variable{ED-s} & \tabvalue{0.53}{0.061} & \tabvalue{0.52}{0.052}& \tabvalue{6.53}{0.176} & \tabvalue{3.53}{0.142}& \tabvalue{14.14}{0.612} & \tabvalue{6.02}{0.815}& \tabvalue{24.11}{0.279} & \tabvalue{10.02}{0.173}\\

             \bottomrule
             \end{tabular} 
             }
             \label{tab:gtcs_training}
\end{table*}

\begin{table*}[!htbp]
  \centering
  \caption{\dataset{NB15}: Runtime (in seconds) for training the ML-NIDS on the \textit{high-end} platform..}
        \resizebox{1.5\columnwidth}{!}{
        \begin{tabular}{c|c ? cc? cc? cc? cc}
             \multicolumn{2}{c?}{Available Data} & \multicolumn{2}{c?}{Limited (100 per class) [N=1000]} & \multicolumn{2}{c?}{Scarce (15\% of \smabb{D}) [N=100]} &  \multicolumn{2}{c?}{Moderate (40\% of \smabb{D}) [N=100]} &  \multicolumn{2}{c}{Abundant (80\% of \smabb{D}) [N=100]} \\ 
              \hline
            Alg. {\tiny CPU} & Design & 
            Complete & Essential &
            Complete & Essential &
            Complete & Essential &
            Complete & Essential \\
             \toprule
             
             \multirow{4}{*}{\rotatebox{90}{\parbox{40pt}\centering
             \begin{tabular}{c}
                  \textbf{RF} \\
                  \tiny{all cores}
             \end{tabular}}}
                & \variable{BD} & \tabvalue{0.80}{0.013} & \tabvalue{0.33}{0.009}& \tabvalue{1.28}{0.027} & \tabvalue{1.71}{0.264}& \tabvalue{3.58}{0.090} & \tabvalue{3.89}{0.074}& \tabvalue{8.23}{0.214} & \tabvalue{8.85}{0.413}\\
                
                & \variable{MD} & \tabvalue{0.90}{0.015} & \tabvalue{0.90}{0.017}& \tabvalue{1.37}{0.027} & \tabvalue{1.23}{0.036}& \tabvalue{3.77}{0.088} & \tabvalue{3.67}{0.067}& \tabvalue{8.62}{0.228} & \tabvalue{9.01}{0.154}\\
                
                & \variable{ED} & \tabvalue{4.84}{0.075} & \tabvalue{4.95}{0.076}& \tabvalue{7.74}{0.086} & \tabvalue{8.02}{0.106}& \tabvalue{18.33}{0.178} & \tabvalue{16.28}{0.163}& \tabvalue{42.41}{0.527} & \tabvalue{38.48}{0.355}\\
                
                & \variable{ED-s} & \tabvalue{6.66}{0.099} & \tabvalue{6.77}{0.101}& \tabvalue{11.53}{0.104} & \tabvalue{11.23}{0.113}& \tabvalue{24.74}{0.207} & \tabvalue{21.48}{0.169}& \tabvalue{54.48}{0.552} & \tabvalue{47.50}{0.441}\\
             \midrule
             
             \multirow{4}{*}{\rotatebox{90}{\parbox{40pt}\centering
             \begin{tabular}{c}
                  \textbf{DT} \\
                  \tiny{one core}
             \end{tabular}}}
                & \variable{BD} & \tabvalue{0.00}{0.001} & \tabvalue{0.00}{0.001}& \tabvalue{0.65}{0.040} & \tabvalue{0.60}{0.074}& \tabvalue{2.06}{0.860} & \tabvalue{1.37}{0.368}& \tabvalue{3.89}{0.671} & \tabvalue{3.70}{0.234}\\
                
                & \variable{MD} & \tabvalue{0.01}{0.000} & \tabvalue{0.01}{0.000}& \tabvalue{0.70}{0.045} & \tabvalue{0.66}{0.070}& \tabvalue{2.19}{0.853} & \tabvalue{2.19}{0.963}& \tabvalue{4.16}{0.788} & \tabvalue{4.32}{0.965}\\
                
                & \variable{ED} & \tabvalue{0.02}{0.001} & \tabvalue{0.02}{0.001}& \tabvalue{3.53}{0.180} & \tabvalue{2.11}{0.114}& \tabvalue{11.31}{4.645} & \tabvalue{7.11}{3.002}& \tabvalue{21.57}{3.615} & \tabvalue{14.09}{2.499}\\
                
                & \variable{ED-s} & \tabvalue{0.04}{0.001} & \tabvalue{0.03}{0.001}& \tabvalue{4.67}{0.184} & \tabvalue{2.71}{0.117}& \tabvalue{14.76}{6.123} & \tabvalue{8.94}{3.804}& \tabvalue{28.26}{4.703} & \tabvalue{17.29}{3.013}\\
             \midrule
             
             \multirow{4}{*}{\rotatebox{90}{\parbox{40pt}\centering
             \begin{tabular}{c}
                  \textbf{LR} \\
                  \tiny{all cores}
             \end{tabular}}}
                & \variable{BD} & \tabvalue{0.05}{0.209} & \tabvalue{0.04}{0.111}& \tabvalue{3.47}{0.537} & \tabvalue{2.94}{0.754}& \tabvalue{12.16}{2.188} & \tabvalue{8.60}{1.147}& \tabvalue{23.88}{3.130} & \tabvalue{21.17}{1.411}\\
                
                & \variable{MD} & \tabvalue{0.12}{0.023} & \tabvalue{0.12}{0.028}& \tabvalue{7.39}{0.210} & \tabvalue{5.66}{0.519}& \tabvalue{19.15}{0.417} & \tabvalue{18.07}{1.515}& \tabvalue{37.93}{0.818} & \tabvalue{35.38}{2.705}\\
                
                & \variable{ED} & \tabvalue{0.16}{0.149} & \tabvalue{0.29}{0.020}& \tabvalue{24.62}{1.390} & \tabvalue{18.28}{0.956}& \tabvalue{97.78}{5.799} & \tabvalue{79.06}{3.903}& \tabvalue{205.70}{13.803} & \tabvalue{172.12}{6.615}\\
                
                & \variable{ED-s} & \tabvalue{0.22}{0.175} & \tabvalue{0.33}{0.024}& \tabvalue{26.40}{1.402} & \tabvalue{19.63}{0.955}& \tabvalue{101.51}{5.853} & \tabvalue{81.64}{3.927}& \tabvalue{213.28}{13.881} & \tabvalue{176.82}{6.698}\\
             \midrule
             
             \multirow{4}{*}{\rotatebox{90}{\parbox{40pt}\centering
             \begin{tabular}{c}
                  \textbf{HGB} \\
                  \tiny{all cores}
             \end{tabular}}}
                & \variable{BD} & \tabvalue{0.33}{0.082} & \tabvalue{0.31}{0.034}& \tabvalue{1.38}{0.091} & \tabvalue{1.18}{0.168}& \tabvalue{2.76}{0.050} & \tabvalue{2.18}{0.085}& \tabvalue{4.57}{0.073} & \tabvalue{3.67}{0.085}\\
                
                & \variable{MD} & \tabvalue{3.36}{0.157} & \tabvalue{3.07}{0.134}& \tabvalue{1.46}{0.261} & \tabvalue{1.06}{0.154}& \tabvalue{2.60}{0.053} & \tabvalue{1.97}{0.051}& \tabvalue{4.15}{0.121} & \tabvalue{3.25}{0.068}\\
                
                & \variable{ED} & \tabvalue{0.73}{0.174} & \tabvalue{0.69}{0.120}& \tabvalue{7.37}{0.444} & \tabvalue{5.61}{0.454}& \tabvalue{14.48}{0.804} & \tabvalue{10.62}{0.624}& \tabvalue{23.76}{1.494} & \tabvalue{16.97}{0.789}\\
                
                & \variable{ED-s} & \tabvalue{0.84}{0.191} & \tabvalue{0.79}{0.139}& \tabvalue{9.74}{0.463} & \tabvalue{7.30}{0.462}& \tabvalue{20.07}{0.917} & \tabvalue{14.35}{0.661}& \tabvalue{35.56}{1.650} & \tabvalue{24.08}{0.949}\\
             \bottomrule
             \end{tabular} 
             }
             \label{tab:nb15_training}
\end{table*}

\begin{table*}[!htbp]
  \centering
  \caption{\dataset{UF-NB15}: Runtime (in seconds) for training the ML-NIDS on the \textit{high-end} platform..}
        \resizebox{1.5\columnwidth}{!}{
        \begin{tabular}{c|c ? cc? cc? cc? cc}
             \multicolumn{2}{c?}{Available Data} & \multicolumn{2}{c?}{Limited (100 per class) [N=1000]} & \multicolumn{2}{c?}{Scarce (15\% of \smabb{D}) [N=100]} &  \multicolumn{2}{c?}{Moderate (40\% of \smabb{D}) [N=100]} &  \multicolumn{2}{c}{Abundant (80\% of \smabb{D}) [N=100]} \\ 
              \hline
            Alg. {\tiny CPU} & Design & 
            Complete & Essential &
            Complete & Essential &
            Complete & Essential &
            Complete & Essential \\
             \toprule
             
             \multirow{4}{*}{\rotatebox{90}{\parbox{40pt}\centering
             \begin{tabular}{c}
                  \textbf{RF} \\
                  \tiny{all cores}
             \end{tabular}}}
& \variable{BD} & \tabvalue{0.78}{0.014} & \tabvalue{0.33}{0.024}& \tabvalue{1.20}{0.030} & \tabvalue{1.70}{0.054}& \tabvalue{2.46}{0.067} & \tabvalue{3.46}{0.084}& \tabvalue{5.67}{0.115} & \tabvalue{7.96}{0.107}\\

& \variable{MD} & \tabvalue{0.90}{0.015} & \tabvalue{0.90}{0.018}& \tabvalue{1.17}{0.026} & \tabvalue{1.41}{0.048}& \tabvalue{2.51}{0.069} & \tabvalue{3.21}{0.062}& \tabvalue{5.76}{0.123} & \tabvalue{7.90}{0.170}\\

& \variable{ED} & \tabvalue{4.74}{0.074} & \tabvalue{5.00}{0.062}& \tabvalue{9.65}{0.065} & \tabvalue{8.98}{0.128}& \tabvalue{14.28}{0.236} & \tabvalue{15.66}{0.156}& \tabvalue{30.22}{0.239} & \tabvalue{36.18}{0.478}\\

& \variable{ED-s} & \tabvalue{6.54}{0.096} & \tabvalue{6.81}{0.088}& \tabvalue{13.41}{0.084} & \tabvalue{12.07}{0.138}& \tabvalue{20.50}{0.364} & \tabvalue{20.57}{0.184}& \tabvalue{41.08}{0.262} & \tabvalue{44.56}{0.518}\\
             \midrule
             
             \multirow{4}{*}{\rotatebox{90}{\parbox{40pt}\centering
             \begin{tabular}{c}
                  \textbf{DT} \\
                  \tiny{one core}
             \end{tabular}}}
& \variable{BD} & \tabvalue{0.00}{0.000} & \tabvalue{0.01}{0.000}& \tabvalue{0.44}{0.033} & \tabvalue{0.02}{0.029}& \tabvalue{1.30}{0.124} & \tabvalue{0.70}{0.111}& \tabvalue{3.57}{0.876} & \tabvalue{1.66}{0.578}\\

& \variable{MD} & \tabvalue{0.01}{0.000} & \tabvalue{0.00}{0.000}& \tabvalue{0.44}{0.037} & \tabvalue{0.30}{0.014}& \tabvalue{1.30}{0.118} & \tabvalue{0.83}{0.077}& \tabvalue{3.52}{0.848} & \tabvalue{2.25}{0.572}\\

& \variable{ED} & \tabvalue{0.02}{0.001} & \tabvalue{0.02}{0.001}& \tabvalue{1.85}{0.058} & \tabvalue{1.38}{0.039}& \tabvalue{5.28}{0.392} & \tabvalue{3.80}{0.307}& \tabvalue{14.13}{3.225} & \tabvalue{10.30}{2.432}\\

& \variable{ED-s} & \tabvalue{0.03}{0.001} & \tabvalue{0.03}{0.001}& \tabvalue{2.85}{0.060} & \tabvalue{1.88}{0.040}& \tabvalue{8.00}{0.585} & \tabvalue{4.97}{0.379}& \tabvalue{20.45}{4.571} & \tabvalue{13.18}{3.021}\\
             \midrule
             
             \multirow{4}{*}{\rotatebox{90}{\parbox{40pt}\centering
             \begin{tabular}{c}
                  \textbf{LR} \\
                  \tiny{all cores}
             \end{tabular}}}
& \variable{BD} & \tabvalue{0.05}{0.142} & \tabvalue{0.27}{0.197}& \tabvalue{4.77}{0.554} & \tabvalue{2.38}{0.678}& \tabvalue{21.43}{3.804} & \tabvalue{11.06}{2.987}& \tabvalue{45.62}{5.266} & \tabvalue{27.78}{4.997}\\

& \variable{MD} & \tabvalue{0.12}{0.024} & \tabvalue{0.12}{0.025}& \tabvalue{6.76}{0.230} & \tabvalue{5.85}{0.800}& \tabvalue{18.95}{1.235} & \tabvalue{18.68}{3.129}& \tabvalue{36.81}{1.202} & \tabvalue{39.99}{4.944}\\

& \variable{ED} & \tabvalue{0.23}{0.141} & \tabvalue{0.32}{0.033}& \tabvalue{31.27}{1.151} & \tabvalue{13.59}{1.249}& \tabvalue{132.16}{10.707} & \tabvalue{53.76}{8.105}& \tabvalue{292.26}{13.293} & \tabvalue{127.41}{14.443}\\

& \variable{ED-s} & \tabvalue{0.28}{0.168} & \tabvalue{0.36}{0.035}& \tabvalue{32.95}{1.159} & \tabvalue{14.79}{1.267}& \tabvalue{136.07}{11.071} & \tabvalue{56.08}{8.263}& \tabvalue{299.48}{13.352} & \tabvalue{131.67}{14.515}\\
             \midrule
             
             \multirow{4}{*}{\rotatebox{90}{\parbox{40pt}\centering
             \begin{tabular}{c}
                  \textbf{HGB} \\
                  \tiny{all cores}
             \end{tabular}}}
& \variable{BD} & \tabvalue{0.23}{0.090} & \tabvalue{0.44}{0.132}& \tabvalue{1.20}{0.105} & \tabvalue{1.14}{0.087}& \tabvalue{2.51}{0.076} & \tabvalue{2.04}{0.066}& \tabvalue{4.18}{0.055} & \tabvalue{3.43}{0.078}\\

& \variable{MD} & \tabvalue{3.37}{0.161} & \tabvalue{3.09}{0.288}& \tabvalue{1.22}{0.071} & \tabvalue{0.99}{0.086}& \tabvalue{2.47}{0.069} & \tabvalue{1.80}{0.053}& \tabvalue{3.90}{0.095} & \tabvalue{3.15}{0.095}\\

& \variable{ED} & \tabvalue{0.60}{0.186} & \tabvalue{0.71}{0.075}& \tabvalue{5.96}{0.324} & \tabvalue{5.00}{0.318}& \tabvalue{13.43}{0.457} & \tabvalue{9.69}{0.373}& \tabvalue{22.14}{0.458} & \tabvalue{16.09}{0.879}\\

& \variable{ED-s} & \tabvalue{0.69}{0.198} & \tabvalue{0.84}{0.090}& \tabvalue{8.16}{0.364} & \tabvalue{6.51}{0.347}& \tabvalue{19.52}{0.649} & \tabvalue{12.91}{0.441}& \tabvalue{33.33}{0.564} & \tabvalue{23.01}{1.345}\\

             \bottomrule
             \end{tabular} 
             }
             \label{tab:ufnb15_training}
\end{table*}

\begin{table*}[!htbp]
  \centering
  \caption{\dataset{CICIDS17}: Runtime (in seconds) for training the ML-NIDS on the \textit{high-end} platform..}
        \resizebox{1.5\columnwidth}{!}{
        \begin{tabular}{c|c ? cc? cc? cc? cc}
             \multicolumn{2}{c?}{Available Data} & \multicolumn{2}{c?}{Limited (100 per class) [N=1000]} & \multicolumn{2}{c?}{Scarce (15\% of \smabb{D}) [N=100]} &  \multicolumn{2}{c?}{Moderate (40\% of \smabb{D}) [N=100]} &  \multicolumn{2}{c}{Abundant (80\% of \smabb{D}) [N=100]} \\ 
              \hline
            Alg. {\tiny CPU} & Design & 
            Complete & Essential &
            Complete & Essential &
            Complete & Essential &
            Complete & Essential \\
             \toprule
             
             \multirow{4}{*}{\rotatebox{90}{\parbox{40pt}\centering
             \begin{tabular}{c}
                  \textbf{RF} \\
                  \tiny{all cores}
             \end{tabular}}}
& \variable{BD} & \tabvalue{0.84}{0.015} & \tabvalue{0.23}{0.008}& \tabvalue{3.33}{0.048} & \tabvalue{0.94}{0.103}& \tabvalue{10.11}{0.157} & \tabvalue{3.47}{0.098}& \tabvalue{24.20}{0.549} & \tabvalue{13.20}{0.341}\\

& \variable{MD} & \tabvalue{0.86}{0.016} & \tabvalue{0.85}{0.017}& \tabvalue{3.73}{0.055} & \tabvalue{1.42}{0.055}& \tabvalue{11.16}{0.151} & \tabvalue{3.58}{0.050}& \tabvalue{26.21}{0.656} & \tabvalue{8.68}{0.156}\\

& \variable{ED} & \tabvalue{6.19}{0.113} & \tabvalue{6.21}{0.120}& \tabvalue{13.40}{0.165} & \tabvalue{12.26}{0.086}& \tabvalue{38.33}{0.424} & \tabvalue{16.47}{0.159}& \tabvalue{89.38}{1.622} & \tabvalue{33.09}{0.369}\\

& \variable{ED-s} & \tabvalue{8.42}{0.178} & \tabvalue{8.41}{0.196}& \tabvalue{21.10}{0.255} & \tabvalue{16.54}{0.090}& \tabvalue{54.73}{0.588} & \tabvalue{24.19}{0.200}& \tabvalue{121.04}{1.665} & \tabvalue{48.20}{0.370}\\
             \midrule
             
             \multirow{4}{*}{\rotatebox{90}{\parbox{40pt}\centering
             \begin{tabular}{c}
                  \textbf{DT} \\
                  \tiny{one core}
             \end{tabular}}}
& \variable{BD} & \tabvalue{0.02}{0.003} & \tabvalue{0.01}{0.000}& \tabvalue{4.85}{0.745} & \tabvalue{0.44}{0.013}& \tabvalue{12.27}{1.452} & \tabvalue{1.26}{0.090}& \tabvalue{24.89}{2.915} & \tabvalue{2.71}{0.499}\\

& \variable{MD} & \tabvalue{0.02}{0.001} & \tabvalue{0.00}{0.000}& \tabvalue{4.46}{0.557} & \tabvalue{0.49}{0.053}& \tabvalue{11.27}{0.826} & \tabvalue{1.30}{0.094}& \tabvalue{22.84}{4.166} & \tabvalue{2.77}{0.563}\\

& \variable{ED} & \tabvalue{0.03}{0.002} & \tabvalue{0.02}{0.001}& \tabvalue{13.35}{1.434} & \tabvalue{2.19}{0.250}& \tabvalue{37.77}{2.280} & \tabvalue{5.65}{0.336}& \tabvalue{81.49}{15.688} & \tabvalue{12.33}{1.905}\\

& \variable{ED-s} & \tabvalue{0.07}{0.003} & \tabvalue{0.04}{0.002}& \tabvalue{18.02}{1.835} & \tabvalue{3.31}{0.336}& \tabvalue{49.31}{2.592} & \tabvalue{8.24}{0.442}& \tabvalue{103.94}{17.805} & \tabvalue{18.16}{2.597}\\
             \midrule
             
             \multirow{4}{*}{\rotatebox{90}{\parbox{40pt}\centering
             \begin{tabular}{c}
                  \textbf{LR} \\
                  \tiny{all cores}
             \end{tabular}}}
& \variable{BD} & \tabvalue{0.07}{0.153} & \tabvalue{0.05}{0.134}& \tabvalue{18.57}{0.568} & \tabvalue{13.21}{0.645}& \tabvalue{60.62}{7.938} & \tabvalue{39.45}{5.54}& \tabvalue{129.09}{9.790} & \tabvalue{112.89}{11.14}\\

& \variable{MD} & \tabvalue{0.15}{0.028} & \tabvalue{0.16}{0.031}& \tabvalue{19.47}{0.844} & \tabvalue{17.17}{3.146}& \tabvalue{48.32}{4.827} & \tabvalue{46.13}{7.626}& \tabvalue{96.15}{5.844} & \tabvalue{96.44}{13.016}\\

& \variable{ED} & \tabvalue{0.37}{0.210} & \tabvalue{0.23}{0.030}& \tabvalue{119.46}{3.840} & \tabvalue{31.92}{3.638}& \tabvalue{376.68}{52.504} & \tabvalue{132.97}{22.547}& \tabvalue{867.38}{68.230} & \tabvalue{275.00}{29.632}\\

& \variable{ED-s} & \tabvalue{0.46}{0.238} & \tabvalue{0.29}{0.034}& \tabvalue{125.06}{3.860} & \tabvalue{34.45}{3.652}& \tabvalue{390.13}{52.424} & \tabvalue{138.92}{22.410}& \tabvalue{894.22}{69.537} & \tabvalue{288.14}{29.984}\\
             \midrule
             
             \multirow{4}{*}{\rotatebox{90}{\parbox{40pt}\centering
             \begin{tabular}{c}
                  \textbf{HGB} \\
                  \tiny{all cores}
             \end{tabular}}}
& \variable{BD} & \tabvalue{0.55}{0.042} & \tabvalue{0.01}{0.012}& \tabvalue{2.75}{0.067} & \tabvalue{0.52}{0.076}& \tabvalue{5.43}{0.124} & \tabvalue{1.28}{0.086}& \tabvalue{9.53}{0.219} & \tabvalue{2.94}{0.486}\\

& \variable{MD} & \tabvalue{3.04}{0.626} & \tabvalue{2.50}{1.382}& \tabvalue{3.08}{0.306} & \tabvalue{1.85}{0.453}& \tabvalue{5.85}{0.768} & \tabvalue{3.40}{0.600}& \tabvalue{9.58}{0.788} & \tabvalue{5.84}{0.525}\\

& \variable{ED} & \tabvalue{1.12}{0.245} & \tabvalue{1.00}{0.077}& \tabvalue{14.80}{0.512} & \tabvalue{8.27}{0.516}& \tabvalue{30.81}{0.884} & \tabvalue{16.83}{0.784}& \tabvalue{48.19}{1.059} & \tabvalue{26.91}{1.091}\\

& \variable{ED-s} & \tabvalue{1.31}{0.249} & \tabvalue{1.20}{0.087}& \tabvalue{23.38}{0.682} & \tabvalue{10.91}{0.543}& \tabvalue{48.96}{0.965} & \tabvalue{23.21}{1.039}& \tabvalue{79.28}{1.198} & \tabvalue{41.13}{1.209}\\

             \bottomrule
             \end{tabular} 
             }
             \label{tab:ids17_training}
\end{table*}

\begin{table*}[!htbp]
  \centering
  \caption{\dataset{CTU13}: Runtime (in seconds) for testing (on \scbb{E}) the ML-NIDS on the \textit{high-end} platform..}
        \resizebox{1.1\columnwidth}{!}{
\begin{tabular}{c|c ? cc? cc? cc? cc}
             \multicolumn{2}{c?}{Available Data} & \multicolumn{2}{c?}{Limited (100 per class)} & \multicolumn{2}{c?}{Scarce (20\% of \scbb{D})} &  \multicolumn{2}{c?}{Moderate (40\% of \scbb{D})} &  \multicolumn{2}{c}{Abundant (80\% of \scbb{D})} \\ 
              \hline
            Alg. & Design & 
            Complete & Essential &
            Complete & Essential &
            Complete & Essential &
            Complete & Essential \\
             \toprule
             
             \multirow{4}{*}{\rotatebox{90}{\parbox{40pt}\centering \textbf{RF} }}
                & \variable{BD} & \tabvaluetemp{0.34} & \tabvaluetemp{0.15} & \tabvaluetemp{0.37} & \tabvaluetemp{0.22} & \tabvaluetemp{0.38} & \tabvaluetemp{0.25} & \tabvaluetemp{0.38} & \tabvaluetemp{0.25} \\
                
                & \variable{MD} & \tabvaluetemp{0.62} & \tabvaluetemp{0.44} & \tabvaluetemp{0.63} & \tabvaluetemp{0.45} & \tabvaluetemp{0.64} & \tabvaluetemp{0.48} & \tabvaluetemp{0.67} & \tabvaluetemp{0.51} \\
                
                & \variable{ED} & \tabvaluetemp{2.01} & \tabvaluetemp{0.95} & \tabvaluetemp{2.06} & \tabvaluetemp{1.02} & \tabvaluetemp{2.09} & \tabvaluetemp{1.09} & \tabvaluetemp{2.11} & \tabvaluetemp{1.13} \\
                
                & \variable{ED-s} & \tabvaluetemp{2.15} & \tabvaluetemp{1.09} & \tabvaluetemp{2.19} & \tabvaluetemp{1.13} & \tabvaluetemp{2.20} & \tabvaluetemp{1.21} & \tabvaluetemp{2.24} & \tabvaluetemp{1.25} \\
             \midrule
             
             \multirow{4}{*}{\rotatebox{90}{\parbox{40pt}\centering \textbf{DT} }}
                & \variable{BD} & \tabvaluetemp{0.18} & \tabvaluetemp{0.01} & \tabvaluetemp{0.19} & \tabvaluetemp{0.02} & \tabvaluetemp{0.19} & \tabvaluetemp{0.02} & \tabvaluetemp{0.19} & \tabvaluetemp{0.02} \\
\bottomrule
             \end{tabular} 
             }
             \label{tab:ctu_testing}
\end{table*}

\begin{table*}[!htbp]
  \centering
  \caption{\dataset{GTCS}: Runtime (in seconds) for testing (on \scbb{E}) the ML-NIDS on the \textit{high-end} platform..}
        \resizebox{1.1\columnwidth}{!}{
        \begin{tabular}{c|c ? cc? cc? cc? cc}
             \multicolumn{2}{c?}{Available Data} & \multicolumn{2}{c?}{Limited (100 per class)} & \multicolumn{2}{c?}{Scarce (20\% of \scbb{D})} &  \multicolumn{2}{c?}{Moderate (40\% of \scbb{D})} &  \multicolumn{2}{c}{Abundant (80\% of \scbb{D})} \\ 
              \hline
            Alg. & Design & 
            Complete & Essential &
            Complete & Essential &
            Complete & Essential &
            Complete & Essential \\
             \toprule
             
             \multirow{4}{*}{\rotatebox{90}{\parbox{40pt}\centering \textbf{RF} }}
                & \variable{BD} & \tabvaluetemp{0.48} & \tabvaluetemp{0.11} & \tabvaluetemp{0.50} & \tabvaluetemp{0.12} & \tabvaluetemp{0.51} & \tabvaluetemp{0.13} & \tabvaluetemp{0.52} & \tabvaluetemp{0.14} \\
                
                & \variable{MD} & \tabvaluetemp{0.56} & \tabvaluetemp{0.20} & \tabvaluetemp{0.55} & \tabvaluetemp{0.19} & \tabvaluetemp{0.55} & \tabvaluetemp{0.19} & \tabvaluetemp{0.56} & \tabvaluetemp{0.19} \\
                
                & \variable{ED} & \tabvaluetemp{2.00} & \tabvaluetemp{0.68} & \tabvaluetemp{1.97} & \tabvaluetemp{0.50} & \tabvaluetemp{1.97} & \tabvaluetemp{0.49} & \tabvaluetemp{1.97} & \tabvaluetemp{0.50} \\
                
                & \variable{ED-s} & \tabvaluetemp{2.14} & \tabvaluetemp{0.86} & \tabvaluetemp{2.07} & \tabvaluetemp{0.64} & \tabvaluetemp{2.07} & \tabvaluetemp{0.62} & \tabvaluetemp{2.08} & \tabvaluetemp{0.61} \\
             \midrule
             
             \multirow{4}{*}{\rotatebox{90}{\parbox{40pt}\centering \textbf{DT} }}
                & \variable{BD} & \tabvaluetemp{0.34} & \tabvaluetemp{0.01} & \tabvaluetemp{0.35} & \tabvaluetemp{0.02} & \tabvaluetemp{0.35} & \tabvaluetemp{0.02} & \tabvaluetemp{0.36} & \tabvaluetemp{0.02} \\
\bottomrule
             \end{tabular} 
             }
             \label{tab:gtcs_testing}
\end{table*}

\begin{table*}[!htbp]
  \centering
  \caption{\dataset{NB15}: Runtime (in seconds) for testing (on \scbb{E}) the ML-NIDS on the \textit{high-end} platform..}
        \resizebox{1.1\columnwidth}{!}{
        \begin{tabular}{c|c ? cc? cc? cc? cc}
             \multicolumn{2}{c?}{Available Data} & \multicolumn{2}{c?}{Limited (100 per class)} & \multicolumn{2}{c?}{Scarce (20\% of \scbb{D})} &  \multicolumn{2}{c?}{Moderate (40\% of \scbb{D})} &  \multicolumn{2}{c}{Abundant (80\% of \scbb{D})} \\ 
              \hline
            Alg. & Design & 
            Complete & Essential &
            Complete & Essential &
            Complete & Essential &
            Complete & Essential \\
             \toprule
             
             \multirow{4}{*}{\rotatebox{90}{\parbox{40pt}\centering \textbf{RF} }}
                & \variable{BD} & \tabvaluetemp{0.32} & \tabvaluetemp{0.21} & \tabvaluetemp{0.31} & \tabvaluetemp{0.23} & \tabvaluetemp{0.32} & \tabvaluetemp{0.23} & \tabvaluetemp{0.32} & \tabvaluetemp{0.24} \\
                
                & \variable{MD} & \tabvaluetemp{0.52} & \tabvaluetemp{0.42} & \tabvaluetemp{0.52} & \tabvaluetemp{0.42} & \tabvaluetemp{0.54} & \tabvaluetemp{0.43} & \tabvaluetemp{0.54} & \tabvaluetemp{0.44} \\
                
                & \variable{ED} & \tabvaluetemp{2.34} & \tabvaluetemp{1.63} & \tabvaluetemp{2.20} & \tabvaluetemp{1.54} & \tabvaluetemp{2.24} & \tabvaluetemp{1.53} & \tabvaluetemp{2.15} & \tabvaluetemp{1.54} \\
                
                & \variable{ED-s} & \tabvaluetemp{2.48} & \tabvaluetemp{1.82} & \tabvaluetemp{2.35} & \tabvaluetemp{1.65} & \tabvaluetemp{2.33} & \tabvaluetemp{1.65} & \tabvaluetemp{2.25} & \tabvaluetemp{1.67} \\
             \midrule

            \multirow{4}{*}{\rotatebox{90}{\parbox{40pt}\centering \textbf{DT} }}
                & \variable{BD} & \tabvaluetemp{0.18} & \tabvaluetemp{0.09} & \tabvaluetemp{0.19} & \tabvaluetemp{0.10} & \tabvaluetemp{0.19} & \tabvaluetemp{0.10} & \tabvaluetemp{0.20} & \tabvaluetemp{0.10} \\
\bottomrule
             \end{tabular} 
             }
             \label{tab:nb15_testing}
\end{table*}

\begin{table*}[!htbp]
  \centering
  \caption{\dataset{UF-NB15}: Runtime (in seconds) for testing (on \scbb{E}) the ML-NIDS on the \textit{high-end} platform..}
        \resizebox{1.1\columnwidth}{!}{
        \begin{tabular}{c|c ? cc? cc? cc? cc}
             \multicolumn{2}{c?}{Available Data} & \multicolumn{2}{c?}{Limited (100 per class)} & \multicolumn{2}{c?}{Scarce (20\% of \scbb{D})} &  \multicolumn{2}{c?}{Moderate (40\% of \scbb{D})} &  \multicolumn{2}{c}{Abundant (80\% of \scbb{D})} \\ 
              \hline
            Alg. & Design & 
            Complete & Essential &
            Complete & Essential &
            Complete & Essential &
            Complete & Essential \\
             \toprule
             
             \multirow{4}{*}{\rotatebox{90}{\parbox{40pt}\centering \textbf{RF} }}
                & \variable{BD} & \tabvaluetemp{0.32} & \tabvaluetemp{0.19} & \tabvaluetemp{0.31} & \tabvaluetemp{0.22} & \tabvaluetemp{0.30} & \tabvaluetemp{0.23} & \tabvaluetemp{0.27} & \tabvaluetemp{0.23} \\
                
                & \variable{MD} & \tabvaluetemp{0.51} & \tabvaluetemp{0.38} & \tabvaluetemp{0.51} & \tabvaluetemp{0.39} & \tabvaluetemp{0.50} & \tabvaluetemp{0.40} & \tabvaluetemp{0.48} & \tabvaluetemp{0.39} \\
                
                & \variable{ED} & \tabvaluetemp{2.48} & \tabvaluetemp{1.47} & \tabvaluetemp{2.20} & \tabvaluetemp{1.42} & \tabvaluetemp{2.13} & \tabvaluetemp{1.43} & \tabvaluetemp{2.00} & \tabvaluetemp{1.45} \\
                
                & \variable{ED-s} & \tabvaluetemp{2.69} & \tabvaluetemp{1.65} & \tabvaluetemp{2.30} & \tabvaluetemp{1.55} & \tabvaluetemp{2.26} & \tabvaluetemp{1.55} & \tabvaluetemp{2.14} & \tabvaluetemp{1.55} \\
             \midrule
             
            \multirow{4}{*}{\rotatebox{90}{\parbox{40pt}\centering \textbf{DT} }}
                & \variable{BD} & \tabvaluetemp{0.17} & \tabvaluetemp{0.07} & \tabvaluetemp{0.15} & \tabvaluetemp{0.08} & \tabvaluetemp{0.17} & \tabvaluetemp{0.08} & \tabvaluetemp{0.19} & \tabvaluetemp{0.08} \\
\bottomrule
             \end{tabular} 
             }
             \label{tab:ufnb15_testing}
\end{table*}

\begin{table*}[!htbp]
  \centering
  \caption{\dataset{CICIDS17}: Runtime (in seconds) for testing (on \scbb{E}) the ML-NIDS on the \textit{high-end} platform..}
        \resizebox{1.1\columnwidth}{!}{
        \begin{tabular}{c|c ? cc? cc? cc? cc}
             \multicolumn{2}{c?}{Available Data} & \multicolumn{2}{c?}{Limited (100 per class)} & \multicolumn{2}{c?}{Scarce (20\% of \scbb{D})} &  \multicolumn{2}{c?}{Moderate (40\% of \scbb{D})} &  \multicolumn{2}{c}{Abundant (80\% of \scbb{D})} \\ 
              \hline
            Alg. & Design & 
            Complete & Essential &
            Complete & Essential &
            Complete & Essential &
            Complete & Essential \\
             \toprule
             
             \multirow{4}{*}{\rotatebox{90}{\parbox{40pt}\centering \textbf{RF} }}
                & \variable{BD} & \tabvaluetemp{0.79} & \tabvaluetemp{0.29} & \tabvaluetemp{0.82} & \tabvaluetemp{0.30} & \tabvaluetemp{0.83} & \tabvaluetemp{0.31} & \tabvaluetemp{0.82} & \tabvaluetemp{0.31} \\
                
                & \variable{MD} & \tabvaluetemp{1.30} & \tabvaluetemp{0.81} & \tabvaluetemp{1.30} & \tabvaluetemp{0.81} & \tabvaluetemp{1.31} & \tabvaluetemp{0.79} & \tabvaluetemp{1.28} & \tabvaluetemp{0.77} \\
                
                & \variable{ED} & \tabvaluetemp{7.19} & \tabvaluetemp{2.73} & \tabvaluetemp{7.04} & \tabvaluetemp{2.65} & \tabvaluetemp{6.96} & \tabvaluetemp{2.65} & \tabvaluetemp{7.04} & \tabvaluetemp{2.68} \\
                
                & \variable{ED-s} & \tabvaluetemp{7.23} & \tabvaluetemp{2.82} & \tabvaluetemp{7.03} & \tabvaluetemp{2.79} & \tabvaluetemp{7.16} & \tabvaluetemp{2.80} & \tabvaluetemp{7.03} & \tabvaluetemp{2.80} \\
             \midrule
             
             \multirow{4}{*}{\rotatebox{90}{\parbox{40pt}\centering \textbf{DT} }}
                & \variable{BD} & \tabvaluetemp{0.61} & \tabvaluetemp{0.25} & \tabvaluetemp{0.65} & \tabvaluetemp{0.14} & \tabvaluetemp{0.62} & \tabvaluetemp{0.14} & \tabvaluetemp{0.62} & \tabvaluetemp{0.14} \\
\bottomrule
             \end{tabular} 
             }
             \label{tab:ids17_testing}
\end{table*}


\begin{table*}
  \centering
  \caption{Runtime (training and testing) on the other platforms (for the \dataset{GTCS} dataset).}

                \begin{subtable}[htbp]{1.5\columnwidth}
        \resizebox{1.0\columnwidth}{!}{
        %
}}
             & \variable{BD} & \scriptsize{0.22} & \scriptsize{0.51} & \scriptsize{0.21} & \scriptsize{0.20} & \scriptsize{3.04} & \scriptsize{0.64} & \scriptsize{1.81} & \scriptsize{0.26} & \scriptsize{9.74} & \scriptsize{0.66} & \scriptsize{5.55} & \scriptsize{0.31} & \scriptsize{23.30} & \scriptsize{0.68} & \scriptsize{13.49} & \scriptsize{0.32} \\
             
             & \variable{MD} & \scriptsize{0.22} & \scriptsize{0.65} & \scriptsize{0.21} & \scriptsize{0.34} & \scriptsize{2.93} & \scriptsize{0.75} & \scriptsize{2.19} & \scriptsize{0.40} & \scriptsize{9.01} & \scriptsize{0.75} & \scriptsize{6.55} & \scriptsize{0.40} & \scriptsize{21.22} & \scriptsize{0.77} & \scriptsize{14.74} & \scriptsize{0.41} \\
             
             & \variable{ED} & \scriptsize{0.82} & \scriptsize{1.95} & \scriptsize{0.85} & \scriptsize{0.82} & \scriptsize{4.41} & \scriptsize{2.17} & \scriptsize{3.26} & \scriptsize{0.97} & \scriptsize{11.28} & \scriptsize{2.13} & \scriptsize{7.97} & \scriptsize{0.94} & \scriptsize{26.94} & \scriptsize{2.16} & \scriptsize{17.74} & \scriptsize{0.95} \\
             
             & \variable{ED-s} & \scriptsize{0.91} & \scriptsize{2.04} & \scriptsize{0.97} & \scriptsize{1.00} & \scriptsize{5.67} & \scriptsize{2.24} & \scriptsize{3.97} & \scriptsize{1.13} & \scriptsize{13.45} & \scriptsize{2.25} & \scriptsize{10.17} & \scriptsize{1.11} & \scriptsize{31.57} & \scriptsize{2.25} & \scriptsize{26.41} & \scriptsize{1.088} \\
             \midrule
                         \multirow{4}{*}{\rotatebox{90}{\parbox{40pt}\centering
             \begin{tabular}{c}
                  \textbf{DT} \\
                  \scriptsize{all cores}
             \end{tabular}}}
            & \variable{BD} & \scriptsize{0.00} & \scriptsize{0.32} & \scriptsize{0.00} & \scriptsize{0.02} & \scriptsize{1.02} & \scriptsize{0.31} & \scriptsize{0.40} & \scriptsize{0.02} & \scriptsize{3.24} & \scriptsize{0.31} & \scriptsize{1.09} & \scriptsize{0.02} & \scriptsize{6.84} & \scriptsize{0.31} & \scriptsize{2.11} & \scriptsize{0.03} \\
             
             & \variable{MD} & \scriptsize{0.00} & \scriptsize{0.32} & \scriptsize{0.00} & \scriptsize{0.02} & \scriptsize{1.01} & \scriptsize{0.31} & \scriptsize{0.41} & \scriptsize{0.03} & \scriptsize{2.76} & \scriptsize{0.32} & \scriptsize{1.13} & \scriptsize{0.03} & \scriptsize{6.33} & \scriptsize{0.31} & \scriptsize{2.51} & \scriptsize{0.03} \\
             
             & \variable{ED} & \scriptsize{0.01} & \scriptsize{1.27} & \scriptsize{0.01} & \scriptsize{0.07} & \scriptsize{1.26} & \scriptsize{1.24} & \scriptsize{0.54} & \scriptsize{0.08} & \scriptsize{3.49} & \scriptsize{1.23} & \scriptsize{1.51} & \scriptsize{0.08} & \scriptsize{7.39} & \scriptsize{1.22} & \scriptsize{3.23} & \scriptsize{0.08} \\
             
             & \variable{ED-s} & \scriptsize{0.02} & \scriptsize{1.23} & \scriptsize{0.01} & \scriptsize{0.06} & \scriptsize{1.67} & \scriptsize{1.19} & \scriptsize{0.64} & \scriptsize{0.06} & \scriptsize{4.97} & \scriptsize{1.18} & \scriptsize{1.97} & \scriptsize{0.06} & \scriptsize{9.87} & \scriptsize{1.18} & \scriptsize{4.21} & \scriptsize{0.07} \\

                 \midrule
             
             \multirow{4}{*}{\rotatebox{90}{\parbox{40pt}\centering
             \begin{tabular}{c}
                  \textbf{LR} \\
                  \scriptsize{all cores}
             \end{tabular}}}
            & \variable{BD} & \scriptsize{0.89} & \scriptsize{0.36} & \scriptsize{0.40} & \scriptsize{0.02} & \scriptsize{2.30} & \scriptsize{0.35} & \scriptsize{3.21} & \scriptsize{0.01} & \scriptsize{5.13} & \scriptsize{0.34} & \scriptsize{7.42} & \scriptsize{0.01} & \scriptsize{9.51} & \scriptsize{0.34} & \scriptsize{14.61} & \scriptsize{0.01} \\
             
             & \variable{MD} & \scriptsize{0.40} & \scriptsize{0.37} & \scriptsize{0.42} & \scriptsize{0.02} & \scriptsize{5.83} & \scriptsize{0.35} & \scriptsize{3.86} & \scriptsize{0.02} & \scriptsize{8.94} & \scriptsize{0.35} & \scriptsize{8.77} & \scriptsize{0.02} & \scriptsize{25.71} & \scriptsize{0.34} & \scriptsize{17.41} & \scriptsize{0.02} \\
             
             & \variable{ED} & \scriptsize{1.51} & \scriptsize{1.43} & \scriptsize{0.79} & \scriptsize{0.06} & \scriptsize{8.17} & \scriptsize{1.38} & \scriptsize{4.93} & \scriptsize{0.06} & \scriptsize{24.78} & \scriptsize{1.36} & \scriptsize{12.04} & \scriptsize{0.06} & \scriptsize{45.98} & \scriptsize{1.35} & \scriptsize{21.05} & \scriptsize{0.06} \\
             
             & \variable{ED-s} & \scriptsize{1.87} & \scriptsize{1.38} & \scriptsize{0.97} & \scriptsize{0.05} & \scriptsize{10.47} & \scriptsize{1.35} & \scriptsize{5.98} & \scriptsize{0.04} & \scriptsize{27.87} & \scriptsize{1.31} & \scriptsize{16.47} & \scriptsize{0.04} & \scriptsize{56.87} & \scriptsize{1.30} & \scriptsize{27.14} & \scriptsize{0.05} \\
             \midrule

                          \multirow{4}{*}{\rotatebox{90}{\parbox{40pt}\centering
             \begin{tabular}{c}
                  \textbf{HGB} \\
                  \scriptsize{one core}
             \end{tabular}}}
                & \variable{BD} & \scriptsize{1.05} & \scriptsize{0.45} & \scriptsize{0.47} & \scriptsize{0.16} & \scriptsize{3.97} & \scriptsize{0.47} & \scriptsize{1.72} & \scriptsize{0.17} & \scriptsize{6.28} & \scriptsize{0.45} & \scriptsize{2.77} & \scriptsize{0.14} & \scriptsize{10.14} & \scriptsize{0.45} & \scriptsize{4.66} & \scriptsize{0.14} \\
             
                 & \variable{MD} & \scriptsize{3.83} & \scriptsize{0.88} & \scriptsize{2.13} & \scriptsize{0.70} & \scriptsize{14.91} & \scriptsize{0.96} & \scriptsize{7.57} & \scriptsize{0.69} & \scriptsize{22.49} & \scriptsize{0.91} & \scriptsize{11.20} & \scriptsize{0.61} & \scriptsize{35.24} & \scriptsize{0.88} & \scriptsize{18.32} & \scriptsize{0.61} \\
                 
                 & \variable{ED} & \scriptsize{1.22} & \scriptsize{1.64} & \scriptsize{0.63} & \scriptsize{0.44} & \scriptsize{8.81} & \scriptsize{1.78} & \scriptsize{4.53} & \scriptsize{0.52} & \scriptsize{14.18} & \scriptsize{1.74} & \scriptsize{6.66} & \scriptsize{0.50} & \scriptsize{22.87} & \scriptsize{1.72} & \scriptsize{10.34} & \scriptsize{0.52} \\
                 
                 & \variable{ED-s} & \scriptsize{1.67} & \scriptsize{1.68} & \scriptsize{0.87} & \scriptsize{0.50} & \scriptsize{11.17} & \scriptsize{1.86} & \scriptsize{5.98} & \scriptsize{0.58} & \scriptsize{17.84} & \scriptsize{1.78} & \scriptsize{9.74} & \scriptsize{0.56} & \scriptsize{29.84} & \scriptsize{1.77} & \scriptsize{14.64} & \scriptsize{0.57} \\
              
             \bottomrule
             \end{tabular} 
             }
            \caption{\textit{Workstation}: Intel Core-i7 10750HQ@2.6GHz (12 cores) with 32GB RAM. The OS is Windows 10.} 
             \label{stab:hardware_workstation}
             
            \end{subtable}

    \begin{subtable}[htbp]{1.5\columnwidth}
        \resizebox{1.0\columnwidth}{!}{
        \begin{tabular}{c|c ? cc|cc? cc|cc? cc|cc? cc|cc}
             \multicolumn{2}{c?}{Available Data} & \multicolumn{4}{c?}{Limited (100 per class)} & \multicolumn{4}{c?}{Scarce (20\% of \smabb{D})} &  \multicolumn{4}{c?}{Moderate (40\% of \smabb{D})} &  \multicolumn{4}{c}{Abundant (80\% of \smabb{D})} \\ 
              \hline
              
            \multicolumn{2}{c?}{Feature Set} & 
            \multicolumn{2}{c|}{Complete} & \multicolumn{2}{c?}{Essential} &
            \multicolumn{2}{c|}{Complete} & \multicolumn{2}{c?}{Essential} &
            \multicolumn{2}{c|}{Complete} & \multicolumn{2}{c?}{Essential} &
            \multicolumn{2}{c|}{Complete} & \multicolumn{2}{c}{Essential} \\ \hline
              
            Alg. {\scriptsize CPU} & Design & 
            Train & Test &
            Train & Test &
            Train & Test &
            Train & Test &
            Train & Test &
            Train & Test &
            Train & Test &
            Train & Test \\
             \toprule
             
             \multirow{4}{*}{\rotatebox{90}{\parbox{40pt}\centering
             \begin{tabular}{c}
                  \textbf{RF} \\
                  \scriptsize{all cores}
             \end{tabular}}}
             & \variable{BD} & \scriptsize{0.29} & \scriptsize{0.29} & \scriptsize{0.25} & \scriptsize{0.67} & \scriptsize{6.88} & \scriptsize{1.34} & \scriptsize{4.42} & \scriptsize{0.69} & \scriptsize{20.43} & \scriptsize{1.38} & \scriptsize{12.63} & \scriptsize{0.70} & \scriptsize{44.47} & \scriptsize{1.35} & \scriptsize{30.25} & \scriptsize{0.75} \\
             
             & \variable{MD} & \scriptsize{1.11} & \scriptsize{0.26} & \scriptsize{0.26} & \scriptsize{1.56} & \scriptsize{6.09} & \scriptsize{1.44} & \scriptsize{4.53} & \scriptsize{0.88} & \scriptsize{18.35} & \scriptsize{1.46} & \scriptsize{12.71} & \scriptsize{0.86} & \scriptsize{39.60} & \scriptsize{1.47} & \scriptsize{30.45} & \scriptsize{0.91} \\
             
             & \variable{ED} & \scriptsize{1.04} & \scriptsize{3.29} & \scriptsize{0.96} & \scriptsize{2.15} & \scriptsize{8.50} & \scriptsize{4.27} & \scriptsize{6.14} & \scriptsize{2.21} & \scriptsize{21.74} & \scriptsize{3.91} & \scriptsize{15.35} & \scriptsize{2.13} & \scriptsize{41.26} & \scriptsize{3.97} & \scriptsize{37.01} & \scriptsize{2.18} \\
             
             & \variable{ED-s} & \scriptsize{1.24} & \scriptsize{3.50} & \scriptsize{1.07} & \scriptsize{2.22} & \scriptsize{10.14} & \scriptsize{4.58} & \scriptsize{8.94} & \scriptsize{2.62} & \scriptsize{29.87} & \scriptsize{4.22} & \scriptsize{19.87} & \scriptsize{2.50} & \scriptsize{56.12} & \scriptsize{4.23} & \scriptsize{50.65} & \scriptsize{2.55} \\
             \midrule
             
             \multirow{4}{*}{\rotatebox{90}{\parbox{40pt}\centering
             \begin{tabular}{c}
                  \textbf{DT} \\
                  \scriptsize{one core}
             \end{tabular}}}
                & \variable{BD} & \scriptsize{0.01} & \scriptsize{0.40} & \scriptsize{0.01} & \scriptsize{0.02} & \scriptsize{1.44} & \scriptsize{0.47} & \scriptsize{0.42} & \scriptsize{0.03} & \scriptsize{3.44} & \scriptsize{0.40} & \scriptsize{1.17} & \scriptsize{0.03} & \scriptsize{7.59} & \scriptsize{0.39} &  \scriptsize{2.50} & \scriptsize{0.03} \\
             
                 & \variable{MD} & \scriptsize{0.00} & \scriptsize{0.40} & \scriptsize{0.01} & \scriptsize{0.03} & \scriptsize{1.26} & \scriptsize{0.47} & \scriptsize{0.45} & \scriptsize{0.03} & \scriptsize{3.15} & \scriptsize{0.41} & \scriptsize{1.29} & \scriptsize{0.03} & \scriptsize{7.18} & \scriptsize{0.40} & \scriptsize{2.76} & \scriptsize{0.03} \\
                 
                 & \variable{ED} & \scriptsize{0.02} & \scriptsize{1.60} & \scriptsize{0.04} & \scriptsize{0.09} & \scriptsize{1.64} & \scriptsize{1.83} & \scriptsize{0.59} & \scriptsize{0.10} & \scriptsize{4.03} & \scriptsize{1.61} & \scriptsize{1.67} & \scriptsize{0.09} & \scriptsize{8.33} & \scriptsize{1.58} & \scriptsize{3.67} & \scriptsize{0.01} \\
                 
                 & \variable{ED-s} & \scriptsize{0.02} & \scriptsize{1.56} & \scriptsize{0.05} & \scriptsize{0.07} & \scriptsize{1.99} & \scriptsize{1.80} & \scriptsize{0.75} & \scriptsize{0.08} & \scriptsize{6.15} & \scriptsize{1.55} & \scriptsize{2.01} & \scriptsize{0.08} & \scriptsize{11.45} &\scriptsize{1.52} & \scriptsize{4.97} & \scriptsize{0.08} \\
                 \midrule
             
             \multirow{4}{*}{\rotatebox{90}{\parbox{40pt}\centering
             \begin{tabular}{c}
                  \textbf{LR} \\
                  \scriptsize{all cores}
             \end{tabular}}}
            & \variable{BD} & \scriptsize{1.25} & \scriptsize{0.40} & \scriptsize{0.03} & \scriptsize{0.02} & \scriptsize{2.84} & \scriptsize{0.46} & \scriptsize{4.11} & \scriptsize{0.02} & \scriptsize{6.78} & \scriptsize{0.41} & \scriptsize{8.75} & \scriptsize{0.02} & \scriptsize{11.08} & \scriptsize{0.39} & \scriptsize{17.61} & \scriptsize{0.02}\\
             
             & \variable{MD} & \scriptsize{0.92} & \scriptsize{0.41} & \scriptsize{0.04} & \scriptsize{0.02} & \scriptsize{5.92} & \scriptsize{0.50} & \scriptsize{5.71} & \scriptsize{0.02} & \scriptsize{15.99} & \scriptsize{0.41} & \scriptsize{11.33} & \scriptsize{0.02} & \scriptsize{32.88} & \scriptsize{0.40} & \scriptsize{22.04} & \scriptsize{0.02} \\
             
             & \variable{ED} & \scriptsize{0.92} & \scriptsize{1.60} & \scriptsize{0.06} & \scriptsize{0.07} & \scriptsize{11.81} & \scriptsize{1.74} & \scriptsize{6.39} & \scriptsize{0.08} & \scriptsize{26.30} & \scriptsize{1.63} & \scriptsize{12.78} & \scriptsize{0.07} & \scriptsize{49.36} & \scriptsize{1.58} & \scriptsize{29.04} & \scriptsize{0.06} \\
             
             & \variable{ED-s} & \scriptsize{1.26} & \scriptsize{1.55} & \scriptsize{0.08} & \scriptsize{0.05} & \scriptsize{14.64} & \scriptsize{1.63} & \scriptsize{9.45} & \scriptsize{0.06} & \scriptsize{33.14} & \scriptsize{1.56} & \scriptsize{17.64} & \scriptsize{0.05} & \scriptsize{59.45} & \scriptsize{1.52} & \scriptsize{37.64} & \scriptsize{0.05} \\
             \midrule
             
             \multirow{4}{*}{\rotatebox{90}{\parbox{40pt}\centering
             \begin{tabular}{c}
                  \textbf{HGB} \\
                  \scriptsize{all cores}
             \end{tabular}}}
                & \variable{BD} & \scriptsize{0.40} & \scriptsize{0.67} & \scriptsize{0.36} & \scriptsize{0.36} & \scriptsize{4.13} & \scriptsize{0.82} & \scriptsize{1.69} & \scriptsize{0.41} & \scriptsize{6.78} & \scriptsize{0.71} & \scriptsize{2.72} & \scriptsize{0.28} & \scriptsize{12.12} & \scriptsize{0.68} & \scriptsize{5.12} & \scriptsize{0.28} \\
                 
                & \variable{MD} & \scriptsize{1.32} & \scriptsize{1.56} & \scriptsize{1.34} & \scriptsize{1.67} & \scriptsize{14.14} & \scriptsize{2.05} & \scriptsize{6.68} & \scriptsize{1.64} & \scriptsize{21.67} & \scriptsize{1.69} & \scriptsize{9.99} & \scriptsize{1.27} & \scriptsize{39.49} & \scriptsize{1.67} & \scriptsize{19.24} & \scriptsize{1.28} \\
                 
                & \variable{ED} & \scriptsize{0.46} & \scriptsize{2.15} & \scriptsize{0.53} & \scriptsize{1.07} & \scriptsize{8.48} & \scriptsize{3.10} & \scriptsize{3.82} & \scriptsize{1.26} & \scriptsize{12.58} & \scriptsize{2.66} & \scriptsize{5.55} & \scriptsize{1.04} & \scriptsize{23.81} & \scriptsize{2.65} & \scriptsize{10.09} & \scriptsize{1.05} \\
                 
                & \variable{ED-s} & \scriptsize{0.67} & \scriptsize{2.22} & \scriptsize{0.77} & \scriptsize{1.28} & \scriptsize{12.47} & \scriptsize{3.40} & \scriptsize{5.01} & \scriptsize{1.51} & \scriptsize{16.97} & \scriptsize{2.75} & \scriptsize{8.40} & \scriptsize{1.18} & \scriptsize{29.01} & \scriptsize{3.00} & \scriptsize{14.57} & \scriptsize{1.18} \\

             \bottomrule
             \end{tabular} 
             }
            \caption{\textit{Desktop}: Intel Core i5-4670@3.2GHz (4 cores) and 8GB of RAM. The OS is Windows 10.} 
             \label{stab:hardware_desktop}
             
            \end{subtable}

                \begin{subtable}[htbp]{1.5\columnwidth}
        \resizebox{1.0\columnwidth}{!}{
        \begin{tabular}{c|c ? cc|cc? cc|cc? cc|cc? cc|cc}
             \multicolumn{2}{c?}{Available Data} & \multicolumn{4}{c?}{Limited (100 per class)} & \multicolumn{4}{c?}{Scarce (20\% of \smabb{D})} &  \multicolumn{4}{c?}{Moderate (40\% of \smabb{D})} &  \multicolumn{4}{c}{Abundant (80\% of \smabb{D})} \\ 
              \hline
              
            \multicolumn{2}{c?}{Feature Set} & 
            \multicolumn{2}{c|}{Complete} & \multicolumn{2}{c?}{Essential} &
            \multicolumn{2}{c|}{Complete} & \multicolumn{2}{c?}{Essential} &
            \multicolumn{2}{c|}{Complete} & \multicolumn{2}{c?}{Essential} &
            \multicolumn{2}{c|}{Complete} & \multicolumn{2}{c}{Essential} \\ \hline
              
            Alg. {\scriptsize CPU} & Design & 
            Train & Test &
            Train & Test &
            Train & Test &
            Train & Test &
            Train & Test &
            Train & Test &
            Train & Test &
            Train & Test \\
             \toprule
             
             \multirow{4}{*}{\rotatebox{90}{\parbox{40pt}\centering
             \begin{tabular}{c}
                  \textbf{RF} \\
                  \scriptsize{all cores}
             \end{tabular}}}
             & \variable{BD} & \scriptsize{0.70} & \scriptsize{2.55} & \scriptsize{0.65} & \scriptsize{3.00} & \scriptsize{40.83} & \scriptsize{5.74} & \scriptsize{23.02} & \scriptsize{4.31} & \scriptsize{124.82} & \scriptsize{6.42} & \scriptsize{71.92} & \scriptsize{3.58} & \scriptsize{263.27} & \scriptsize{6.27} & \scriptsize{160.68} & \scriptsize{4.22} \\
             
             & \variable{MD} & \scriptsize{0.62} & \scriptsize{3.42} & \scriptsize{1.20} & \scriptsize{3.55} & \scriptsize{35.81} & \scriptsize{5.92} & \scriptsize{25.22} & \scriptsize{4.90} & \scriptsize{123.11} & \scriptsize{6.62} & \scriptsize{72.77} & \scriptsize{4.45} & \scriptsize{243.23} & \scriptsize{6.67} & \scriptsize{158.24} & \scriptsize{4.82} \\
             
             & \variable{ED} & \scriptsize{3.76} & \scriptsize{11.17} & \scriptsize{4.82} & \scriptsize{9.09} & \scriptsize{52.74} & \scriptsize{16.45} & \scriptsize{30.36} & \scriptsize{12.90} & \scriptsize{152.04} & \scriptsize{18.23} & \scriptsize{77.40} & \scriptsize{12.23} & \scriptsize{298.37} & \scriptsize{18.76} & \scriptsize{192.61} & \scriptsize{12.15} \\
             
             & \variable{ED-s} & \scriptsize{4.06} & \scriptsize{14.20} & \scriptsize{6.74} & \scriptsize{11.11} & \scriptsize{66.11} & \scriptsize{20.18} & \scriptsize{38.45} & \scriptsize{14.43} & \scriptsize{169.45} & \scriptsize{18.76} & \scriptsize{86.45} & \scriptsize{14.73} & \scriptsize{397.41} & \scriptsize{19.87} & \scriptsize{234.10} & \scriptsize{14.07} \\
             \midrule
             
             \multirow{4}{*}{\rotatebox{90}{\parbox{40pt}\centering
             \begin{tabular}{c}
                  \textbf{DT} \\
                  \scriptsize{one core}
             \end{tabular}}}
                & \variable{BD} & \scriptsize{0.02} & \scriptsize{0.92} & \scriptsize{0.01} & \scriptsize{0.11} & \scriptsize{4.46} & \scriptsize{0.93} & \scriptsize{1.88} & \scriptsize{0.11} & \scriptsize{11.68} & \scriptsize{0.93} & \scriptsize{4.13} & \scriptsize{0.08} & \scriptsize{25.95} & \scriptsize{0.91} & \scriptsize{7.31} & \scriptsize{0.09} \\
             
                 & \variable{MD} & \scriptsize{0.02} & \scriptsize{1.32} & \scriptsize{0.01} & \scriptsize{0.11} & \scriptsize{3.03} & \scriptsize{0.93} & \scriptsize{1.06} & \scriptsize{0.08} & \scriptsize{8.31} & \scriptsize{1.16} & \scriptsize{2.93} & \scriptsize{0.08} & \scriptsize{30.29} & \scriptsize{0.93} & \scriptsize{6.64} & \scriptsize{0.09} \\
                 
                 & \variable{ED} & \scriptsize{0.05} & \scriptsize{4.22} & \scriptsize{0.02} & \scriptsize{0.24} & \scriptsize{5.04} & \scriptsize{5.46} & \scriptsize{1.45} & \scriptsize{0.27} & \scriptsize{17.66} & \scriptsize{5.09} & \scriptsize{4.03} & \scriptsize{0.26} & \scriptsize{38.17} & \scriptsize{4.99} & \scriptsize{8.96} & \scriptsize{0.27} \\
                 
                 & \variable{ED-s} & \scriptsize{0.07} & \scriptsize{4.93} & \scriptsize{0.03} & \scriptsize{0.19} & \scriptsize{7.01} & \scriptsize{5.29} & \scriptsize{1.98} & \scriptsize{0.22} & \scriptsize{23.17} & \scriptsize{5.29} & \scriptsize{5.45} & \scriptsize{0.21} & \scriptsize{55.64} & \scriptsize{5.11} & \scriptsize{13.14} & \scriptsize{0.22} \\
                 \midrule
             
             \multirow{4}{*}{\rotatebox{90}{\parbox{40pt}\centering
             \begin{tabular}{c}
                  \textbf{LR} \\
                  \scriptsize{all cores}
             \end{tabular}}}
            & \variable{BD} & \scriptsize{3.69} & \scriptsize{1.27} & \scriptsize{3.43} & \scriptsize{0.05} & \scriptsize{9.12} & \scriptsize{1.55} & \scriptsize{13.16} & \scriptsize{0.05} & \scriptsize{18.15} & \scriptsize{1.51} & \scriptsize{28.03} & \scriptsize{0.05} & \scriptsize{34.66} & \scriptsize{1.38} & \scriptsize{50.68} & \scriptsize{0.05} \\
             
             & \variable{MD} & \scriptsize{4.43} & \scriptsize{1.59} & \scriptsize{2.38} & \scriptsize{0.07} & \scriptsize{30.05} & \scriptsize{1.78} & \scriptsize{19.44} & \scriptsize{0.12} & \scriptsize{54.52} & \scriptsize{1.49} & \scriptsize{47.54} & \scriptsize{0.11} & \scriptsize{134.30} & \scriptsize{1.65} & \scriptsize{92.40} & \scriptsize{0.07} \\
             
             & \variable{ED} & \scriptsize{2.49} & \scriptsize{5.10} & \scriptsize{2.37} & \scriptsize{0.20} & \scriptsize{40.57} & \scriptsize{5.18} & \scriptsize{22.53} & \scriptsize{0.36} & \scriptsize{87.90} & \scriptsize{5.38} & \scriptsize{40.22} & \scriptsize{0.21} & \scriptsize{154.10} & \scriptsize{5.90} & \scriptsize{75.10} & \scriptsize{0.28} \\
             
             & \variable{ED-s} & \scriptsize{5.01} & \scriptsize{5.71} & \scriptsize{4.44} & \scriptsize{0.20} & \scriptsize{45.74} & \scriptsize{5.62} & \scriptsize{29.78} & \scriptsize{0.15} & \scriptsize{101.21} & \scriptsize{5.31} & \scriptsize{49.65} & \scriptsize{0.15} & \scriptsize{198.45} & \scriptsize{5.61} & \scriptsize{101.61} & \scriptsize{0.22} \\
             \midrule
             
             \multirow{4}{*}{\rotatebox{90}{\parbox{40pt}\centering
             \begin{tabular}{c}
                  \textbf{HGB} \\
                  \scriptsize{all cores}
             \end{tabular}}}
            & \variable{BD} & \scriptsize{2.45} & \scriptsize{2.99} & \scriptsize{1.59} & \scriptsize{1.88} & \scriptsize{22.27} & \scriptsize{3.68} & \scriptsize{5.63} & \scriptsize{1.92} & \scriptsize{35.46} & \scriptsize{3.54} & \scriptsize{16.23} & \scriptsize{1.74} & \scriptsize{66.47} & \scriptsize{3.46} & \scriptsize{31.81} & \scriptsize{1.79} \\
             
             & \variable{MD} & \scriptsize{5.83} & \scriptsize{8.81} & \scriptsize{6.40} & \scriptsize{7.41} & \scriptsize{68.72} & \scriptsize{7.90} & \scriptsize{38.10} & \scriptsize{7.77} & \scriptsize{148.65} & \scriptsize{8.58} & \scriptsize{69.24} & \scriptsize{7.26} & \scriptsize{262.43} & \scriptsize{8.55} & \scriptsize{120.57} & \scriptsize{7.23} \\
             
             & \variable{ED} & \scriptsize{1.91} & \scriptsize{8.52} & \scriptsize{1.89} & \scriptsize{3.88} & \scriptsize{29.13} & \scriptsize{12.74} & \scriptsize{19.86} & \scriptsize{6.02} & \scriptsize{71.74} & \scriptsize{10.22} & \scriptsize{31.64} & \scriptsize{5.53} & \scriptsize{130}.34 & \scriptsize{10.91} & \scriptsize{48.33} & \scriptsize{6.33} \\
             
             & \variable{ED-s} & \scriptsize{3.84} & \scriptsize{9.34} & \scriptsize{3.98} & \scriptsize{5.30} & \scriptsize{39.47} & \scriptsize{13.50} & \scriptsize{21.68} & \scriptsize{8.88} & \scriptsize{99.87} & \scriptsize{12.37} & \scriptsize{40.58} & \scriptsize{6.63} & \scriptsize{169.64} & \scriptsize{12.22} & \scriptsize{69.14} & \scriptsize{7.07} \\

             \bottomrule
             \end{tabular} 
             }
            \caption{\textit{Laptop}: Intel Core i5-430M@2.5GHz (4 cores) and 8GB of RAM. The OS is Windows 10.} 
             \label{stab:hardware_laptop}
            
            \end{subtable}
            
                \begin{subtable}[htbp]{1.5\columnwidth}
        \resizebox{1.0\columnwidth}{!}{
        \begin{tabular}{c|c ? cc|cc? cc|cc? cc|cc? cc|cc}
             \multicolumn{2}{c?}{Available Data} & \multicolumn{4}{c?}{Limited (100 per class)} & \multicolumn{4}{c?}{Scarce (20\% of \smabb{D})} &  \multicolumn{4}{c?}{Moderate (40\% of \smabb{D})} &  \multicolumn{4}{c}{Abundant (80\% of \smabb{D})} \\ 
              \hline
              
            \multicolumn{2}{c?}{Feature Set} & 
            \multicolumn{2}{c|}{Complete} & \multicolumn{2}{c?}{Essential} &
            \multicolumn{2}{c|}{Complete} & \multicolumn{2}{c?}{Essential} &
            \multicolumn{2}{c|}{Complete} & \multicolumn{2}{c?}{Essential} &
            \multicolumn{2}{c|}{Complete} & \multicolumn{2}{c}{Essential} \\ \hline
              
            Alg. {\scriptsize CPU} & Design & 
            Train & Test &
            Train & Test &
            Train & Test &
            Train & Test &
            Train & Test &
            Train & Test &
            Train & Test &
            Train & Test \\
             \toprule
             
             \multirow{4}{*}{\rotatebox{90}{\parbox{40pt}\centering
             \begin{tabular}{c}
                  \textbf{RF} \\
                  \scriptsize{all cores}
             \end{tabular}}}
             & \variable{BD} & \scriptsize{0.57} & \scriptsize{1.68} & \scriptsize{0.61} & \scriptsize{0.78} & \scriptsize{12.46} & \scriptsize{2.26} & \scriptsize{8.34} & \scriptsize{1.07} & \scriptsize{37.92} & \scriptsize{2.39} & \scriptsize{24.72} & \scriptsize{1.12} & \scriptsize{88.37} & \scriptsize{2.53} & \scriptsize{53.90} & \scriptsize{1.23} \\
             
             & \variable{MD} & \scriptsize{0.42} & \scriptsize{2.12} & \scriptsize{0.51} & \scriptsize{1.13} & \scriptsize{11.91} & \scriptsize{2.43} & \scriptsize{8.70} & \scriptsize{1.43} & \scriptsize{34.03} & \scriptsize{2.60} & \scriptsize{24.18} & \scriptsize{1.44} & \scriptsize{81.65} & \scriptsize{2.78} & \scriptsize{57.79} & \scriptsize{1.55} \\
             
             & \variable{ED} & \scriptsize{1.91} & \scriptsize{5.83} & \scriptsize{1.77} & \scriptsize{2.57} & \scriptsize{16.14} & \scriptsize{6.75} & \scriptsize{11.71} & \scriptsize{3.21} & \scriptsize{42.35} & \scriptsize{6.92} & \scriptsize{31.25} & \scriptsize{3.30} & \scriptsize{100.05} & \scriptsize{7.25} & \scriptsize{70.28} & \scriptsize{3.48} \\
             
             & \variable{ED-s} & \scriptsize{2.10} & \scriptsize{6.23} & \scriptsize{1.98} & \scriptsize{3.13} & \scriptsize{20.64} & \scriptsize{7.14} & \scriptsize{14.74} & \scriptsize{3.77} & \scriptsize{49.43} & \scriptsize{7.33} & \scriptsize{39.78} & \scriptsize{3.90} & \scriptsize{124.84} & \scriptsize{7.58} & \scriptsize{89.21} & \scriptsize{3.93} \\
             \midrule
             
             \multirow{4}{*}{\rotatebox{90}{\parbox{40pt}\centering
             \begin{tabular}{c}
                  \textbf{DT} \\
                  \scriptsize{one core}
             \end{tabular}}}
                & \variable{BD} & \scriptsize{0.06} & \scriptsize{0.92} & \scriptsize{0.00} & \scriptsize{0.02} & \scriptsize{2.47} & \scriptsize{0.91} & \scriptsize{0.64} & \scriptsize{0.03} & \scriptsize{7.75} & \scriptsize{0.91} & \scriptsize{2.17} & \scriptsize{0.04} & \scriptsize{16.50} & \scriptsize{0.90} & \scriptsize{4.97} & \scriptsize{0.04} \\
             
                 & \variable{MD} & \scriptsize{0.00} & \scriptsize{0.92} & \scriptsize{0.00} & \scriptsize{0.03} & \scriptsize{2.48} & \scriptsize{0.91} & \scriptsize{0.92} & \scriptsize{0.04} & \scriptsize{6.96} & \scriptsize{0.89} & \scriptsize{2.60} & \scriptsize{0.05} & \scriptsize{14.57} & \scriptsize{0.90} & \scriptsize{5.80} & \scriptsize{0.06} \\
                 
                 & \variable{ED} & \scriptsize{0.01} & \scriptsize{3.65} & \scriptsize{0.00} & \scriptsize{0.09} & \scriptsize{3.04} & \scriptsize{3.64} & \scriptsize{1.25} & \scriptsize{0.13} & \scriptsize{8.11} & \scriptsize{3.60} & \scriptsize{3.28} & \scriptsize{0.12} & \scriptsize{18.31} & \scriptsize{3.60} & \scriptsize{7.38} & \scriptsize{0.14} \\
                 
                 & \variable{ED-s} & \scriptsize{0.04} & \scriptsize{5.58} & \scriptsize{0.01} & \scriptsize{0.08} & \scriptsize{3.87} & \scriptsize{3.55} & \scriptsize{1.87} & \scriptsize{0.10} & \scriptsize{11.74} & \scriptsize{3.53} & \scriptsize{4.84} & \scriptsize{0.11} & \scriptsize{24.15} & \scriptsize{3.57} & \scriptsize{9.87} & \scriptsize{0.11} \\
                 \midrule
             
             \multirow{4}{*}{\rotatebox{90}{\parbox{40pt}\centering
             \begin{tabular}{c}
                  \textbf{LR} \\
                  \scriptsize{all cores}
             \end{tabular}}}
            & \variable{BD} & \scriptsize{1.35} & \scriptsize{0.97} & \scriptsize{1.46} & \scriptsize{0.07} & \scriptsize{6.43} & \scriptsize{0.97} & \scriptsize{5.39} & \scriptsize{0.02} & \scriptsize{15.31} & \scriptsize{0.98} & \scriptsize{20.02} & \scriptsize{0.07} & \scriptsize{32.00} & \scriptsize{0.99} & \scriptsize{36.10} & \scriptsize{0.06} \\
             
             & \variable{MD} & \scriptsize{0.86} & \scriptsize{1.00} & \scriptsize{1.00} & \scriptsize{0.07} & \scriptsize{10.77} & \scriptsize{0.99} & \scriptsize{9.80} & \scriptsize{0.10} & \scriptsize{28.53} & \scriptsize{1.02} & \scriptsize{21.66} & \scriptsize{0.10} & \scriptsize{57.55} & \scriptsize{1.01} & \scriptsize{42.93} & \scriptsize{0.09} \\
             
             & \variable{ED} & \scriptsize{1.03} & \scriptsize{3.96} & \scriptsize{0.95} & \scriptsize{0.19} & \scriptsize{21.85} & \scriptsize{3.88} & \scriptsize{6.75} & \scriptsize{0.13} & \scriptsize{68.23} & \scriptsize{3.89} & \scriptsize{16.02} & \scriptsize{0.21} & \scriptsize{140.39} & \scriptsize{3.97} & \scriptsize{55.37} & \scriptsize{0.28} \\
             
             & \variable{ED-s} & \scriptsize{1.94} & \scriptsize{4.14} & \scriptsize{1.12} & \scriptsize{0.12} & \scriptsize{27.41} & \scriptsize{3.93} & \scriptsize{9.99} & \scriptsize{0.19} & \scriptsize{81.41} & \scriptsize{3.94} & \scriptsize{21.58} & \scriptsize{0.44} & \scriptsize{167.64} & \scriptsize{3.91} & \scriptsize{75.45} & \scriptsize{0.06} \\
             \midrule
             
             \multirow{4}{*}{\rotatebox{90}{\parbox{40pt}\centering
             \begin{tabular}{c}
                  \textbf{HGB} \\
                  \scriptsize{all cores}
             \end{tabular}}}
            & \variable{BD} & \scriptsize{29.21} & \scriptsize{12.76} & \scriptsize{0.74} & \scriptsize{1.47} & \scriptsize{184.45} & \scriptsize{5.16} & \scriptsize{123.33} & \scriptsize{13.44} & \scriptsize{92.49} & \scriptsize{6.11} & \scriptsize{15.50} & \scriptsize{4.31} & \scriptsize{260.53} & \scriptsize{8.88} & \scriptsize{24.33} & \scriptsize{3.52} \\
             
             & \variable{MD} & \scriptsize{3.35} & \scriptsize{26.66} & \scriptsize{4.26} & \scriptsize{56.50} & \scriptsize{167.48} & \scriptsize{39.03} & \scriptsize{41.76} & \scriptsize{16.00} & \scriptsize{504.85} & \scriptsize{37.33} & \scriptsize{578.36} & \scriptsize{34.91} & \scriptsize{526.04} & \scriptsize{59.27} & \scriptsize{278.97} & \scriptsize{14.13} \\
             
             & \variable{ED} & \scriptsize{3.53} & \scriptsize{18.40} & \scriptsize{14.80} & \scriptsize{8.24} & \scriptsize{55.38} & \scriptsize{66.50} & \scriptsize{180.47} & \scriptsize{28.68} & \scriptsize{552.68} & \scriptsize{31.09} & \scriptsize{426.50} & \scriptsize{5.37} & \scriptsize{129.22} & \scriptsize{42.36} & \scriptsize{854.91} & \scriptsize{45.37} \\
             
             & \variable{ED-s} & \scriptsize{4.94} & \scriptsize{26.63} & \scriptsize{16.41} & \scriptsize{11.61} & \scriptsize{67.98} & \scriptsize{40.61} & \scriptsize{207.58} & \scriptsize{38.90} & \scriptsize{597.84} & \scriptsize{27.56} & \scriptsize{504.68} & \scriptsize{34.52} & \scriptsize{187.54} & \scriptsize{20.51} & \scriptsize{904.65} & \scriptsize{29.39} \\

             \bottomrule
             \end{tabular} 
             }
            \caption{\textit{Low-end}: a Virtual Machine that is set up to use only 4 cores and at most 40\% of the frequency and 8GB of RAM of the Workstation platform. The OS is Ubuntu 20.04.} 
             \label{stab:hardware_lowend}
             
            \end{subtable}

         \label{tab:hardware}   
\end{table*}

\end{document}